\documentclass[12pt,openright,twoside]{report}

\usepackage{times}
\usepackage{amsbsy,amssymb}
\usepackage{graphicx,color}
\usepackage{fancyheadings}

\def\one{{\mathchoice {\rm 1\mskip-4mu l} {\rm 1\mskip-4mu l} {\rm
1\mskip-4.5mu l} {\rm 1\mskip-5mu l}}}
\newcommand{\ket}[1]{|{#1}\rangle}
\newcommand{\bra}[1]{\langle{#1}|}

\def\fu{\mathfrak{u}}

\def\fI{\mathfrak{I}}

\def\fh{\mathfrak{h}}

\def\fr{\mathfrak{r}}

\def\fsu{\mathfrak{su}}

\def\fso{\mathfrak{so}}
\def\fsu{\mathfrak{su}}

\def\cO{\cal O}
\def\cE{\cal E}
\def\cA{\cal A}
\def\cB{\cal B}
\def\cH{\cal H}
\def\cS{\cal S}
\def\cl{\ell}

\def\hh{\hat{h}}
\def\hE{\hat{E}}
\def\hO{\hat{O}}
\def\hA{\hat{A}}
\def\hW{\hat{W}}

\def\hH{\hat{H}}
\def\hM{\hat{M}}
\def\hQ{\hat{Q}}

\def\oS{\overrightarrow{S}}
\def\oT{\overrightarrow{T}}

\begin{document}


\thispagestyle{empty} 
 
\begin{center} 
\vspace*{\stretch{1}} 
 
\Large 
\textbf{Quantum Computation, Complexity, and Many-Body Physics}
\vspace{2cm} 

\textmd{Rolando D. Somma}\\ 
\vspace{3cm} 
 
\normalsize

Instituto Balseiro, S. C. de Bariloche, Argentina, and Los Alamos National
Laboratory, Los Alamos, USA.\\

\vspace{2cm}

August 2005 \\ 
 
\vspace*{\stretch{2}} 
\end{center} 
 
\normalsize 
 
\vspace{2cm}

\vspace{2cm} 
 
\begin{minipage}[c]{0.4\textwidth} 
\centering 
Dr. Gerardo Ortiz \\PhD Advisor\\ 
\end{minipage}%
 
\cleardoublepage

\thispagestyle{empty}
\newenvironment{dedication}
	{\cleardoublepage \thispagestyle{empty} \vspace*{\stretch{1}} \begin{flushright} \em}
	{\end{flushright} \vspace*{\stretch{3}} \clearpage}
		\begin{dedication}
		{\large To Janine and Isabel}
		\end{dedication}
		\thispagestyle{empty} \cleardoublepage
		

\pagenumbering{roman} 
\setcounter{page}{1}
\chapter*{Abstract}

By taking advantage of the laws of physics it is possible to revolutionize the
way we communicate (transmit), process or even store information. It is now
known that quantum computers, or computers  built from quantum mechanical
elements, provide new resources to solve certain problems and perform certain
tasks more efficiently than today's conventional computers. However, on the
road to a complete understanding of the power of quantum computers there are
intermediate steps that need to be addressed. The primary focus of this thesis
is the understanding of the possibilities and limitations of the
quantum-physical world in the areas of quantum computation and quantum
information processing.

First I investigate the simulation of quantum systems on a quantum computer
(i.e., a quantum simulation) constructed of two-level quantum elements or
qubits. For this purpose, I present algebraic mappings that allow one to
efficiently obtain physical properties and compute correlation functions of
fermionic, anyonic, and bosonic systems with such a computer. By studying the
amount of resources required for a quantum simulation, I show that the
complexity of preparing a quantum state which contains the desired information
is crucial at the time of evaluating the advantages of having a quantum
computer over a conventional one. As a small-scale demonstration of the
validity of these results, I show the simulation of a fermionic system using a
liquid-state nuclear magnetic resonance (NMR) device.

Remarkably, the conclusions obtained in the area of quantum simulations can be
extended to general quantum computations by means of the notion of generalized
entanglement. This is a generalization based on the idea that quantum
entanglement (i.e., the existence of non-classical correlations) is a concept
that depends on the accesible information, that is, relative to the observer.
Then I present a wide class of quantum computations that can be efficiently
simulated on a conventional computer and where quantum computers cannot be
claimed to be more powerful. The idea is that a quantum algorithm, performed by
applying a restricted set of gates which do not create generalized entangled
states relative to small (polynomially-large) sets of observables, can be
imitated using a similar amount of resources with a conventional computer.
However, a similar statement cannot be obtained when generalized entangled
states (relative to these sets) are involved, because this purely quantum
phenomena cannot be easily reproduced by classical-information methods.

Finally, I show how these concepts developed from an information-theory point
of view can be used to study other important problems in many-body physics. To
begin with, I exploit the notion of Lie-algebraic purity to identify and
characterize the quantum phase transitions present in the Lipkin-Meshkov-Glick
model and the spin-1/2 anisotropic XY model in a transverse magnetic field. The
results obtained show how generalized entanglement leads to useful tools for
distinguishing between ordered and disordered phases in quantum systems.
Moreover, I discuss  how the concept of general mean field hamiltonians
naturally emerges from these considerations and show that these can be exactly
diagonalized by using a conventional computer.

In brief, in this thesis I apply several topics developed in the context of
quantum information theory to study the complexity of obtaining relevant
physical properties of quantum systems with a quantum computer, and to study
different physical processes in quantum  many-body systems.

\chapter*{Acknowledgements}

Many people have contributed in one way or another to my PhD thesis.  Without
them, this work would have been impossible.  It is time then, to express my
gratitude to each one who participated in this long journey.

I would first like to thank my family,  starting with a special thanks to my
wife Janine and our little miracle Isabel, for bringing happiness every
morning, which allows me to enjoy and continue with this life. Also, I'm very
thankful to my parents and siblings for their limitless support and for having
provided and sustained the basis  that guides me every day. Only God can
explain how much I love everyone of them.

On the scientific side, I want to give special thanks to my PhD advisor and
friend, Dr. Gerardo Ortiz, for his dedication and for giving me the tools to
realize this work. His constant help, teachings, and his own personal passion
for science have been the main reasons for my achievements during this time as
a PhD student. 

Also, I'm very thankful to everyone in the quantum information group at Los
Alamos National Laboratory for sharing their knowledge and for their 
dedication. They are the reason for many important results obtained in this
thesis. In particular, I want to thank my collaborators Howard Barnum, Many
Knill, Raymond Laflamme, Camille Negrevergne, and Lorenza Viola, for their help
and contributions. Throughout my years at Los Alamos, they have treated me as
an equal as a scientist, something that is priceless. 

In the same way, I want to thank my professors and administrators at the
Instituto Balseiro that, although most of my PhD studies were not performed in
Argentina, they were a constant source of support. In particular, I want to
thank Dr. Armando A. Aligia for having helped me in many scientific and
pedagogical aspects of this thesis. Also, I want to thank Carlos Balseiro,
Ra\'{u}l Barrachina, Manuel C\'{a}ceres, Daniel Dom\'{\i}nguez, Jim Gubernatis,
Armando F. Guillermet, Karen Hallberg, Eduardo Jagla, Lisetta, Marcela
Margutti, Hugo Montani and Juan Pablo Paz. Each one of them has contributed in
one way or another to this work.

Finally, I want to thank Andr\'{e}s, Seba, Sequi, the `star team', and all my
friends from Buenos Aires and Bariloche, and the Ortiz, the Batistas, the
Dalvits, the Machados, and every member of the Argentinean community in New
Mexico. Through their support and sense of humor, we have  entertained
ourselves over the years.


\tableofcontents 

\newpage

\setcounter{page}{1}
\pagenumbering{arabic} 

\setcounter{chapter}{0}

\chapter{Introduction} 
\label{chapter1}
\begin{quote}
{\em ...there is plenty of room to make computers smaller...nothing that I can see in
the physical laws... 

R. P . Feynman, Caltech (1959).}
\end{quote}

During the last few decades, the theory of Quantum Information Processing (QIP)
has acquired great importance because it has been shown that information based
on quantum mechanics provides new resources that go beyond the traditional
`classical information'. It is now known that certain quantum mechanical
systems, named quantum computers (QCs), can be used to easily solve certain
problems which are difficult to solve using  today's conventional or classical
computers CCs. Having a QC would allow one to communicate in secret~\cite{BB84}
(quantum cryptography), perform a variety of search algorithms~\cite{Gro97},
factor large numbers~\cite{Sho94}, or simulate efficiently some physical
systems~\cite{OGK01,SOG02}. Additionally, it would allow us to break security
codes used, for example, to secure internet communications, optimize a large
variety of scheduling problems, etc., which make of quantum information an
exciting and relevant subject. Consequently,  the science of quantum
information is mainly focused on better understanding the  foundations of
quantum mechanics (which are different of classical mechanics) and the physical
realization of quantum controllable physical devices. While the first allows
clever and not so obvious ways of taking advantage of the quantum
world, the latter will let us achieve our most important goal: the building of a
QC.
\\

When one looks for the word {\em information} in the dictionary one finds  many
definitions: i) a message received and understood, ii) knowledge acquired
through study or experience, iii) propagated signal through a given channel,
iv) broadcasted news, and more. Information is then the basis of all human
knowledge and we usually base our behavior on it. It always requires a physical
representation to be able to use it, propagate it, or store it, such as a
telephone, a computer disk, etc. Depending on the physical representation,
information can be {\em classical} or {\em quantum}. 

We define as {\em classical information} the one that is manipulated and stored by
today's CCs. In {\em classical information theory} the basic unit
is the {\it bit}. A bit's state can be in one of two states represented by the
numbers 0 and 1, which constitute the logical basis. A possible physical
representation of a bit is given by a system in which the state is determined by
the distribution of, for example, electrical charge. The idea is then to
process information through the manipulation of the state of a set of bits
(i.e., bit sequence) by performing elementary gates. These gates are
different processes that depend on the particular physical realization of
the CC.

\begin{figure}[hbt]
\begin{center}
\includegraphics[height=5.5cm]{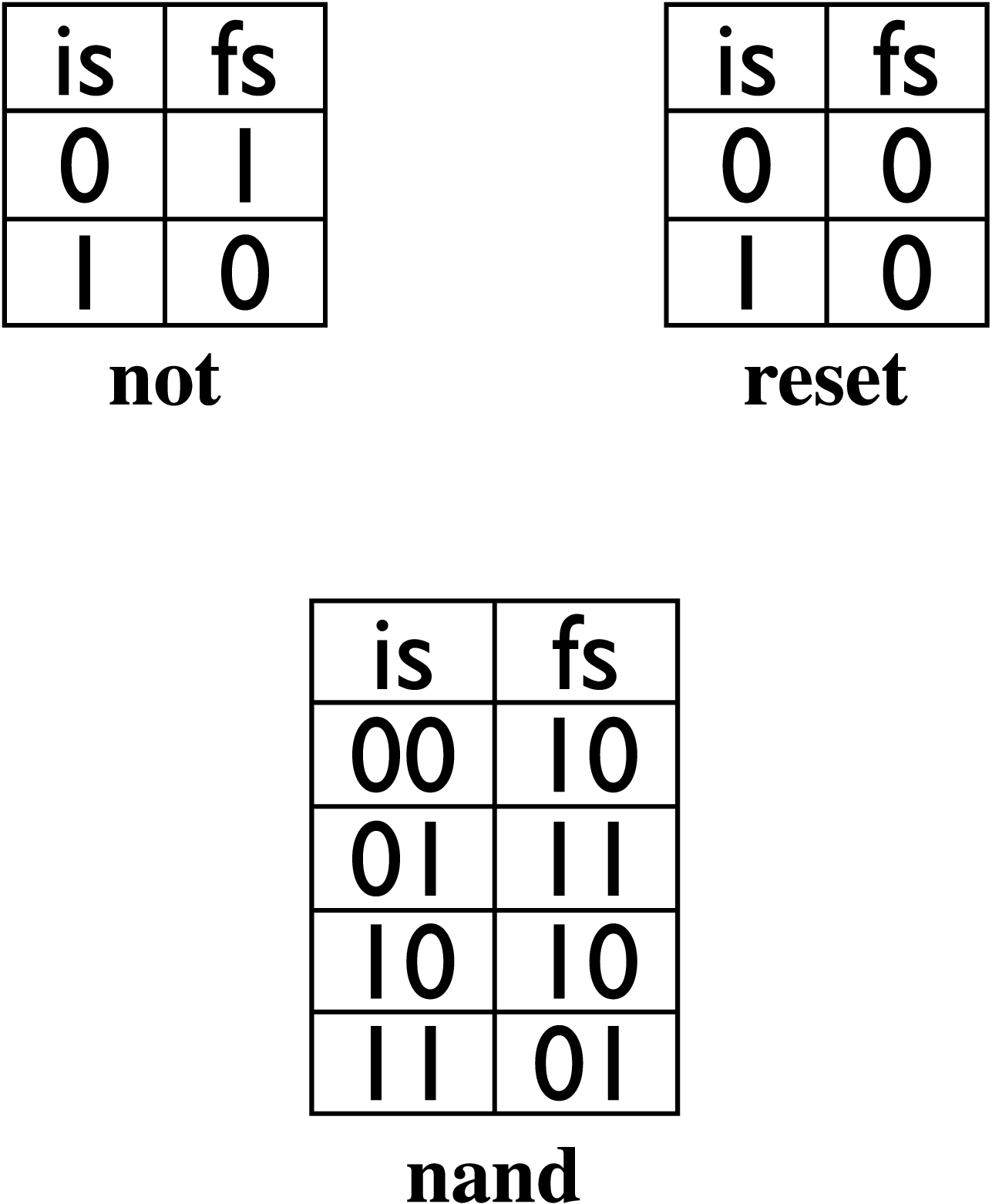}
\end{center}
\caption{Logical action of the single bit gates {\bf not} and {\bf reset}, and
the two-bit gate {\bf nand}. Here is and fs denote the initial and final bit
states, respectively.}
\label{fig1-1}
\end{figure}

Examples of one-bit gates are the {\bf not} and {\bf reset} gates, and of
two-bit gates is the {\bf nand} gate. Their action over logical initial states
are shown in Fig.~\ref{fig1-1}. They suffice for implementing arbitrary state
transformations. That is, any classical algorithm can be implemented through a
circuit that consists of applying these elementary gates to a bit sequence.
In fact, this method is used by today's computers, where the
{\em program} sets up a particular order for performing the elementary gates and
the {\em chips} implement them physically. Finally, one reads the final state where
the required information is supposed to be encoded (e.g., the solution to
a problem).

The idea of quantum information processing is similar to that of classical
information but under the laws of the quantum world. One defines {\em quantum
information} as the one which is stored and manipulated by physical devices obeying
the laws of quantum physics; that is, satisfying the Schr\"{o}dinger evolution
equation
\begin{equation}
\label{schroev}
i \hbar \frac{d}{dt} \ket{\psi} = H \ket{\psi},
\end{equation}
where $H$ is the Hamiltonian describing the interactions that one manipulates to
perform the desired evolution, and
$\ket{\psi}$ is some pure state (i.e., wave function) of the system.
Such devices constitute {\em quantum
computers}. 

\begin{figure}[hbt]
\begin{center}
\includegraphics*[height=5cm]{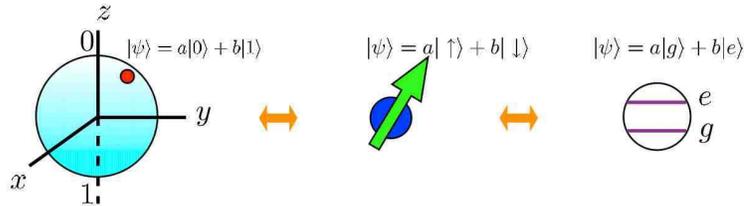}
\end{center}
\caption{Different physical representations of the single qubit state
$\ket{\psi}=a\ket{0} + b\ket{1}$. 
The red dot on the surface of the sphere (left) represents such a linear
combination of states.
Here $\ket{g}$ and $\ket{e}$ denote the
ground and an excited state of a certain atom.}
\label{fig1-2}
\end{figure}

In the conventional model of {\em quantum information theory} (QIT) 
the basic
unit is the quantum bit or {\it qubit}. A qubit's pure state can be in any
superposition of the logical states and is expressed as $a \ket{0} + b
\ket{1}$, where the complex numbers $a$ and $b$ are the probability amplitudes
of being in the states $\ket{0}$ and $\ket{1}$, respectively. They  are
normalized to the unity: $|a|^2 + |b|^2=1$. A possible physical representation
of a qubit is given by any two-level quantum system (Fig.~\ref{fig1-2}) such as
a spin-1/2, where the state represented by $\ket{0}$ ($\ket{1}$) corresponds to
the state with the spin pointing up (down), or a single atom. Bohm's
rule~\cite{Boh51} tells one that such state corresponds to having
probabilities $|a|^2$ and $|b|^2$ of being in the state with the spin pointing
up and down, respectively.

Due to the superposition principle of quantum physics, a pure state of a set of
$N$ qubits (register) is expressed as $a_0 \ket{0 \cdots 00} + a_1 \ket{0
\cdots 01} + \cdots + a_{2^N-1} \ket{1 \cdots 11}$. Again, the complex
coefficients $a_i$ are the corresponding probability amplitudes, and $\sum_i
|a_i|^2 =1 $. The idea is then to perform computation
by executing a quantum algorithm that consists of performing a set of
elementary gates in a given order (i.e., a quantum circuit). The action of
these {\em quantum gates} is rigorously discussed in Chap.~\ref{chapter2} and
requires some previous knowledge in linear algebra. As in classical
information, these gates involve single qubit and two-qubit operations. In
order to preserve the features of quantum physics, these operations must be
reversible (i.e., unitary operations), and are usually performed by making
the register interact with external oscillating electromagnetic fields.

The great advantages of having a QC are two-fold: First, working at the quantum
level allows one to make these computers extremely small and, if
scalable\footnote{A computer is said to be scalable if the number of resources
needed scale almost linearly with the problem size.}~\cite{Div95}, it
would allow one to process a large number of qubits at the same time. Second, a
computer ruled by the laws of quantum physics should contain certain features
that go beyond those of classical information, since the latter can be
considered a limit of the first one\footnote{Any classical algorithm can be
simulated efficiently with a QC~\cite{NC00}}. For example, one immediately
observes that the superposition principle gives one more freedom when
manipulating quantum information, in the sense that many different logical
states can be carried simultaneously (parallelism).

To analyze the computational complexity to solve a certain problem one
needs to determine the total amount of physical resources required, such as
bits or qubits, the number of operations performed or number of elementary
gates, the number of times that the algorithm is executed, etc. While nobody
knows yet the power of quantum computation, certain
algorithms~\cite{Sho94,Gro97} suggest that QCs are more powerful than their
classical analogues. All these algorithms share the feature that they not only
make use of the superposition principle (which is not sufficient to claim  that
a QC is more efficient), but also of the non-classical correlations between
different quantum elements in the QC. (Interference phenomena also plays an
important role in the efficiency of quantum algorithms.) Such
correlations are inherent to quantum systems~\cite{Sch35,EPR35} and do
not exist in classical systems. They are usually referred as {\em quantum
entanglement} (QE), an emerging field of QIT.

\begin{figure}[hbt]
\begin{center}
\includegraphics*[height=6cm]{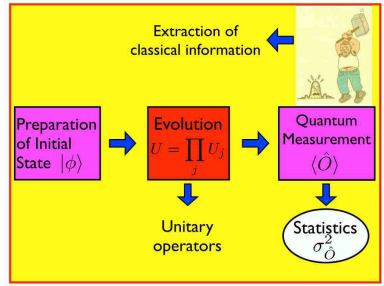}
\end{center}
\caption{Circuit representation of a quantum algorithm. A pure state
$\ket{\psi}$ is evolved by applying elementary gates. The evolution 
obeys Schr\"{o}dinger's
equation (Eq.~\ref{schroev}). After the evolution, a measurement
collapses the evolved state with statistics given by Bohm's rule.}
\label{fig1-3}
\end{figure}

In order to access the quantum
information, one needs to perform a {\em measurement}. This is defined as the
extraction of some classical information from the quantum register. Due to
the features of quantum physics, after a measurement process the state of the register is
{\em collapsed} into the logical state corresponding to the outcome, with
statistics given by Bohm's rule (Fig.~\ref{fig1-3}). This process
could {\it destroy} the efficiency of the computation. For example, if after
the execution of the quantum algorithm the state of two qubits is $a_0 \ket{00}
+ a_1 \ket{01} + a_2 \ket{10} + a_3 \ket{11}$, a measurement process in the
logical basis has the effect of collapsing the state to $\ket{00}$ with
probability $|a_0|^2$, to $\ket{01}$ with probability $|a_1|^2$,  to $\ket{10}$
with probability $|a_2|^2$, and to $\ket{11}$ with probability $|a_3|^2$.
Therefore, a single measurement does not give the whole information about the
state of the register and usually one needs to run the quantum algorithm
repeatedly many times to obtain more accurate statistics to recover the state of the
register. This is a main difference with the case of
classical information, where such measurement or
convertion is not necessary. 

\begin{figure}[hbt]
\begin{center}
\includegraphics*[height=5cm]{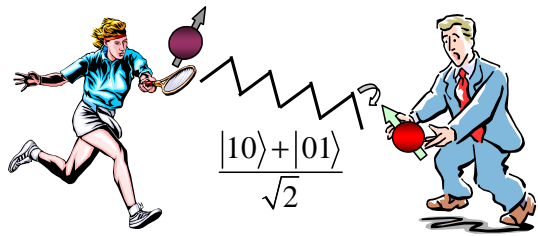}
\end{center}
\caption{Maximally entangled two-qubit state. The quantum correlations cannot be
represented by any classical state.}
\label{fig1-4}
\end{figure}

The simplest case of an entangled state is the pure two-qubit state
(Fig.~\ref{fig1-4})
\begin{equation}
\label{siment}
\ket{\psi}=\frac{1}{\sqrt{2}}[\ket{10} + \ket{01}],
\end{equation}
or similar states obtained by flipping or by changing the phase of a single
qubit. Equation~\ref{siment} states that if one qubit is measured and projected
in the logical basis, the other qubit is automatically projected in the same
basis. Moreover, if the outcome of a measurement performed in one qubit is 0
(1), the outcome of a post-measurement performed on the other qubit will be 1
(0). Remarkably, similar results are obtained for such state if the measurements
are performed in a basis other than the logical one. These correlations between
outcomes cannot be explained by a classical theory.

During the last few years, several authors~\cite{EHK04,GK,SOK03,Val02,Vid03}
have started to study the relation between different definitions and measures
of QE and quantum complexity. Naturally, they mostly agreed that whenever the
QE of the evolved state in a quantum computation is small enough, such
algorithms can be simulated with the same efficiency on a CC. However, the lack
of a unique computable measure of entanglement that could be applied to any
quantum state and quantum system is the main reason why the power of QCs
is still not fully understood.

One of the purposes of my thesis is to show the computational
complexity (i.e.., the number of resources and operations needed) to solve
certain problems with QCs and to compare it with the corresponding classical
complexity. In particular, I will first focus on the study of the simulation of
physical systems by quantum networks or quantum simulations
(QSs)~\cite{OGK01,SOG02,SOK03}. As noticed by R. P. Feynman~\cite{Fey82} and Y.
Manin, the obvious difficulty with deterministically solving a quantum
many-body problem (e.g., computing some correlation functions) on a CC is the
exponentially large basis set needed for its simulation. Known exact
diagonalization approaches like the Lanczos method suffer from this exponential
catastrophe. For this reason, it is expected that using a computer constructed
of distinct quantum mechanical elements (i.e., a QC) that
`imitates\footnote{In general, the QC used to perform a QS is built of quantum
elements that are different in nature of those that compose the system to be
simulated. However, this is not  a drawback in a simulation because  usually
one can perform a one-to-one association between the quantum states of the QC
and the quantum states of the physical system to be simulated.}' the physical system to be
simulated (i.e., simulates the interactions) would overcome this difficulty.

The results obtained by studying the complexity of QSs can be extended to
understand the complexity of solving other problems. For example, different 
quantum search algorithms~\cite{Gro97} admit a Hamiltonian representation and
can be equivalently considered as a particular QS. But most importantly, I
will show how these simulations lead to the definition of a general measure of
quantum (pure state) entanglement, so called {\em generalized entanglement}
(GE), which can be applied to any quantum state regardless of its nature.
Remarkably, this measure is crucial when analyzing the efficiency-related
advantages of having a QC.


This report arises from the studies and novel results obtained together with my
colleagues at {\em Los Alamos National Laboratory} (USA) and {\em Instituto
Balseiro} (Argentina) during my PhD studies. For a better understanding, the
main results are presented in chronological order. In
Chap.~\ref{chapter2}, I analyze the problem of simulating different finite
physical systems on a QC using deterministic quantum algorithms. The
corresponding computational complexity is also studied. As a proof of
principles, I present the experimental simulation of a particular fermionic
many-body system on a liquid-state nuclear-magnetic-resonance (NMR) QC. 

In Chap.~\ref{chapter3}, I introduce the concept of quantum generalized
entanglement, a notion that goes beyond the traditional quantum entanglement
concept and makes no reference to a particular subsystem decomposition. I show
that important results are obtained whenever a Lie algebraic setting exists
behind the problem under consideration. In particular, I apply this novel
approach to the study of quantum correlations in different quantum systems,
regardless of their nature or particle statistics, including different spin
and fermionic systems.

In Chap.~\ref{chapter4} I compare the effort of simulating certain quantum
systems with a QC or a CC. In particular, I show that the concept of
generalized entanglement is crucial to the efficiency of a quantum algorithm
and can be used as a resource in quantum computation.  Moreover, generalized
entanglement allows one to make a connection between QIT and many-body physics 
by studying different problems in quantum mechanics, such as the
characterization of quantum phase transitions in matter or the study of
integrable quantum systems. These results are presented in Chap.~\ref{chapter5}.

Finally, in Chap.~\ref{chapter6}, I present the conclusions, open questions,
and future directions related to this subject.

\chapter{Simulations of Physics with Quantum Computers}
\label{chapter2}
Since Richard P. Feynman conjectured that an arbitrary discrete quantum system
may be {\em simulated} by another one~\cite{Fey82}, the simulation of quantum
phenomena became a fundamental problem that a quantum computer (QC), i.e., a
system of universally controlled distinct quantum elements, may potentially
solve in a more efficient way than a classical computer (CC). The main problem
with the simulation of a quantum system  on a CC is that the dimension of the 
associated Hilbert space grows exponentially with the volume of the system to
be simulated. For example, the classical simulation of a system composed of $N$
qubits requires, in general, an amount of computational operations (additions and
products of complex numbers) that is proportional to $D=2^N$, where $D$ is the
dimension of the Hilbert space given by the number of different logical states
$\ket{i_1 i_2 \cdots i_N}$, with $i_j=\{0,1\}$. Nevertheless, a QC allows one
to imitate the evolution of the corresponding quantum system by cleverly
controlling and manipulating its elements. This process is called a {\em
quantum simulation} (QS). It is expected then that the number of resources
required for the QS increases linearly (or at most, polynomially) with the
volume of the system to be simulated~\cite{AL97}. If this is the case, we say
that the QS can be performed efficiently. 

\begin{figure}[hbt]
\begin{center}
\includegraphics*[angle=270,width=13.0cm,scale=1.0]{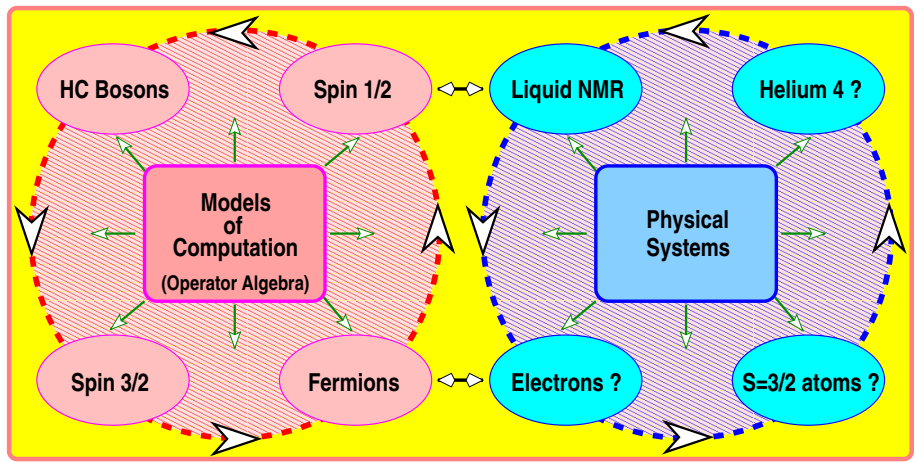}
\end{center}
\caption{Relationship between different models of computation
(with their associated operator algebras) and different physical systems.
Question marks refer to the present lack of a quantum computer 
device using the corresponding elementary physical components indicated 
in the box. Diamond-shaped arrows represent the natural connection 
between physical system and operator language, while arrows on the
circle indicate the existence of isomorphisms of $*$-algebras,
therefore, the corresponding simulation of one physical system by
another.}
\label{fig2-1}
\end{figure}

To be able to perform a QS, it is necessary to make a connection between the
operator algebra associated to the system and the operator algebra which
defines  the {\em model} of quantum computation~\cite{OGK01}. The existence of
one-to-one mappings between different algebras of operators and one-to-one
mappings between different Hilbert spaces~\cite{BO01,SOK03}, is a necessary 
requirement to simulate a physical system using a QC built on the basis of
another system (Fig.~\ref{fig2-1}). For example, one can simulate a fermionic
system on  a liquid-state nuclear magnetic resonance (NMR) QC by making use of
the Jordan-Wigner transformation~\cite{JW28} that maps fermionic operators onto
the Pauli (spin-1/2) operators. Although these mappings can usually be 
performed efficiently, this is not sufficient to establish that any system can
be simulated efficiently on a QC. It is then necessary to prove that all steps
involved in the QS, including the initialization, evolution, and  measurement,
can be performed efficiently~\cite{SOG02}.

This chapter will explore the theoretical and experimental
issues associated with the simulations of physical phenomena on QCs. In
Sec.~\ref{sec2-1}, I  start by describing different models of quantum
computation. In particular, I rigorously introduce the {\em conventional model} 
by
means of the Pauli operators, where a natural set of elementary gates (i.e.,
set of universal operations) is obtained. This model, roughly
described in Chap.~\ref{chapter1}, is the one generally needed for the
practical implementation of a QS. In Sec.~\ref{sec2-2}, I present a class of
quantum algorithms (QAs) in the language of the conventional model, for the
computation of relevant physical properties of quantum systems, such as
correlation functions, energy spectra, etc. In Sec.~\ref{sec2-3}, I explain how
the QS of quantum physical systems obeying fermionic, anyonic, and bosonic
particle statistics, can be performed on a QC described by the conventional
model, presenting some mappings between the different operator algebras. As an
application, in Sec.~\ref{sec2-4} I show the QS (imitated by a classical
computer) of a particular fermionic system: The two-dimensional fermionic Hubbard model. It
is expected that such simulation gives an insight into the limitations of
quantum computation, showing that certain issues remain to be solved to assure
that a QC is more powerful than a CC (Sec.~\ref{sec2-5}). In Sec.~\ref{sec2-7},
I describe the experimental implementation on an NMR QC of the QS of another
fermionic system: The Fano-Anderson model. For this purpose, an elementary
introduction to the physical processes on an NMR setting is described in
Sec.~\ref{sec2-6}.
Finally, I summarize  in Sec.~\ref{sec2-8}.

\section{Models of Quantum Computation}
\label{sec2-1}
When performing a quantum computation, the quantum elements
which constitute the QC can be universally controlled and manipulated by
modulating and changing their interactions. This quantum control model assumes
then the existence of a control Hamiltonian $H_P$, which describes these
interactions. The control possibilities are used to implement specific quantum
gates, allowing one, for example, to represent the time evolution of the physical
system to be simulated~\cite{OGK01}. 

In order to define a model of quantum computation it is necessary to give a
physical setting together with its initial state, an algebra of operators
associated to the system, a set of controllable Hamiltonians necessary to
define a set of elementary gates, and a set of measurable operators (i.e.,
observables). In this way, many different models of quantum computation can be
described, but for historic reasons and practical purposes I will  focus mostly
on the {\em conventional model}~\cite{NC00}.

\subsection{The Conventional Model of Quantum Computation}
\label{sec2-1-1}
As mentioned in Chap.~\ref{chapter1}, in the conventional model of quantum
computation, the fundamental unit of information is the quantum bit or {\em
qubit}. A qubit's pure state $\ket{\sf a}=a \ket{0} + b \ket{1}$ (with $a,b \in
\mathbb{C}$ and $|a|^2 + |b|^2=1$), is a linear superposition of the logical
states $\ket{0}$ and $\ket{1}$, and can be represented by the state of a
two-level quantum system such as a spin-1/2. Assigned to each qubit are the
identity operator $\one$ (i.e., the no-action operator) and the Pauli operators
$\sigma_x$, $\sigma_y$, and $\sigma_z$. In the logical single-qubit basis
$\mathcal{B}= \{ \ket{0}, \ket{1} \}$, these are
\begin{equation}
\label{pauli1}
\one =\pmatrix{1 & 0 \cr 0 & 1}, \ \sigma_x=\pmatrix{0 & 1 \cr 1& 0},
\ \sigma_y=\pmatrix{0 & -i \cr i& 0}, \
\sigma_z=\pmatrix{1 & 0 \cr 0& -1}.
\end{equation}
Because of its action over the logical states, the operator $\sigma_x$ is
usually referred as the {\em flip} operator:
\begin{equation}
\label{flip} 
\sigma_x \left\{ \matrix{ \ket{0} \rightarrow \ket{1} \ , \cr
\ket{1} \rightarrow \ket{0} \ .} \right.
\end{equation}
For practical purposes in this thesis, 
it is also useful to define the raising (+) and
lowering (-) Pauli operators $\sigma_\pm= \frac{1}{2} (\sigma_x \pm i
\sigma_y)$, and the eigenstates of the flip operator
$\ket{+}=\frac{1}{\sqrt{2}}[\ket{0} + \ket{1}]$ and 
$\ket{-}=\frac{1}{\sqrt{2}}[\ket{0} - \ket{1}]$, satisfying
\begin{equation}
\sigma_x \ket{\pm} = \pm \ket{\pm}.
\end{equation}

The Pauli operators form the $\fsu(2)$ Lie algebra and satisfy the commutation
relations ($\mu,\nu,\lambda = \{x,y,z\}$)
\begin{equation} 
\label{comm1} 
\left[\sigma_\mu, \sigma_\nu \right]= 
i2 \epsilon_{\mu \nu \lambda} \sigma_\lambda,
\end{equation} 
where $[A,B]=AB -BA$ and $\epsilon_{\mu \nu \lambda}$ is the total
anti-symmetric Levi-Civita symbol. They constitute a complete set of local
observables, that is, a basis for the $2 \times 2 $ dimensional Hermitian
matrices with $\sigma_\mu = (\sigma_\mu)^\dagger$. The symbol $\dagger$
denotes the corresponding complex conjugate transpose. 

\begin{figure}
\begin{center}
\includegraphics*[angle=0,width=7cm,scale=1.0]{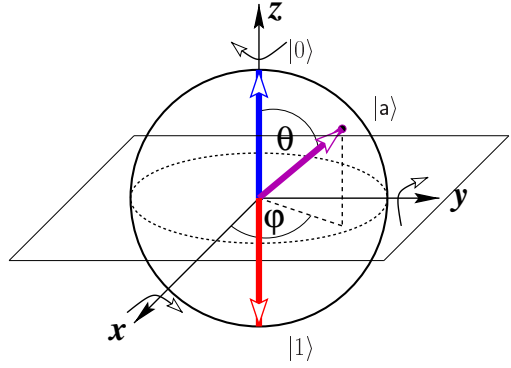}
\end{center}
\caption{Bloch-Sphere representation of a one qubit state parametrized as
$\ket{\sf a}=\cos(\theta/2) \ket{0} +e^{i\varphi}\sin(\theta/2) \ket{1}$. 
The curved arrows denote rotations $R_\mu$ along the corresponding axis. The
(arrow) color convention is:  $\ket{0} \rightarrow$ blue; $\ket{1} \rightarrow$
red; other linear combinations $\rightarrow$ magenta.}
\label{fig2-2}
\end{figure}

\begin{figure}
\begin{center}
\includegraphics*[width=9cm]{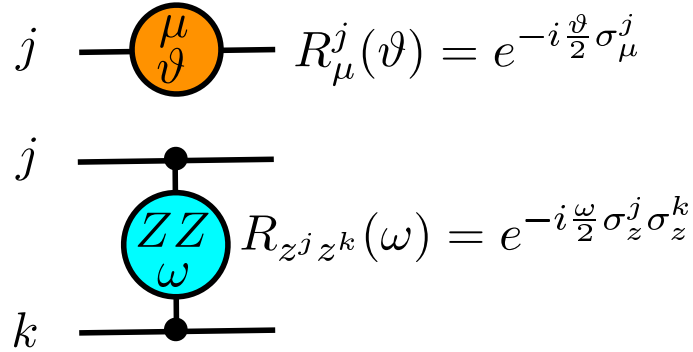}
\end{center}
\caption{Circuit representation of the elementary gates in the conventional
model. The top picture indicates a single-qubit rotation while the bottom one
indicates the two-qubit Ising gate. Any quantum algorithm can be represented by
a circuit composed of these elementary gates.}
\label{fig2-3}
\end{figure}

Any qubit's pure state can be represented as a point on the surface of the unit
sphere (Bloch-sphere representation) by parametrizing the state as $\ket{\sf a}
=a \ket{0} + b \ket{1} = \cos (\theta/2) \ket{0} + e^{i \varphi} \sin
(\theta/2) \ket{1}$  (Fig.~\ref{fig2-2}). In order to process a single qubit, a
complete set of single-qubit gates has to be given. These operations constitute
then, any rotation in the Bloch-sphere representation, which are given by the
operators $R_\mu(\vartheta) = e^{-i (\vartheta/2) \sigma_\mu}
=\cos(\vartheta/2) \one -i \sin (\vartheta/2) \sigma_\mu$;  that is, a rotation
by an angle $\vartheta$ along the $\mu$ axis. These rotations are unitary
(reversible) operations, satisfying $R_\mu (\vartheta) [R_\mu
(\vartheta)]^\dagger = \one$ (i.e., no-action), where $R_\mu(\vartheta)^\dagger
\equiv R_\mu (-\vartheta)$. This reversibility property allows one to perform
these gates with no thermodynamical cost. In Fig.~\ref{fig2-3}, I present these
elementary single-qubit gates in their circuit representation.

Similarly, a pure state of a $N$-qubit register (quantum register) is
represented as the {\em ket} $\ket{\psi}=\sum_{n=0} ^{2^N-1} a_n \ket{n}$,
where $\ket{n}$ is a product of states of each qubit in the logical (or other)
basis,
e.g., its binary representation ($\ket{0}\equiv \ket{0_10_2 \cdots0_N},
\ket{1}\equiv\ket{0_10_2\cdots0_{N-1}1_N},\ket{2}\equiv\ket{0_10_2\cdots1_{N-1}0_N}$,
etc.), and $\sum_{n=0}^{2^N-1} |a_n|^2=1$ ($a_n \in \mathbb{C}$). Assigned to
the $j$th qubit of the 
quantum register are, together with the identity operator $\one^j$,
the local Pauli operators $\sigma_\mu^j$ (with $\mu=x$,
$y$, or $z$); that is
\[
\sigma^j_\mu = \overbrace{\one \otimes \one \otimes \cdots \otimes
\underbrace{\sigma_\mu}_{j^{th}\ \mbox{factor}} \otimes \cdots
\otimes \one}^{n\ \mbox{factors}} \ ,
\]
where $\otimes$ represents a Kronecker tensorial product. Their matrix
representation in the basis ordered as $\mathcal{B} =\{ \ket{0_1 \cdots
0_{N-1}0_N}, \ket{0_1 \cdots 0_{N-1}1_N}, \cdots, \ket{1_1 \cdots 1_{N-1}1_N}
\}$ is just the matrix tensor product of the corresponding $2 \times 2$
matrices defined by Eq.~\ref{pauli1}. For two different qubits,  these
operators commute:
\begin{equation}
\label{comm2}
\left[ \sigma_\mu^j , \sigma_\nu^k \right] =0 \ \forall j \neq k.
\end{equation}

In order to describe a generic operation on the quantum register, it is also
necessary to consider products of the Pauli operators $\sigma_\mu^j$. 
Remarkably, every unitary (reversible) operation acting on the quantum register
can be decomposed in terms of single qubit rotations $R_\mu^j (\vartheta) =
e^{-i \frac{\vartheta}{2} \sigma_\mu^j}$ and two-qubit gates, such as the Ising
gate $R_{z^jz^k}(\omega) = e^{-i \frac{\omega}{2} \sigma_z^j \sigma_z^k} =
\cos(\omega/2) \one -
i \sin (\omega/2) \sigma_z^j \sigma_z^k$, with $\omega \in \mathbb{R}$
(\cite{BBC95,DiV95}). The operations $R_{z^jz^k}(\omega)$ are also unitary,
satisfying $R_{z^jz^k}(\omega)[R_{z^jz^k}(\omega)]^\dagger=\one$, with
$[R_{z^jz^k}(\omega)]^\dagger \equiv R_{z^jz^k}(-\omega)$. Together with the
single-qubit rotations they define a {\em universal set of quantum gates}.
Their quantum circuit representation is shown in Fig~\ref{fig2-3}. 

Every (logical) state of the quantum register has associated a mathematical
object denoted as {\em bra} in the following way
\begin{equation}
\ket{n} \leftrightarrow \bra{n},
\end{equation}
and can be linearly extended for a general state $\ket{\psi}$ by conjugation
as
\begin{equation}
\sum_n a_n \ket{n} \leftrightarrow \sum_n a_n^* \bra{n},
\end{equation}
where $a_n^*$ denotes the complex conjugate of $a_n$. The product between a
logical bra and a logical ket defines the inner-product in the associated
Hilbert space given by
\begin{equation}
\label{product}
\bra m \ket{n} = \langle m \ket{n} = \delta_{mn},
\end{equation}
with $\delta_{mn}$ being the Kronecker delta. In this vectorial space, two
vectors (states) $\ket{\psi}$ and $\ket{\phi}$ are orthogonal if their {\em
overlap}, that is, their inner product $\langle\psi \ket{\phi}$ given by
Eq.~\ref{product}, vanishes. 
Moreover, the bra-ket notation allows one to represent every (Pauli) operator.
For example, the single qubit flip operator $\sigma_x $ is represented as
$\sigma_x = \ket{0} \bra{1} + \ket{1} \bra{0}$. This notation is very useful
when computing, for example, expectation values.

A {\em measurement} is defined as the action that gives some classical
information about the state of the quantum register. In quantum mechanics, a
measurement is considered to be a probabilistic process that {\em collapses}
the actual quantum state of the system~\cite{Per98}. For example, a measurement of the
polarization in the logical basis of every qubit (i.e., the measurement of the
observables $\sigma_z^j$) when the state of the register is $\ket{\psi} =\sum_n
a_n \ket{n}$, projects  it onto a certain logical state $\ket{m}$ with
probability $|a_m|^2$ (Bohm's rule). 
This is a von Neumann measurement. In particular, 
in a general {\em von Neumann}
measurement of an observable
$\hM$ ($\equiv \hM^\dagger$), the probability $p_m$
that the outcome $m$ is obtained,
where $m$ is one possible eigenvalue of $\hM$, is given by~\cite{NC00}
\begin{equation}
p_m = \bra{\psi} \hM^\dagger_m \hM_m \ket{\psi},
\end{equation}
where $\hM_m$ is the projector onto the subspace of states with quantum number
$m$. Moreover, if $m$ is the actual outcome, the state after the measurement is
given by
\begin{equation}
\ket{\psi'} = \frac{\hM_m}{\sqrt{\bra{\psi} \hM_m^\dagger \hM_m \ket{\psi}}}
 \ket{\psi}.
\end{equation}
For example, when measuring the operator $\sigma_x$ for a single qubit state
$\ket{0}$, the two possible outcomes are $m=\pm1$ (i.e., the eigenvalues of
$\sigma_x$). Since $\ket{0} = \frac{1}{\sqrt{2}}[\ket{+} + \ket{-}]$, the
corresponding probabilities are
\begin{equation}
p_1 = p_{-1} = 1/2,
\end{equation}
and if the outcome is $+1$ ($-1$), the state is projected onto $\ket{+}$
($\ket{-}$).
Therefore, to obtain accurate information about the actual state of the quantum
system, one needs  to prepare many copies of the state and perform many
different measurements.

The expectation value of a measurement outcome is the expectation of the
outcomes of many measurement repetitions. It can also be expressed in the bra-ket
notation. If the state of
the quantum system is $\ket{\psi}$, the expectation value of $\hM$ is given by 
\begin{equation}  
\label{expectation}  
\langle \hM \rangle = \bra{\psi} \hM\ket{\psi}.  
\end{equation}  
In the conventional model, any observable $\hM$ can be written as a combination
(sums and/or products) of the identity and Pauli operators. Therefore, if
$\ket{\psi}$ is known, the expectation $\langle \hM \rangle$ can be
algebraically computed by obtaining first the state $\hM \ket{\psi}$, and by
projecting it onto the bra $\bra{\psi}$ using the inner-product relations of
Eq.~\ref{product}. For example, if a two-qubit state is given by
$\ket{\psi}=\frac{1}{\sqrt{2}} [\ket{0_10_2}+\ket{1_11_2}]$, then
\begin{equation}
\label{expexamp1}
\bra{\psi} \sigma_z^1 \ket{\psi} = \frac{1}{2} \left( \bra{0_10_2} + \bra{1_11_2} 
\right)\left( \ket{0_10_2} - \ket{1_11_2}  \right) =0,
\end{equation}
and
\begin{equation}
\label{expexamp2}
\bra{\psi} \sigma_x^1 \sigma_x^2 \ket{\psi} = \frac{1}{2} 
\left( \bra{0_10_2} + \bra{1_11_2} 
\right) \left( \ket{1_11_2} + \ket{0_10_2}  \right) =1.
\end{equation}
Equations~\ref{expexamp1} and \ref{expexamp2} have been obtained by noticing
that $\sigma_z^1 \ket{1_1 1_2} = - \ket{1_1 1_2}$, $\sigma_x^1 \sigma_x^2
\ket{0_1 0_2} = \ket{1_1 1_2}$, and $\sigma_x^1 \sigma_x^2 \ket{1_1 1_2} =
\ket{0_1 0_2}$, together with Eq.~\ref{product}.

Nevertheless, certain quantum computations and QSs are done by evolving mixed states instead of pure 
states.  A quantum register in a probabilistic mixture of pure states can be
described in the bra-ket notation by a density matrix $\rho=\sum_s p_s \rho_s$,
with $\rho_s=\ket{\psi_s}\bra{\psi_s}$ representing  the quantum register being
in the pure state $\ket{\psi_s}$, with probability $p_s$ ($\sum_s p_s=1; p_s\ge
0$). Equivalently, every density operator $\rho$ can also be written as a combination (sums
and/or products)  of the Pauli operators $\sigma_\alpha^j$ ($\alpha=x,y,z$) and
the identity operator $\one$. These mixed states are useful when performing
quantum computation with devices such as the NMR QC, where the state of the
quantum register is approximated by the average state of an ensemble of
molecules at room temperature; that is, an extremely mixed state.
The expectation value of a measurement outcome over a mixed state is given by
\begin{equation}
\label{mixedmeasur}
\langle \hM \rangle = {\sf Tr} (\rho \hM),
\end{equation}
where $\hM$ is the measured observable, $\rho$ the density operator of the mixed
state, and ${\sf Tr}$ denotes the trace.

In brief, the conventional model allows one to describe every step in a QS by
means of Pauli operators. The idea is to represent any quantum algorithm
(QA) as a circuit composed
of elementary single and two-qubit gates, together with the measurement
process.  The complexity of a QA is then determined by the amount of
resources required, given by the number of qubits needed, the number
of universal single and two-qubit operations (Fig.~\ref{fig2-3}), and the
number of measurements needed to obtain an accurate result (e.g., the number of
times that the algorithm needs to be performed). For this purpose,  a procedure
to decompose an arbitrary operation in terms of elementary gates has to be
explained.  In the following subsection, I present some useful techniques and
examples.

\subsection{Hamiltonian Evolutions}
\label{sec2-1-2}
When simulating a physical system on a QC it is necessary, in general, to
perform a Hamiltonian (unitary) evolution to the 
quantum register~\cite{OGK01,SOG02},
of the form 
\begin{equation}
U(t)=e^{-i H t} =
\one -i H t + \frac{1}{2} (-i H t)^2 + \cdots,
\end{equation}
where $H=H^\dagger$ is a physical Hamiltonian  and $t$ is a real parameter
(e.g., time). A common $H$ is given by
\begin{equation}
H=H_{x}+H_y=\bar{\alpha} \ \sigma_x^1 \left( \prod\limits_{i=2}^{j-1} \sigma_z^i
\right)
\sigma^j_x+\bar{\beta} \ \sigma_y^1 \left( \prod\limits_{i=2}^{j-1} \sigma_z^i
\right)
\sigma^j_y \ ,
\end{equation}
where $\bar{\alpha}$ and $\bar{\beta}$ are real numbers. From Eqs.~\ref{comm1}
and
\ref{comm2} one obtains
$[H_{x},H_{y}]=0$, and therefore, $U(t)=e^{-iH_{x}t} e^{-iH_{y}t}$.

To decompose $U(t)$ into single and two-qubit operations, the
following steps can be taken. First, the unitary operator
\begin{equation}
U_{1}=e^{i\frac{\pi}{4}\sigma_{y}^1}=\frac{1}{\sqrt{2}}
\left[{\one} +i \sigma_y^1\right]
\end{equation}
takes $\sigma_{z}^{1} \rightarrow \sigma_x^1$, i.e., $U_1^\dagger
\sigma_z^1 U_1 = \sigma_x^1$, so $U_1^{\dagger} e^{i{\bar{\alpha}}
\sigma_z^1} U_1 = e^{i{\bar{\alpha}}\sigma_x^1}$. Second, the
operator
\[
U_2=e^{i{\pi\over 4}\sigma_z^1\sigma_z^2}= \frac{1}{\sqrt{2}}
\left [{\one}+ i \sigma_z^1 \sigma_z^2 \right]
\]
takes $\sigma^1_x \rightarrow \sigma^1_y\sigma^2_z$, so $U_2^\dagger
e^{i{\bar{\alpha}} \sigma_x^1} U_{2} = e^{i{\bar{\alpha}} \sigma_y^{1}
\sigma_z^2}$. Then,
\[
U_3=e^{i{\pi\over 4}\sigma_z^1\sigma_z^3}
\]
takes $\sigma^1_y\sigma^2_z \rightarrow -\sigma^1_x \sigma^2_z
\sigma^3_z$. By successively similar steps the required
string of operators can be easily built: $\sigma_x^1\sigma_z^2 \cdots\sigma_z^{j-1}
\sigma^j_x$ and also $\exp[i{\bar{\alpha}}\sigma_x^1\sigma_z^2
\cdots\sigma_z^{j-1} \sigma^j_x]$ (up to a global irrelevant phase):
\begin{equation}
U_{k}^{\dagger}\cdots U_{2}^{\dagger}U_{1}^{\dagger}
e^{i{\bar{\alpha}}\sigma_z^1} U_{1} U_{2} \cdots U_{k} =
\exp \left[i{\bar{\alpha}}\sigma_x^1\sigma_z^2 \cdots\sigma_z^{j-1} \sigma^j_x
\right]
\ 
\end{equation}
where the integer $k$ scales linearly with $j$.
The evolution $e^{-iH_{y}t}$ can be decomposed similarly so $U(t)$ is
decomposed as the product of both
decompositions. 

\subsection{Controlled Operations}
\label{sec2-1-3}
\begin{figure}
\begin{center}
\includegraphics*[width=9cm]{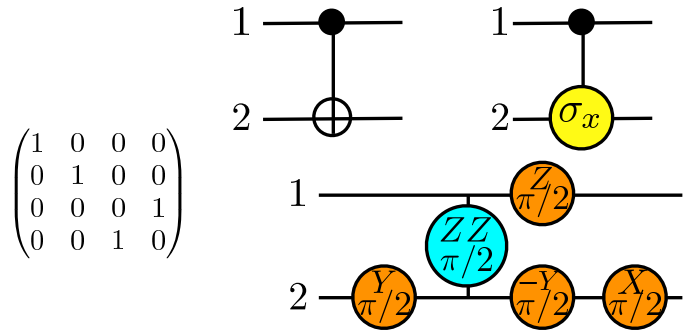}
\end{center}
\bigskip
\caption{CNOT gate decomposition and its matrix representation. The
control qubit is 1. Note that the last circuit realizes the CNOT matrix
operation up to a global phase $e^{-i \frac{\pi}{4}}$.}
\label{fig2-4}
\end{figure}

Alternatively, one can use the well known Controlled-Not, or CNOT, gate
instead of the two-qubit Ising gate. Its action on a pair of qubits (1 and 2)
is 
\[
\mbox{CNOT}
\left\{\begin{array}{c}
\ket{0_10_2}\rightarrow\ket{0_10_2} \ , \ \ \ket{0_11_2}\rightarrow\ket{0_11_2} \ ,   \\
\ket{1_10_2}\rightarrow\ket{1_11_2} \ , \ \ \ket{1_11_2}\rightarrow\ket{1_10_2} \ .   \\
\end{array}\right. 
\]
Here, qubit 1 is the control qubit (the controlled operation on its state
$\ket{1_1}$ is represented by a solid circle in Fig.~\ref{fig2-4}). If the
state of qubit 1 is $\ket{0_1}$ nothing happens (identity operation) but if its
state is $\ket{1_1}$, the state of qubit 2 is flipped. The decomposition of the
CNOT unitary operation into single and two-qubit operations is
\begin{equation}
\label {cnotdec}
 \mbox{CNOT: }
e^{i \frac{\pi}{4}} e^{-i \frac{\pi}{4} \sigma_{z}^{1}}
 e^{-i \frac{\pi}{4} \sigma_{x}^{2}}
 e^{ i \frac{\pi}{4}\sigma_{z}^{1}\sigma_{x}^{2}},
 \end{equation}
which was obtained by noticing that $i e^{-i \frac{\pi}{2}
\sigma_x^2} \equiv \sigma_x^2$, i.e., the spin-flip operator 
acting on qubit 2 (Eq.~\ref{pauli1}): 
\begin{equation}
\sigma_x^2 \left\{ \matrix{ \ket{\phi_1 0_2} \rightarrow \ket{s_1 1_2} \ , \cr
\ket{\phi_1 1_2} \rightarrow \ket{s_1 0_2} \ . } \right.
\end{equation}
By using the techniques described in
Sec.~\ref{sec2-1-2}, the CNOT operation in terms of single and two-qubit Ising
gates is
\begin{equation}
\label{cnotdec2}
 \mbox{CNOT: }
e^{i \frac{\pi}{4}} e^{-i \frac{\pi}{4} \sigma_{z}^{1}}
 e^{-i \frac{\pi}{4}\sigma_{x}^{2}}
 e^{ i \frac{\pi}{4}\sigma_{y}^{2}}
 e^{-i \frac{\pi}{4}\sigma_{z}^{1}\sigma_{z}^{2}}
 e^{-i \frac{\pi}{4}\sigma_{y}^{2}} \ .
 \end{equation}
The circuit representation of this decomposition is shown in
Fig.~\ref{fig2-4}. Five elementary single and two-qubit Ising gates are required
to perform the CNOT gate.

\begin{figure}
\begin{center}
\includegraphics*[width=8.5cm]{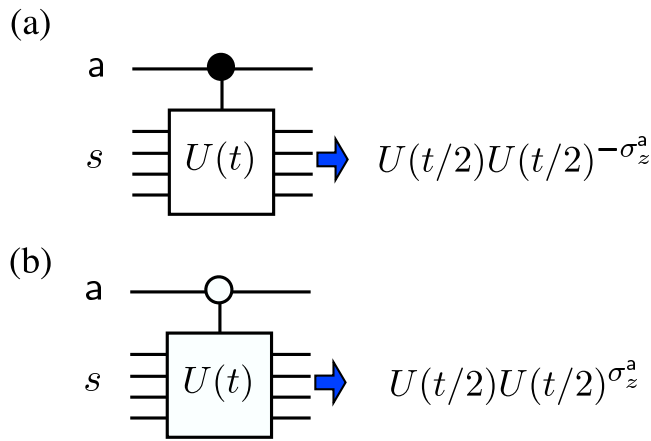}
\end{center}
\caption{(a) CU operation with the state of the control qubit ${\sf
a}$ being in
$\ket{1}_{\sf a}$ and (b) CU' operation controlled 
with the state $\ket{0}_{\sf a}$.}
\label{fig2-5}
\end{figure}

These results can be extended to other controlled unitary operations like the
CU operation defined as
\begin{equation}
\mbox{CU: }
\ket{0}_{\sf a} \bra{0} \otimes \one_s + \ket{1}_{\sf a} \bra{1} \otimes U_s.
\end{equation}
The unitary operation above performs the transformation $U_s$ (also unitary)
over a
set of qubits $s$ if the state of the control qubit ${\sf a}$ is $\ket{1}$, and
does not act otherwise. For the transformation
$U_s \equiv U(t)=e^{-i \hQ t}$, with $\hQ = \hQ^\dagger$, the operational representation of
the CU gate is
\begin{equation}
U(t/2)U(t/2)^{-\sigma_z^{\sf a}},
\end{equation}
where $U(t/2)^{-\sigma_z^{\sf a}} \equiv e^{i \hQ \otimes \sigma_z^{\sf a}}$
[Fig.~\ref{fig2-5}(a)].
Equivalently, one can define another controlled operation, CU',
on the state $\ket{0}_{\sf a}$ [Fig.~\ref{fig2-5}(b)]:
$U(t/2)U(t/2)^{\sigma_z^{\sf a}}$.

Controlled operations are widely used in quantum
algorithms. In general, their decomposition into single and two-qubit gates
require a large number of these elementary operations, so they should be
avoided when possible.

\section{Deterministic Quantum Algorithms}
\label{sec2-2}
In a QS, a QC performs certain tasks which are expected to give some
information about the physical system being simulated. These tasks are
communicated by means of a {\em program} or {\em quantum algorithm} (QA), which
can be schematically represented as a quantum circuit. In this section, I
present a particular type of QA that can be used to obtain relevant properties
of a quantum physical system $s$, using the conventional model (Sec.~\ref{sec2-1}).
Nevertheless, the same techniques can be used to simulate physical systems with
other particle statistics (e.g., fermionic or bosonic systems), if they can be
described by Pauli operators after an algebraic mapping.

A deterministic QA is based on three different steps: i) the preparation of a
pure initial state, ii) its evolution,  and iii) the measurement of certain
property of the evolved state, in which the result of the algorithm is encoded.
To preserve the features of the quantum world, the evolution step is performed
through a unitary operation and the measurement step is described by a certain
observable (i.e., Hermitian operator). Here, I present only the class of QAs
that allows one to determine, in a register of $N$ qubits, physical correlation
functions of the form
\begin{equation} 
\label{correl1}
G=\bra{\phi} \hat{W}_s \ket{\phi },
\end{equation} 
where $\hat{W}_s$ is a 
unitary (reversible) operator associated to the system to be simulated; 
that is, $\hat{W}_s\hat{W}_s^\dagger = \one$ ($s$ refers to the system).

Indirect measurement techniques can be used to obtain such correlation functions on a
QC. In addition to the qubits used to {\it represent} the physical system $s$ to be
simulated (i.e., the qubits-system) extra qubits, called {\em ancillas}, are
required. These ancillas constitute the probes that contain information about
the qubits-system.  In the following section I describe different measurement 
techniques.
\subsection{One-Ancilla Qubit Measurement Processes}
\label{sec2-2-1}
In this case a single ancilla qubit allows one to obtain the correlation functions
of Eq.~\ref{correl1}, with $\hat{W}_s=U^\dagger V$, and $U$, $V$ unitary
operators acting on $s$~\cite{OGK01}. For
this purpose, the ancilla qubit ${\sf a}$ is first initialized in the state
$\ket{+}_{\sf a}=\frac{1}{\sqrt{2}}(\ket{0}_{\sf a}+\ket{1}_{\sf a})$ by
applying, for example,  the unitary Hadamard gate to the state $\ket{0}_{\sf
a}$\footnote{The Hadamard gate in terms of single qubit rotations is
$i e^{-i \frac{\pi}{4} \sigma_y}  e^{-i \frac{\pi}{2} \sigma_z}$}. 
Second, one makes it interact with the qubits-system, initially in certain
pure state $\ket{\phi}$, through two controlled  unitary operations $\tilde{V}$
and $\tilde{U}$, associated to the $V$ and $U$ operations, respectively.  The
first operation   $\tilde{V}$ evolves the system by $V$ if the ancilla is in
the state $\ket{1}$: $\tilde{V}=\ket{0}_{\sf a}\bra{0}\otimes
\one_s+\ket{1}_{\sf a}\bra{1}\otimes V$.  The second operation $\tilde{U}$
evolves the system by $U$ if the ancilla state is $\ket{0}$:
$\tilde{U}=\ket{0}_{\sf a}\bra{0} \otimes U+\ket{1}_{\sf a}\bra{1}\otimes
\one_s$. Notice that $\tilde{V}$ and $\tilde{U}$ are reversible and commute
with each other. 

\begin{figure}
\begin{center}
\includegraphics*[width=7.0cm]{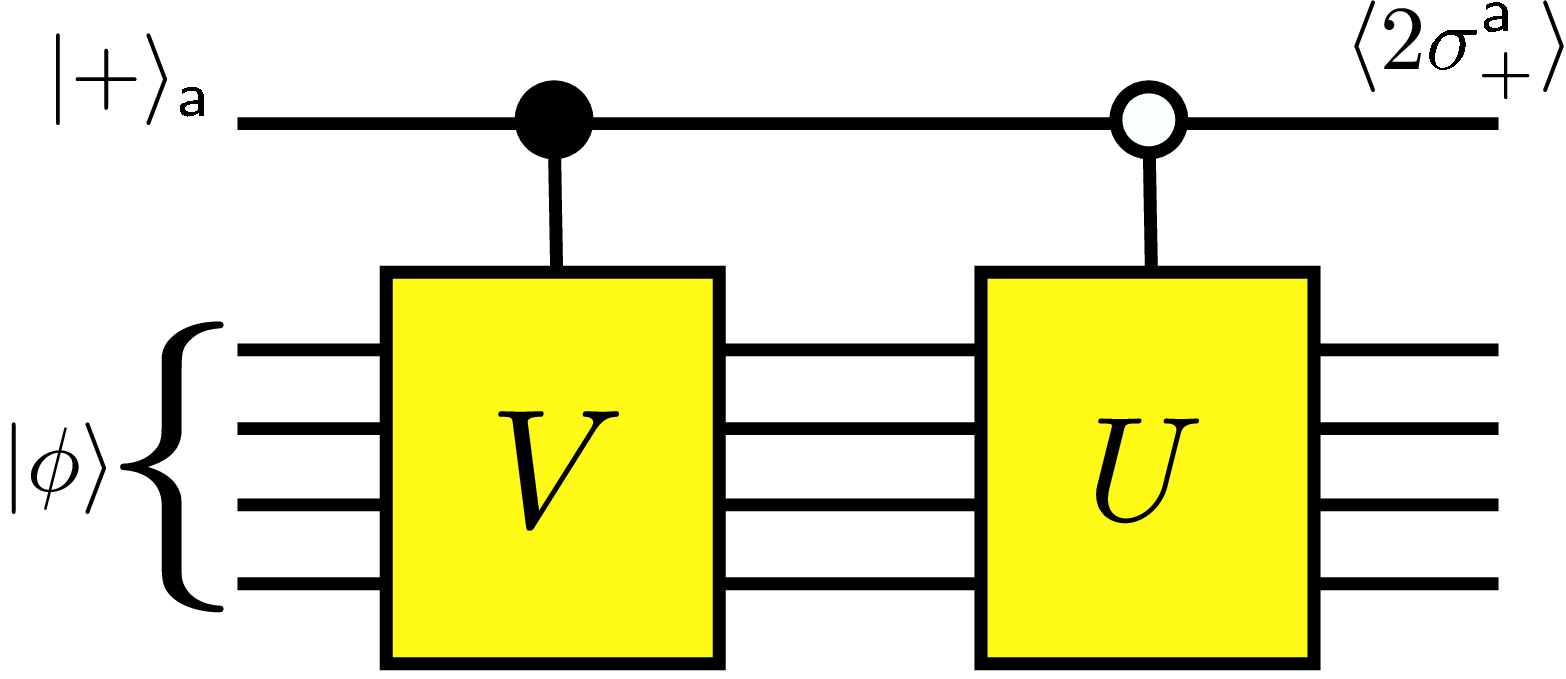}
\end{center}
\caption{Quantum algorithm for the evaluation of the correlation
$G=\bra{\phi} U^\dagger V \ket{\phi}$.}
\label{fig2-6}
\end{figure}

After such evolution, the final state of the quantum register, $\ket{ \psi_f}$,
is
\begin{equation}
\ket{\psi_f} = \tilde{V} \tilde{U} \ket{+}_{\sf a} \ket{\phi}=
\frac{1}{\sqrt2} \left[ \ket{0}_{\sf a} \otimes U \ket{\phi} +
\ket{1}_{\sf a} \otimes V \ket{\phi} \right].
\end{equation}
Interestingly, the expectation value of the Pauli operator $2 \sigma_+^{\sf a}=
\sigma_x^{\sf a} + i \sigma_y^{\sf a}$ associated with the ancilla qubit,
in this state, gives the desired correlation function:
\begin{equation}
G= \langle \psi_f | \sigma_x^{\sf a} + i \sigma_y^{\sf a} | \psi_f \rangle,
\end{equation}
where I have used the orthogonality property (Sec.~\ref{sec2-1-1}), that is,
$\langle 0 \ket{1}_{\sf a} = \langle 1 \ket{0}_{\sf a}=0$, $\langle 0
\ket{0}_{\sf a} = \langle 1 \ket{1}_{\sf a}=1$, and the action of the operators
$\sigma_\mu^{\sf a}$ over the state of the ancilla qubit given by
Eq.~\ref{pauli1}. The corresponding
circuit for this quantum algorithm is shown in Fig.~\ref{fig2-6}. Due to the
probabilistic nature of quantum measurements, the desired
expectation value is obtained with variance ${\cal O}(1)$ for each instance.
That is, in order to get an accurate value for $\langle 2 \sigma_+^{\sf
a}\rangle$, repetition must be used to reduce the variance below what is
required (Sec.~\ref{sec2-1-1}).

Nevertheless, sometimes it is necessary to compute the expectation value of an
operator $\hW$ 
of the form
\begin{equation}
\hat{W}=\sum\limits_{i=1}^Ma_{i} \ U^{\dagger}_{i}V_{i} \ ,
\end{equation}
where $U_i$ and $V_i$ are unitary operators, $a_{i}\geq 0 \in \mathbb{R}$ (with
no loss of generality).   In principle, this
expectation value can be computed by preparing $M$ different circuits as the
one represented in Fig.~\ref{fig2-6}, such that each algorithm computes
$\langle U^{\dagger}_iV_i \rangle$. However, in most practical cases, the
preparation of the initial state $\ket{\phi}$ is a very difficult task. This
difficulty can then be reduced  by using a particular QA that requires only one
circuit, but with $L$ ancilla qubits, where $L=J+1$ and $J \ge \log_2 M$. Such QA
has been described in Ref.~\cite{SOG02}. The idea is to extend the results
described above using controlled operations with respect to different ancilla
qubits.

\subsection{Quantum Algorithms and Quantum Simulations}
\label{sec2-2-2}
Based on the indirect-measurement methods described in Sec.~\ref{sec2-2-1}, I
now present certain QAs for QSs. These are useful for
obtaining relevant properties of
quantum systems, like the evaluation of the correlation function
\begin{equation}
\label{greenfunction}
G(t)= \langle \phi | T^{\dagger} A^\dagger_i T B_j |\phi \rangle .
\end{equation}
Here, $A_i$ and $B_j$ are unitary operators (any operator can be
decomposed in a unitary operator basis as $A=\sum\limits_i \alpha_i
A_i$, $B=\sum\limits_j \beta_j B_j$), $T=e^{-iHt}$ is the time
evolution operator of a time-independent Hamiltonian $H$ associated to the
physical system to be simulated, and
$\ket{\phi}$ is a particular state of the physical system.
Notice that Eq.~\ref{greenfunction} is a particular case of Eq.~\ref{correl1}. 
In particular, the evaluation of spatial
correlation functions can be obtained by replacing the evolution operator $T$
by the space translation operator. 

\begin{figure}
\begin{center}
\includegraphics*[width=10cm]{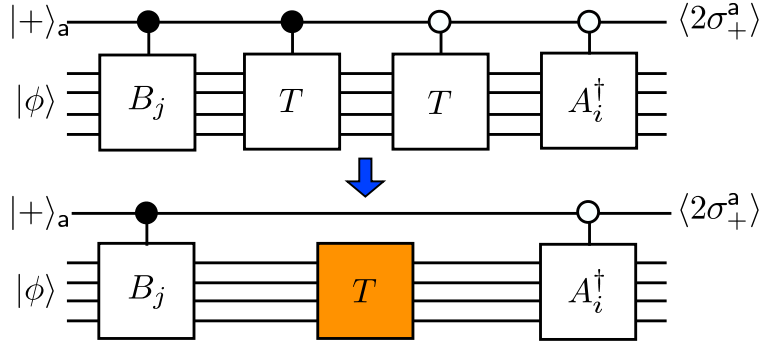}
\end{center}
\caption{Quantum algorithm for the computation of spatial and time-correlation
functions. In this case, $\langle 2 \sigma_+^{\sf a}\rangle = \bra{\phi}
T^\dagger A^\dagger_i T B_j \ket{\phi}$. Notice the simplification achieved by
reducing the controlled-T operations. The same convention of Fig.~\ref{fig2-6}
has been used.}
\label{fig2-7}
\end{figure}

The quantum circuit for the evaluation of $G(t)$ is shown in Fig.~\ref{fig2-7}.
It is equivalent to the one shown in Fig.~\ref{fig2-6} by choosing  $U^\dagger=
T^\dagger A_i$ and $V=TB_j$. This particular selection allows one to reduce the
complexity of the problem in the sense that the operation $T$ need not to be
controlled by the state of the ancilla qubit (\cite{SOG02}). As mentioned in
Sec.~\ref{sec2-1-3}, controlled operations require a large  number of elementary 
gates, so they must be avoided when possible.

In brief, the computation of $G(t)$ is performed as follows: First, the ancilla
qubit ${\sf a}$ is prepared in the state $\ket{+}_{\sf a}$, and the system is
prepared in the state $\ket{\phi}$. Second, a controlled evolution on the
state $\ket{1}_{\sf a}$, given by $\mbox{C-B}= \ket{0}_{\sf a} \langle 0 | \otimes
\one_s + \ket{1}_{\sf a} \langle 1 | \otimes B_j$, is performed. Third, the time
evolution $T$ is performed. Fourth, a controlled evolution on the state
$\ket{0}_{\sf a}$, given by  $\mbox{C-A}= \ket{0}_{\sf a} \langle 0 | \otimes A_i +
\ket{1}_{\sf a} \langle 1 | \otimes \one_s$, is performed. Finally the
observable $\langle 2\sigma_+^{\sf a} \rangle=  \langle \sigma_x^{\sf a} +i
\sigma_y^{\sf a}\rangle =G(t)$ is measured. 

Sometimes one is interested in obtaining the
spectrum (eigenvalues) of a given observable $\hat{Q}$ (
Hermitian operator), associated with some physical property of the system to be
simulated. Techniques for getting spectral information can be used on the
quantum Fourier transform~\cite{Kit95,CEM98} and can be applied to physical
problems~\cite{AL99}.
Nevertheless, the idea here is to use the methods developed in
Sec.~\ref{sec2-2-1}.

For some Hermitian operator $\hat{Q}$, such as the Hamiltonian $H$ of the
system to be simulated, a common type of problem is the computation of its
eigenvalues or, at least, the lowest eigenvalue (related, for example, to the
ground state of the system). For this purpose, the qubits-system needs to be
initialized in a certain state $\ket{\phi}$, that has a non-zero overlap with the
eigenstates corresponding to the eigenvalues that need to be
computed. Such a state can
always be decomposed as a linear combination of eigenstates of $\hat{Q}$,
\begin{equation}
\label{initspec}
\ket{\phi}=\sum_n \gamma_n \  \ket{\psi_n},
\end{equation}
where $\gamma_n$ are complex coefficients, $\ket{\psi_n}$ are the eigenstates
of $\hat{Q}$, and $\lambda_n$ the corresponding eigenvalues. If interested in
computing $\lambda_m$, it is then required that $|\gamma_m| \neq 0$ in
Eq.~\ref{initspec}.

\begin{figure}
\begin{center}
\includegraphics*[width=10.0cm]{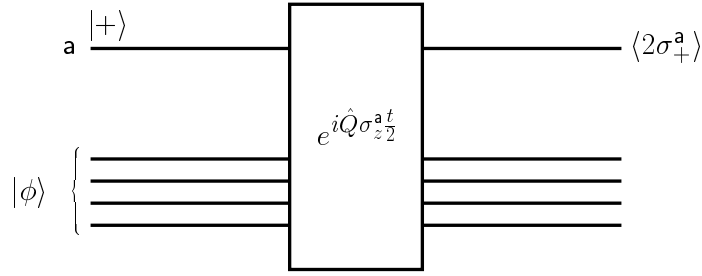}
\end{center}
\caption{Quantum algorithm for the computation of the spectrum of an observable
$\hQ$. In this case, $\langle 2 \sigma_+^{\sf a} \rangle = \bra{\phi} e^{-i \hQ
t} \ket{\phi}$.}
\label{fig2-8}
\end{figure}

For this purpose, the correlation function 
\begin{equation}
\label{spectra}
S(t)=\bra{\phi} U(t) \ket{\phi},
\end{equation}
with $U(t)=e^{-i \hat{Q} t}$, needs to be computed for different values of the
real parameter $t$ (usually related with time). For a particular $t$, the
measurement of $S(t)$ can be performed using the one-ancilla method
(Sec.~\ref{sec2-2-1}) as follows.
First, the initial state 
$\ket{+}_{\sf a} \otimes \ket{\phi}$ is prepared. 
Second, the unitary evolution  $\exp[i \hat{Q} \sigma_z^{\sf a} t/2]$
is performed. Finally, the expectation $\langle 2 \sigma_+^{\sf a} \rangle
=S(t)$ is measured. The circuit representation of this QA is shown in
Fig.~\ref{fig2-8}. It is equivalent to the one shown in Fig.~\ref{fig2-6} by
replacing $U^\dagger =V=  e^{-i \hat{Q} t/2}$.

For a particular value of $t$, the function $S(t)$ is
\begin{equation}
\label{spectra2}
S(t)  = \sum_n |\gamma_{n}|^{2} \ 
e^{-i{\lambda}_{n}t} .
\end{equation}
Then, the eigenvalues $\lambda_n$ can be obtained by 
performing a classical Fourier transform to Eq.~\ref{spectra2} (i.e., 
$\tilde{S}(\lambda)=\int  S(t) e^{i\lambda t}dt$)
\begin{equation}
\label{FFT}
\tilde{S}(\lambda)=\sum_n
2 \pi |\gamma_{n}|^{2}{\delta} ({\lambda}-{\lambda}_{n}) \ .
\end{equation}
However, $S(t)$ is only obtained for a discrete set of values of $t$. It is
needed, instead, to calculate the corresponding discrete Fourier transform (see
Appendix~\ref{appA}) to obtain information about the $\lambda_n$'s.

\section{Quantum Simulations of Quantum Physics}
\label{sec2-3}
In the most general case, a QS requires the simulation of systems with diverse
degrees of freedom, like fermions, anyons, bosons, etc. The associated Hilbert
spaces (space of states) differ from the one defined for the conventional
model. For example, in the case of fermionic systems, fermionic states are
governed by Pauli's exclusion principle. Then, at most a single spinless (or
two spin-1/2) fermion can occupy a certain (atomic) quantum state at the same
time. Therefore, all the features associated with the physical system to be
simulated must be preserved when transforming its operators to the operators
describing the computational model of the QC. 

In this section, I present isomorphic mappings that allow one to simulate
arbitrary quantum systems, regardless of their particle statistics, by using
the QAs defined for the conventional model (Sec.~\ref{sec2-2}). Fortunately,
such mappings can be easily performed without breaking the efficiency of a
QA.

\subsection{Simulations of Fermionic Systems}
\label{sec2-3-1}
The systems considered here consist mainly of a lattice with $N$ modes (sites),
where spinless fermions can hop between sites. These results can be easily
extended for the case of spin-1/2 fermions or higher spin fermions.

In the second quantization representation, the (spinless) fermionic
operators $c^{\dagger}_j$ and $c^{\;}_j$ are defined as the creation
and annihilation operators of a fermion in the $j$-th mode 
($j=1,\cdots,N$), respectively. Due to the Pauli's exclusion principle and the
antisymmetric nature of the  fermionic wave function under the
permutation of two fermions, the fermionic algebra is given by the
following anticommutation relations
\begin{equation}
\label{fermcom}
\{ c_i,c_j \}=0 ,  \mbox{ } \{ c^{\dagger}_i,c_j \} = \delta_{ij}
\end{equation}
where $\{,\}$ denotes the anticommutator (i.e., $\{ A,B \} = AB+BA$).

The Jordan-Wigner transformation~\cite{JW28} is the isomorphic mapping that
allows the description of a fermionic system by the conventional model. It is
performed in the following way:
\begin{eqnarray}
\label{JW}
c_j \rightarrow \left( \prod\limits_{l=1}^{j-1} -\sigma_z^l \right) \sigma_-^j,\\
\label{JW2}
c^{\dagger}_j \rightarrow \left( \prod\limits_{l=1}^{j-1} -\sigma_z^l \right) 
\sigma_+^j ,
\end{eqnarray}
where the Pauli operators $\sigma_{\mu}^i$ were previously introduced in
Sec~\ref{sec2-1}.
If these operators satisfy
the $\fsu(2)$ commutation relations (Eqs.~\ref{comm1} and \ref{comm2}), 
the operators
$c^{\dagger}_j$ and $c^{\;}_j$ obey the anticommutation relations of 
Eqs. \ref{fermcom}.
This is an
isomorphic mapping between operator  algebras and is independent of the
Hamiltonian of the fermionic system to be simulated. 

Different
Hamiltonians establish different  connections (connectivity) between
fermionic modes. Historically, Eqs. \ref{JW} and \ref{JW2} correspond
to lattices in one space dimension. Nevertheless, it is also valid for
lattice systems in any dimension, when the set of modes $j$ is 
countable. In particular, the  set of all ordered $p$-tuples of
integers can be placed in one-to-one correspondence with the set of
integers. For example, the simulation of a two dimensional fermionic
lattice system can be done by re-mapping each mode $(l,m)$ into a new
set of modes as $j=m+(l-1)N_x$, where  $[l=1 \cdots N_y]$ and   $[m=1
\cdots N_x]$ are integer numbers that refer to the position of a site
in the lattice,  and $N_x$ and $N_y$ are the number of sites (modes) in
the $x$ and $y$ direction, respectively. 

In order to compute physical properties of a fermionic system on a QC described
by the conventional model, every step of the quantum simulation has to be
expressed in terms of Pauli operators. For a (spinless)  fermionic system with
$N$ modes, a QC must contain, besides the ancilla qubit ${\sf a}$, $N$ qubits
to represent the system. In the following, I describe how certain fermionic
initial states can be prepared and how they can be evolved under a particular
fermionic Hamiltonian evolution.

\subsubsection{Preparation of Initial Fermionic States}
Associated to each fermionic mode, there are two levels which correspond 
to the mode being
empty or being occupied by a spinless fermion. The state-state mapping is then
trivial. Basically, the logical state $\ket{1_j}$ is associated to the
$j$th mode if it is empty, and the logical state $\ket{0_j}$ (up to a phase)
is associated if
the $j$th mode is occupied. In this way, the vacuum or no-fermion state  $\ket{\sf
vac}$, which satisfies $c_j \ket{ {\sf vac}}=0 \mbox{ }\forall j$, is mapped to
the logical $N$-qubit state $\ket{1_1 1_2 \cdots 1_N}$. 

However, when simulating a fermionic system, more complex states need to be
prepared. A general state $\ket{\psi}$  of $N_e$ fermions is a linear
combination of Slater determinants (i.e., fermionic product states),
\begin{equation}
\label{genferm}
\ket{\psi}=\sum\limits_{\alpha=1}^L g_\alpha \ \ket{\phi_\alpha} , 
\end{equation}
where the Slater determinants $\ket{\phi_\alpha}$ are
\begin{equation} 
\label{slater1}
\ket{\phi_\alpha} = \prod \limits_{j=1}^{N_e} c^{\dagger}_j \ \ket{\sf vac}.
\end{equation}
Due to the anticommutation relations of Eqs.
\ref{fermcom}, the fermionic operators satisfy
\begin{equation}
c^\dagger_i c^\dagger_j = -c^\dagger_j
c^\dagger_i \mbox{ if } i\neq j,
\end{equation}
implying that the Slater determinants $\ket{\phi_\alpha}$ are antisymmetric
wave functions under the permutation of an even number of fermions.

Every state $\ket{\phi_\alpha}$ can be prepared on a QC made of qubits,
by noticing that
the quantum gate (i.e., unitary operator)
\begin{equation}
\label{Um}
U_m=e^{i \frac{\pi}{2} (c^{\;}_m+c^\dagger_m)}
\end{equation}
creates a particle in the $m$-th mode when acting on the vacuum state. In other
words, $U_m \ket{{\sf vac}} = e^{i \frac{\pi}{2}} c^\dagger_m \ket{\sf vac}$.
Then, making use of the Jordan-Wigner transformation (Eqs. \ref{JW} and
\ref{JW2}), the operators $U_m$ in the spin language are
\begin{equation}
\tilde{U}_m=e^{i \frac{\pi}{2} \sigma_x^m \prod\limits_{j=1}^{m-1} -\sigma_z^j}.
\end{equation}
The operators $\tilde{U}_m$ can easily be decomposed into elementary single and
two-qubit gates as described in Sec.~\ref{sec2-1-2}. The successive application
of $N_e$ similar unitary operators to the state $\ket{1_1 1_2 \cdots 1_N}$
generates the mapped state $\ket{\phi_\alpha}$, up to an irrelevant global
phase.

The general fermionic state of Eq.~\ref{genferm} can be prepared by using $L$
ancilla qubits, performing unitary controlled-$\tilde{U}_m$ evolutions on the state of the
ancillas, and finally, performing a measurement (projecting) on the ancillas.
For example, if one is interested in preparing the state
$\ket{\psi}=\frac{1}{\sqrt{2}}[\ket{\phi_1}+\ket{\phi_2}]$, one needs to add an
extra ancilla to the system. This ancilla is prepared in the state $\ket{+}_{\sf
a}$ and a controlled evolution to obtain the state $\frac{1}{\sqrt{2}}
[\ket{0}_{\sf a} \otimes
\ket{\phi_1} + \ket{1}_{\sf a} \otimes\ket{\phi_2}]$, is performed later.
If the Hadamard gate is applied to the ancilla, this state evolves into
\begin{equation}
\ket{0}_{\sf a} \otimes \frac{1}{\sqrt{2}} [\ket{\phi_1}+\ket{\phi_2}]
+ \ket{1}_{\sf a} \otimes \frac{1}{\sqrt{2}} [\ket{\phi_1}+\ket{\phi_2}].
\end{equation}
Therefore, the ancilla qubit is measured and projected, with probability 1/2,
into $\ket{0}_{\sf a}$ or $\ket{1}_{\sf a}$. If the former is obtained, the
desired state is prepared. However, if the ancilla is projected into
$\ket{1}_{\sf a}$, the whole method needs to be applied again from the begining.

In general, the probability of successful preparation of $\ket{\psi}$  
(Eq.~\ref{genferm}) using
this method is $1/L$. Then, the order of $L$ trials need to be performed before
a successful preparation. A detailed description of this method can be found in
Ref.~\cite{OGK01}.

Nevertheless, an important case consists of the preparation of Slater
determinants (product state) $\ket{\phi_\beta}$ in a different basis mode than
the one given before:
\begin{equation}
\label{slater2}
\ket{\phi_\beta}=\prod\limits_{j=1}^{N_e} d^\dagger_j \ \ket{\sf vac} .
\end{equation}
The fermionic operators $d^\dagger_j$'s are sometimes related to the operators
$c^\dagger_j$ through the following canonical transformation
\begin{equation}
\label{unitmap}
\overrightarrow{d}^\dagger = e^{i\bar{M}} \overrightarrow{c}^\dagger,
\end{equation}
with
$\overrightarrow{d}^\dagger=(d^\dagger_1,d^\dagger_2,\cdots,d^\dagger_N)$,
$\overrightarrow{c}^\dagger=(c^\dagger_1,c^\dagger_2,\cdots,c^\dagger_N)$,
and $\bar{M}$ being a $N \times N$ Hermitian matrix. (Sometimes the operators
$d^\dagger$ are combinations of both, the creation and annihilation operators
$c^\dagger_i$ and $c_i$.)

Thouless's
theorem states that one Slater determinant
evolves into the other as 
\begin{equation}
\ket{\phi_\beta}=U\ket{\phi_\alpha},
\end{equation}
where the unitary fermionic operator 
\begin{equation}
\label{thoulessprep}
U=e^{-i \overrightarrow{c}^\dagger \bar{M},
\overrightarrow{c}}
\end{equation}
can be written in terms of Pauli operators using the
Jordan-Wigner transformation (Sec.\ref{sec2-3-1}), and can also be decomposed
into elementary gates as described in Sec.~\ref{sec2-1-2}. 

In brief, the described fermionic product states can be prepared on a QC
described by the conventional model, if the Jordan-Wigner transformation is
performed. Interestingly, the preparation can be done efficiently: the number
of elementary single-qubit and two-qubit gates required scales polynomially
with the system size $N$. In Chap.~\ref{chapter4}, I present another class of
fermionic states that can also be prepared efficiently. 

\subsubsection{Fermionic Evolutions}
The evolution of a quantum state is the second step in the realization of a
QA. The goal is to decompose a generic evolution into the
elementary gates $R_{\mu}(\vartheta)\mbox{ and } R_{z^{j}z^{k}}(\omega)$
(Sec.~\ref{sec2-1}).
Sometimes, the evolution step is associated to a Hermitian operator $\tilde{H}$
which is, for example, the Hamiltonian $H$ of the fermionic system to be simulated in terms of
Pauli operators after Eqs.~\ref{JW} and \ref{JW2} have been performed.
In this case, the corresponding evolution unitary operator is
$\tilde{U}(t)=e^{-i \tilde{H}t}$ (i.e, the solution to the Schr\"{o}dinger's evolution equation).

In general, a fermionic Hamiltonian can be decomposed as
$H=K+V$ ,where $K$ represents the kinetic energy of the fermions
and $V$ their potential energy. Usually, $[K,V] \neq 0$ and the decomposition
of $\tilde{U}(t)$ in terms of elementary gates
is a complicated task.  To
avoid this difficulty this operator is approximated by, for example,
using a first order Trotter decomposition~\cite{Suz93}. That is,
\begin{eqnarray}
\label{trotter}
\tilde{U}(t)&=&\prod\limits_{g=1}^{\cal N} \tilde{U}(\Delta t), \\
\tilde{U}(\Delta t)&=& e^{i \tilde{H} \Delta t}= e^{ i (\tilde{K}+
\tilde{V}) \Delta t}=  e^{i \tilde{K} \Delta t}
e^{i \tilde{V} \Delta t} + \mathcal{O}(\Delta t^2),
\end{eqnarray}
where $\tilde{K}$ and $\tilde{V}$ are the terms $K$ and $V$ in Pauli operators,
respectively.
Therefore, for $\Delta t \rightarrow 0$, $\tilde{U}(\Delta t) \sim e^{i
\tilde{K} \Delta t}
e^{i \tilde{V} \Delta t}$.

The potential energy $V$ is usually a sum of commuting diagonal terms,
and the decomposition of $e^{i \tilde{V} \Delta t}$ into elementary gates is
simple. However, the kinetic energy $K$  is usually a sum of
noncommuting hopping terms of the form  $c^\dagger_j c^{\;}_k + c^\dagger_k
c^{\;}_j$ (bilinear fermionic operators), and its decomposition is again
approximated. A typical kinetic term  $e^{i (c^\dagger_j
c^{\;}_k +c^\dagger_k c^{\;}_j)\Delta t}$ ($j<k$), when mapped onto the
spin language gives
\begin{equation}
\label{decomp2}
e^{-\frac{i}{2}  (\sigma_x^j \sigma_x^k +\sigma_y^j \sigma_y^k) 
\prod\limits_{l= j+1}^{k-1}(-\sigma_z^l) } = e^{-\frac{i}{2}   \sigma_x^j
\sigma_x^k  \prod\limits_{l= j+1}^{k-1}(-\sigma_z^l) } e^{-\frac{i}{2}  
\sigma_y^j \sigma_y^k \prod\limits_{l= j+1}^{k-1}(-\sigma_z^l) }.
\end{equation}
The decomposition of each term on the right hand side of Eq. 
\ref{decomp2} into  elementary single and two-qubit gates was previously
discussed in Sec.~\ref{sec2-1-2}.
The amount of elementary gates required depends on
the distance $|j-k|$, and scales polynomially with that distance. Moreover,
since $H$ represents a physical system, it is a linear combination of a
polynomially large (with $N$) amount of fermionic operators.  Then, $\tilde{U}(t)$ can be
performed efficiently by applying a polynomially large amount of elementary
gates.
In the same way, the unitary operation $U=e^{-i \overrightarrow{c}^\dagger M
\overrightarrow{c}}$ of Eq.~\ref{thoulessprep} can also be efficiently
implemented.

Obviously, the accuracy of approximating $\tilde{U}(t)$ using the Trotter decomposition
increases as $\Delta t$ decreases. Then, a large amount of gates might  be
required to perform the desired evolution with small errors. To overcome this
problem, one could use a Trotter approximation of higher order in  $\Delta
t$~\cite{Suz93}. All these approximation methods do not destroy the efficiency
of the QA. Moreover, the evolution step induced by fermionic physical
Hamiltonians with higher order products of creation and annihilation operators
can also be efficiently implemented using the same techniques.

\subsection{Simulations of Anyonic Systems}
\label{sec2-3-2}
The concepts described in Sec.~\ref{sec2-3-1} can be extended to other and more
general particle statistics, namely {\em hard-core} anyons~\cite{BO01}. These
are particles that also obey the Pauli's exclusion principle: At most one
(spinless) anyon can occupy a single mode. Assigned to each mode of the lattice
are the creation and annihilation anyonic operators $a^{\dagger}_j$ and
$a^{\;}_j$, respectively. Their commutation relations are given by ($j \leq j'$)
\begin{eqnarray}
\label{anyoncom1}
[ a^{\;}_j,a^{\;}_{j'} ]_\theta &=& 
[ a^\dagger_j,a^\dagger_{j'} ]_\theta=0  
 \ , \nonumber\\
{[}a^{\;}_j,a^\dagger_{j'}{]}_{-\theta}&=&\delta_{jj'} (1-(e^{-i
\theta}+1)n_{j'}) \ , \\
{[} n_j, a^\dagger_{j'} ]&=& \delta_{jj'} 
a^\dagger_{j'}  \ , \nonumber
\end{eqnarray}
where $n_{j'}=a^\dagger_{j'} a^{\;}_{j'}$ is the number operator, 
$[A,B ]_\theta = AB - e^{i \theta} BA$, 
and $\theta$ is the statistical angle. In particular, 
$\theta=\pi$  mod($2\pi$) represents canonical spinless fermions, while
$\theta=0$ mod($2\pi$) represents hard-core bosons. 

In order to simulate this problem on a QC described by the conventional model,
the following isomorphic mapping between algebras can be performed:
\begin{eqnarray}
a^{\dagger}_j &=& \prod\limits_{l<j} \left[\frac {e^{-i \theta} +1}{2} + 
\frac {e^{-i \theta} -1}{2} \sigma_z^l \right] \ \sigma_+^j , \nonumber \\
\label{anyonmap}
a_j &=& \prod\limits_{l<j} \left[\frac {e^{i \theta} +1}{2} + 
\frac {e^{i \theta} -1}{2} \sigma_z^l \right] \ \sigma_-^j ,  \\
n_j &=& \frac{1}{2} (1+ \sigma_z^j) ,\nonumber
\end{eqnarray}
where the Pauli operators $\sigma_{\mu}^i$ were introduced in
Sec.~\ref{sec2-1-1}. 
Since they satisfy the commutation relations of Eqs.~\ref{comm1} and 
\ref{comm2}, the
commutation relations for the anyonic operators (Eqs.
\ref{anyoncom1}) are satisfied.

As in the fermionic case (Sec.~\ref{sec2-3-1}), an anyonic evolution operator
can be written in terms of Pauli operators using Eq.~\ref{anyonmap}, and can be
decomposed into single and two-qubit elementary gates. Therefore, the  same
procedure described in the previous section can be followed. 

Anyon statistics have
fermion and hard-core boson statistics as limiting cases, satisfying always the
Pauli's exclusion principle. In the next section this hard-core condition is
relaxed and the important case of canonical bosons is considered.

\subsection{Simulations of Bosonic Systems}
\label{sec2-3-3}
Quantum computation is based on the manipulation of quantum systems
that possess a finite number of degrees of freedom (e.g., qubits). From
this point of view, the simulation of bosonic systems appears to be
impossible, since the non-existence of an exclusion principle implies
that the Hilbert space used to represent bosonic quantum states on a lattice is
infinite-dimensional; that is, there is no limit to the number of
bosons that can occupy a given mode $j$. However, sometimes it is necessary
to simulate and study properties such that the use of
the complete Hilbert space is unnecessary, and only a finite  sub-basis of
states is sufficient.  This is the case for $N$-mode (e.g., $N$ sites) lattice systems with
interactions given by the boson-preserving Hamiltonian
\begin{equation}
\label{bosonhamilt}
H= \sum \limits_{j,j'=1}^N \alpha_{jj'} \ b^{\dagger}_j b^{\;}_{j'} +
\beta_{jj'} \ \hat{n}_j \hat{n}_{j'},
\end{equation}
where the operators $b^{\dagger}_j$ ($b^{\;}_j$) create (destroy) a
boson  at site $j$, and $\hat{n}_j=b^{\dagger}_j b^{\;}_j$ is the number
operator; that is
\begin{eqnarray}
\nonumber
b^{\dagger}_j \ket{n_1,n_2,\cdots,n_j,\cdots,n_N} &=& \sqrt{n_j+1} \ 
\ket{n_1,n_2,
\cdots,n_j+1,\cdots,n_N}, \\
\nonumber
b^{\;}_j \ket{n_1,n_2,\cdots,n_j,\cdots,n_N}&= & \ \ \ \sqrt{n_j} \ \ \ \ \ 
\ket{n_1,n_2,
\cdots,n_j-1,\cdots,n_N}, \\
\label{boso}
\hat{n}_j \ket{n_1,n_2,\cdots,n_j,\cdots,n_N} &= & \ \ \ \ \ n_j
\ \ \ \ \ \ \ket{n_1,n_2,\cdots,n_j,\cdots,n_N},
\end{eqnarray}
where the bosonic state $\ket{n_1,n_2,\cdots,n_j,\cdots,n_N}$ represents a
quantum state with $n_j$ bosons in the $j$-th mode (site).

The space dimension of the lattice is encoded in the parameters $\alpha_{jj'}$
and $\beta_{jj'}$ of the Hamiltonian.  Since $H$ contains pairs of creation and
annihilation operators, the total number of bosons $N_P$ in the system is
preserved and the idea is to work in this finite sub-basis of states (where
the dimension of the associated Hilbert space depends on the magnitude of $N_P$). 

The corresponding bosonic commutation relations (in an
infinite-dimensional Hilbert space) are~\cite{CDG98}
\begin{equation}
\label{bosoncom}
[b_j,b_{j'}]=0 , [b_j,b^{\dagger}_{j'}]=\delta_{jj'}.
\end{equation}
However, if the operators $b^{\dagger}_j$ are restricted to the finite
basis of  states represented by $\{ \ket{n_1,n_2,\cdots,n_N }$ with
$\max(n_i)= N_P \}$, that is, $N_P$ is the maximum number of bosons per
site, they acquire the following matrix representation (see Eqs.
\ref{boso})
\begin{equation}
\label{bosonprod}
\bar{b}^{\dagger}_j =\one \otimes \cdots \otimes \one \otimes 
\underbrace{\hat {b}^{\dagger}}_{j^{th}\mbox{ factor}} \otimes
\one \otimes \cdots \otimes \one \\
\end{equation}
where the symbol $\otimes$ indicates the usual tensorial product between matrices,
and the $(N_P+1) \times (N_P+1)$ dimensional matrices $\one$ and
$\hat{b}^{\dagger}$  are given by
\begin{equation}
\label{bosonrep}
\one =\pmatrix {1&0&0&\cdots&0 \cr  0&1&0&\cdots&0\cr 0&0&1&\cdots&0 \cr
\vdots&\vdots&\vdots&\cdots &\vdots\cr 0&0&0&\cdots&1} 
\mbox{ , }
\hat{b}^{\dagger} = \pmatrix{0 & 0 & 0 &\cdots &0 & 0 \cr 1 & 0 & 0 & 
\cdots & 0& 0 \cr 0& \sqrt{2} & 0 &\cdots &0 & 0 \cr \vdots & \vdots & \vdots
&\cdots &\vdots &\vdots \cr 0 & 0 & 0 &\cdots &\sqrt{N_P} & 0}.
\end{equation}
It is important to note that in this finite basis, the commutation
relations of the $\bar{b}^{\dagger}_i$ differ from the standard
bosonic ones (Eq. \ref{bosoncom}) \cite{BO02} 
\begin{equation}
\label{bosoncom2}
[\bar{b}^{\;}_i,\bar{b}^{\;}_j]=0 , \mbox{ } 
[\bar{b}^{\;}_i,\bar{b}^{\dagger}_j]=\delta_{ij} \left[ 1-
\frac{N_P+1}{N_P!} (\bar{b}^{\dagger}_i)^{N_P}(\bar{b}^{\;}_i)^{N_P}
\right] ,
\end{equation}
and clearly $(\bar{b}^{\dagger}_j)^{N_P+1}=0$.

Since the goal is to simulate the bosonic system on a QC described by the
conventional model, a corresponding mapping between both operator algebras must
be given. Nevertheless, Eqs. \ref{bosoncom2}  imply that the linear span of
the  operators  $\bar{b}^{\dagger}_j$ and $\bar{b}^{\;}_j$ is not closed under
the commutator,  and a mapping between the bosonic operators and the
Pauli operators like the  Jordan-Wigner transformation (Sec.~\ref{sec2-3-1})
is not 
possible. Therefore, such isomorphic mapping needs to be found by first mapping
quantum bosonic states onto quantum logical states in the conventional model
(i.e., a Hilbert space mapping).

\begin{figure}
\begin{center}
\includegraphics*[width=7.5cm]{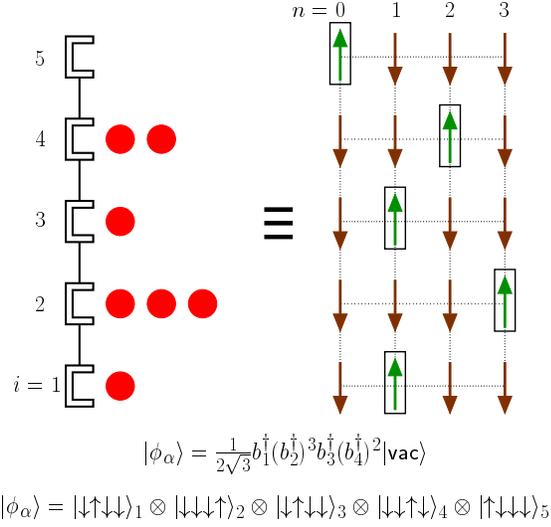}
\end{center}
\caption{Mapping of the bosonic state $\ket{\phi_\alpha}$, of a chain with 5
sites and 7 bosons ($N_P=3$), into a four-spin-1/2 (or four-qubit) state. The convention
is $\ket{\uparrow_j} \equiv \ket{0_j}$ and $\ket{\downarrow_j} 
\equiv \ket{1_j}$.}
\label{fig2-9}
\end{figure}

The idea is to
start by considering only the $j$th mode in the chain. Since this
mode can be occupied with at most $N_P$ bosons, it is possible to
associate an $(N_P+1)$-qubit quantum state to each particle number
state, in the following way:
\begin{eqnarray}
\label{bosonmap}
|0\rangle_j &\leftrightarrow& |0_0 1_1 1_2 \cdots 
1_{N_P} \rangle_j \nonumber \\
|1\rangle_j &\leftrightarrow& |1_0 0_1 1_2 \cdots 
1_{N_P} \rangle_j \nonumber \\
|2\rangle_j &\leftrightarrow& |1_0 1_1 0_2 \cdots 
1_{N_P} \rangle_j  \\
\vdots && \vdots \nonumber \\
|N_P\rangle_j &\leftrightarrow& |1_0 1_1 1_2
\cdots 0_{N_P} \rangle_j \nonumber
\end{eqnarray} 
where $\ket{n}_j$ denotes a quantum state with $n$ bosons in $j$th mode.
Therefore,  $N(N_P+1)$ qubits for the simulation (where $N$
is the total number of modes) are needed. 
An example of this mapping for a quantum state with 7 bosons in a
chain of 5 sites, where the maximum number of bosons per site is $N_P=3$, is
shown in Fig.~\ref{fig2-9}.

By definition (see Eqs. \ref{boso}, \ref{bosonprod}, and \ref{bosonrep}) 
$\bar{b}^{\dagger}_j \ \ket{n}_j= \sqrt{n+1} \ \ket{n+1}_j$, and this operator
in the conventional model maps to
\begin{equation}
\label{bosonmap2}
\bar{b}^{\dagger}_j \rightarrow \tilde{b}^{\dagger}_j =
\sum \limits_{n=0}^{N_P-1} \sqrt{n+1} \ \sigma_-^{n,j}
\sigma_+^{n+1,j} ,
\end{equation}
where a pair $(n,j)$ refers to the $n$th qubit in the chain of qubits
representing the $j$th bosonic mode. The Pauli creation and annihilation
operators $\sigma_\pm^k$ were previously defined in Sec.~\ref{sec2-1}. The operator 
$\bar{b}^{\dagger}_j$ acts then on the $(N_P+1)$-qubit chain representing the
$j$th bosonic mode as
\begin{equation}
\tilde{b}^{\dagger}_j \ket{ 1_0 \cdots  1_{n-1}  0_n 1_{n+1} \cdots 1_{N_P} }_j =
\sqrt{n+1} \  \ket{ 1_0 \cdots 1_n 0_{n+1} 1_{n+2} \cdots 1_{N_P} }_j,
\end{equation}
so its matrix representation in this basis
is analogous to the matrix representation of $\bar{b}^{\dagger}_j$ in the
basis of bosonic states. 

Similarly, the number operator is mapped as
\begin{equation}
\label{bosonmap2d}
\bar{n}_j \rightarrow \tilde{n}_j =\sum \limits _{n=0}^{N_P} n \
\frac{\sigma_z^{n,j}+1}{2} ,
\end{equation}
so its action over the corresponding logical states is
\begin{equation}
 \tilde{n}_j \ket{ 1_0 \cdots  1_{n-1}  0_n 1_{n+1} \cdots 1_{N_P} }_j = n \ 
\ket{ 1_0 \cdots 1_n 0_{n+1} 1_{n+2} \cdots 1_{N_P} }_j. 
\end{equation}
Since the commutator $[\tilde{b}^{\dagger}_j,
\sum_{n=0}^{N_P} \sigma_z^{n,j}]=0$  the operators $\tilde{b}^\dagger_j$
($\tilde{b}_j$) always keep states within the same subspace.

The Hamiltonian of Eq. \ref{bosonhamilt} in terms of Pauli operators is then
\begin{equation}
\label{bosonhamilt2}
\tilde{H}=\sum\limits_{j,j'=1}^N \alpha_{jj'} \ \tilde{b}^{\dagger}_j
\tilde{b}^{\;}_{j'} + \beta_{jj'} \  \tilde{n}_j \tilde{n}_{j'}  , 
\end{equation}
where the operators $\tilde{b}^{\dagger}_j$ ($\tilde{b}^{\;}_j$) are given by
Eqs. \ref{bosonmap2}, and $\tilde{n}_j$ by Eq. \ref{bosonmap2d}. In this way,
physical properties of the bosonic system such as the  spectrum of $H$ can be
obtained using a QC made of qubits.  The same methods can be used
when simulating any other type of boson-preserving quantum system.

\subsubsection{Preparation of Initial Bosonic States}
As in the fermionic case, the most general bosonic state of an $N$-mode lattice
system with a maximum of $N_P$ bosons per site can be written as a  linear
combination of bosonic product states like
\begin{equation}
\label{bosprodstate}
\ket{\phi_\alpha}= {\sf K} (b^\dagger_1)^{n_1}(b^\dagger_2)^{n_2} \cdots
(b^\dagger_N)^{n_N} \ \ket{\sf vac} ,
\end{equation}
where ${\sf K}$ is a normalization factor, $n_j$ is the number of  bosons at
site $j$ ($\max(n_j)=N_P$), and $\ket{\sf vac}$ is vacuum or no-boson state,
that is, $b_j \ket{\sf vac} =0 \ \forall j$. 

Using the mapping
described in Eq. \ref{bosonmap}, the vacuum state in the conventional model maps
as 
\begin{equation}
\ket{\sf vac} \rightarrow \ket{ 0_0 1_1 \cdots 1_{N_P}}_1 \otimes \cdots
\otimes \ket{ 0_0 1_1 \cdots 1_{N_P}}_N ,
\end{equation}
and the product state of Eq.~\ref{bosprodstate} maps as 
\begin{equation}
\ket{\phi_\alpha} = \ket{ 1_0 \cdots 0_{n_1} \cdots 1_{N_P}}_1 \otimes \cdots
\otimes \ket{ 1_0 \cdots 0_{n_N} \cdots 1_{N_P}}_N
\end{equation}
(see Fig.~\ref{fig2-9} for an example). 

The preparation of the mapped bosonic state $\ket{\phi_\alpha}$ on a QC made of
qubits is then performed by flipping the states of the corresponding qubits
from a fully polarized state (i.e., the logical state with all qubits in 
$\ket{1}$), using for example the flip operations $\sigma^{n,j}_x$. 
Nevertheless, more general bosonic states like
\begin{equation}
\label{bosstate}
\ket{\psi}=
\sum\limits_{\alpha=1}^L g_{\alpha} \ \ket{\phi_\alpha}
\end{equation}
can also be realized as in the fermionic case.  Again, the idea is to add $L$
ancillas (extra qubits), perform controlled evolutions on their states, and
finally perform measurements on the state of the ancillas. The state is
successfully prepared with probability $1/L$~\cite{OGK01}.

\subsubsection{Bosonic Evolutions}
Again, the idea is to represent certain bosonic unitary evolution operator
$\tilde{U}(t)=e^{-i\tilde{H}t}$, where $\tilde{H}$ is some boson-preserving
Hermitian operator such as the Hamiltonian of the system to be simulated 
(Eq.~\ref{bosonhamilt}), in terms of Pauli operators (Eq.~\ref{bosonhamilt2}).
Usually, a first order Trotter approximation~\cite{Suz93} also needs to be
performed  to  separate those terms in $H$ that do not commute. 

\begin{figure}
\begin{center}
\includegraphics*[width=7.4cm]{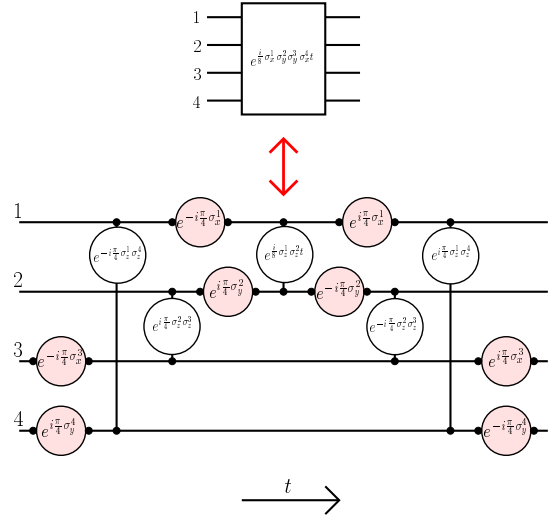}
\end{center}
\caption{Decomposition of the unitary operator $\tilde{U}(t)=e^{\frac{it}{8} \sigma_x^1
\sigma_y^2 \sigma_y^3 \sigma_x^4}$ into elementary gates as described in
Sec.~\ref{sec2-1-2}. The labeling convention is $(0,1)\equiv 1$, 
$(1,1)\equiv 2$, $(0,2)\equiv 3$, and $(1,2)\equiv 4$.}
\label{fig2-10}
\end{figure}

In general, $H=K+V$ (Eq.~\ref{bosonhamilt}), where $K$ is a kinetic term and $V$ a potential term. 
The kinetic term is a linear combination of terms like 
$b^{\dagger}_k b^{\;}_l + b^{\dagger}_l b^{\;}_k$. Therefore, a single-step
evolution operator $e^{i (b^{\dagger}_k b^{\;}_l + b^{\dagger}_l b^{\;}_k)
\Delta t}$ is mapped onto Pauli operators as
\begin{eqnarray}
\nonumber
\exp \left[ i \vartheta \sum\limits_{n,n'=0}^{N_P-1} 
\sqrt{(n+1)(n'+1)} \ [(\sigma_x^{n,k}\sigma_x^{n+1,k}
+\sigma_y^{n,k}\sigma_y^{n+1,k}) 
(\sigma_x^{n',l}\sigma_x^{n'+1,l} +\right. \\
\label{decomp3}
\left.
\sigma_y^{n',l}\sigma_y^{n'+1,l}) 
+ (\sigma_x^{n,k}\sigma_y^{n+1,k}
-\sigma_y^{n,k}\sigma_x^{n+1,k})(\sigma_x^{n',l}\sigma_y^{n'+1,l}
-\sigma_y^{n',l}\sigma_x^{n'+1,l})]  \right] ,
\end{eqnarray}
where $\vartheta=\Delta t/8$ and $N_P$ is the maximal number of bosons per
site.  The terms in the  exponent of Eq. \ref{decomp3} commute with each other,
so the decomposition into elementary gates can be done using the
methods described in Sec.~\ref{sec2-1-2}. As an example, 
consider a system of two sites with maximal one boson per site ($N_P=1$).  Thus,
$2(1+1)=4$ qubits are needed for the simulation, and  Eq. \ref{bosonmap2}
implies that
$\tilde{b}^\dagger_1 = \sigma_-^{0,1}\sigma_+^{1,1}$ and
$\tilde{b}^\dagger_2 = \sigma_-^{0,2}\sigma_+^{1,2}$. The mapped bosonic
operator $e^{i
(b^\dagger_1 b^{\;}_2  + b^\dagger_2 b^{\;}_1) \Delta t}$ in terms of Pauli
operators is
\begin{eqnarray}
\label{bosexamp}
\exp (i \vartheta \sigma_x^{0,1}\sigma_x^{1,1}\sigma_x^{0,2}\sigma_x^{1,2})
\times
\exp (i \vartheta \sigma_x^{0,1}\sigma_x^{1,1}\sigma_y^{0,2}\sigma_y^{1,2})
\times \ \ \ \ \ \ \ \ \ \ 
\\
\nonumber
\exp (i \vartheta \sigma_y^{0,1}\sigma_y^{1,1}\sigma_x^{0,2}\sigma_x^{1,2})
\times
\exp (i \vartheta \sigma_y^{0,1}\sigma_y^{1,1}\sigma_y^{0,2}\sigma_y^{1,2})
\times
\exp (i \vartheta \sigma_y^{0,1}\sigma_x^{1,1}\sigma_y^{0,2}\sigma_x^{1,2})
\times
\\
\nonumber
\exp (-i \vartheta \sigma_y^{0,1}\sigma_x^{1,1}\sigma_x^{0,2}\sigma_y^{1,2})
\times
\exp (-i \vartheta \sigma_x^{0,1}\sigma_y^{1,1}\sigma_y^{0,2}\sigma_x^{1,2})
\times
\exp (i \vartheta \sigma_x^{0,1}\sigma_y^{1,1}\sigma_x^{0,2}\sigma_y^{1,2}),
\end{eqnarray}
and the decomposition of each of the terms in Eq. \ref{bosexamp} in terms of
single and two-qubit elementary gates can be done, again, using the methods
described in Sec.~\ref{sec2-1-2}. An example of 
the decomposition of the term $\exp \left (\frac{it}{8}
\sigma_x^{0,1}\sigma_y^{1,1}\sigma_y^{0,2}\sigma_x^{1,2} t\right )$,
where the qubits were relabeled as $(n,j) \equiv n+2j -1$ (e.g.,
$(0,1)\rightarrow 1$) is shown in Fig.~\ref{fig2-10}.

Contrary to the fermionic case, the number of elementary operations involved in
the decomposition is not related to the distance between sites, $|k-l|$.
Nevertheless,  a physical bosonic operator $H$, such as the  Hamiltonian of
Eq. \ref{bosonhamilt}, involves a polynomially large number (with respect to
$N$) of bosonic terms.  Therefore, the corresponding evolution $\tilde{U}(t)=e^{-i
\tilde{H} t}$ 
can be
efficiently performed on a QC by applying a polynomially large number of
elementary single and two-qubit gates. Again, when using approximate methods
like the Trotter decomposition, the number of operations needed increases with
the desired accuracy. However, such an approximation does not destroy the
efficiency of the simulation.

\section{Applications: The 2D fermionic Hubbard model}
\label{sec2-4}
To clarify the methods described previously, in this section I present, as an
example, the QS of the finite two-dimensional fermionic Hubbard model by using a CC that
imitates a QC; that is, a {\em quantum simulator}. Since this is a classical
simulation,  the CC must keep track of an exponentially large number of quantum
states, associated with the Hilbert space of the quantum system. Nevertheless,
this simulation provides a good example for understanding the advantages and
results that can be obtained when using a real QC.

\begin{figure}
\begin{center}
\includegraphics*[width=7.0cm]{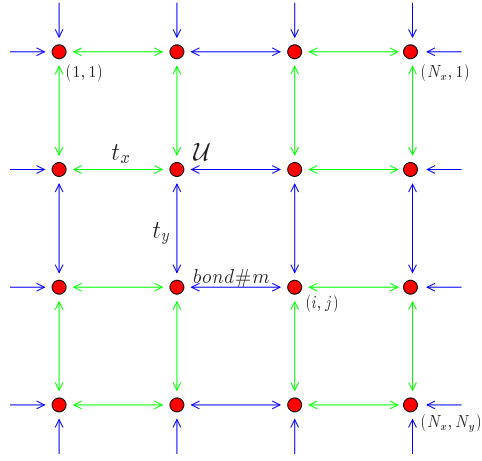}
\end{center}
\caption{Two-dimensional lattice in the Hubbard model.}
\label{fig2-11}
\end{figure}

The physical system to be simulated consists of a rectangular lattice
(Fig.~\ref{fig2-11}), with
$N_x \times N_y$ sites, where spin-1/2 fermions hop from site to site under the
interaction Hamiltonian
\begin{equation}
\label{hubbard}
H=-\sum\limits_{(i,j);\sigma}
[ t_x c^{\dagger}_{(i,j);\sigma}c_{(i+1,j);\sigma}^{\;}
  + t_y c^{\dagger}_{(i,j);\sigma}c_{(i,j+1);\sigma}^{\;}
    +H.C. ]
+{\cal U}\sum\limits_{(i,j)}n_{(i,j);\uparrow}n_{(i,j);\downarrow} \ ,
\end{equation}
where the operators $c^\dagger_{(i,j);\sigma}$ ($c_{(i,j);\sigma}$) create
(annihilate) a fermion with spin component denoted by $\sigma$ ($=\uparrow$ or
$\downarrow$), at the site located at $(x=i,
y=j)$. $t_x$ and $t_y$ are the hopping terms in the $x$ and $y$ directions,
respectively, and $n_{(i,j);\sigma} = c^{\dagger}_{(i,j);\sigma}
c_{(i,j);\sigma}^{\;}$ is the corresponding
number operator. ($H.C.$ denotes the Hermitian conjugate).
Periodic boundary conditions (PBC) are
assumed: $[(i,j);\sigma] \equiv=[(i+N_x,j);\sigma] \equiv
[(i,j+N_y);\sigma]$.

To use the QA for the obtention of the spectrum of $H$, described in
Sec.~\ref{sec2-2-2}, it is necessary first to map the fermionic operators onto
Pauli operators using, for example, the Jordan-Wigner transformation
(Sec.~\ref{sec2-3-1}). Considering that these are spin-1/2 fermions,  the QA
needs $2(N_x \times N_y)$ qubits to represent the system (qubits
system), plus the ancilla qubit.  

Assuming that one is mainly interested in obtaining the lowest energy of
Eq.~\ref{hubbard}, the prepared initial state should be the ground state of
Eq.~\ref{hubbard}. However, since no algebraic methods exist to  exactly
diagonalize  Eq.~\ref{hubbard} for large $N_x$ and $N_y$, such a state is not
known, and therefore impossible to prepare. Nevertheless, the ground state of
the associated mean-field Hamiltonian
\begin{eqnarray}
\nonumber
H_{MF}=-\sum\limits_{(i,j);\sigma}
[t_x c^{\dagger}_{(i,j);\sigma}c_{(i+1,j);\sigma}^{\;}
  + t_y c^{\dagger}_{(i,j);\sigma}c_{(i,j+1);\sigma}^{\;} +H.C.] + \\
{\cal U}\sum\limits_{(i,j)}
[\langle n_{(i,j);\uparrow}\rangle n_{(i,j);\downarrow}
 +n_{(i,j);\uparrow}\langle n_{(i,j);\downarrow}\rangle 
 -\langle n_{(i,j);\uparrow} \rangle \langle n_{(i,j);\downarrow}
 \rangle] \ ,
\end{eqnarray}
is known to be a fermionic product state (Slater determinant) $\ket{\phi_\beta}$,
and its corresponding state in the conventional model can be efficiently
prepared by using the methods described in Sec.~\ref{sec2-3-1}; that is, it can be
prepared by applying a polynomially large (with respect to $N_x \times N_y$) set
of elementary gates to the fully polarized state. For finite small lattices,
$\ket{\phi_\beta}$ is a good approximation to the ground state of
Eq.~\ref{hubbard} and can be used to obtain the ground state energy.

The second step of the QA is to apply the unitary operator $\tilde{U}(t)= e^{i
\tilde{H} \sigma_z^{\sf a} t/2}$ using single and two-qubit gates (see
Fig.~\ref{fig2-8} for $\hQ \equiv \tilde{H}$)),  where, in this case, 
$\tilde{H}$ is the Hamiltonian of Eq.~\ref{hubbard} in terms of Pauli
operators, $t$ is a real (fixed) parameter, and $ \sigma_z^{\sf a}$ is the
Pauli operator associated with the ancilla qubit.  Since $H$ is a linear
combination of non-commuting terms (Eq.~\ref{hubbard}), the operator
$\tilde{U}(t)=
\prod_j \tilde{U}(\Delta t)$ can be approximated by using, for example, the
first order Trotter decomposition~\cite{Suz93}. That is, $H = K_\uparrow +
K_\downarrow + V$, where $K_\sigma$ denotes the kinetic energy associated with
the fermions of spin $\sigma$ and $V$ denotes the potential energy. Then,
\begin{equation}
\tilde{U}(\Delta t) = e^{-i \tilde{H} \sigma_z^{\sf a} \Delta t/2} \sim
e^{-i \tilde{K}_\uparrow \sigma_z^{\sf a} \Delta t/2} \times e^{-i \tilde{K}_\downarrow 
\sigma_z^{\sf a} \Delta t/2} \times  e^{-i \tilde{V} \sigma_z^{\sf a} \Delta
t/2} ,
\end{equation}
where $\tilde{K}_\sigma$ and $\tilde{V}$ are the corresponding terms in Pauli
operators. Also, the same approximation can be used to decompose each term 
$e^{-i \tilde{K}_\sigma \sigma_z^{\sf a} \Delta t/2}$. Such approximation leads
to operators that can easily be decomposed in terms of elementary gates by
using the methods described in Sec.~\ref{sec2-1-2}.

\begin{figure}
\begin{center}
\includegraphics*[angle=270,width=15cm]{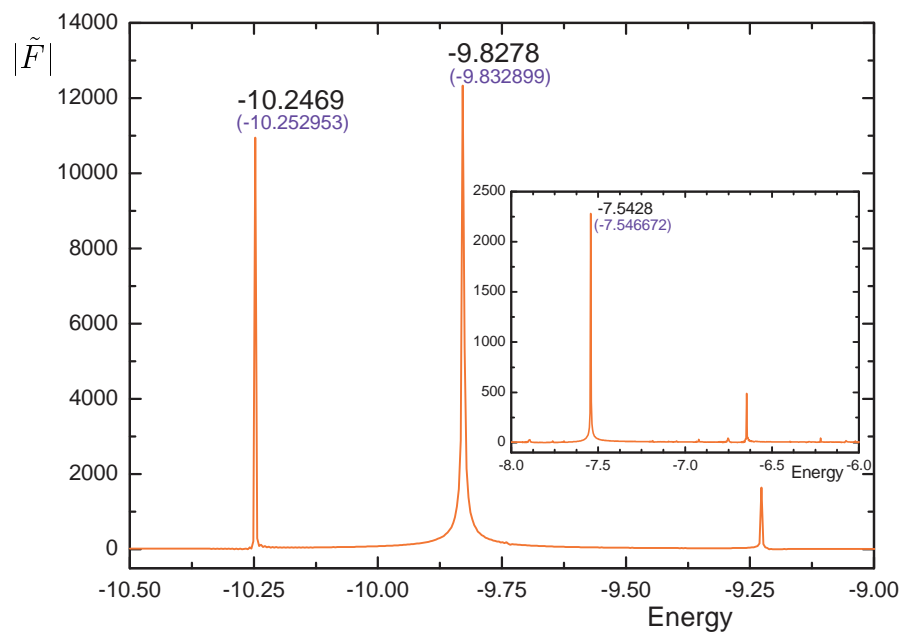}
\end{center}
\caption{Energy spectrum of the Hubbard model obtained simulating the QA of
Fig.~\ref{fig2-8} on a CC.
The lattice has $4\times 2$ sites (which requires 16+1 qubits). Here,
$t_x=t_y=1$, and ${\cal U}=4$. The time steps used in the Trotter approximation
to prepare the initial state and apply the evolution are $\Delta t_1= \Delta t_2
=0.05$, respectively. The numbers in brackets are the results obtained from the
exact diagonalization using the Lanczos method.}
\label{fig2-12}
\end{figure}

The energy spectrum of the Hubbard model for a $4 \times 2$ lattice is shown in
Fig.~\ref{fig2-12}. It has been obtained after running the classical simulation
for many different  values of $t$, and performing the Fourier transform on the
data. The peaks show the eigenvalues and the results are compared to those of
an exact diagonalization method.

The algorithm described allows one to easily obtain  the energy spectra of a finite
small lattice. Nevertheless, the spectra of a large lattice cannot be
efficiently obtained using the same methods. The problem is that the ground
state of the mean-field approximation differs more from the ground state of
Eq.~\ref{hubbard} as the system increases, and the QA needs to be
performed an exponentially large number of times (with respect to $N_x \times
N_y$) to obtain the desired property~\cite{SOG02}.

\section{Quantum Algorithms: Efficiency and Errors}
\label{sec2-5}
A QA for a physical simulation is considered efficient if the
total number of operations involved for the initial state preparation, the
evolution, and the measurement process, scales at most
polynomially with the system
size and with $1/\epsilon$,   where $\epsilon$ is the maximal tolerable error in the
measurement of  a relevant property. 

While the decomposition of the operator $U(t)=e^{-iHt}$ can be done efficiently
(e.g., when
using the Trotter  approximation) if $H$ is a physical operator, the
preparation of a general initial state could be inefficient. Such inefficiency
would arise, for example, if  the state $\ket{\psi}$ defined in 
Eq.~\ref{genferm} or Eq.~\ref{bosstate} is a linear combination of an
exponentially large number of elementary product states. 
In this case,  $L\sim
x^N$, with $N$ the number of modes in the system and $x>1$, so an exponentially
large number of trials need to be performed before successful preparation.
However,  if $L \le \mbox{ poly}(N)$,
it can be prepared efficiently. This
construction generalizes to more general coherent states (see
Chap.~\ref{chapter4}). 

The three main reasons for the existence of errors $\epsilon$ in the outcome of
the quantum computation are gate imperfections, the use of the Trotter
approximation in the evolution operator, and the statistics in measuring the
polarization of the ancilla qubit (Sec.~\ref{sec2-2-2}). Gate imperfections are
very common in quantum information because, contrary to classical information,
quantum gates are usually dominated by a continuous parameter. This problem 
can be solved by using quantum error correction methods and fault tolerant
quantum computation~\cite{Ste96,Kit97,NC00,Got97}. According to the accuracy
threshold theorem, provided that the physical gates have sufficiently low
error, it is possible to quantum compute accurately in an efficient way.

The type of error introduced by the discretization of the evolution operator
$U(t)$ (e.g., by using the Trotter decomposition or other approximations) is
very similar to the error obtained when performing a classical simulation, such
as when using Monte Carlo methods. This error can be estimated by a detailed
analysis of the discretization and can also be arbitrarily reduced in an
efficient way. 

Finally, when using the QAs described in Sec.~\ref{sec2-2-2}, the step
corresponding to the measurement process can also be performed efficiently
because it only involves the measurement of a single (ancilla) qubit,
regardless of the number of qubits needed for the simulation. Nevertheless,
repeatedly many same-simulations need to be performed to get an accurate value
of such measurement. This is an inherent property of the quantum mechanichs
where a single measurement projects the quantum state (Sec.~\ref{sec2-1-1}) and
does not give sufficient information. If the relevant signal at the end of the
quantum computation is small, like when obtaining the spectra of the
two-dimensional
Hubbard model on a large lattice (Sec.~\ref{sec2-4}), it is necessary to run
the algorithm a larger number of times and the efficiency can be destroyed. 

In brief, a QC can only be more efficient than a CC when simulating quantum
physical systems if the three main steps of the corresponding QA can be
performed efficiently. For example, the evaluation of certain correlation
functions over a quantum state that can be easily prepared, can be efficiently
done with a QC. In general (i.e., for non-integrable Hamiltonians), there is no
known way to evaluate such correlation efficiently with a
CC~\cite{SOG02,SOK03}.

\section{Experimental Implementations of Quantum Algorithms}
\label{sec2-6}
In this chapter, I have shown that if a large QC existed today, some
simulations of quantum systems could be performed more efficiently on it
than on a
CC. Nevertheless, I did not discuss how the corresponding QAs could be
experimentally implemented. Although numerous proposals for implementing
quantum information processors (QIPs) are found in the
literature~\cite{CZ95,CLK00,KLM01},  only few of them have been successfully
implemented to process more than one qubit. In particular, liquid-state NMR
devices allow one to simulate several systems by manipulating, nowadays, up to
ten qubits~\cite{RBC04}.

The physical implementation of a large scale QC still remains one of the
most important challenges for today's physicists. The problem is a QC should be
designed such that the interaction between its constituents and the environment
is small enough to keep coherence of the quantum state. But if such interaction
is too small, the manipulation and control processes using external sources
becomes impracticable. For this reason, quantum decoherence has been one of the
most important subjects of study during the last decade. In general,
decoherence phenomena is hard to predict due to the infinite degrees of
freedom associated with the environment. Nevertheless, a QC is reliable whenever
the time required to perform a certain task is much smaller than the corresponding
decoherence time.

In this section, I describe the experimental setting of a liquid-state NMR QIP
and show how such devices can be used to execute the QAs
described previously. Later on, I will show the experimental NMR
simulation of a particular fermionic system, where some correlation functions
and energy spectra have been obtained.

\subsection{Liquid-State NMR Quantum Information Processor}
\label{sec2-6-1}
Liquid-state NMR methods allow one to physically implement a slightly different
version of the conventional model of quantum computation, with respect to the
initial state preparation and the measurement process. In this set-up the
quantum register is represented by the average state of the nuclear spin-1/2 of
an ensemble of identical molecules. Each nuclear spin is a two-level physical
system and can then be considered a possible qubit. Thus, the idea is to
perform single and two-qubit elementary gates by external radio-frequency (rf)
pulses that interact with the nuclear spin state. In the following, I present a
basic analysis about how these processors can be used as possible QCs. 

\begin{figure}
\begin{center}
\includegraphics*[width=7cm]{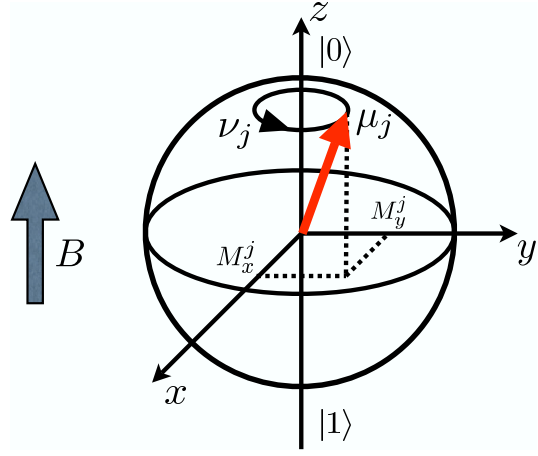}
\end{center}
\caption{Bloch's sphere representation of a single nuclear spin-1/2 state
precessing around the quantization axis determined by the external magnetic
field $B$. The precession frequency is given by $\nu_j = \mu_j B$, with $\mu_j$
the magnetic moment of the $j$th nucleus. Due to the chemical environment, each
nucleus precesses at a different Larmor frequency $\nu_j$.}
\label{fig2-13}
\end{figure}

In a liquid NMR setting, the molecules are placed in a strong magnetic field
$B(\hat{z}) \simeq 10$ T, so that the spin of the $j$-th nucleus of a
single molecule precesses at its Larmor frequency $\nu_j$
(Fig.~\ref{fig2-13}).  In the frame rotating with the $j$th spin, its qubit
state can then be rotated by sending rf pulses in the XY
plane at the resonant frequency $\nu_r \approx \nu_j$. If the duration of this
pulse is $\delta t$, the corresponding evolution operator in the rotating frame
is~\cite{LKC02} 
\begin{equation} 
U_j =  e^{-i H_j \delta t} = e^{-i {\sf A}
(\cos(\varphi) \sigma_x^j  + \sin(\varphi) \sigma_y^j) \delta t} ,
\end{equation}  
where ${\sf A}$ is the amplitude of the RF-pulse and $\varphi$ is its phase
(i.e., orientation) in the XY plane ($\hbar=1$). Then, one can induce single
spin rotations\footnote{One actually is restricted to 90 and 180 degrees
rotations for experimental calibration issues.} around any axis in that plane by
adjusting $\delta t$ and $\varphi$.

Single-qubit rotations around the $z$ axis can be implemented with no
experimental imperfection or physical duration simply by changing the
phase of the abstract rotating frame with which one is working. One has then
to keep track of all these phase changes with respect to a reference
phase associated with the spectrometer. Nevertheless, these phase
tracking calculations are linear with respect to the number of pulses
and spins, and can be efficiently done on a classical
computer. Together with the rotations around any axis in the XY plane, the
$z$ rotations can generate any single qubit rotation on the Bloch
sphere.

\begin{figure}
\begin{center}
\includegraphics*[width=9cm]{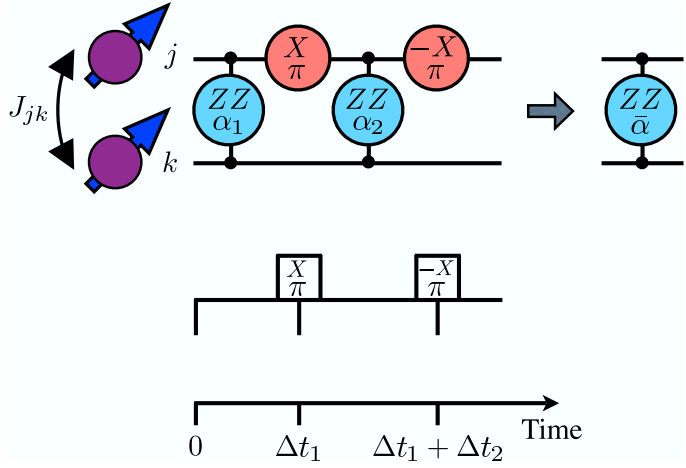}
\end{center}
\caption{Circuit representation for the refocusing scheme to control
$J$ couplings. The Ising-like coupling $J_{jk}$ between spins can be controlled
by performing flips on one of the spins at times $t_1 =\Delta t_1$ and $t_2 =
t_1 + \Delta t_2$, respectively. The effective coupling is
$\bar{\alpha}=\alpha_1 - \alpha_2 = J_{jk} (\Delta t_1 - \Delta t_2)$, and
vanishes when $\Delta t_1 = \Delta t_2$.}
\label{fig2-14}
\end{figure}

Two-qubit gates, like the Ising gate $R_{z^jz^k}(\omega)$
(Sec.~\ref{sec2-1-1}), can be performed by taking advantage of the spin-spin
interactions (i.e., nuclei interaction) present in the molecule, and then
achieve universal control. To first order in perturbation, this interaction,
named the $J$-coupling, has the form
\begin{equation}
\label{zzinteraction}
H_{j,k} = \frac{J_{jk}}{4} \sigma_z^j \sigma_z^k ,
\end{equation}
where $j,k$ denote the corresponding pair of qubits and $J_{jk}$ is
their coupling strength. Under typical NMR operating conditions, these
interaction terms are small enough to be neglected when performing
single-qubit rotations with rf pulses of short duration. Nevertheless,
between two pulses they are driving the evolution of the system. By
cleverly designing a pulse sequence, i.e., a succession of pulses and
free evolution periods, one can easily apply two-qubit gates on the
state of the system. Indeed, the so-called {\em refocusing techniques'
principle} consists of  performing an arbitrary Ising gate by flipping
one of the coupled spins ($\pi$-pulse), as shown in Fig~\ref{fig2-14}. 
The interaction evolutions before and after the
refocusing pulse compensate, leading to the effective evolution
\begin{equation}
U^{\sf eff}_{j,k} = e^{i \frac{\pi}{2} \sigma_x^j} e^{-i
\frac{J_{jk}}{4}\sigma_z^j \sigma_z^k \delta t_2} e^{-i \sigma_x^j
\pi/2} e^{-i\frac{J_{jk}}{4} \sigma_z^j \sigma_z^k \delta t_1} = e^{-i
\frac{\bar{\alpha}}{4} \sigma_z^j \sigma_z^k} ,
\end{equation}
where the effective coupling strength $\bar{\alpha}= J_{jk} (\delta
t_1-\delta t_2)$ is being determined by the difference between the
durations $\delta t_1$ and $\delta t_2$.

Although the physics on a single molecule has been analyzed,
liquid-state NMR uses an ensemble of about
$10^{23}$ molecules in a solution maintained at room temperature
($\simeq 300K$). For typical values of the magnetic field, this
thermal state is extremely mixed. Clearly, this is not the usual state
in which one initializes a quantum computation since qubits are nearly
randomly mixed.  Nevertheless, known NMR methods~\cite{LKC02} can be
used to prepare the so-called \emph{pseudo-pure state} ($\rho_{\sf
pp}$)\footnote{Even though efficient techniques to prepare a pseudo-pure state
exist in theory~\cite{SV98}, they are very hard to implement in practice, and one
instead uses non-efficient methods that suffer an exponential decay of the
observed signal with respect to the number of qubits in the pseudo-pure state.}
\begin{equation}
\label{pseudopure}
\rho_{{\sf pp}} = \frac{(1-\bar{\epsilon})}{2^N} \one + \bar{\epsilon}\rho_{{\sf
pure}} ,
\end{equation} 
where $\one$ is the identity operator,
$\rho_{{\sf pure}}$ is a density operator that describes a pure
state, and $\bar{\epsilon}$ is a small real constant (i.e., $\bar{\epsilon}$ decays
exponentially with the number of atoms in the solution due to the Boltzmann's
distribution).

Under the action of any unitary evolution $U$, this state evolves
as
\begin{equation}
\label{ppstate}
\rho_{\sf pp}^{\sf final} = U \rho_{\sf pp} U^\dagger =
\frac{(1-\bar{\epsilon})}{2^N}\one + U \bar{\epsilon}\rho_{{\sf pure}} U^\dagger .
\end{equation}
The first term in Eq. \ref{ppstate} did not change because the
identity operator is invariant under any unitary
transformation. Therefore, performing quantum computation on the
ensemble is equivalent to performing quantum computation over the initial
state represented only by $\rho_{{\sf pure}}$.  

At the end of the computation, the orthogonal components of the sample
polarization in the XY plane, $M_x = {\sf Tr} (\rho_{\sf pp}^{\sf final}
\sum_{i=1}^N \sigma_x^i)$, and $M_y = {\sf Tr}(\rho_{\sf pp}^{\sf final}
\sum_{i=1}^N \sigma_y^i)$ are measured (Eq.~\ref{mixedmeasur}). 
Note that the invariant component of
$\rho_{\sf pp}^{\sf final}$ does not contribute to the signal since ${\sf
Tr}(\one\sigma_{x,y}^j)=0$. Because the polarization of each single spin,
$M_x^j= {\sf Tr} (\rho_{\sf pp}^{\sf final} \sigma_x^j)$ and $M_y^j= {\sf Tr}
(\rho_{\sf pp}^{\sf final} \sigma_y^j)$, precesses at its own Larmor frequency
$\nu_j$, a Fourier transformation of the temporal recording (called FID, for
Free Induction Decay) of the total magnetization needs to be performed. By
doing so, one obtains the expectation value of the polarization of each spin
(averaged over all molecules in the sample).

Summarizing, a liquid-state NMR setting allows one to initialize a
register of qubits in a pseudo-pure state, apply any unitary
transformation to this state by sending controlled rf pulses or by leaving
free
interaction periods, and measure the expectation value of some quantum
observables (i.e., the spin polarization). Hence, these systems can be
used as QIPs.

\section{Applications: The Fano-Anderson Model}
\label{sec2-7}
I now present the experimental QS of the fermionic one-dimensional (1D)
Fano-Anderson model using a liquid-state NMR~\cite{NSO05}, by manipulating the state of the
spin nuclei as described in Sec.~\ref{sec2-6-1}. Such simulation shows then
how reliable these experimental methods are and how well the elementary gates
(Sec.~\ref{sec2-1-1}) can be implemented using NMR techniques.

\begin{figure}
\begin{center}
\includegraphics*[width=6cm]{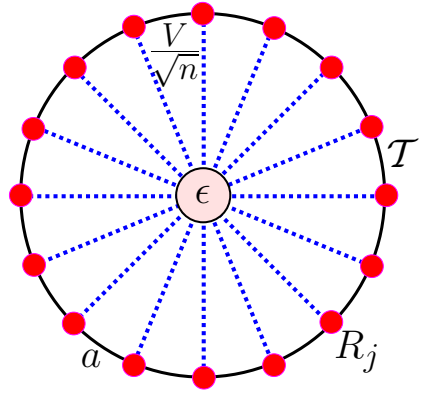}
\end{center}
\caption{Fermionic Fano-Anderson model. Fermions can hop between nearest
neighbor sites (exterior circles) and between a site and the impurity (centered
circle), with hopping matrix elements $\tau$ and $V/{\sqrt{n}}$, respectively.
The energy of localization in the impurity is $\epsilon$.}
\label{fig2-15}
\end{figure}

The 1D fermionic Fano-Anderson model consists of an
$n$-sites  ring with an impurity in the center (Fig.~\ref{fig2-15}), 
where spinless fermions can hop between nearest-neighbors sites with
hopping matrix element (overlap integral) $\tau$, or between a site and
the impurity with matrix element $V/\sqrt{n}$. Taking the
single-particle energy of a fermion in the impurity to be $\epsilon$,
and considering the translational invariance of the system, the
Fano-Anderson Hamiltonian can be written in the wave vector
representation as~\cite{OGK01}
\begin{equation}
\label{Hamilt2}
H=\sum_{l=0}^{n-1} \varepsilon_{k_l} c^{\dagger}_{k_l}c_{k_l}^{\;}+
\epsilon b^{\dagger}b + V(c^{\dagger}_{k_0}b+b^{\dagger}c_{k_0}^{\;}),
\end{equation}
where the fermionic operators $c^{\dagger}_{k_l}$ and $b^\dagger$
($c_{k_l}^{\;}$ and $b$) create (destroy) a spinless fermion in the
conduction mode $k_l$ and in the impurity, respectively.  Here, the
wave vectors (modes) are $k_l=\frac{2\pi l}{n}$ ($l=[0,..,n-1]$) and the
energies per mode are $\varepsilon_{k_l} = -2 \tau \cos k_l$.

In this form, the Hamiltonian in Eq.~\ref{Hamilt2} is almost diagonal
and can be exactly solved: There are no interactions between fermions
in different modes $k_l$, except for the mode $k_0$, which interacts
with the impurity. Therefore, the relevant physics comes from this
latter interaction, and its spectrum can be exactly obtained by
diagonalizing a $2 \times 2$ Hermitian matrix, regardless of $n$ and
the number of fermions in the ring, $N_e$. Nevertheless, its simulation
in a liquid-state NMR QIP is the first step in QSs of
quantum many-body problems and constitutes a proof of the principles described
throughout this thesis.

\begin{figure}
\begin{center}
\includegraphics*[width=7.4cm]{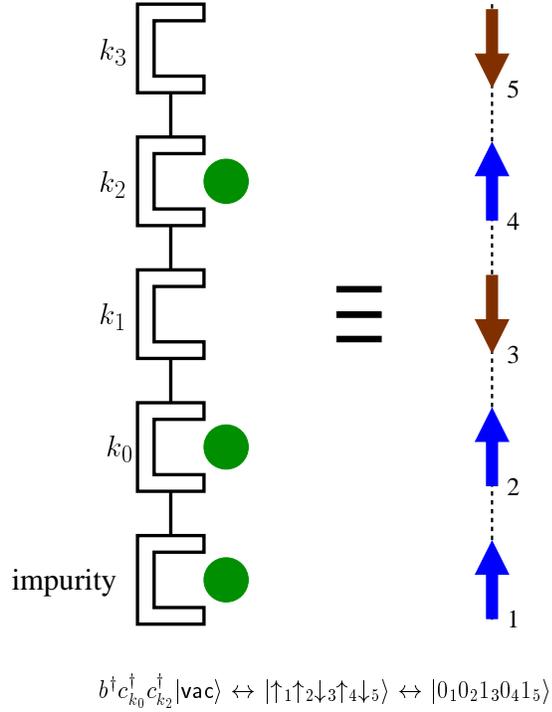}
\end{center}
\caption{Mapping of the fermionic product state $b^\dagger c^\dagger_{k_0}
c^\dagger_{k_2} \ket{\sf vac}$ into a five-qubit state, using the Jordan-Wigner
transformation. The convention is $\ket{\uparrow_j} \equiv \ket{0_j}$ (filled)
and
$\ket{\downarrow_j} \equiv \ket{1_j}$ (empty).}
\label{fig2-16}
\end{figure}

In order to 
successfully simulate this system in a liquid-state NMR QIP, 
the fermionic operators need to be mapped onto the Pauli operators
(Sec.~\ref{sec2-3-1}).
This
is done by using the following Jordan-Wigner
transformation
\begin{equation}
\label{jwmap2}
\matrix {b = \sigma_-^1 & b^\dagger = \sigma_+^1 \cr c^{\;}_{k_0} = -
\sigma_z^1 \sigma_-^2 & c^{\dagger}_{k_0} = - \sigma_z^1 \sigma_+^2
\cr \vdots & \vdots \cr 
c^{\;}_{k_{n-1}} = \left ( \prod_{j=1}^n -\sigma_z^j \right )
\sigma_-^{n+1}  & \qquad c^{\dagger}_{k_{n-1}} = \left ( \prod_{j=1}^n
-\sigma_z^j \right )\sigma_+^{n+1} . }
\end{equation}
In this language, a logical state $\ket{0_j}$ (with $\ket{0}\equiv
\ket{\uparrow}$ in the usual spin-1/2 notation) corresponds to having a
spinless fermion in either the impurity, if $j=1$, or in the mode
$k_{j-2}$, otherwise (Fig.~\ref{fig2-16}). (Again, the fermionic vacuum
state $\ket{{\sf vac}}$ maps onto $\ket{{\sf \widehat{vac}}} = \ket{1_1 1_2
\cdots 1_{n+1}}$.)

The algorithms described in Sec.~\ref{sec2-2-2} can be used, for example, 
to evaluate 
the probability amplitude of
having a fermion in mode $k_0$ at time $t$, if initially ($t=0$)
the quantum state is the Fermi sea state with $N_e$ fermions; that is,
$\ket{\sf FS}=\prod\limits_{l=0}^{N_e-1} c^\dagger_{k_l} \ket{{\sf
vac}}$. This probability is given by the modulus square of the
following dynamical correlation function: 
\begin{equation}
\label{correlation}
G(t)= \bra{\sf FS} b(t) b^\dagger(0) \ket{\sf FS} \ ,
\end{equation}
where $b(t) = T^\dagger b(0) T$, $T=e^{-i Ht}$ is the time evolution
operator, and $b^\dagger(0)=b^\dagger$. Basically, $G(t)$ is the
overlap between the quantum state $ b^\dagger(0) \ket{\sf FS}$, which
does not evolve, and the state $b^\dagger(t) \ket{\sf FS} $, which does
not vanish unless the evolved state $T\ket{\sf FS}$ already contains a
fermion in the impurity site ($(b^\dagger(t))^2 =(b^\dagger(0))^2= 0$),
i.e., contains the fermion which initially was in the $k_0$ mode.
In terms of Pauli operators (see Eq. \ref{jwmap2}), this correlation
function reduces to a two-qubit problem~\cite{OGK01}:
\begin{equation}
\label{correl2}
G(t)= \bra{\phi} \bar{T}^\dagger \sigma_-^1 \bar{T} \sigma_+^1
\ket{\phi} \ ,
\end{equation}
where $\bar{T} = e^{-i \bar{H} t}$ is an evolution operator arising
from the interaction terms in Eq. \ref{Hamilt2}, with
\begin{equation}
\label{Hamilt4}
\bar{H} = \frac{\epsilon}{2} \sigma_z^1 + \frac{\varepsilon_{k_0}}{2}
\sigma_z^2 + \frac{V}{2} (\sigma_x^1 \sigma_x^2 + \sigma_y^1
\sigma_y^2) \ ,
\end{equation}
and $\ket{\phi} = \ket{1_1 0_2}$ in the logical basis (i.e., the
initial state with one fermion in the $k_0$ mode).

\begin{figure}
\begin{center}
\includegraphics*[width=14cm]{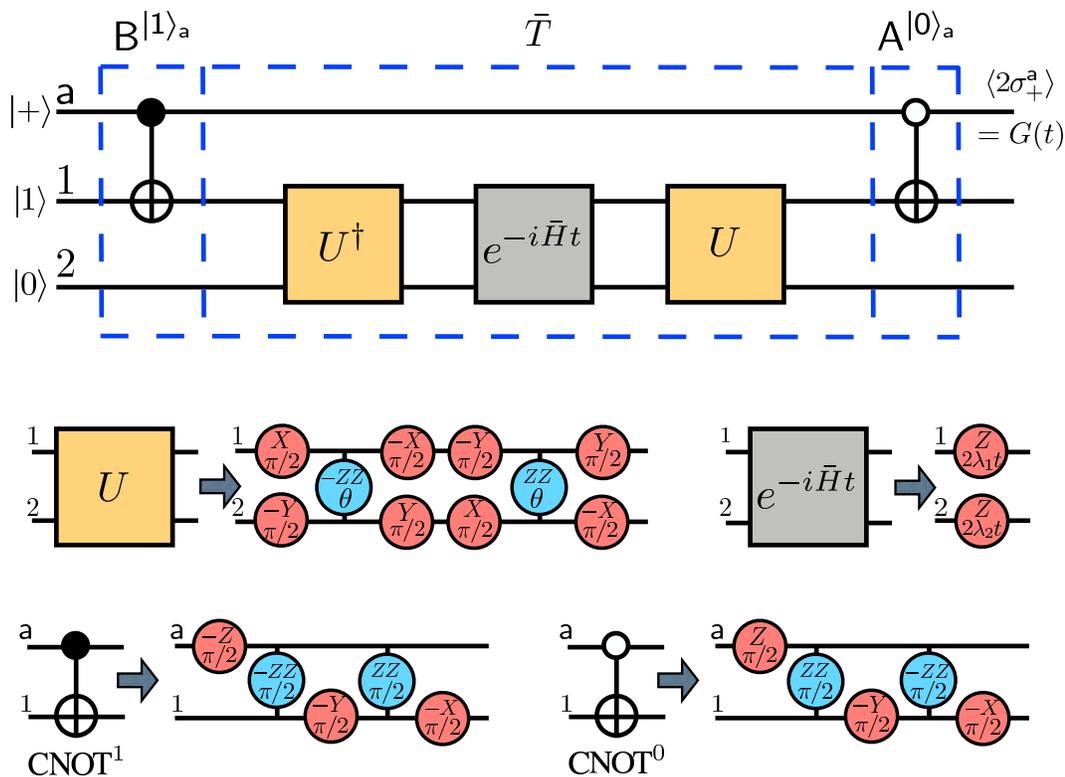}
\end{center}
\caption{Quantum circuit for the evaluation of $G(t)$ (Eq.~\ref{correlation}) in
terms of elementary gates directly able to be implemented
with liquid-state NMR methods.
The controlled operations ${\sf B}^{\ket{1}_{\sf a}}$ and
${\sf A}^{\ket{0}_{\sf a}}$ correspond to the operations C-B and C-A of
Sec.~\ref{sec2-2-2}, respectively.
 }
\label{fig2-17}
\end{figure}

In order to use the quantum circuit depicted in Fig.~\ref{fig2-7}, all 
operators in Eq. \ref{correl2} must be unitary. Because of the symmetries of
$\bar{H}$, such as the global $\pi/2$-$z$ rotation that maps $(\sigma_x^j ,
\sigma_y^j ) \rightarrow (\sigma_y^j , -\sigma_x^j )$, leaving the state
$\ket{\phi}$ invariant (up to a phase factor), then
$\langle \phi | \bar{T}^\dagger \sigma_x^1 \bar{T} \sigma_y^1 \ket{\phi}=
\langle \phi | \bar{T}^\dagger \sigma_y^1 \bar{T} \sigma_x^1 \ket{\phi}=0$
and 
$\langle \phi | \bar{T}^\dagger \sigma_x^1 \bar{T} \sigma_x^1 \ket{\phi}=
\langle \phi | \bar{T}^\dagger \sigma_y^1 \bar{T} \sigma_y^1 \ket{\phi}$.
Therefore, Eq. \ref{correl2} can be written in terms of unitary operators as
\begin{equation}
\label{correl3}
G(t) = \bra{\phi}  e^{i \bar{H} t} \sigma_x^1 
e^{-i \bar{H} t}  \sigma_x^1 \ket{\phi}.
\end{equation}
Figure~\ref{fig2-17} shows the quantum circuit used to obtain $G(t)$.  It
is derived from Fig.~\ref{fig2-7} by making the following identifications:
$T \rightarrow e^{-i \bar{H} t}$, $ A_i \rightarrow \sigma_x^1$, and $
B_j \rightarrow \sigma_x^1$. The corresponding
controlled operations C-A and C-B
transform into the well-known controlled-not (CNOT) gates
(Sec.~\ref{sec2-1-3}).  All
the unitary operations appearing in Fig.~\ref{fig2-17} were decomposed into
elementary NMR gates (single qubit rotations and Ising interactions).
In particular, the decomposition of $e^{-i \bar{H} t}$ can be found in
Ref.~\cite{OGK01}, obtaining
\begin{equation}
\label{decomp}
e^{-i \bar{H} t} = U e^{-i \lambda_1 \sigma_z^1 t} e^{-i \lambda_2
\sigma_z^2 t} U^\dagger \ ,
\end{equation}
where $\lambda_{1(2)} = \frac{1}{2} (E \mp \sqrt{\Delta^2 +V^2})$, with
$E= \frac{\epsilon + \varepsilon_{k_0}}{2}$, and  $\Delta=
\frac{\epsilon - \varepsilon_{k_0}}{2}$. The unitary operator $U$ is
decomposed as (Fig.~\ref{fig2-17})
\begin{equation}
\label{Adecomp}
U = e^{i \frac{\pi}{4} \sigma^2_x} e^{-i \frac{\pi}{4} \sigma^1_y}
e^{-i \frac{\theta}{2} \sigma^1_z \sigma^2_z} e^{i
\frac{\pi}{4}\sigma^1_y} e^{i \frac{\pi}{4} \sigma^1_x} e^{-i
\frac{\pi}{4}\sigma^2_x} e^{-i \frac{\pi}{4} \sigma^2_y} e^{i
\frac{\theta}{2}\sigma^1_z \sigma^2_z} e^{-i \frac{\pi}{4} \sigma^1_x}
e^{i \frac{\pi}{4} \sigma^2_y} ,
\end{equation}
with the parameter $\theta$ satisfying $\cos \theta =
1/\sqrt{1+\delta^2}$, and $\delta=(\Delta+\sqrt{\Delta^2+V^2})/V$. 

The CNOT gates C-A and C-B
can also be decomposed into elementary gates, as explained in
Sec.~\ref{sec2-1-2}.  
Therefore, $G(t)$ can be obtained using an NMR QIP by
applying the appropriate rf pulses (Sec.~\ref{sec2-6-1}).  Remarkably, only
three qubits are required for the simulation (Fig.~\ref{fig2-17}): The
ancilla qubit ${\sf a}$, one qubit representing the impurity site
(qubit-1), and one qubit representing the $k_0$ mode (qubit-2).

Similarly, the algorithm depicted in Fig.~\ref{fig2-8} can be used if interested
in obtaining the spectrum of the Hamiltonian $H$ of Eq.~\ref{Hamilt2},
replacing $\hat{Q} \rightarrow H$. In particular, when
$n=1$ (one site plus the impurity), Eq.~\ref{Hamilt2} in terms of Pauli
operators reduces to $\tilde{H}=
\frac{\epsilon + \varepsilon_{k_0}}{2} +\bar{H}$, with $\bar{H}$
defined in Eq.~\ref{Hamilt4}. In this case,
the two eigenvalues $\lambda_i$ ($i=1,2$) of the one-particle subspace
can be extracted from the correlation function
\begin{equation}
\label{fanospectrum}
S(t)= \bra{\phi}  e^{-i \tilde{H} t} \ket{\phi} =  e^{-i (\epsilon +
\varepsilon_{k_0}) t} \bra{\phi} e^{-i \bar{H} t} \ket{\phi} ,
\end{equation}
which can be obtained by measuring the polarization of the ancilla qubit after
the quantum circuit shown in Fig.~\ref{fig2-8} has been applied. Since
$\ket{\phi} =\ket{1_1 0_2}$ is not an eigenstate of $H$, it has a non-zero
overlap with the two one-particle eigenstates, called $\ket{{\sf 1P}_i}$ (see
Appendix~\ref{appB}).

\begin{figure}
\begin{center}
\includegraphics*[width=10cm]{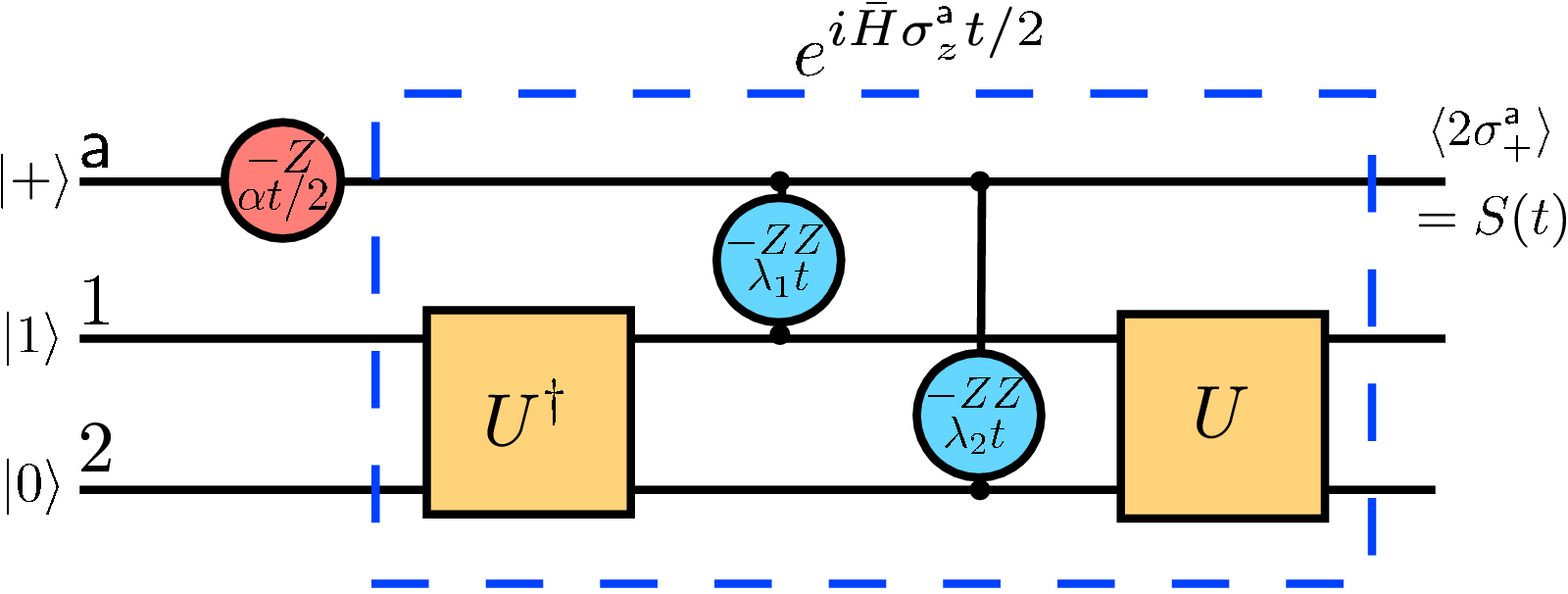}
\end{center}
\caption{Quantum circuit for the evaluation of $S(t)$ [Eq.~\ref{fanospectrum}].
The parameters $\lambda_1$ and $\lambda_2$ are defined in Sec.~\ref{sec2-7}, and
$\alpha=(\epsilon + \varepsilon_{k_0})/2$. The decomposition of the operator $U$
in NMR gates is shown in Fig.~\ref{fig2-17}.}
\label{fig2-18}
\end{figure}

Again, the operator $e^{i \tilde{H} \sigma_z^{\sf a} t/2}$ (Fig.~\ref{fig2-8})
needs to be decomposed into elementary gates for its implementation in
an NMR QIP.  Noticing that $[ \sigma_z^{\sf a} , \tilde{H} ]= [ \sigma_z^{\sf
a} , U ]=0$, then
\begin{equation}
e^{i \tilde{H} \sigma_z^{\sf a} t/2} = U e^{i \lambda_1 \sigma_z^1
\sigma_z^{\sf a} t/2} e^{i \lambda_2 \sigma_z^2 \sigma_z^{\sf a} t/2}
U^\dagger e^{i (\epsilon + \varepsilon_{k_0})
\sigma_z^{\sf a} t/2},
\end{equation}
where the unitary operator $U$ is decomposed as in Eq. \ref{Adecomp}.
Figure~\ref{fig2-18} shows the corresponding circuit in terms of elementary
gates. Again, qubits 1 and 2
represent the impurity site and the $k_0$ mode, respectively.
${\sf a}$ denotes the ancilla qubit.  Since the
idea is to perform a DFT on the results obtained from the measurement
(see Appendix~\ref{appB}), this circuit needs to be applied for several
values of the parameter $t$ (Sec.~\ref{sec2-1-3}).

\subsection{Experimental Protocol and Results}
\label{sec2-7-1}
\begin{figure}
\begin{center}
\includegraphics*[width=11cm]{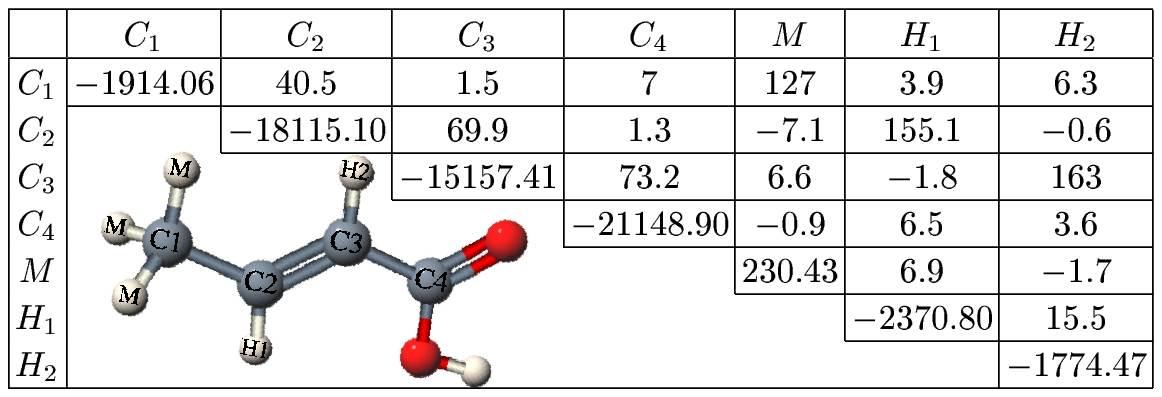}
\end{center}
\caption{The transcrotonic acid molecule is a seven-qubit register. The methyl
group is used as a single qubit~\cite{KLM00}. The table shows the values of the
chemical shifts (main diagonal) and the $J$ couplings (off-diagonal) between every
pair of nuclei or qubits, in hertz.}
\label{fig2-19}
\end{figure}

The NMR setting used for the evaluation of $G(t)$ and $S(t)$ is  based on an
ensemble
solution of trans-crotonic acid and methanol dissolved in acetone
(Fig.~\ref{fig2-19}). This
molecule can be used as a seven-qubit register, where methanol is used to
perform rf-power selection and accurately calibrate the rf pulses. In this way,
the decoherence time ($T_2^*$) is in the range from several hundreds
of milliseconds to more than a second, allowing one to perform around 1000
single-qubit gates and around 100 two-qubit (Ising) gates.

The relevant nuclei of the molecule are denoted as $\mbox{C}_1$,  $\mbox{C}_2$,
$\mbox{C}_3$, and $\mbox{C}_4$, corresponding to the Carbon atoms, $\mbox{H}_1$
and $\mbox{H}_2$ corresponding to the Hydrogen atoms, and M corresponding to
the methyl group. Although this is a seven-qubit register, the simulation of
the Fano-Anderson model requires only three qubits. Considering different
practical issues, such as the nuclei-nuclei interaction,  the selection has
been made as follows: The spin nucleus $\mbox{C}_1$ represents qubit-1, the
spin nucleus $\mbox{M}$ represents qubit-2, and the spin nucleus $\mbox{C}_2$
represents the ancilla qubit ${\sf a}$.

The idea is then to apply the desired elementary gates by sending appropriate rf
pulses. Nevertheless, designing a pulse sequence to implement exactly the
desired unitary transformation would require very long refocusing schemes to
cancel out all the unwanted naturally occurring $J$ couplings (free
evolutions).  Then, the overall duration of the pulse sequence increases and
decoherence effects could destroy the signal. Therefore, a pulse sequence
compiler is used to approximate the evolution and numerically optimize the
delays between pulses to minimize the error introduced by such approximation.

The approximate evolution is then applied to the corresponding initial state
$\rho_{\sf init}=\frac{1}{2}\left[ (\one^{\mbox{C}_2} + \sigma_x^
{\mbox{C}_2})\mbox{\bf 1}^{\mbox{C}_1} \mbox{\bf 0}^{\mbox{M}} \right]$, which
is the density operator corresponding to the pure state $\ket{+}_{\sf a}
\ket{1_1 0_2}$. Since the identity part $\one^{\mbox{C}_2}$ is not an
observable, the pseudo-pure state $\rho'_{\sf init}= \sigma_x^
{\mbox{C}_2}\mbox{\bf 1}^{\mbox{C}_1} \mbox{\bf 0}^{\mbox{M}}$, with $\mbox{\bf
1} \equiv\ket{1}\bra{1}$ and $\mbox{\bf 0} \equiv\ket{0}\bra{0}$, can be used
equivalently. Then $\rho'_{\sf init}$, which is a deviation of
the completely mixed state due to the high temperature of the ensemble
(Sec.~\ref{sec2-6-1}),
can easily be prepared
using already developed NMR techniques~\cite{LKC02}.

As mentioned in Sec.~\ref{sec2-2-2}, the desired result of the QA is encoded in
the polarization $\langle 2 \sigma_+^{\mbox{C}_2}\rangle$ of the nucleus
$\mbox{C}_2$ (i.e., ancilla qubit). This component precesses at the
$\mbox{C}_2$-Larmor frequency $\nu_{\mbox{C}_2}$. To measure it, a Fourier
transformation on the measured {\em free induction decay} FID 
(i.e., the decay in the polarization due to the contribution of different Larmor
frequencies)
must be performed, integrating only the peak
located at $\nu_{\mbox{C}_2}$. Nevertheless, the absolute value of this signal
is irrelevant since it depends on
many experimental parameters such as the solution concentration, the
probe sensitivity, and the gain of the amplifier. The relevant quantity
is its intensity relative to a reference signal given by the
observation of the initial state $\rho_{\sf init}$. To get a good
signal-to-noise ratio, each experiment (or {\it scan}) was done several
times and the corresponding experimental data were added.
Moreover, to average over small magnetic fluctuations occurring within
the duration of the whole experiment the scans of the
reference experiment (i.e., the measurement of the reference signal)
are interlaced
with scans of the actual complete pulse sequence. To increase the
spatial homogeneity of the field over the sample, 
several automated shimming periods, consisting of fine tuning small
additional coils located around the sample, have been inserted. 

\begin{figure}
\begin{center}
\includegraphics*[width=12cm]{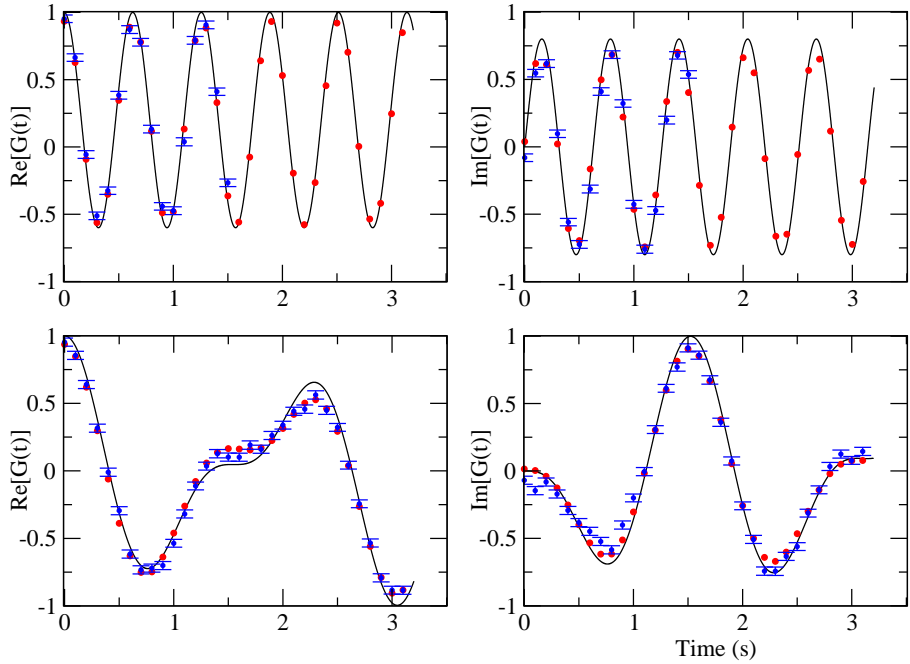}
\end{center}
\caption[Real part of $G(t)$.]{Real and imaginary parts of the
correlation function $G(t)$ of Eq.~\ref{correlation}.  The top panels
show the results when the parameters in Eq.~\ref{Hamilt2} are
$\varepsilon_{k_0} = -2, \epsilon=-8 ,V=4$. The corresponding
parameters $\lambda_1, \lambda_2, \theta$ 
can be determined using
Eqs. \ref{decomp} and \ref{Adecomp}.  The bottom panels show the
results for $\varepsilon_{k_0} = -2, \epsilon=0 , V=4$.  The (black)
solid line is the analytic solution, the red circles are obtained by
the numerical simulation (including the refocusing pulses), and the
blue circles with the error bars are experimental data.}
\label{fig2-20}
\end{figure}

In Fig.~\ref{fig2-20}, I show the experimental results obtained for the evaluation of $G(t)$
for $\varepsilon_{k_0}=-2 \mbox{, } \epsilon=-8, \ V=4, $ and
$\varepsilon_{k_0}=-2\mbox{, } \epsilon=0\mbox{, } V=4$, and for different
values of $t$.  The duration of the optimized pulse sequences from the
beginning of the initialization step to the beginning of the data acquisition,
was 97 ms. For comparison, the analytical form of $G(t)$, as well as the
simulated data points (i.e., data points obtained by simulating the quantum
algorithm on a conventional computer) are also shown in the figure.

\begin{figure}
\begin{center}
\includegraphics*[width=11cm]{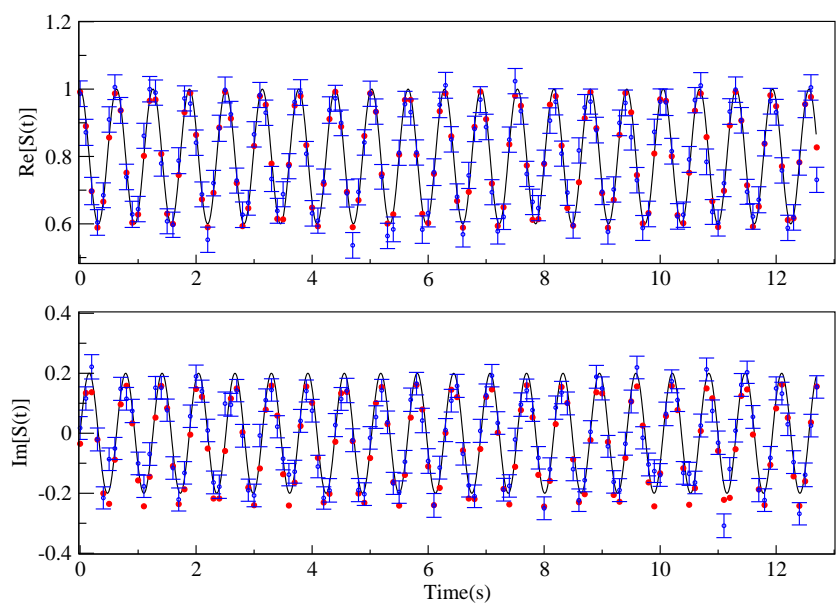}
\end{center}
\caption{Real and imaginary parts of $S(t)$, for 
$\varepsilon_{k_0}=-2, \ \epsilon=-8$, and $V=0.5$ in Eq.
\ref{Hamilt2}. The (black) solid line corresponds to the analytic
solution. The red circles correspond to the numerical simulation (using
refocusing pulses) and the blue circles with the error bars are
experimental data. $S(t)$ has been measured using the network of Fig.
\ref{fig2-18} with $\alpha = (\epsilon +\varepsilon_{k_0})/2$. }
\label{fig2-21}
\end{figure}

\begin{figure}
\begin{center}
\includegraphics*[width=13.cm]{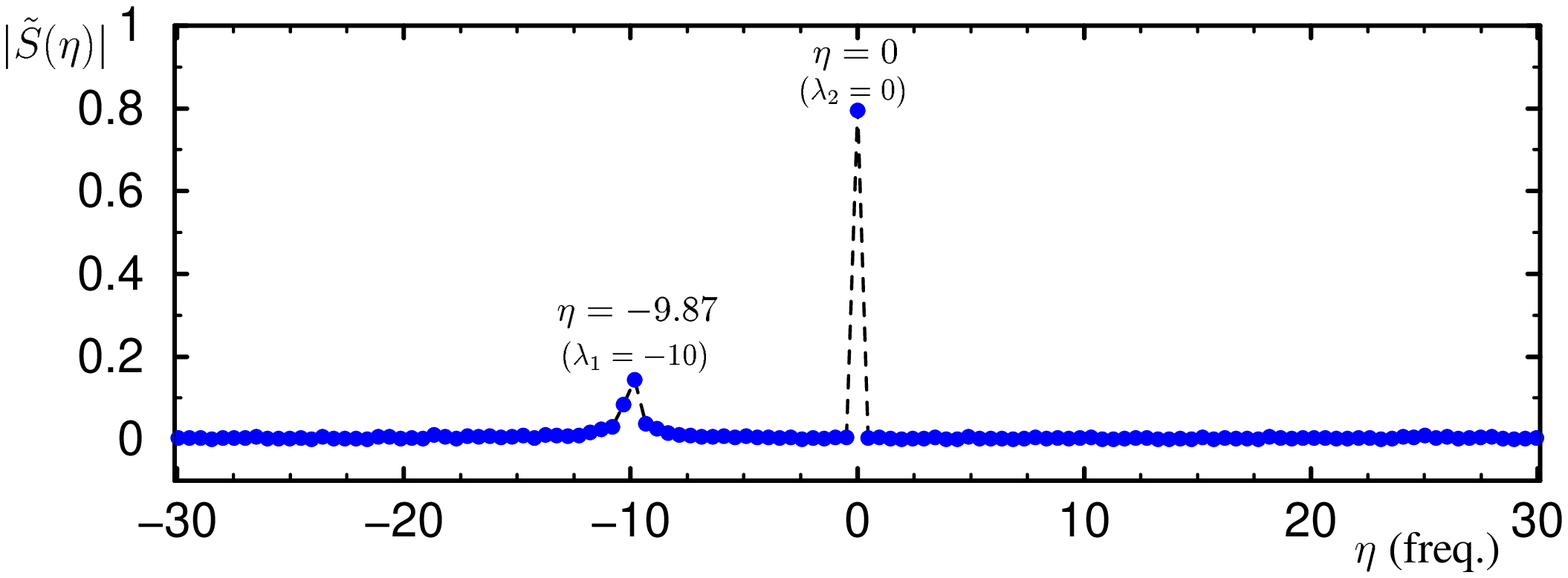}
\end{center}
\caption{Discrete Fourier transform of the real part of the
experimental data of Fig.~\ref{fig2-21}.  The position of the two
peaks corresponds to the two eigenvalues of the Hamiltonian of Eq.
\ref{Hamilt2} for $\varepsilon_{k_0}=-2, \epsilon=-8$, and $V=0.5$.
Numbers in parentheses denote the exact solution. The size of the dots
representing experimental points is the error bar (see Appendix~\ref{appB}).
An upper bound to the error in the frequency domain is
$\approx 0.5$, which was determined by the resolution of the spectrum
due to the time sampling of the simulation.}
\label{fig2-22}
\end{figure}

In Fig.~\ref{fig2-21}, I present the experimental results obtained for the
evaluation of $S(t)$, to obtain the corresponding eigenvalues for the
Hamiltonian, with  $\varepsilon_{k_0}=-2, \ \epsilon=-8$, and $V=0.5$. The pulse
sequence applied is the one corresponding to the quantum circuit shown in
Fig.~\ref{fig2-18} with the corresponding refocusing pulses.  In
Fig.~\ref{fig2-21}, I also show the analytical and simulated data
points. The DFT of the experimental data is shown in Fig.~\ref{fig2-22}, 
revealing the expected peaks at the frequency corresponding to the two
one-particle eigenvalues of Eq.~\ref{Hamilt2}, for the above parameters.

The close agreement between the experimental results and the corresponding
simulations, using the refocusing pulses, suggests that the main contribution
to errors comes from the incomplete refocusing scheme used in the optimization
procedure. Therefore, increasing the number of refocusing pulses might have led
to more accurate results even if they would have increased the overall duration
of the pulse sequence.

\section{Summary}
\label{sec2-8}
Throughout this chapter, I have addressed several broad issues associated with the
simulation of physical phenomena with QCs. In particular, I
presented efficient ways to map the algebras of operators associated to the
physical system to be simulated onto the algebra of Pauli operators. These
mappings are sufficient to establish the equivalence of the different physical
models to a universal model of quantum computation: The conventional model. 

I also explored various issues associated with simulations (Sec.~\ref{sec2-3}),
remarking that every step in a QA must be performed efficiently to have an
efficient simulation. Although a QC can simulate some quantum physical systems
more efficiently than its classical analogue, I showed that many challenges still
remain to prove this statement in a general way (Sec.~\ref{sec2-5}). As an
example, I presented the QS of the two-dimensional fermionic Hubbard model,
where there is no known way to obtain its ground state energy efficiently
using a QC.

Finally, I described the experimental implementation of a QA using a
liquid-state NMR QIP (Sec.~\ref{sec2-7}). This experiment allows one to
understand  the advantages and disadvantages of simulating physical systems
with today's QCs, and, in particular, to understand the power of quantum
computation.

\chapter{Quantum Entanglement as an Observer-Dependent Concept}
\label{chapter3}
\begin{quote}
{\em ``...Maximal knowledge of a total (quantum) system does not necessarily
include total knowledge of all its parts, not even when these are fully
separated from each other and at the moment are not influencing each other at
all.''

E. Schr\"{o}dinger (1935).}
\end{quote}

Quantum Entanglement (QE) is referred to the existence of certain correlations in a
quantum system that have no classical interpretation.  This concept was first
introduced by E. Schr\"{o}dinger~\cite{Sch35} as the essence of quantum mechanics, and is
responsible for many counterintuitive physical processes like the violation of
the local realism. Naturally, it has been the main focus of philosophical
discussions since the early days of quantum mechanics,  and it is now
known that entanglement is the defining resource that allows one to perform certain
protocols like quantum cryptography, quantum teleportation, and even more
efficient computation. For this reason, QE has been one of
the most important subjects of study in QIT during recent
years. Nevertheless, its analysis requires a good understanding of the
conceptual foundations of the quantum theory. For this purpose, a historical
introduction to the subject is given in the following sections.
\\
\\
{\large{ \bf The EPR Paradox}}

In 1935, Einstein, Podolsky and Rosen designed an experiment, the so called EPR
paradox~\cite{EPR35}, to prove that quantum mechanics was an incomplete
description of physical reality. Neither they nor others agreed that
the (probabilistic)
outcome of a measurement performed on a quantum system was not uniquely
determined by its quantum state $\ket{\psi}$. They believed, instead, that the
result of a measurement was a property associated with the quantum system right
before the measurement was performed. 

\begin{figure}[hbt]
\begin{center}
\includegraphics[width=9.0cm]{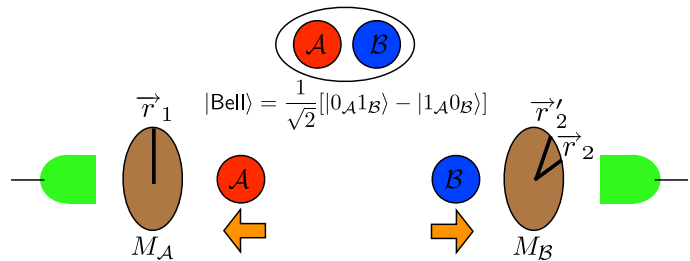}
\end{center}
\caption{Bohm's representation of J. Bell's experiment. Two qubits in a pure
quantum state are separated and their polarizations are locally measured along
three different directions. The measurement outcomes violate the
Bell inequality of Eq.~\ref{ineq}. Here $M$ denotes the polarizers.}
\label{fig3-1}
\end{figure}

To prove this statement, they considered a system composed of two quantum
particles where, although the position and momentum of each particle were not
well defined (uncertainty), the sum of their positions and the difference of
their momenta were. For pedagogical purposes, I consider here
a simpler version of this
experiment by means of the conventional model~\cite{Boh51} where the qualitative results are
equivalent to those of the EPR paradox. This simplified model consists of two
qubits initially prepared in some state (Fig.~\ref{fig3-1})
\begin{equation}
\label{belexp}
\ket{\sf Bell}= \frac{1}{\sqrt{2}} \left[ \ket{ 0_{\cA} 1_{\cB}}- 
\ket{1_{\cA}0_{\cB}} \right].
\end{equation}
[Such a state represents, for example, the singlet state (total spin 0)
of two-spin 1/2.] Assuming that a measurement (in the logical basis) is
performed on qubit $\cA$, quantum mechanics says that the outcome of such
measurement projects the state $\ket{\sf Bell}$ onto the state $\ket{0_{\cA}
1_{\cB}}$
with probability 1/2, and onto the state $\ket{1_{\cA} 0_{\cB}}$ with the same
probability (Sec.~\ref{sec2-1-1}). Thus, a measurement on qubit $\cA$
projects also the state of qubit ${\cB}$, such that the outcome of a later
measurement performed on it can be predicted with certainty.

What worried Einstein and others is the fact that, in principle, both particles
(qubits) could be very far apart in physical space. Therefore, a person located
close to qubit $\cA$ could gain information about the state of 
qubit ${\cB}$ almost immediately, violating the principle of locality; that is,  some
information could be propagated at infinitely large speed. If such were the
case, the existence of these quantum correlations would be against the
relativity theory, and the understanding of the physical world would be
completely different. Nevertheless, a deeper analysis shows that such violation
does not exist and that quantum mechanics is right. Below, I present the
experiments suggested by Bell to show that quantum correlations exist in nature
and cannot be explained by means of classical theory.
\\
\\
{\large{ \bf Bell's Inequalities}}

The idea of existence of hidden variables which determine the outcome of a
measurement in a quantum system was proved to be incompatible with quantum
mechanics by J. Bell~\cite{Bel64} in 1964. Basically, he proposed a similar
experiment to the one described above but such that the measurements could be
performed in any direction, that is, in a possible basis different to the
logical one. For example, when working with photons, such measurements could be
performed by using polarizers rotated by different angles.

Assuming that qubit ${\cA}$ is measured in some basis denoted by the vector
$\overrightarrow{r}_1$ and qubit ${\cB}$ is measured in the basis denoted
by  $\overrightarrow{r}_2$, such measurements project the state of either qubit
in the corresponding direction with eigenvalues $\pm 1$ (Sec.~\ref{sec2-1-1}).
In other words, the state after the measurement of the $j$th qubit can be
represented as a vector pointing in the direction $\overrightarrow{r}_j$ (+1)
or in the opposite direction $-\overrightarrow{r}_j$ (-1).

The idea of Bell was to obtain the function
$A(\overrightarrow{r}_1,\overrightarrow{r}_2)$, which denotes the average
product of the outcomes of the corresponding measurements for different
directions $\overrightarrow{r}_j$. It can be proved that if
$\overrightarrow{r}_1=\overrightarrow{r}_2$, then
$A(\overrightarrow{r}_1,\overrightarrow{r}_1)=-1$ for a Bell state of the form
of Eq.~\ref{belexp}. This is because such state can be written in the same form for
other locally rotated basis. For example
\begin{equation}
\frac{1}{\sqrt{2}} \left[ \ket{1_{\cA}0_{\cB}} - \ket{0_{\cA}1_{\cB}} \right ] =
\frac{1}{\sqrt{2}} \left[ \ket{+_{\cA}-_{\cB}} - \ket{-_{\cA}+_{\cB}} \right ] ,
\end{equation}
where $\ket{\pm_j}=\frac{1}{\sqrt{2}}[\ket{0_j}\pm\ket{1_j}]$  are the
eigenstates of the Pauli flip operator $\sigma_x^j$ with eigenvalue $\pm1$.
Then, a local measurement of the Bell state in the $x$ direction shows the same
results as a measurement in the $z$ direction (logical basis).

Equivalently, for the Bell state $\ket{\sf Bell}$ one obtains
$A(\overrightarrow{r}_1,-\overrightarrow{r}_1)=+1$, and for arbitrary
directions, $A(\overrightarrow{r}_1,\overrightarrow{r}_2)=-\overrightarrow{r}_1
. \overrightarrow{r}_2$; that is, the usual scalar product
between vectors in real space. Interestingly, such a result is associated with the
existence of correlations of quantum nature and cannot be obtained by any
classical theory, including hidden variables.

To show this, J. Bell first assumed that the complete state of both qubits was
characterized by the existence of uncontrollable hidden variables, denoted by
$\lambda$. He also assumed that the measurement outcome of qubit ${\cB}$ was
independent of the orientation $\overrightarrow{r}_1$ 
where qubit ${\cA}$ is measured (i.e.,
locality). Therefore, there exist two functions $M_{\cA}(\overrightarrow{r}_1,
\lambda)$ and $M_{\cB}(\overrightarrow{r}_2,
\lambda)$, which correspond to both measurements outcomes, with values
\begin{equation}
M_{\cA}(\overrightarrow{r}_1,\lambda)= \pm 1 \mbox{ ;  }
M_{\cB}(\overrightarrow{r}_2,\lambda)= \pm 1.
\end{equation}
In particular, both functions satisfy the desired result if 
$\overrightarrow{r}_1=\overrightarrow{r}_2$ (see Eq.~\ref{belexp}):
\begin{equation}
\label{bellcorrel}
M_{\cA}(\overrightarrow{r}_1,\lambda)= -
M_{\cB}(\overrightarrow{r}_1,\lambda) \ \ \forall \lambda.
\end{equation}
Nevertheless, if $p(\lambda)$ is the probability distribution for the hidden
variable, with $\int p(\lambda) d\lambda =1$, the average of the product of both
outcomes should be
\begin{equation}
A(\overrightarrow{r}_1,\overrightarrow{r}_2) = \int p(\lambda)
M_{\cA}(\overrightarrow{r}_1,\lambda) M_{\cB}(\overrightarrow{r}_2,\lambda) d \lambda,
\end{equation}
and from Eq.~\ref{bellcorrel},
\begin{equation}
\label{bell1}
A(\overrightarrow{r}_1,\overrightarrow{r}_2) =- \int p(\lambda)
M_{\cA}(\overrightarrow{r}_1,\lambda) M_{\cA}(\overrightarrow{r}_2,\lambda) d \lambda.
\end{equation}
Considering another unit vector $\overrightarrow{r}'_2$ and considering that the
three integers $M_{\cA}=\pm 1$ satisfy
\begin{equation}
M_{\cA}(\overrightarrow{r}_1) [M_{\cA}(\overrightarrow{r}_2) -
M_{\cA}(\overrightarrow{r}'_2)] \equiv \pm [1 -M_{\cA}(\overrightarrow{r}_2)
M_{\cA}(\overrightarrow{r}'_2)],
\end{equation}
and
after some simple
algebraic manipulations of Eq.~\ref{bell1}, together with $[M_{\cA}
(\overrightarrow{r}_1,\lambda)]^2=1$, the following Bell inequality is obtained:
\begin{equation}
\label{ineq}
\left| A(\overrightarrow{r}_1,\overrightarrow{r}_2) -A(\overrightarrow{r}_1,
\overrightarrow{r}'_2)\right|\le 1+ A(\overrightarrow{r}_2,\overrightarrow{r}'_2).
\end{equation}

Remarkably, this Bell inequality  sometimes is violated in a
quantum mechanical system due to the existence of non-classical correlations.
For example, if the vectors $\overrightarrow{r}_1$ and $\overrightarrow{r}_2$
are orthogonal, and $\overrightarrow{r}'_2$ lies between them making a $45^0$
angle (Fig.~\ref{fig3-1}), then for the quantum state of Eq.~\ref{belexp}
\begin{equation}
\label{ineq2}
A(\overrightarrow{r}_1,\overrightarrow{r}_2)=0 \ , \ \ 
A(\overrightarrow{r}_1,\overrightarrow{r}_3)=
A(\overrightarrow{r}_2,\overrightarrow{r}_3)=-\cos (\pi/4),
\end{equation}
and
\begin{equation}
\cos (\pi/4) \nleq 1 - \cos (\pi/4).
\end{equation}
\\
\\
{\large{ \bf Interpretation}}

The violation of Bell's inequalities was experimentally proved in the 1980s
using photon atomic transitions~\cite{AGR82}. The results obtained were in excellent
agreement with the predictions of quantum mechanics, assuring that no
hidden variable theory could overcome the problem of nonlocality. However, an
analysis of the measurement process tells one that the existence of non-classical
correlations does not imply a violation of causality, as Einstein, et. al., 
stated. In other words, the
almost instant influence of the measurement outcome of one qubit in the
measurement outcome of the other, cannot be regarded as the existence of some
sort of
communication (i.e., travelling information)
between both qubits: The outcome of the measurement (projection) is completely
random.
\\

Many different analysis about the nonclassical properties of quantum systems are
discussed in almost any book about quantum mechanics. In most
cases, the problem of measurement, which is not analyzed in detail in the
present work, is also considered. In this chapter, I mainly focus
on the study of these nonclassical correlations which define
QE. In the first place, I review the usual (or traditional) notion 
to prepare for a more general concept denoted as generalized entanglement, which
is also presented here. Finally, I present some examples to show how generalized
entanglement can be used in a more general framework.

\section{Quantum Entanglement}
\label{sec3-1}
The standard setting for studying entanglement usually involves a quantum
system $\cS$ composed of two distinguishable subsystems $\cA$ (for Alice) and
$\cB$ (for Bob), with Hilbert spaces denoted by ${\cal H_A}$ and ${\cal H_B}$,
respectively. Then the total Hilbert space for the joint system is ${\cal
H_S=H_A \otimes H_B}$, with $\otimes$ the usual tensor product.
A quantum state $\rho$ (i.e., its density operator)
is said to be separable if and only if
\begin{equation}
\label{separable}
\rho= \sum_s p_s \rho_s^{\cA} \otimes \rho_s^{\cB} ,
\end{equation}
where $\sum_s p_s =1$, with $p_s$ ($\ge 0$) being the corresponding probabilities.
The density operators $\rho_s^{\cA}=\ket{\psi^s_{\cA}}\bra{\psi^s_{\cA}}$ and
$\rho_s^{\cB}=\ket{\psi^s_{\cB}}\bra{\psi^s_{\cB}}$, which describe pure states 
of Alice and Bob, respectively,
need not describe orthogonal states for
different integers $s$.
A quantum state of the joint system is then separable whenever it can be
written as a probabilistic (convex) combination or a mixture of product (separable) states.
For example, if both $\cA$ and $\cB$ are single qubits, the pure state
\begin{equation}
\ket{\psi} =\frac{1}{2} \left[\ket{0_{\cA}0_{\cB}} + \ket{0_{\cA}1_{\cB}} +
\ket{1_{\cA}0_{\cB}} + \ket{1_{\cA}1_{\cB}} \right]
\end{equation}
is separable because it can be rewritten as
\begin{equation}
\ket{\psi} =
\frac{\ket{0}_{\cA}+\ket{1}_{\cA}}{\sqrt{2}} \otimes
\frac{\ket{0}_{\cB}+\ket{1}_{\cB}}{\sqrt{2}}.
\end{equation}

Interestingly, separable states can be prepared by means of local operations
(e.g., rotations, local measurements, etc.) and classical communication between
the different parties of the system. Then, they do not possess correlations of
quantum nature and cannot be distinguished from classical states. This
property leads to the following definition: A quantum state $\rho$ of a
composite system is said to be entangled if it is not separable and unentangled
otherwise. For example, the Bell  state of Eq.~\ref{belexp} is entangled because there is
no basis where it can be written as a product state. Naturally, this definition
of entanglement applies not only to bipartite systems but to all systems
composed of many (i.e., more than two)  distinguishable subsystems.

Finding a decomposition like Eq.~\ref{separable} 
for a given quantum state $\rho$ of a joint system, or even showing that
$\rho$ describes an entangled state (i.e., no such decomposition exists)
is, in general, a very difficult task. In fact, algebraic methods to check if
$\rho$ is entangled or not only exist for the case of bipartite systems, and for
Alice and Bob being single qubits or two level subsystems. However, if
the state is pure  (i.e., $\rho=\ket{\psi} \bra{\psi}$), 
simple methods to check separability exist.

\subsection{Separability and von Neuman Entropy}
\label{sec3-1-1}
The Schmidt decomposition of a pure quantum state $\ket{\psi}$ provides a
useful tool to check its separability.  When using this decomposition, any state
$\ket{\psi}$ of a bipartite quantum system can be written as~\cite{Per98}
\begin{equation}
\label{schmidt1}
\ket{\psi} = \sum\limits_{j=1}^R c_j \ket{\phi^j_{\cA}} \otimes
\ket{\phi^j_{\cB}} ,
\end{equation}
where the pure states $\ket{\phi^j_{\cA}}$ and $\ket{\phi^j_{\cB}}$ are
orthonormal states of subsystems $\cA$ and $\cB$, respectively; that is
\begin{equation}
\langle \phi^j_{\cA} \ket{\phi^{j'}_{\cA}}=
\langle \phi^j_{\cB} \ket{\phi^{j'}_{\cB}}= \delta_{jj'} \ \forall j,j' \in [1
\cdots R].
\end{equation}
Without loss of generality, the coefficients $c_j \ne 0$ of Eq.~\ref{schmidt1} are
real (i.e., the phases have been absorbed in the corresponding states) and, from
the normalization of $\ket{\psi}$, they
satisfy $\sum_{j=1}^R (c_j)^2=1$. If $d_a$ and $d_b$ denote the dimensions
of the Hilbert spaces ${\cal H_A}$ and ${\cal H_B}$, respectively, the integer
$R$ satisfies
\begin{equation}
R \le \min \{d_a, d_b \}.
\end{equation}

Obviously, the pure state $\ket{\psi}$ is entangled whenever $R >1$ in the
decomposition and unentangled otherwise ($R=1$). Moreover,  $\ket{\psi}$ is
entangled whenever it looks mixed from the point of view of either observer
(Alice or Bob). In other words, when obtaining the reduced density operator of
one subsystem by tracing out the other, such density operator describes a mixed
state when $\ket{\psi}$ is non-separable or entangled. This is only valid for
pure states of the joint system. For example, the reduced
density operator $\rho^{\cA}$ associated to Alice is
\begin{equation}
\rho^{\cA}= {\sf Tr}_{\cB} (\rho) =
\sum\limits_{j=1}^R \bra{\phi^j_{\cB}} \psi \rangle \bra{\psi}
 \phi^j_{\cB}\rangle = \sum\limits_{j=1}^R (c_j)^2 \ket{\phi^j_{\cA}}
\bra{\phi^j_{\cA}} .
\end{equation}
Then, if $(c_j)^2 <1 \ \forall j$ (or equivalently $R>1$), the reduced density
operator $\rho^{\cA}$ describes a mixed state and $\ket{\psi}$ is entangled.
Similar results are obtained from the point of view of Bob.

A useful way to quantify the entanglement of a pure bipartite state
$\ket{\psi}$ is given by the von Neuman entropy $E_{\cS}$ of either
$\rho^{\cA}$ or  $\rho^{\cB}$. Its definition is 
\begin{equation}
\label{vonneu1}
E_{\cS}(\psi)= - {\sf Tr} \left( \rho^{\cA} . \log_2 \rho^{\cA} \right) =
- {\sf Tr} \left( \rho^{\cB} . \log_2 \rho^{\cB} \right) =
-\sum\limits_{j=1}^R (c_j)^2 \log_2 (c_j)^2.
\end{equation}
$E_{\cS}$ is zero (minimal) for a product state and takes its maximum value 
($E_{\cS}=1$) for maximally entangled states, such as the Bell state of
Eq.~\ref{belexp}. Moreover, $E_{\cS}$ remains invariant or decreases under any 
operation on $\ket{\psi}$ performed locally by Alice or Bob. As I will show,
this is an important property for a measure of entanglement.

All the concepts described in this section can be naturally extended for
multipartite systems. Then a pure quantum state $\ket{\psi}$ is entangled whenever it
looks mixed to, at least, one of the observers associated to one of the parties. A
von Neuman measure can be defined accordingly, by measuring the entanglement of
every single party with the rest of the system.

\subsection{Mixed-State Entanglement and the Concurrence}
\label{sec3-1-2}
As mentioned, checking the separability of a mixed quantum state  is a difficult task because
of the many equivalent ways that the density operator can be decomposed.
Nevertheless, certain (classical) algorithms to calculate the entanglement of a mixed state
exist, but their complexity scales with the dimension $d_S$ of the Hilbert
space ${\cal H}$ associated to the joint quantum system. Such dimension is
known to scale exponentially with the system size. Thus, these algorithms can
only be applied to study the
entanglement of small systems. 

For a mixed state of a bipartite system $\rho= \sum_s p_s\ket{\psi_s}
\bra{\psi_s}$, that is, the system being in the pure state $\ket{\psi_s}$ with
probability $p_s <1$, the entanglement of formation $E(\rho)$ is defined as
\begin{equation}
\label{formation}
E(\rho) = \min \left[ \sum_s p_s E_{\cS}(\psi_s) \right],
\end{equation}
where $E_{\cS}(\psi_s)$ is the von Neuman entropy defined in Eq.~\ref{vonneu1}. 
Each possible decomposition of $\rho$ corresponds to a certain amount of
entanglement, so the minimum needs to be obtained.
(Equation~\ref{formation} is trivially extended for multipartite
systems.) For
example, the mixed state of a two-qubit system
\begin{equation}
\rho= \frac{1}{2} \ket{\sf Bell}_1 \bra{\sf Bell}_1 +
\frac{1}{2} \ket{\sf Bell}_2 \bra{\sf Bell}_2 ,
\end{equation}
where $\ket{\sf Bell}_1=\ket{0_{\cA} 0_{\cB}}+ \ket{1_{\cA} 1_{\cB}}$ and
$\ket{\sf Bell}_2=\ket{0_{\cA} 0_{\cB}}- \ket{1_{\cA} 1_{\cB}}$ are maximally
entangled states (i.e., Bell states),
seems to be maximally entangled. However, $\rho$ is actually separable and thus
unentangled :
\begin{equation}
\rho=\frac{1}{2}  \ket{0_{\cA} 0_{\cB}} \bra{0_{\cA} 0_{\cB}} +
\frac{1}{2}  \ket{1_{\cA} 1_{\cB}} \bra{1_{\cA} 1_{\cB}},
\end{equation}
that is, a mixture of product states.

As mentioned, the entanglement for a mixed state of two qubits ($\cA$
and $\cB$) can be exactly computed without calculating the minimal value of
Eq.~\ref{formation}~\cite{Woo98}. 
For this purpose, the time reversal
operation needs first to be applied. That is, if $\rho$ is the state of the 
two-qubit system,
then
\begin{equation}
\tilde{\rho} = (\sigma_y^{\cA} \sigma_y^{\cB}) \rho^* 
(\sigma_y^{\cA} \sigma_y^{\cB}),
\end{equation}
where $\rho^*$ is the complex conjugate of $\rho$, is the corresponding
spin-flipped state. Remarkably, the entanglement of formation
(Eq.~\ref{formation}) for this system
can be expressed as
\begin{equation}
E(\rho) = {\cal E}(C(\rho)),
\end{equation}
where the real function ${\cal E}(C)$ is monotonically increasing with $C$, and
$C(\rho)$ is denoted as the {\em concurrence} of $\rho$~\cite{Woo98}. (Due to
this property, one can use the concurrence as a measure of entanglement in the
two-qubit system, instead.) Defining the operator
\begin{equation}
R = \sqrt{ \sqrt{\rho} \tilde{\rho} \sqrt{\rho}},
\end{equation}
the concurrence is defined by
\begin{equation}
C(\rho) = \max \{ 0, \lambda_1 - \lambda_2 - \lambda_3 - \lambda_4 \},
\end{equation}
where the $\lambda_j$s are the eigenvalues of $R$ in decreasing order
($\lambda_j \ge 0$). Then, $C(\rho)$ takes its maximum value [$C(\rho)=1$]
for Bell states (Eq.~\ref{belexp}) and vanishes for any separable (mixed)
state.

\subsection{Measures of Quantum Entanglement}
\label{sec3-1-3}
A good measure of entanglement ${\sf E}(\rho)$ for a multipartite quantum
system, such as the entanglement of formation of Eq.~\ref{formation}, needs to
satisfy certain requirements~\cite{Vid00}. First,
such a measure must take its minimum value, ${\sf E}(\rho)=0$, whenever $\rho$
describes a separable state. Second, unitary local operations, local
measurements, and classical communication between different parties in the
system (usually referred to LOCC operations) cannot increase ${\sf E}(\rho)$.
It is reasonable
that LOCC operations do not transform, for example, a separable state into a
nonseparable one. Moreover, local measurements project the state lowering its
entanglement. 

Other requirements also have to be considered when defining a good measure of
entanglement. These include continuity, convexity, additivity, and
subadditivity. For example, if two density operators describe almost the same
state, they must have similar amounts of entanglement. That is, 
${\sf E}(\rho)$ must be a continuous function of $\rho$. Also, the entanglement
of a linear combination of two density operators $\rho_1$ and $\rho_2$, that is,
$\rho= p \rho_1 + (1-p) \rho_2$ (with $0<p<1$) must satisfy
\begin{equation}
\label{convex}
{\sf E}(\rho) \le p {\sf E}(\rho_1) + (1-p) {\sf E}(\rho_2).
\end{equation}
A function satisfying Eq.~\ref{convex} is said to be convex. Basically,
the reason for such
convexity is because an operator like $\rho$ tends to have less entanglement than the
corresponding operators $\rho_1$ and $\rho_2$.

It is not very clear how important the requirements of additivity and
subadditivity are. The additivity property states that if the system is composed
of $m$ identical subsystems, all being in the state given by the density
operator $\tau$, then
\begin{equation}
{\sf E}(\rho) = {\sf E} (\otimes_{i=1}^m \tau) = m {\sf E}(\tau).
\end{equation}
The subadditivity property states, however, that if the density operator $\rho$
of the total (multipartite) system can be expressed as $\rho=\tau_1 \otimes
\tau_2$, then
\begin{equation}
{\sf E}(\rho) \le {\sf E}(\tau_1) + {\sf E}(\tau_2).
\end{equation}
In fact, the entanglement of formation for multipartite systems, as defined by
Eq~\ref{formation}, does not satisfy the additivity property.

\section{Generalized Entanglement}
\label{sec3-2}
The main purpose of {\em generalized entanglement} (GE) is to extend the concepts of
traditional entanglement to a more general setting by defining new measures
that can be applied to any quantum system, even when a nontrivial subsystem
decomposition exists. In this way, GE is considered as an
observer-dependent property of a quantum state and, as in the usual case, it is
determined by a preferred set of observables of the quantum system under study.
It is expected that this new concept allows for a better understanding of
non-classical correlations in quantum mechanics. 

In this section, I rigorously define GE~\cite{BKO03,BKO04} in terms of reduced
states and I analyze the important case where the preferred set of observables
belong to a certain Lie algebra (i.e., an algebra closed under commutation). Some
examples to show how GE works in different quantum
systems will be presented and, in particular, I will
show how the traditional notion can
be recovered.

\subsection{Generalized Entanglement: Definition}
\label{sec3-2-1}
Assume that $\fh$ is a finite set of observables; that is, $\fh=\{\hat{O}_1,
\hat{O}_2,\cdots, \hat{O}_M\}$, with $\hat{O}_j=\hat{O}_j^\dagger$ the
Hermitian operators that linearly map quantum states of a Hilbert space $\cH$
into quantum states of the same space.   Such set of observables is usually
intrinsically associated to the quantum system $\cS$ under study and depends on
its nature, control access, superselection rules (e.g., particle number
conservation), etc. For a quantum state with
corresponding density matrix $\rho=\sum_s p_s \ket{\psi_s}\bra{\psi_s}$ ($p_s
\ge 0$; $\sum_s p_s=1$), its $\fh$-state (reduced state) is defined to be a
linear functional $\Lambda$ on the operators of $\fh$ according to 
\begin{equation}
\Lambda ({\hO}_j)= {\sf Tr}(\rho {\hO}_j)=\langle {\hO}_j \rangle  \ \forall j
\in [1 \cdots M],
\end{equation}
with ${\hO}_j \in \fh$, and $ \langle {\hO}_j \rangle$ its expectation value
over the state $\rho$. In particular, if $\fh$ denotes the set of all linearly
independent observables associated to the quantum system, the set of
$\fh$-states completely determine the state of the system, since $\rho$ can be
exactly recovered from those expectations.

Considering the set of all $\fh$-states, it can be shown that such set is
closed under convex (or probabilistic) combination. In other words, if
$\Lambda_k$ are $\fh$-states associated to different density matrices $\rho_k$,
then $\sum_k p_k \Lambda_k$ is also a possible $\fh$-state for any
probabilistic distribution $p_k$ ($p_k \ge 0$, $\sum_k p_k=1$). If the set $\fh$
is compact then all $\fh$-states can be obtained as combinations of extremal
states. Extremal states are those defined as the reduced states that cannot be
written as a convex combination or mixture of two or more $\fh$-states. In general, extremal
states are least uncertainty states so they are also referred to $\fh$-pure
states.

With these definitions and properties, a pure state $
\ket{\psi}$ (i.e., $\rho=\ket{\psi}\bra{\psi}$)
is said to be generalized unentangled relative to the
distinguished set of observables $\fh$ if its reduced state, that is, the state
defined by the expectations of the operators in $\fh$, is pure or extremal.
Otherwise, $\ket{\psi}$ is said to be generalized entangled
relative to $\fh$.

Although such definition can be applied to any set of observables, relevant
physical applications are usually found when the set $\fh$ is a Lie
algebra of operators. In the following, I describe this important case while
giving a
short description of Lie algebra theory.

\subsection{Generalized Entanglement and Lie Algebras}
\label{sec3-2-2}
I now focus on the case when $\fh=\{\hat{O}_1, \hat{O}_2,\cdots,
\hat{O}_M\}$ is a finite real $M$-dimensional
Lie algebra of observables acting irreducibly on
${\cal H}$ (the Hilbert space), with bracket given by
\begin{equation}
\left[ \hat{O}_j, \hat{O}_{j'} \right]=i \left( \hat{O}_j \hat{O}_{j'} - 
\hat{O}_{j'} \hat{O}_j \right) = i \sum\limits_{k=1}^M f^{jj'}_k \hat{O}_k \ .
\end{equation}
(Here, no distinction between the abstract Lie algebra isomorphic to $\fh$, and
the concrete matrix Lie algebra $\fh$ of observables acting on the Hilbert
space ${\cal H}$ is made.) The corresponding induced Lie group of $\fh$ is given
by the map $X \rightarrow e^{iX}$, with $X \in \fh$. Also, $\fh$ is usually
assumed to
be semi-simple, that is, with no commutative ideals\footnote{An ideal of $\fh$ is a
subalgebra $\fI \subset \fh$ invariant under the commutation with any element of $\fh$.}.

The projection map of a quantum state, with density matrix $\rho$, onto $\fh$ can
be uniquely defined by the trace inner product as
\begin{equation}
\label{projecmap}
\rho \rightarrow {\cal P}_\fh (\rho)=\sum\limits_{j=1}^M \langle \hat{O}_j
\rangle \hat{O}_j,
\end{equation}
where $\langle \hat{O}_j \rangle = {\sf Tr} (\rho \hat{O}_j)$ is the corresponding
expectation value. Notice that only the $\fh$-state associated to $\rho$ must
be known to build ${\cal P}_\fh$. Similarly, the relative purity or
$\fh$-purity is defined as the squared length of the projection; that is
\begin{equation}
\label{purity}
P_\fh (\rho) = {\sf Tr} \left( {\cal P}^2_\fh (\rho) \right),
\end{equation}
and if the operators are Schmidt orthogonal satisfying ${\sf Tr} \left(
\hat{O}_j \hat{O}_k \right)= \delta_{jk}$, then
\begin{equation}
\label{purity2}
P_\fh (\rho)= {\sf K} \sum\limits_{j=1}^N \langle \hat{O}_j \rangle^2.
\end{equation}
Here ${\sf K}$ is a real constant for normalization requirements.
The $\fh$-purity as defined by Eq.~\ref{purity2}
is a group invariant function:
\begin{equation}
\label{groupinv}
P_\fh (\rho)=P_\fh (e^{iX} \rho e^{-iX}); \ \ X\in \fh.
\end{equation}

Interesting consequences from the definition of the relative purity are
obtained when the matrix Lie algebra $\fh$ acts irreducibly on the Hilbert
space ${\cal H}$ associated to the system, that is, the representation of $\fh$
is irreducible. In such a case, a pure quantum state $\ket{\psi}$ is extremal or
$\fh$-pure if and only if the $\fh$-purity takes its maximum
value~\cite{BKO03}. Such states are the so-called {\em generalized coherent
states} (GCSs)
of $\fh$ and are constructed as
\begin{equation}
\label{gencoh}
\ket{\sf GCS}=\exp \left[i (\sum\limits_{j=1}^M \zeta_j \hat{O}_j) \right] 
\ket{\sf ref},
\end{equation}
where $\zeta_j \in \mathbb{R}$, and $\ket{\sf ref}$ is a reference state
corresponding to the highest (or lowest) weight state of $\fh$. GCSs are a
generalization of the traditional coherent states of the harmonic
oscillator or the radiation field~\cite{SZ02}, where in those cases $\ket{\sf
ref}$ is the vacuum or no-excitation state. (The first term on the right hand
side of Eq.~\ref{gencoh} is a general displacement operator.)

In general, the reference state $\ket{\sf ref}$ 
is a well defined object in finite semi-simple Lie algebras, after
a particular Cartan-Weyl (CW) decomposition is performed. 
In a CW basis, the algebra is written
as 
\begin{equation}
\label{cardecom}
\fh =\{\fh_D, \fh_+, \fh_- \}. 
\end{equation}
Here, $\fh_D=\{\hh_1,\hh_2,\cdots,\hh_r\}$,
with $\hh_j=\hh_j^\dagger$, is a Cartan subalgebra (CSA) of $\fh$ defined as
the biggest set of commuting operators in $\fh$, and the integer $r$ defines
the rank of $\fh$. The sets $\fh_+=\{\hE_{\alpha_1}, \hE_{\alpha_2},\cdots,
\hE_{\alpha_l} \}$ and $\fh_-=\{ \hE_{-\alpha_1}, \hE_{-\alpha_2},\cdots,
\hE_{-\alpha_l} \}$, with $\hE_{\alpha_j}=\left( \hE_{-\alpha_j}
\right)^\dagger$, are usually referred to the sets of raising and lowering
operators, respectively. By notation, $r+2l =M$, the dimension of $\fh$.
The corresponding commutation relations are
\begin{eqnarray}
\label{cartancom}
\left[ \hh_k, \hh_{k'} \right] &=&0 \  , \\
\label{raiscom}
\left[ \hh_k, \hE_{\alpha_j} \right] &=&\alpha_j^k  \hE_{\alpha_j} \ 
 ,\\
\label{lowcom}
\left[ \hE_{\alpha_j}, \hE_{-\alpha_j} \right] &=&\sum\limits_{k=1}^r \alpha_j^k 
\hh_k \  ,\\
\left[ \hE_{\alpha_j}, \hE_{\alpha_{j'}} \right]& = &N_{jj'} \hE_{\alpha_j +
\alpha_{j'}} \ \ \forall \alpha_j \neq -\alpha_{j'} \ ,
\end{eqnarray}
where the vectors $\alpha_j=(\alpha_j^1, \alpha_j^2, \cdots, \alpha_j^r) \in 
\mathbb{R}^r$
are defined as the {\em roots} of the algebra, and $N_{jj'}$ depends on
the corresponding roots. 

Equation~\ref{cartancom} states that the operators in
$\fh_D$ can be simultaneously diagonalized. Their eigenstates in a given
representation are the corresponding {\em weight states}. A weight state 
$\ket{\phi^p}$ of $\fh$ satisfies the eigenvalue equation
\begin{equation}
h_k \ket{\phi^p} = e^p_k \ket{\phi^p} \ \forall k\in [1 \cdots r],
\end{equation}
where the eigenvalues $e^p_k \in \mathbb{R}$ are defined as the {\em weights}.
Moreover, the vectors $\overrightarrow{e}^p =(e_1^p, e_2^p, \cdots, e_r^p) \
\in \mathbb{R}^r$ are defined as the {\em weight vectors}.

Due to the commutation relations of Eq.~\ref{raiscom}, a raising operator acting over a
weight state either annihilates it (i.e., $\hE_{\alpha_j} \ket{\phi^p} = 0$)
or maps it to a different weight state:
\begin{equation}
\hh_k \hE_{\alpha_j} \ket{\phi^p} = 
(\hE_{\alpha_j}\hh_k + \alpha_j^k \hE_{\alpha_j})\ket{\phi^p} =
(\alpha_j^k + e_k^p)\hE_{\alpha_j} \ket{\phi^p}.
\end{equation}
Similar results are obtained for the lowering operators $\hE_{-\alpha_j}$ from
Eq.~\ref{lowcom}. The simplest case is given by the algebra $\fsu(2)=\{
\sigma_z, \sigma_+, \sigma_- \}$, where the spin-1/2 representation is given
by the Pauli matrices defined in Eq.~\ref{pauli1}.

A weight vector $\overrightarrow{e}^p = (e_1^p, e_2^p, \cdots, e_r^p)$  is said
to be {\em positive} if the first non-zero component is positive. Then, a
weight vector $\overrightarrow{e}^p$ is said to be larger  than a weight
vector  $\overrightarrow{e}^{p'}$ if they differ on a positive weight vector.
In particular, the root vectors $\alpha_j= (\alpha_j^1, \alpha_j^2, \cdots,
\alpha_j^r)$ in Eq.~\ref{cardecom}  are weight vectors of a  representation of
$\fh$ called {\em adjoint representation} (Appendix~\ref{appC}). With no loss
of generality, these vectors  have been chosen to be positive (or equivalently,
$-\alpha_j$ to be negative) in the CW decomposition.

The definition of positivity naturally extends to any weight vector in any
representation of $\fh$. Therefore, a {\em highest weight state} $\ket{\sf HW}$ of
$\fh$ is defined as the state with highest weight vector $\overrightarrow{e}^p$
in the representation.
Similarly, a {\em lowest weight state} $\ket{\sf LW}$ is defined as the state
with lowest weight
vector in the representation. These states are annihilated by every raising and
lowering operator, respectively: 
\begin{equation}
\hE_{\alpha_j} \ket{\sf HW}= \hE_{-\alpha_j} \ket{\sf LW}=0.
\end{equation}
Both, $\ket{\sf HW}$ and $\ket{\sf LW}$, are possible reference states
$\ket{\sf ref}$ to generate the GCSs or generalized unentangled states of $\fh$
(Eq.~\ref{gencoh}). With no loss of generality, I will mainly use  $\ket{\sf
HW}$ as the reference state. Since its weight vector is positive, this state 
is then the unique ground state of a Hamiltonian
\begin{equation}
H = \sum\limits_{k=1}^r \gamma_k \hh_k \ ; \ \gamma_k \in \mathbb{R},
\end{equation}
where the sign of the coefficients $\gamma_k$ depends on the weight.
Therefore, every GCS is the unique ground state of some Hamiltonian in
$\fh$~\cite{BKO03}.

A common uncertainty measure for the algebra $\fh$ is given by
\begin{equation}
\label{uncer}
(\Delta F)^2 = \sum\limits_{j=1}^M \langle \hO_j^2 \rangle - \langle \hO_j
\rangle ^2.
\end{equation}
Remarkably, GCSs of $\fh$ are also minimal uncertainty states. 
When $\fh$ acts irreducibly on ${\cal H}$, the first term on the rhs of
Eq.~\ref{uncer} is a Casimir, i.e. invariant, while the second term is the
$\fh$-purity. Because GCSs have maximum purity, the corresponding uncertainty is
minimal.

It is important to realize that the relationships just mentioned between
maximal purity, generalized coherence, and generalized unentanglement
established for a pure state relative to an irreducibly represented Lie algebra
$\fh$ do not automatically extend to the case where $\fh$ acts reducibly on
${\cal H}$. In fact, for a semi-simple Lie algebra $\fh$, a generic finite
dimensional representation of $\fh$ may be decomposed as a direct sum of
irreducibly invariant subspaces, $\cH = \oplus_{\cl} \cH_{\cl}$,  with each of
the $ \cH_{\cl}$ being in turn the direct sum of its weight spaces. Then, the
algebra acts irreducibly on each $\cH_{\cl}$, with the corresponding
irreducible representation (irrep). In particular, every irrep appearing in the
decomposition has a highest (and lowest) weight state, and for each of these irreps
there is a manifold of GCSs constructed as the orbit of a highest weight state
for that irrep. Therefore, these GCSs will not satisfy in general the
extremality property that defines generalized unentangled states. Indeed, the
extremal weight vectors, which correspond to generalized unentangled states,
need not all have the same length ($\fh$-purity). Also, the minimal uncertainty
property could be lost. Maximal purity remains then as a sufficient, though no
longer a necessary, condition for generalized unentanglement. Nevertheless, all
the statements obtained for the irreducible case still apply for the examples
and problems presented in this thesis.

\subsection{Generalized Entanglement and Mixed States}
\label{sec3-2-3}
For mixed states on ${\cal H}$, the direct generalization of the
squared length of the projection onto $\fh$ as in Eq. (\ref{purity}) does
{\em not} give a GE measure with well-defined monotonicity properties
under appropriate generalizations of the LOCC set
transformations as explained in Sec.~\ref{sec3-1-3}~\cite{BKO03}. 
A proper extension of the quadratic
purity measure defined by Eq.~\ref{purity} for pure states to
mixed states may be naturally obtained via a standard convex roof
construction.  If $\rho = \sum\limits_s p_s \ket{\psi_s}\bra{\psi_s}$,
with $\sum\limits_s p_s =1$ and $\sum\limits_s p_s^2 < 1$, the latter
is obtained by calculating the maximum $\fh$-purity (minimum
entanglement) over all possible convex decompositions $\{p_s,
\ket{\psi_s} \}$ of the density operator $\rho$ as a pure-state
ensemble. In general, similarly to what happens for most mixed-state
entanglement measures, the required extremization makes the resulting
quantity very hard to compute. Nevertheless, the cases studied in this thesis
are always related with pure states and no extremization process is
required.

\section{Relative Purity as a Measure of Entanglement in
Quantum Systems}
\label{sec3-3}
In order to understand its meaning as a measure of entanglement for pure
quantum states, I now apply the definition of relative purity to the study of
non-classical correlations in different physical systems. First, I will focus
on spin systems, showing that for particular subsets of observables, the
$\fh$-purity can be reduced to the usual notion of entanglement: The pure
quantum states that can be written as a product of states of each party are
generalized unentangled in this case.  
For this purpose, and because they will also be needed in future cases, 
I introduce the following representative quantum states for $N$ spins of
magnitude $S$:
\begin{eqnarray}
\label{statedef}
\ket{{\sf F}_S^N}&=&\ket{S,S, \cdots ,S} \;,\\ \nonumber
\ket{{\sf W}_S^N} &= &\frac{1}{\sqrt{N}} \sum\limits_{i=1}^N \ket{S,\cdots,S ,
(S-1)_i , S, \cdots, S}\; ,\\ \nonumber
\ket{{\sf GHZ}_S^N}& = &\frac{1}{\sqrt{2S+1}} \sum\limits_{l=0}^{2S}
\ket{S-l,S-l,\cdots,S-l} \;,
\end{eqnarray}
where $\ket{S_1,S_2,\cdots,S_N}=\ket{S_1}_1 \otimes
\ket{S_2}_2 \otimes \cdots \otimes \ket{S_N}_N$ is a product state, and $\ket{S_i}_i$ 
denotes the state of the $i$th party with $z$-component of the 
spin equal to $S_i$ (defining the relevant computational basis for
the $i$th subsystem).

However, for other physically natural choices of observable sets a different
notion of QE is obtained. In this way, one can go beyond the usual
(distinguishable) subsystem
partition.  For this reason, I will focus later on the study of the
$\fh$-purity as a measure of entanglement for fermionic systems. This is a good
starting point to understand how such a measure can be applied to quantum
systems obeying different particle statistics and/or described by different
operator languages.

\subsection{Two-Spin Systems}
\label{sec3-3-1}
\begin{figure}[hbt]
\begin{center}
\includegraphics[width=10.5 cm]{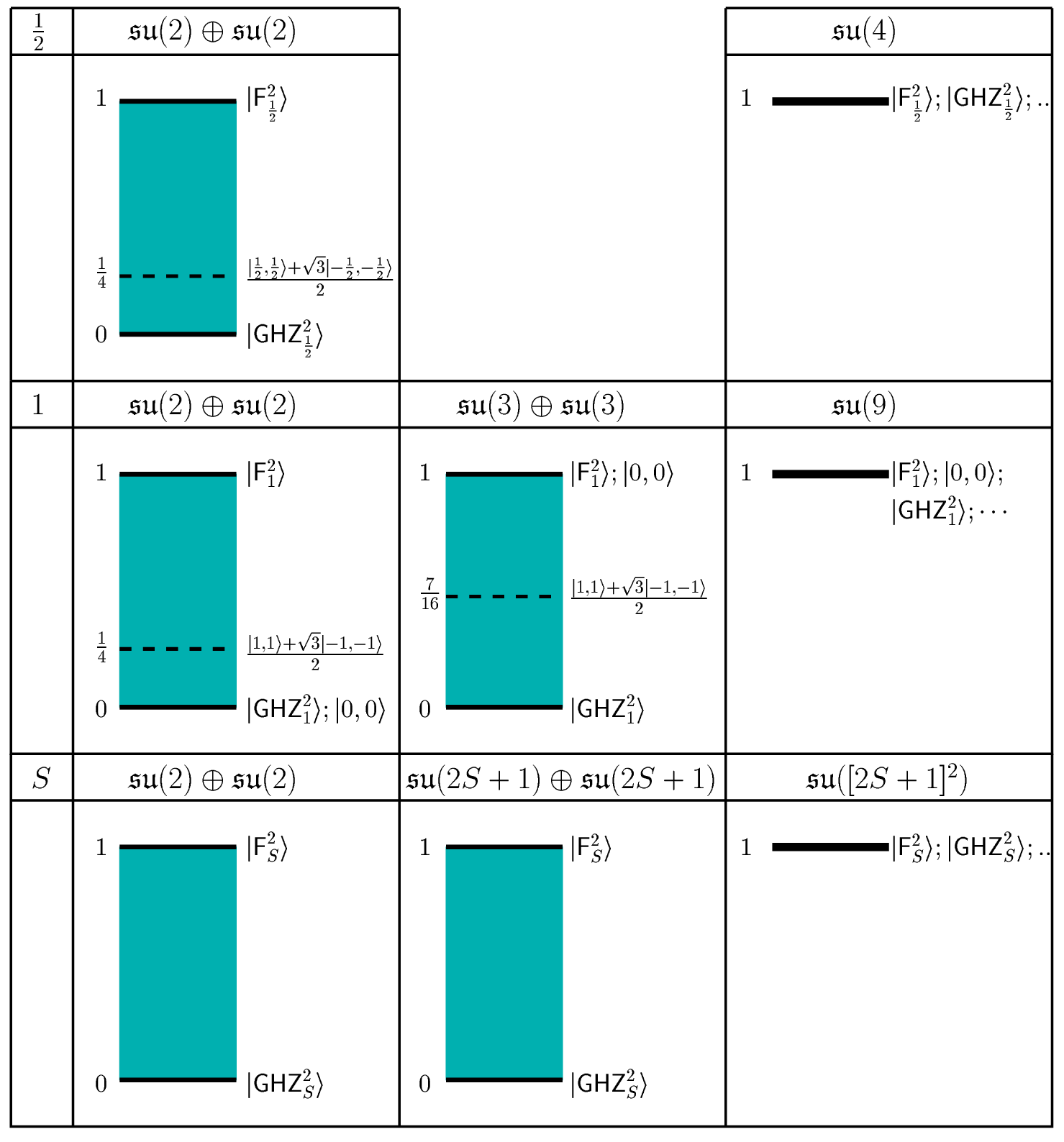}
\end{center}
\caption{Purity relative to different possible algebras for a two-spin-$S$ system.
The quantum states $\ket{{\sf GHZ}_S^2}$ and $\ket{{\sf F}_S^2}$ are defined in
Eqs.~\ref{statedef}.}
\label{fig3-2}
\end{figure}

The simplest system to study a measure of entanglement is composed of two
spins. 
For simplicity, I begin by studying the GE of a two-qubit system (two-spin-1/2, 
$\cA$ and $\cB$), where the most general pure quantum state can be
written as  $\ket{\psi}=a \ket{0_{\cA}0_{\cB}} + b \ket{0_{\cA}1_{\cB}} +c
\ket{1_{\cA}0_{\cB}} + d \ket{1_{\cA}1_{\cB}} $, with the complex numbers $a$,
$b$, $c$,  and $d$ satisfying  $|a|^2+|b|^2+|c|^2+|d|^2=1$.  For a spin-1/2
system the associations $\ket{0}=\ket{\uparrow}=\ket{\frac{1}{2}}$ and 
$\ket{1}=\ket{\downarrow}=\ket{-\frac{1}{2}}$ are commonly considered. The
traditional measures of pure state entanglement in this case are well
understood (Sec.~\ref{sec3-1}),
indicating that the  Bell states $\ket{\sf
GHZ_{\frac{1}{2}}^2}=\frac{1}{\sqrt{2}} \left[ \ket{0_{\cA} 0_{\cB}} +
\ket{1_{\cA} 1_{\cB}} \right]$~\cite{EPR35}  (and their local spin  rotations
such as the state of Eq.~\ref{belexp}) are maximally entangled
with respect to the local Hilbert space  decomposition ${\cal H}_{\cA}
\otimes{\cal H}_{\cB}$.  On the other hand, calculating the purity
relative to the (irreducible) Lie algebra of all {\em local 
observables} $\fh=\fsu(2)_{\cA}\oplus \fsu(2)_{\cB} =\{ \sigma_{\alpha}^i;
\ i: {\cA},{\cB};\ \alpha=x,y,z\}$ classifies the pure two-qubit  
states in the same way as the traditional measures (Fig.~\ref{fig3-2}).
Here, the operators $\sigma_{\alpha}^{\cA}=\sigma_\alpha
\otimes \one $ and $\sigma_\alpha^{\cB}=\one \otimes \sigma_\alpha$ are the
Pauli operators acting on qubits $\cA$ and $\cB$, respectively. 
In this case, Eq.~\ref{purity}  simply gives 
\begin{equation}
P_{\fh} (\ket{\psi})=\frac{1}{2}\sum\limits_{i,\alpha} \langle 
\sigma_{\alpha}^i \rangle^2 \;, 
\end{equation} 
where Bell's states are maximally entangled ($P_{\fh}=0$) and product
states of  the form $\ket{\psi}=\ket{\phi_{\cA}}_{\cA} \otimes
\ket{\phi_{\cB}}_{\cB}$
(GCSs of the algebra $\fh=\fsu(2)_{\cA} \oplus \fsu(2)_{\cB}$) are generalized
unentangled,  with maximum purity.  Therefore, the normalization factor ${\sf
K} =
1/2$ may be obtained by setting $P_{\fh}=1$ in such a product state. 
As explained in Sec.~\ref{sec3-2-2}, $P_{\fh}$ is invariant
under group operations, i.e., local rotations in this case. Since
all GCSs of $\fh$ belong to the same orbit generated by the application
of group operations to a particular product state (a reference state
like $\ket{0_{\cA}0_{\cB}} \equiv \ket{\frac{1}{2},\frac{1}{2}}$), 
they all consistently have maximum $\fh$-purity ($P_{\fh}=1$). 

Another important insight may be gained by calculating the purity relative to
the algebra of {\em all} observables,
$\fh=\fsu(4)=\{\sigma_{\alpha}^i,\sigma_{\alpha}^{\cA} \otimes
\sigma_{\beta}^{\cB}; \mbox{ }i={\cA},{\cB}; \mbox{ }\alpha,\beta=x,y,z\}$ in
the case of two-qubit system. One finds that {\em any} two-qubit pure state
$\ket{\psi}$ (including the Bell's states of Eq.~\ref{belexp}) is
generalized unentangled ($P_{\fh}=1$; See Fig.~\ref{fig3-2}).  This property is
a manifestation of the relative nature of GE, since considering the set of all
observables as being physically accessible is equivalent to not making any
preferred subsystem decomposition. Accordingly, in this case any pure quantum
state becomes a GCS of $\fsu(4)$.

In Fig.~\ref{fig3-2} I also show the GE for systems of two parties
of spin-$S$ relative to different algebras.  It is observed that the purity
reduces again to the traditional concept of entanglement for higher
spin if it is calculated relative to the (irreducible) Lie algebra of
{\em all} local observables $\fh=\fsu(2S+1)_{\cA}\oplus \fsu(2S+1)_{\cB}$. For
example, when  interested in distinguishing product states from 
entangled states in a two-spin-1 system, the
purity relative to the (irreducible) algebra  $\fh=\fsu(3)_{\cA} \oplus
\fsu(3)_{\cB} = \{ \lambda^{\cA}_\alpha \otimes \one^{\cB} , \one^{\cA}  \otimes
\lambda^{\cB}_\alpha \mbox{ } (1 \leq \alpha \leq 8) \}$, 
needs to be calculated. Here, the
$3
\times 3$ Hermitian and traceless matrices $\lambda_i$ are the well
known Gell-Mann matrices~\cite{Geo99}:
\[
\begin{array}{l}
\lambda_1= \frac{1}{\sqrt{2}} 
\pmatrix {0 & 1& 0 \cr 1 & 0 & 0 \cr 0 & 0 & 0 \cr} \mbox{ ; }\;
\lambda_2= \frac{1}{\sqrt{2}} 
\pmatrix {0 & -i & 0 \cr i & 0 & 0 \cr 0 & 0 & 0 \cr} \\
\lambda_3= \frac{1}{\sqrt{2}} 
\pmatrix {1 & 0& 0 \cr 0 & -1 & 0 \cr 0 & 0 & 0 \cr} \mbox{ ; }\;
\lambda_4= \frac{1}{\sqrt{2}} 
\pmatrix {0 & 0& 1 \cr 0 & 0 & 0 \cr 1 & 0 & 0 \cr} \\
\lambda_5= \frac{1}{\sqrt{2}} 
\pmatrix {0 & 0& -i \cr 0 & 0 & 0 \cr i & 0 & 0 \cr} \mbox{ ; }\;
\lambda_6= \frac{1}{\sqrt{2}} 
\pmatrix {0 & 0& 0 \cr 0 & 0 & 1 \cr 0 & 1 & 0 \cr} \\
\lambda_7= \frac{1}{\sqrt{2}} 
\pmatrix {0 & 0& 0 \cr 0 & 0 & -i \cr 0 & i & 0 \cr} \mbox{ ; }\;
\lambda_8= \frac{1}{\sqrt{6}} 
\pmatrix {1 & 0& 0 \cr 0 & 1 & 0 \cr 0 & 0 & -2 \cr} \;,
\end{array}
\]
which satisfy 
${\sf Tr}[ \lambda_\alpha \lambda_\beta ] = \delta_{\alpha,\beta}$.  
In this basis, the spin-1 states are represented by the 
3-dimensional vectors
\begin{equation}
\ket{1} = \pmatrix {1 \cr 0 \cr 0 \cr} \mbox{ ; } 
\ket{0} = \pmatrix {0 \cr 1 \cr 0 \cr} \mbox{ and } 
\ket{-1} = \pmatrix {0 \cr 0 \cr 1 \cr}\; .
\end{equation}
Then, the relative purity for a generic pure state $\ket{\psi}$ becomes
\begin{equation}
\label{purityspin1}
P_{\fh} (\ket{\psi})= \frac{3}{4} \sum \limits_{\alpha=1}^8 
\sum \limits_i \langle \lambda^i_\alpha \rangle ^2\;,
\end{equation}
where $\langle \lambda^i_\alpha \rangle$ denotes the expectation  value
of $\lambda^i_\alpha$ in the state $\ket{\psi}$. In this way, product
states like $\ket{\psi}=\ket{\phi_{\cA}}_{\cA} \otimes \ket{\phi_{\cB}}_{\cB}$ are
generalized unentangled ($P_{\fh}=1$) and states like $\ket{{\sf
GHZ}_1^2}$ (and states connected through local spin unitary
operations),  are maximally entangled in this algebra ($P_{\fh}=0$).

Different results are obtained if the purity is calculated relative to
a {\em subalgebra of local observables}. For example, the two-spin-1
product  state $\ket{0,0}= \ket{0}\otimes\ket{0}$, where both spins have
zero projection along $z$, becomes generalized entangled relative to the
(irreducible) algebra $\fsu(2)_{\cA} \oplus \fsu(2)_{\cB}$ of local spin
rotations, which is  generated by $\{ S_{\alpha}^i;\mbox{ }
i:{\cA},{\cB};\mbox{ } \alpha=x,y,z\}$, the spin-1 angular momentum operators
$S_\alpha$ for each spin being given by
\begin{eqnarray}
\label{pauli2}
S_x= \frac{1}{\sqrt{2}}
\pmatrix {0 & 1 & 0 \cr 1 & 0 & 1\cr 0 & 1 & 0\cr}, \mbox{ } 
S_y=\frac{1}{\sqrt{2}}
\pmatrix {0 & -i & 0 \cr i & 0 & -i \cr 0 & i & 0 \cr}, \mbox{ } 
S_z=
\pmatrix {1 & 0 & 0\cr 0  & 0 & 0\cr 0 & 0 & -1 \cr} \;. 
\end{eqnarray}
Notice that access to local angular momentum observables suffices to
operationally characterize the system as describable in terms of two
spin-1 particles (by imagining, for instance, the performance of a
Stern-Gerlach-type of experiment on each particle). Thus, even when a
subsystem decomposition can be naturally identified from the beginning
in this case, states which are manifestly separable (unentangled) in the 
standard sense may exhibit GE (see also Appendix~\ref{appD}). On the 
other hand, this is physically quite natural in the example, since 
there are no SU(2) $\times$ SU(2) group operations (local rotations from
exponentiating $\fh =\fsu(2)_{\cA} \oplus \fsu(2)_{\cB}$) 
that are able to transform the state $\ket{0,0}$ into the unentangled 
product state $\ket{1,1}$. In fact, $\ket{0,0}$ is maximally entangled with
respect to $\fsu(2)_{\cA} \oplus \fsu(2)_{\cB}$ (i.e., $P_{\fh}=0$).

The examples described in this section together with other examples of
states of bipartite quantum systems are shown in
Fig.~\ref{fig3-2}.  It is clear that calculating the purity
relative to different algebras gives information about different types
of quantum correlations present in the system.

\subsection{$N$-Spin Systems}
\label{sec3-3-2}
The $\fh$-purity distinguishes pure product states from entangled ones if it is
calculated relative to the (irreducible) algebra of local observables
$\fh=\bigoplus \limits_{j=1}^N \fsu (2S+1)_i$ (see Appendix~\ref{appD}) because
of the group invariance of the relative purity (Eq.~\ref{groupinv}), which, in
this case, constitutes all local rotations. Here $j$ denotes every subsystem
(spin) of the total system. For example,  in the previous section I denoted the
subsystem 1 to be $\cA$ (or Alice) and the subsystem 2 to be ${\cB}$ (or Bob). 
Therefore, product states are GCSs and generalized unentangled relative to the
set $\fh$ of all local observables.

For example, the usual concept 
of QE in an $N$-qubit quantum state can be 
recovered if the purity is calculated relative to the local algebra 
$\fh=\bigoplus\limits_{j=1}^N \fsu(2)_j = \{ \sigma_x^1 , \sigma_y^1,
\sigma_z^1, \cdots , \sigma_x^N , \sigma_y^N , \sigma_z^N \}$, where
the Pauli operators $\sigma_\alpha^j$ ($\alpha=x,y,z$) were introduced in
Sec.~\ref{sec2-1-1}.
The local purity is then
\begin{equation} 
\label{purity3}
P_{\fh} (\ket{\psi}) = \frac{1}{N} \sum\limits_{\alpha=x,y,z}
\sum\limits_{j=1}^N \langle \sigma_\alpha^j \rangle^2 \;,
\end{equation}
where the normalization factor ${1}/{N}$ was obtained by
setting $P_{\fh}=1$ in any product state of the form $\ket{\psi} =
\ket{\phi_1}_1 \otimes \ket{\phi_2}_2 \otimes \cdots \otimes
\ket{\phi_N}_N$ (i.e., a GCS of this algebra).  With this definition, 
states like $\ket{{\sf GHZ}_{\frac{1}{2}}^N}$, $[(\ket{01}
-\ket{10})/\sqrt{2}]^{\otimes n}$ (with
obvious notations), and the well known cluster states $\ket{\Phi}_C$ introduced
in Ref.~\cite{BR01}, are maximally
entangled ($P_{\fh}=0$). Also, Eq.~\ref{purity3} can be shown to be equivalent
to the measure of QE introduced by Meyer and Wallach in Ref.~\cite{MW02}.

\begin{figure}[hbt]
\begin{center}
\includegraphics[width=8.5cm]{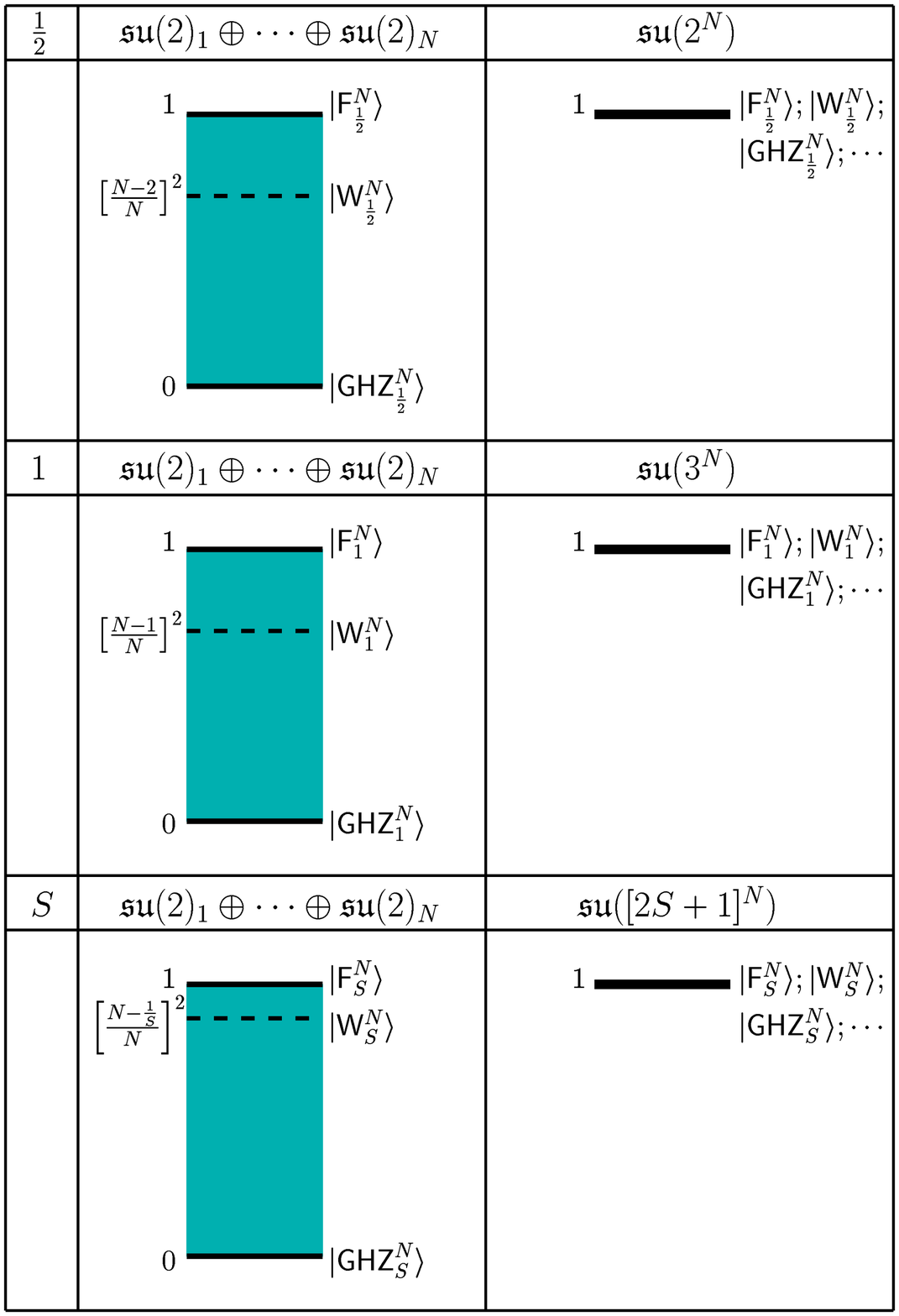}
\end{center}
\caption{Purity relative to different algebras for a $N$-spin-$S$ system. The
quantum states $\ket{{\sf GHZ}_S^N}$, $\ket{{\sf W}_S^N}$, and 
$\ket{{\sf F}_S^N}$ are defined in Eqs.~\ref{statedef}.}
\label{fig3-3}
\end{figure}

In Fig.~\ref{fig3-3}, I present some examples of the purity relative
to the local algebra $\fh=\bigoplus\limits_{j=1}^N \fsu(2)_j$ for a 
$N$-spin-$S$ system.  I also show the purity relative to the algebra of
all observables $\fsu([2S+1]^N)$,  where any pure quantum state is a
GCS, thus generalized unentangled ($P_{\fh}=1$).

\subsection{Fermionic Systems}
\label{sec3-3-3}
The case of fermionic systems is important because it shows how the concept of
GE can be widely used.
The system considered here consists of
$N$ (spinless) fermionic modes $j$, where each mode is described in terms of
canonical creation  and annihilation operators $c^{\dagger}_j$,
$c^{\;}_j$, respectively, satisfying   the anti-commutation
rules of Eqs.~\ref{fermcom}.
For instance, different modes could be associated with different sites in
a  lattice, or to delocalized momentum modes related to the spatial modes
through a Fourier transform (i.e., wave vectors).  
In general, for any $N \times N$ unitary
matrix $V=||v_{ij}||$, any transformation $c^{\;}_j \mapsto \sum_j v_{ij}
c^{\;}_j$ maps the original modes into  another possible set of
fermionic modes (Bogolubov transformation~\cite{BR86}).  

The commutation relations of quadratic fermionic operators can be obtained
using Eqs.~\ref{fermcom}, finding that
\begin{equation}
[c^{\dagger}_i c^{\;}_j, c^{\dagger}_k c^{\;}_l ] = \delta_{jk}
c^{\dagger}_i c^{\;}_l - \delta_{il} c^{\dagger}_k c^{\;}_j \;.
\end{equation}
Thus, the set of bilinear fermionic operators $\{ c^\dagger_j
c^{\;}_{j'}; \; 1 \leq j,j' \leq N \}$ provides a realization of the
Lie algebra  $\fu(N)$\footnote{A basis for the matrix Lie algebra $\fu(N)$ is
given by all real $N \times N$ matrices.} in the $2^N$-dimensional Fock space
${\cal H}_{Fock}$ of the system.  The latter is constructed as the
direct sum of subspaces ${\cal H}_n$  corresponding to a fixed fermion
number $n=0,\ldots,N$, with dim(${\cal H}_n ) = N!/[n! (N-n)!]$.   Here,
it is more convenient to express $\fu(N)$ as the linear span of  
a Hermitian, orthonormal operator basis, which can be chosen as
\begin{equation}
\fu(N)=\left\{
\begin{array}{cl}
(c^{\dagger}_j c^{\;}_{j'} + c^{\dagger}_{j'} c^{\;}_j) & 
\mbox{  with } 1\leq j<j' \leq N \cr
i(c^{\dagger}_j c^{\;}_{j'} - c^{\dagger}_{j'} c^{\;}_j) & 
\mbox{  with } 1\leq j<j' \leq N \cr 
\sqrt{2}(c^{\dagger}_j c^{\;}_j - 1/2 ) & \mbox{  with }1 \leq j \leq N
\end{array}
\right. \;,
\label{uNbasis}
\end{equation}
(the large left curly 
bracket means ``is the span of'').  The action of $\fu(N)$ on ${\cal
H}_{Fock}$ is reducible, because any operator in $\fu(N)$ conserves
the total number of fermions $n=\langle \sum_{j=1}^N c^\dagger_j c^{\;}_j
\rangle$. It turns out that the irrep decomposition of $\fu(N)$ is
identical to the direct sum into fixed-particle-number subspaces
${\cal H}_n$, each irrep thus appearing with multiplicity one.

Using Eq.~\ref{purity}, the $\fh$-purity of a generic pure
many-fermion state relative to $\fu(N)$ is 
\begin{equation}
\label{unpurity}
 P_{\fh}(\ket{\psi}) = \frac{2} {N} \sum \limits_{j<j'=1}^N  
\Big[ \langle c^{\dagger}_j c^{\;}_{j'} +
 c^{\dagger}_{j'} c^{\;}_j \rangle^2  -  \langle c^{\dagger}_j c^{\;}_{j'} -
 c^{\dagger}_{j'} c^{\;}_j \rangle^2 \Big] +\frac{4}{N} 
\sum\limits_{j=1}^N \langle
 c^{\dagger}_j c^{\;}_j -1/2 \rangle^2\;.
\end{equation}
For reasons that will become clear
shortly, the normalization factor was chosen to be ${\sf K}= 2/N$. 
In this case, the fermionic product states (Slater
determinants) of the form 
\begin{equation}
\ket{\phi}= \prod\limits_m c^\dagger_m
\ket{\sf vac}, 
\end{equation}
with $\ket{\sf vac}$ denoting the reference state 
with no fermions and $m$ labeling a particular set of modes, are 
the GCSs of the $\fu(N)$ algebra~\cite{Gil74,Per85}. 

Because a Slater
determinant carries a well defined number of particles, each GCS
belongs to an irrep space ${\cal H}_n$ for some $n$; states with
different $n$ belonging to different orbits under $\fu (N)$.  A fixed
GCS has maximum $\fh$-purity when compared to any other state
within the same irrep space.  Remarkably, it also turns out that
any GCS of $\fh=\fu(N)$ gives rise to a reduced state which is
extremal (thus generalized unentangled) regardless of $n$; the
$\fh$-purity assuming the same (maximum) value in each irrep. Using
this property, the normalization factor ${\sf K}={2}/{N}$ was
calculated by setting $P_{\fh}=1$ in an arbitrary Slater determinant.
Thus, the purity relative to the $\fu(N)$ algebra is a good measure of
entanglement in fermionic systems, in the sense that $P_{\fh}=1$ in
any fermionic product state, and $P_{\fh}<1$ for any other state,
irrespective of whether the latter has a well defined number of fermions
or not.  

Due to the invariance of $P_{\fh}$ under group
transformations (Eq.~\ref{groupinv}), the property of a state
being generalized unentangled is independent of the specific set of
modes that is chosen. That is, if $\ket{\phi}=\prod_j c^\dagger_j \ket{\sf vac}$
is generalized unentangled with respect to $\fu(N)$, so is the state
$\ket{\phi'} = \prod_m c^\dagger_m \ket{\sf vac}$, with $c^\dagger_m = \sum_j
v_{mj} c^\dagger_j$. Therefore, the $\fu(N)$-purity is a measure of entanglement
that goes beyond a particular subsystem decomposition in this case and only
distinguishes fermionic product states from those which are not.  

For example, if the system has only $N=4$ sites (modes), then a fermionic state
like $\ket{\phi}=\frac{1}{\sqrt{2}}(c^\dagger_1   c^\dagger_2 + c^\dagger_3  
c^\dagger_4) \ket{\sf vac}$ is maximally entangled relative to the algebra
$\fu(4)$ (that is, $P_{\fu(4)}=0$) because there is no basis where it
can be written as a fermionic product state. However, the state $\ket{\phi} =
\frac{1}{\sqrt{2}}(c^\dagger_1 c^\dagger_2 +  c^\dagger_1 c^\dagger_4) \ket{\sf
vac}$ is unentangled with respect to $\fu(4)$ (i.e.,
$P_{\fu(4)}=1$) because it can also be written in a certain basis as
the fermionic product state $\ket{\phi}= c^\dagger_1 c^\dagger_l \ket{\sf vac}$,
with $c^\dagger_l = \frac{1}{\sqrt{2}} (c^\dagger_2 + c^\dagger_4)$.

\section{Summary}
\label{sec3-4}
In this Chapter I have introduced a generalization of entanglement which
provides a unifying framework for defining entanglement in an arbitrary physical
setting. For this purpose, I first presented the usual notion of entanglement
which referres to a particular subsystem decomposition of the total system. The
usual concept can be naturally extended to a more general case by means of
generalized entanglement. The latter is regarded as an observer-dependent
property of quantum states, being definable relative to any physically relevant
set of observables for the system under study.

In particular, I implemented some steps for the purpose of associating the
theory of entanglement with the theory of generalized coherent states in the
Lie algebraic setting. I have shown that whenever the preferred set of
observables constitutes a Lie algebra acting irreducibly on the Hilbert space
associated with the system under study, GCSs are unentangled relative to such a
set, and possess minimal uncertainty. Such properties are also obtained in the
usual framework with respect to local observables. That is, the usual notion
of QE is recovered from GE when choosing the algebra
of all local observables, corresponding to a particular subsystem
decomposition.

Finally, some useful examples were presented to realize the dynamics of this
innovative approach.


\chapter{Generalized Entanglement as a Resource in Quantum Information}
\label{chapter4}

Because of the interesting non-classical features of entanglement, physicists
have been studying this quantum mechanical property almost since the early
years of quantum mechanics~\cite{Sch35,EPR35}. Entanglement is considered
nowadays to be a fundamental resource in quantum information processing, where
it can be exploited to perform certain tasks like quantum cryptography, quantum
teleportation, quantum simulations, etc., which are difficult or impossible in
a classical setting.  Nonetheless, it is not yet fully understood how or when
QE can be used to perform more efficient computation. In Chap.~\ref{chapter2},
for example, I have shown that certain properties related with the simulation
of physical systems, can be obtained more efficiently using a QC than a CC. As
I will explain later, this is valid only if entangled states are involved in
the simulation. However, if no entanglement is created at any step of a
deterministic QA (i.e., preparation of a pure initial state, evolution, and
measurement), the quantum state carried over the whole process remains a
product state (unentangled) and such simulation can be performed on a CC using
the same amount of resources. 

On the other hand, it is also known that the creation of entanglement is not a
sufficient condition to claim that there is no classical  algorithm able to
perform the simulation efficiently~\cite{GK,Vid03}.  In fact, certain QSs which
involve entangled states can be imitated on a CC with the same
efficiency. So, which problems can be solved more efficiently using a QC?
The answer to this question is not yet known: The lack of a
general theory of QE that can be applied to any system and any pure or mixed
quantum state is the principal reason behind the difficulty of understanding
and comparing
quantum with classical complexity in a general case. 

In Chap.~\ref{chapter3}, I introduced  the notion of GE which constitutes a
main step towards the developing of a general theory of QE. In the following
sections I will show how this novel concept leads to a better understanding
about when QE can be used as a resource for more efficient computation. In
particular, I will show that the creation of generalized entangled states 
relative to every small dimensional Lie algebra is a necessary condition for a
QA to be more efficient than the corresponding (known) classical one. This
result goes beyond the idea of creating entangled states in the traditional way
to gain efficiency.

This chapter is organized as follows: First, I present how QE can be exploited
to perform interesting protocols in quantum information processing. Second, I
focus on the study of the relation between the concept of GE with classical and
quantum complexity, showing that a wide class of problems in quantum mechanics
can be easily solved with a CC. The problem of efficient state preparation is
also considered.

\section{Quantum Entanglement and Quantum Information}
\label{sec4-1}
Here I present some examples of how the usual notion of
QE can be exploited to perform interesting tasks in quantum
information. These tasks involve the well known processes of quantum
cryptography and quantum teleportation. For simplicity, the setting used here
usually involves two qubits (or many copies of it), $\cA$ and $\cB$ for
Alice and Bob, which are initially prepared in some maximally entangled Bell
state.

\subsection{Quantum Cryptography}
\label{sec4-1-1}
Quantum cryptography~\cite{BB84} is the process that allows Alice and Bob to
share a secret conversation through the encryption of their messages. Such an idea
has been widely used in classical protocols. In a classical framework, the
secret conversation between Alice and Bob is done by sending and receiving
arrays of $0$s and
$1$s. The encryption and decryption of these arrays are done by using a key
which is assumed to be known only by them. This key is a random string of $0$s
and $1$s and its length is usually the same that the length of the message to
be sent. For example, assume that Alice wants to send Bob the binary number
$`10'$. Then Alice creates a random key, such as the array $`11'$, and sends
to Bob the encrypted message $`10' + `11' = `01'$ (i.e., sum of binary numbers). Bob
then receives the encrypted message and by using the key it proceeds to its
decryption by adding them: $`01' + `11' = `10'$, corresponding to the original message.
This cryptography method is trivially extended to longer messages.

The type of encryption described is more secure if the key used is changed in
every sent message. Otherwise, a potential eavesdropper  $\cE$ (for Eve) could
easily gain information about the key and therefore `hear' the conversation
between $\cA$ and $\cB$. Nevertheless, more secure efficient methods exist in a
quantum setting by using entangled states. 
The reason is simple: If Eve tries
to figure out the key, which is now an entangled state, she needs to perform a
measurement, and therefore destroys the entangled state by projecting it into
some other unentangled one.

\begin{figure}[hbt]
\begin{center}
\includegraphics[width=11.0cm]{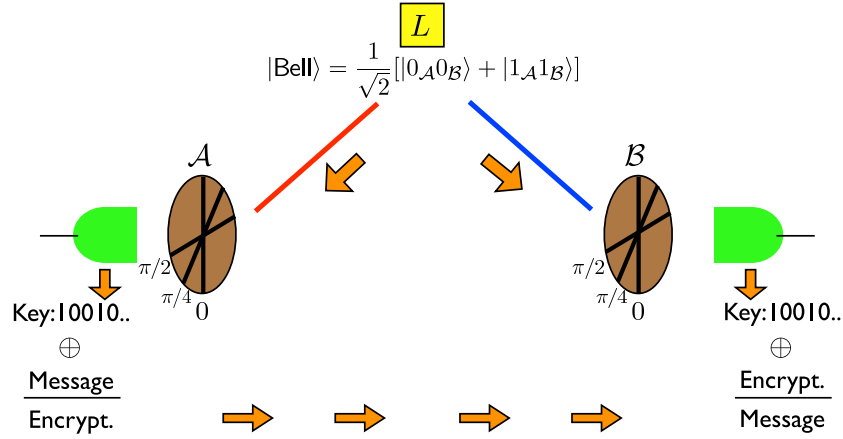}
\end{center}
\caption{Quantum cryptographic protocol. A source $L$ emits pairs of maximally
entangled photons. Alice and Bob measure their polarization in random bases and
they only keep those measured in the same direction. In this way they build a
key used to encrypt their message.}
\label{fig4-1}
\end{figure}

In more detail, the scheme for quantum cryptography consists in a source that
creates maximally entangled two-qubit Bell states $\ket{\sf
Bell}=\frac{1}{\sqrt{2}}[\ket{0_{\cA} 0_{\cB}} + \ket{1_{\cA} 1_{\cB}}]$ (e.g.,
entangling photon polarizations, etc.), where one qubit is given to Alice and
the other is given to Bob (Fig.~\ref{fig4-1}). In addition, each party can measure the state of the
corresponding qubit in a different basis, if they want, from the logical one. (As
mentioned in Chap.~\ref{chapter3}, such freedom is necessary to take advantage of the quantum
correlations.) The source then emits many copies of the Bell state and the
parties measure their corresponding qubits several times and different
orientations. After the measurements are performed, Alice and Bob have a
classical conversation where they tell each other about the chosen
orientations but they never talk about the results of the measurements. Because
of the nature of the Bell state, whenever they choose the same orientation, the
results of the corresponding measurements are the same. For example, if Alice
measures $\ket{1_{\cA}}$ in the logical basis and Bob measures in the same
basis, Alice knows that Bob measured the qubit state
$\ket{1_{\cB}}$. In this way, they only keep the results obtained whenever the
measurement is performed in the same orientation. Such results constitute
the key, which was never discussed between them, for the encryption and
decryption of the message.

Although the scheme presented seems simple and successful, an eavesdropper
could still hear the conversation by, for example, stealing entangled qubits
used to build up the key. Many different cryptography protocols have been
designed for different situations and can be found in the literature. Here
I only discussed the basics of quantum cryptography to show that entangled
states play an important role in information processes.  

\subsection{Quantum Teleportation}
\label{sec4-1-2}
Quantum teleportation~\cite{BBC93} consists of the process of teleporting the
state of a distant quantum system by means of local quantum operations,
including measurements, and classical communication (LOCC). The simplest case
considers the teleportation of the state of a single qubit. For this purpose
assume that Alice possesses two qubits denoted by ${\cA}_1$ and ${\cA}_2$, and
Bob possesses only one, denoted by $\cB$. Moreover, assume that qubits ${\cA}_2$
and $\cB$ are in a maximally entangled state which was given by an external
source. Such an entangled state can then be used to teleport the state of qubit
${\cA}_1$ to qubit $\cB$.

The idea is the following: The global three-qubit state of Alice and Bob
together is given by $\ket{\psi} = (a \ket{0_{{\cA}_1}} + b\ket{1_{{\cA}_1}})
\otimes \frac{1}{\sqrt{2}} (\ket{0_{{\cA}_2} 0_{\cB} } + \ket{1_{{\cA}_2}
1_{\cB}})$. This state can also be written as
\begin{eqnarray}
\label{telep}
\ket{\psi}= \frac{1}{2} \left[ \ket{\phi^+} (a \ket{0_{\cB}} + b\ket{1_{\cB}}) +
\ket{\phi^-} (a \ket{0_{\cB}} - b\ket{1_{\cB}}) + \right.\\
\nonumber
 \left.\ket{\xi^+} (a \ket{1_{\cB}} + b\ket{0_{\cB}}) + 
  \ket{\xi^-} (a \ket{1_{\cB}} - b\ket{0_{\cB}}) \right],
\end{eqnarray}  
where Alice's states $\ket{\phi^\pm}$ and $\ket{\xi^\pm}$ are the Bell states
\begin{eqnarray}
\label{telep2}
\ket{\phi^\pm} &=& \frac{1}{\sqrt{2}} (\ket{0_{{\cA}_1} 0_{{\cA}_2}} \pm
\ket{1_{{\cA}_1} 1_{{\cA}_2}}) \mbox{   and} \\
\nonumber
\ket{\xi^\pm} &=& \frac{1}{\sqrt{2}} (\ket{0_{{\cA}_1} 1_{{\cA}_2}} \pm
\ket{1_{{\cA}_1} 0_{{\cA}_2}}),
\end{eqnarray}
respectively. Remarkably, in this basis (Eq.~\ref{telep2}) the state of Bob's
qubit looks very similar to the state to be teleported (i.e., the state of
${\cA}_1$). Then, Alice performs a measurement in her two qubits projecting
them into one of the four possible Bell states. (Such a measurement can be done
using local operations only but I do not describe it here.). 

The state obtained by Alice corresponds to one of
the four possibilities shown in Eq.~\ref{telep2}. Alice then contacts Bob and
tells him about the result of the measurement, after which Bob acts on its qubit
to recover the state of ${\cA}_1$. For example, if Alice projects its two-qubit
state into $\ket{\xi^+}$, then she tells this result (by means of a classical
communication) to Bob. He thereby transforms the state of his qubit by using a
flip operation transforming 
\begin{equation}
a\ket{1_{\cB}} + b\ket{0_{\cB}} \rightarrow a\ket{0_{\cB}} + b\ket{1_{\cB}},
\end{equation}
which is the desired teleported state. Similar operations can be done for any
other state measured by Alice.

\section{Quantum Entanglement and Quantum Computation}
\label{sec4-2}
Perhaps the most important practical case where QE can be used as a
resource is in computational tasks. As mentioned before, parallelism is one of
the properties of the quantum world that needs to be exploited in a quantum
computation, and such a property is naturally associated to QE.
Shor's factoring algorithm~\cite{Sho94} constitutes a nice example where entangled states
are used to find the prime factors of a whole number. It has been shown that
using a QC, this algorithm can be efficiently performed; that is,
with a  number of steps that scales at most polynomially with the integer
$P$ to be factorized. However, it is not known how to perform such an algorithm
efficiently on a CC: In order to find the factors of $P$ it is
necessary to divide it by all the whole numbers between 1, and $P/2$, and so on,
constituting a time consuming task.

To demonstrate Shor's algorithm, I present the factorization of the number
$P=15$. (Although the solution to this problem is immediately found, $P=15=3
\times 5$, this method can be extended to the factorization of larger numbers
like $P'=12319$, where finding the solution is more complicated: $P'=97 \times
127$.) 
As I will show, the problem of finding the factors of $P$ is equivalent to the
order-finding problem.
Thus, a random whole number $m$ between 1 and $P-1$ needs to be chosen first.
Assume that this number is $m=8$. Second, a QA needs to be performed to find
the {\em order} $n$ of $m$, modulo $P$. Such order is defined as the least positive
integer such that $m^n=1(\mbox{ mod } P)$, where $a=b(\mbox{ mod } P)$ if $a-b$
is divisible by $P$. 

Then, the first step of the QA consists of performing $r$
Hadamard gates to the initial $(r+4)$-qubit
polarized state $\ket{0_1 \cdots 0_r} \otimes
\ket{0_a 0_b 0_c 0_d}$. The number $r$ of extra qubits has to be big enough to
find the order $n$ with high accuracy~\cite{Sho94}. Here, $r=11$. The other 4 qubits are
necessary to encode information about the factors of $15$ (i.e., $2^4=16>15$). 
 
After the Hadamard gates have been applied 
(i.e., the gates that transform $\ket{0} \rightarrow \ket{+}$), 
the evolved state is
\begin{equation}
\ket{\psi} = \frac{1}{\sqrt{2^r}}\sum\limits_{j=0}^{2^r-1} \ket{j} \otimes
\ket{0_a 0_b 0_c 0_d},
\end{equation}
where $\ket{j}$ is the state corresponding to the binary decomposition of the
integer $j$ in the logical basis; for example, $\ket{j=3}=\ket{0_1 \cdots
1_{r-1} 0_r}$. The next step of the algorithm is to apply a unitary operation
that transforms
\begin{equation}
\ket{j} \otimes \ket{0_a 0_b 0_c 0_d} \rightarrow \ket{j} \otimes \ket{m^j
(\mbox{ mod }P)} .
\end{equation}
This operation can be efficiently performed using controlled operations on the
state of the $r$ qubits, but I will not explain it here. After this evolution,
the evolved state is
\begin{eqnarray}
\label{shor2}
\ket{\psi} &=& \frac{1}{\sqrt{2^r}}\sum\limits_{j=0}^{2^r-1} \ket{j} \otimes
\ket{m^j (\mbox{ mod }P)} \\
\nonumber
&=& \frac{1}{\sqrt{2^r}} \left[ \ket{0}\otimes \ket{1} +
\ket{1} \otimes \ket{8} +  \ket{2} \otimes\ket{4} + \ket{3} \otimes \ket{2} +
\cdots \right],
\end{eqnarray}
where again I have chosen the integer representation of the states in the
logical basis. The state of Eq.~\ref{shor2} is a highly entangled state between
the $r=11$ qubits and qubits $a$, $b$, $c$, and $d$.

By defining the states $\ket{\mu_q}$ for qubits $a,b,c,d$, as
\begin{equation}
\ket{\mu_q}= \frac{1}{\sqrt{n}} \sum\limits_{k=0}^{n-1} \exp \left[ \frac{-i 2
\pi q k}{n} \right] \ket{m^k (\mbox{ mod }P)},
\end{equation}
where again $n$ is the not yet known order, Eq.~\ref{shor2} can be written as
\begin{equation}
\ket{\psi} = \frac{1}{\sqrt{n2^r}} \sum\limits_{q=0}^{n-1}
\sum\limits_{j=0}^{2^r-1} \exp \left[ i \frac{2 \pi qj}{n} \right] \ket{j}
\otimes \ket{\mu_q}.
\end{equation} 
A (inverse) quantum Fourier transform~\cite{NC00} performed to the $r$ qubits in the
above state maps it to
\begin{equation}
\frac{1}{\sqrt{2^r}} 
\sum\limits_{j=0}^{2^r-1}\exp \left[ i \frac{2 \pi qj}{n} \right] \ket{j}
\rightarrow
\ket{\overline{K=q/n}},
\end{equation}
where the state $\ket{\overline{K=q/n}}$ of the $r$ qubits is a product state which
depends on the coefficient $q/n$ as in the usual Fourier transform. This
transformation can also be efficiently implemented in a quantum circuit.
Therefore, the transformed state reads
\begin{equation}
\label{shor3}
\ket{\psi} = \frac{1}{\sqrt{n}} \sum\limits_{q=0}^{n-1} \ket{\overline{K=q/n}} \otimes
\ket{\mu_q}.
\end{equation}
In this way, the register of $r$ qubits is measured in the logical basis to
project the state $\ket{\psi}$ and obtain one possible state
$\ket{\overline{K=q/n}} \otimes \ket{\mu_q}$. After applying the same algorithm
several times and by applying a continued-fractions algorithm, it is
possible to obtain the order $n$ from the statistics of the measured states
$\ket{\overline{K=q/n}}$.  In this case one would obtain $n=4$, such that
$8^4=4096=1 (\mbox{ mod }15)$. Such a solution could be obtained if the result of
the measurement in the register of $r$ qubits gives, for example,
$\ket{\overline{K=q/n}}=\ket{1536}$ (in the binary representation), 
where I have used again the integer
representation of states in the logical basis. If this is the case,
$q/n=\frac{1536}{2^{11}}=\frac{3}{4}$, resulting in $n=4$. From this
information, one can obtain that $15=3 \times 5$ as it is explained in
Ref.~\cite{Sho94}.

A factoring QA such as the one presented is more efficient than a classical one
only because entangled states have been created at some step. Otherwise, no
advantages can be gained, in general, from using a deterministic QC. As a
simple example, suppose (with no loss of generality) that the initial pure
state of a register of $N$ qubits is the usual polarized state $\ket{0_1 \cdots
0_N}$, and that the QA never does create QE.
That is, the set of unitary
gates $U_j$ applied to the initial state are only single qubit rotations
transforming product states into product states. Therefore,
\begin{equation}
U_j = e^{i \theta_j \sigma^{\alpha(j)}_{\pi(j)} } = \cos \theta + i \sin \theta
\sigma^{\alpha(j)}_{\pi(j)},
\end{equation}
where $\pi(j)=[1 \cdots N]$ denotes the qubit rotated at the $j$th step, and
$\alpha(j)=[x,y,z]$ is the axis of rotation. The
evolved state is then 
\begin{equation}
\prod_j U_j \ket{0_1 \cdots 0_N}.
\end{equation}
After the evolution is performed, certain qubit(s) is(are) measured in, for
example, the logical basis. As usual, this is the last step of a deterministic
QA (Sec.~\ref{sec2-2}). If the $k$th qubit is
measured, the result obtained is
\begin{equation}
\label{unentalg}
\langle \sigma_z^k \rangle =\bra{0_1 \cdots 0_N} \left(\prod_j U^\dagger_j
\right)
\sigma_z^k \left(\prod_j U_j \right) \ket{0_1 \cdots 0_N}.
\end{equation}

The fact is that such QA can be simulated classically and the expectation
value of Eq.~\ref{unentalg} can be computed in a CC with the same amount of
resources (i.e., same efficiency).
The idea is to only keep track of the information about $[\theta_j,
\alpha(j)] $  if $\pi(j)=k$, because the other qubits are not measured.
If the QA is performed efficiently,
only a polynomially large
number of pairs  $[\theta_j, \alpha(j)] $ are kept. A classical algorithm
to obtain the result of Eq.~\ref{unentalg} would consist of updating at each
step the state of the $k$th qubit only. For example, if at the $j$th step the state
is $a_j \ket{0_k} + b_j \ket{1_k}$ and the gate $U_{j+1}= e^{i
\frac{\pi}{4} \sigma_x^k}$ is performed after, the evolved state becomes
\begin{equation}
[\cos(\pi/4) + i \sin(\pi/4) \sigma_x^k] (a_j \ket{0_k} + b_j \ket{1_k} )=
a_{j+1} \ket{0_k} + b_{j+1} \ket{1_k},
\end{equation}
and since $\sigma_x^k$ is the corresponding flip operation,
$a_{j+1}=\frac{1}{\sqrt{2}}(a_j+ib_j)$ and $b_{j+1}=\frac{1}{\sqrt{2}} (ia_j +
b_j)$. Updating the values of the complex coefficients $a_j$ and
$b_j$ can be easily done with a conventional computer. At the end of the total
evolution the state of the $k$th qubit is $a_M \ket{0_k} + b_M \ket{1_k}$.
Therefore, the result of Eq.~\ref{unentalg} is
\begin{equation}
\langle \sigma_z^k \rangle = |a_M|^2 - |b_M|^2,
\end{equation}
which can be efficiently computed with a CC.

However, if some of the gates $U_j$ create entanglement, there is no known way,
in general, to classically evaluate Eq.~\ref{unentalg} efficiently. Naturally,
the elementary two-qubit gates, like the $R_{z^jz^k}(\omega)$ or the CNOT gate
introduced in Sec.~\ref{sec2-1-3}, are responsible for creating entanglement. For
example, if the initial state is the product (unentangled) state 
$\frac{1}{\sqrt{2}} \ket{0+1} \otimes \ket{0}$, applying the CNOT gate evolves
it into $\frac{1}{\sqrt{2}}(\ket{00}+\ket{11})$, which is known to be one of
the Bell states, that is, maximally entangled. Nonetheless, I will show in the next
section that creation of QE is not a sufficient condition to claim that
Eq.~\ref{unentalg} cannot be obtained efficiently by using a classical
algorithm.

\section{Efficient Classical Simulations of Quantum Physics}
\label{sec4-3}
The purpose of this section is to generalize the concepts described previously,
presenting a wide set of problems which can be solved efficiently with both, a
QC or a classical one. To show this, I will exploit the notion of GE, which was
presented in Sec.~\ref{sec3-2}. Again, an algorithm is said to be efficient  with
respect to a certain variable $N$, whenever the amount of resources required to
perform it is bounded by a polynomial in $N$. Otherwise, the algorithm is said
to be inefficient. It is important to remark that here I only make a
comparison between efficient and inefficient algorithms, with the previous
definition, and do not compare the total amount of resources needed. For
example, if a QA can be performed in $N$ operations and
the corresponding classical one in $N^2$ operations then, despite being more
convenient to use the former, both are considered efficient.

In most cases, the type of problems one encounters in quantum mechanics are
related with the evaluation of a certain expectation value or correlation function
\begin{equation}
\label{exp1}
\langle \hW \rangle = \bra{\phi} \hat{W} \ket{\phi},
\end{equation}
where $\hW$ is an operator acting on
a finite quantum system ${\cal S}$ in some state
$\ket{\phi} \in \cH$. ($\cH$ is the Hilbert space associated to $\cS$.) In
Sec.~\ref{sec2-2}, I presented some QAs that allow one to compute 
Eq.~\ref{exp1} in a QC described by the conventional model,
whenever the operator algebra associated to the
system $\cS$ could be mapped onto the Pauli operators (Sec.~\ref{sec2-3}). 
The main steps of those QAs consist of the preparation of the state
$\ket{\psi}=\ket{+}_{\sf a} \otimes \ket{\phi}$, from some initial (boot-up)
reference state like $\ket{0}_{\sf a} \otimes \ket{0 \cdots 0}$,  then an
evolution, and finally the measurement of the ancilla qubit ${\sf a}$.
Similarly, in the rest of this section I assume that the algebra of
operators used to describe the quantum computational model and/or the algebra
associated to the system $\cS$ are not necessarily described by the
conventional model, but a one-to-one efficient mapping between both exists.

Since I am mainly interested in comparing the efficiencies of obtaining
Eq.~\ref{exp1} using a QC or a classical (conventional) one, the efficiency is
defined relative to a certain variable $N$ which here is, in general, the
number of quantum mechanical elements that compose the system $\cS$. I  assume
that  there exists a QA that evaluates Eq.~\ref{exp1} efficiently, that is,
using at most $\mbox{poly}(N)$ resources. These are the so-called {\em bounded
polynomial quantum algorithms} (BPQA). This is usually valid whenever the QC
can be efficiently initialized and every step, including the preparation of 
$\ket{\phi}$, can be efficiently performed. These results were previously
discussed in Sec.~\ref{sec2-5}.

The most common classical algorithm to evaluate Eq.~\ref{exp1} consists of
writing the matrix representation of $\hW$, and obtaining the corresponding
matrix element. This is usually a hard task because of the dimension of the
Hilbert space $\cH$ growing exponentially with $N$. However, in some cases this
complexity can be highly reduced. The simplest case is when the state
$\ket{\phi}$ of Eq.~\ref{exp1} is known to be an eigenstate of $\hW$, so that
the corresponding expectation is simply the known eigenvalue. 

Nonetheless, a more
general result can be obtained in a Lie algebraic framework. For this purpose, 
I start by pointing out that, with no loss of generality, the state
$\ket{\phi} \in \cH$ of Eq.~\ref{exp1} can be obtained by transforming any
reference state $\ket{\sf ref} \in \cH$; that is
\begin{equation}
\label{expstate}
\ket{\phi}= U \ket{\sf ref},
\end{equation}
where $U^\dagger = U^{-1}$ (unitary). Moreover, the reference state can be
chosen to be the highest (or lowest) weight state $\ket{\sf HW}$ of certain
finite semi-simple Lie algebra $\fh=\{ \hO_1, \hO_2, \cdots, \hO_M \}$, with
$\hO_j = \hO_j^\dagger$ (Sec.~\ref{sec3-2-2}). In this case, $\fh$ also admits
a Cartan-Weyl decomposition: $\fh=\{ \fh_D, \fh_+, \fh_- \}$, with $\fh_D$ the
Cartan subalgebra (CSA) of $\fh$, and $\fh_+$ and $\fh_-$ the set of raising and lowering
operators, respectively (Sec.~\ref{sec3-2-2}).

Now suppose that the transformation $U$ of Eq.~\ref{expstate} is a group operation
induced by $\fh$; that is
\begin{equation}
\label{unitdec}
U = \exp \left[ i \sum\limits_{j=1}^M \zeta_j \hO_j \right] .
\end{equation}
Then $\ket{\psi}$ is a GCS and generalized unentangled state
(extremal) of $\fh$. Moreover, suppose that $\hW$ of Eq.~\ref{exp1} can
be decomposed as
\begin{equation}
\label{observdec}
\hW = \sum\limits_{j=1}^M \kappa_j \hO_j,
\end{equation}
where $\kappa_j \in \mathbb{C} \ \forall j$, and $\hO_j \in \fh$. 
In general, $\hW$ does not have the form of Eq.~\ref{observdec} and it belongs
to a  Lie algebra other than $\fh$. Then, $\fh$ must be considered as the
Lie algebra that contains both. (The important case when $\hW \in H \equiv e^{i \fh}$,
i.e. the group induced by $\fh$, is discussed in the next section.)

In this way, computing classically Eq.~\ref{exp1} is equivalent to evaluating
\begin{equation}
\label{exp2}
\langle \hW \rangle = \sum\limits_{j=1}^M \kappa_j
\bra{\sf HW} e^{- i \sum\limits_{j=1}^M \zeta_j \hO_j } \hO_j
e^{ i \sum\limits_{j=1}^M \zeta_j \hO_j } \ket{\sf HW}.
\end{equation}
The operators
\begin{equation}
\label{transop}
\hO'_j = e^{- i \sum\limits_{j=1}^M \zeta_j \hO_j } \hO_j 
e^{ i \sum\limits_{j=1}^M \zeta_j \hO_j }= \sum\limits_{j'=1}^M
\nu_{jj'} \hO_{j'}
\end{equation}
are also elements in $\fh$ because they are transformed by a group operation.
The coefficients $\nu_{jj'}$ of Eq.~\ref{transop} can be computed by means of the
representation theory. A {\em matrix representation} of $\fh$ is the mapping $\Phi :
\fh \rightarrow \mathbb{C}^{2p} $ such that
\begin{eqnarray}
\label{repres}
\Phi(\hO_j) &= &\bar{O}_j ,\\
\nonumber
\Phi ( [\hO_j, \hO_{j'}] ) &=& \left[ \Phi(\hO_j) , \Phi(\hO_{j'}) \right] =
\left[ \bar{O}_j, \bar{O}_{j'} \right],
\end{eqnarray}
where $[A,B]$ is the commutator and
the complex matrices $\bar{O}_j$ are $p \times p$ dimensional. A 
representation $\Phi$ is said to be {\em faithful} when maps linearly independent
elements in $\fh$ to linearly independent $p \times p$ matrices. If the Lie
algebra $\fh$ is compact, one can always find a faithful representation
satisfying the inner product
\begin{equation}
\label{innerrep}
{\sf Tr} [\bar{O}_j \bar{O}_{j'}] = \delta_{jj'},
\end{equation}
with ${\sf Tr}[\bar{Q}]$ being the trace of $\bar{Q}$. For example, the well known adjoint
representation (Appendix~\ref{appC}), which associates an $M \times M$ matrix to every
operator in $\fh$ ($M$ is the dimension of $\fh$), is faithful and satisfies
Eq.~\ref{innerrep}. Nevertheless, smaller dimensional faithful representations
($p<M$) can usually be found. 
Then, if $\Phi$ is faithful,
\begin{equation}
\label{transop2}
\bar{O}'_j = e^{- i \sum\limits_{j=1}^M \zeta_j \bar{O}_j } \bar{O}_j 
e^{ i \sum\limits_{j=1}^M \zeta_j \bar{O}_j }= \sum\limits_{j'=1}^M
\nu_{jj'} \bar{O}_{j'},
\end{equation}
which is the matrix representation of Eq.~\ref{transop}. Since this is a matrix operation,
the matrices $\bar{O}'_j$ of Eq.~\ref{transop2} can be
classically computed with ${\cO}(p^2)$ computational
operations\footnote{Here ${\cO}$ denotes
the order of required operations.} (addition and multiplication of complex
numbers). Each coefficient
$\nu_{jj'}$ is then classically obtained by using the inner product equation, that is
\begin{equation}
\nu_{jj'} ={\sf Tr}[\bar{O}'_j \bar{O}_{j'}],
\end{equation}
which also requires ${\cO}(p^2)$
computational operations\footnote{ 
To obtain $\nu_{jj'}$ at accuracy
$\epsilon$ the order of $\mbox{poly}(1/\epsilon)$ operations are needed.
Nevertheless, in this section
I consider that the calculation is done at the computational accuracy given by
the particular type of CC used.}.

Similarly, in the CW basis,
\begin{equation}
\label{transop3}
\hO'_j = \sum\limits_{k=1}^r \gamma_{jk} \hh_k + \sum\limits_{j'=1}^{l}
\iota_{jj'} \hE_{\alpha_{j'}} + \iota_{jj'}^* \hE_{-\alpha_{j'}}\,
\end{equation}
where $\hh_k \in \fh_D$ (CSA), 
$E_{\alpha_{j'}} \in \fh_+$, and $E_{-\alpha_{j'}} \in \fh_-$
(Sec.~\ref{sec3-2-2}). The
coefficients $\gamma_{jk}$ and $\iota_{jj'}$ of Eq.~\ref{transop3} can also be
obtained classically with ${\cO}(p^2)$ computational operations by the trace inner product:
The operators $\hh_k$, $\hE_{\alpha_{j'}}$, and
$\hE_{-\alpha_{j'}}$ are usually a
simple linear combinations of the operators $\hO_j$.

By definition, $\ket{\sf HW}$ is an eigenstate of the operators in the CSA, 
that is
\begin{equation}
\hh_k \ket{\sf HW} = e_k \ket{\sf HW},
\end{equation}
where the eigenvalues $e_k$ are assumed to be known. Then,
\begin{equation}
\bra{\sf HW} \hE_{\alpha_{j'}} \ket{\sf HW} =
\bra{\sf HW} \hE_{-\alpha_{j'}} \ket{\sf HW} =0 \ \ \forall j'.
\end{equation}
In other words,
Eq.~\ref{exp2} can be classically computed by only keeping the elements of
$\hO'_j$ (Eq.~\ref{transop}) belonging to $\fh_D$:
\begin{equation}
\label{exp3}
\bra{{\sf HW}} \hO'_j \ket{{\sf HW}} = \sum\limits_{k=1}^r \gamma_{jk} \bra{\sf HW} \hh_k
\ket{\sf HW} = \sum\limits_{k=1}^r \gamma_{jk} e_k,
\end{equation}
and
\begin{equation}
\label{exp4}
\langle \hW \rangle = \sum\limits_{j=1}^M \sum\limits_{k=1}^r \kappa_j
\gamma_{jk} e_k.
\end{equation}
Since the coefficients $e_k$ are assumed to be known, and the coefficients
$\kappa_j$ and $\gamma_{jk}$ can be classically obtained with ${\cO}(p^2)$, 
Eq.~\ref{exp4} can be computed with ${\cO}(Mrp^2)$ computational operations. If
$M \le \mbox{poly}(N)$, 
then $r\le \mbox{poly}(N)$. Also, the existence of
the adjoint representation (Appendix~\ref{appC}) guarantees that one can always find $p$ such
that $p \le M \le \mbox{poly}(N)$. 
Therefore, Eq.~\ref{exp4} can be efficiently
computed with a conventional computer. A more detailed proof about the number of
operations required, including the approximation of the exponential matrix, is
given in the Appendix~\ref{appaux1}.

In brief, when the state $\ket{\phi}$ of Eq.~\ref{exp1} is a GCS
of a polynomially large dimensional Lie algebra $\fh$, and the
operator $\hat{W}$ is an element of $\fh$, Eq.~\ref{exp1} can be
efficiently computed with both, a classical and a quantum device.

However, in some cases $\hW$ or $U$ (Eq.~\ref{expstate}) are associated with an
exponentially large dimensional Lie algebra and the corresponding classical 
simulation is inefficient.
Nevertheless, it
is important to remark that for mixed states of the form 
\begin{equation}
\label{expstate2}
\rho = \sum\limits_{s=1}^L p_s \rho_s,
\end{equation}
where $\rho_s = \ket{\phi_s} \bra{\phi_s}$, 
and $\ket{\phi_s}$ being GCSs and generalized unentangled states
of a polynomially large dimensional Lie algebra,
the computation of
\begin{equation}
\langle \hW \rangle = {\sf Tr}(\rho \hW),
\end{equation}
can also be efficiently performed with a CC if $L \le \mbox{poly}(N)$.

\subsection{Higher Order Correlations}
\label{sec4-3-1}
The previous results can be extended to the computation of higher order
correlations. For simplicity, I start by studying the case of two-body
correlations of the form
\begin{equation}
\label{exp6}
\langle \hW \rangle = \bra{\phi} \hW_1 \hW_2 \ket{\phi},
\end{equation}
where $\hW$ is the product of the observables $\hW_1=\hW_1^\dagger$  and
$\hW_2= \hW_2^\dagger$. Again, $\hW_1 $ and $\hW_2$ are considered to be
elements of a Lie
algebra $\fh=\{ \hO_1, \cdots, \hO_M \}$, and $\ket{\phi}$ is assumed to be a
GCS of $\fh$; that is 
\begin{equation}
\ket{\phi}=U \ket{\sf HW},
\end{equation}
with $U= \exp \left[ i \sum_j \zeta_j \hO_j \right]$, and $\ket{\sf HW}$ the
highest weight state of $\fh$. In this way, Eq.~\ref{exp6} reads
\begin{equation}
\bra{\phi} \hW_1 \hW_2 \ket{\phi} = \bra{\sf HW} U^\dagger
\hW_1 \hW_2 U \ket{\sf HW}.
\end{equation}
Since $UU^\dagger= \one$,
\begin{equation}
\bra{\phi} \hW_1 \hW_2 \ket{\phi} = \bra{\sf HW} \hW'_1 \hW'_2 \ket{\sf HW},
\end{equation}
where $\hW'_i=U^\dagger \hW_i U$.

Equation~\ref{exp6} can be efficiently computed with a classical computer if the
dimension $M$ of $\fh$ scales at most polynomially with the variable $N$; i.e.
$M \le \mbox{poly}(N)$. To show this, I first decompose the corresponding
operators as
\begin{eqnarray}
\hW_1 = \sum\limits_{j=1}^M \kappa^1_j \hO_j, \\ 
\hW_2= \sum\limits_{j=1}^M \kappa^2_j \hO_j,
\end{eqnarray}
which transform under the action of $U$ as
\begin{equation}
\hW'_i=U^\dagger \hW_i U = \sum_{j=1}^M  \kappa^i_j \hO'_j \ ; \ i=1,2.
\end{equation}
In a CW basis of $\fh$, the transformed observables $
\hO'_j=U^\dagger \hO_j U$ can be decomposed as in Eq.~\ref{transop3}. Such
decomposition can be done efficiently with a CC (i.e., computing
the coefficients $\gamma_{jk}$ and $\iota_{jj'}$) by working in low-dimensional
faithful representation
of $\fh$
(Appendix~\ref{appC}). Then,
\begin{equation}
\langle \hW_1 \hW_2 \rangle= \sum\limits_{i,j=1}^M \kappa_i^1 \kappa_{j}^2
\bra{\sf HW} \hO'_i \hO'_{j} \ket{\sf HW},
\end{equation}
with
\begin{eqnarray}
\nonumber
\bra{\sf HW} \hO'_i \hO'_{j} |&{\sf HW} \rangle= \sum\limits_{k,k'=1}^r
\gamma_{ik}
\gamma_{jk'} \langle \hh_k \hh_{k'} \rangle +
\sum\limits_{i',j'=1}^{l} \left[\iota_{ii'}\iota_{jj'} \langle \hE_{\alpha_{i'}}
\hE_{\alpha_{j'}} \rangle   \right. \\
\nonumber
& +  \left. 
\mbox{C.C.} + \iota_{ii'} \iota^*_{jj'} \langle \hE_{\alpha_{i'}}
\hE_{-\alpha_{j'}} \rangle + \iota^*_{ii'} \iota_{jj'} \langle \hE_{-\alpha_{i'}}
\hE_{\alpha_{j'}} \rangle \right]  \\
\label{exp7}
&+ \sum\limits_{k=1}^r \sum\limits_{j'=1}^{l} \left[ \gamma_{ik} \iota_{jj'} 
\langle \hh_k \hE_{\alpha_{j'}} \rangle +   \gamma_{ik} \iota^*_{jj'}  
\langle \hh_k \hE_{-\alpha_{j'}} \rangle + \mbox{C.C.}
\right] ,
\end{eqnarray}
where C.C. denotes the complex conjugate, and the expectation values are now
taken over the highest weight state, i.e. $\langle \hA \rangle = \bra{\sf HW}
\hA \ket{\sf HW}$. Since raising operators annihilate $\ket{\sf HW}$ 
only a few expectations do not vanish in
Eq.~\ref{exp7}. These are
\begin{eqnarray}
\langle \hh_k \hh_{k'} \rangle &=& e_k e_{k'} ,\\
\langle E_{\alpha_{j'}} E_{-\alpha_{j'}} \rangle & = & 
\langle E_{-\alpha_{j'}} E_{\alpha_{j'}} + \sum\limits_{k=1}^r \alpha_{j'}^k \hh_k
\rangle= \sum\limits_{k=1}^r \alpha_{j'}^k e_k ,
\end{eqnarray}
where the coefficients $\alpha_j^k$ are the components of the root vector
$\alpha_j$ (Sec.~\ref{sec3-2-2}) and $e_k$ are the known
weights of $\ket{\sf HW}$.

In brief, Eq.~\ref{exp6} takes the form
\begin{equation}
\label{exp5}
\bra{\phi} \hW_1 \hW_2 \ket{\phi} = \sum\limits_{i,j=1}^M \kappa_i^1
\kappa_{j}^2 \left[ 
\sum\limits_{k,k'=1}^r \gamma_{ik} \gamma_{jk'} e_k e_{k'} +
\sum\limits_{j'=1}^{l} \sum\limits_{k=1}^r 
\iota_{ij'} \iota^*_{jj'} \alpha_{j'}^k
e_k \right],
\end{equation}
which can be efficiently computed with a CC if $M \le
\mbox{poly}(N)$.

Higher order correlation functions of the form
\begin{equation}
\label{ho01}
\langle \hW \rangle = \bra{\phi} \hW^p  \cdots \hW^1 \ket{\phi}
\end{equation}
can also be computed efficiently  for a fixed integer $R$, whenever there is a
polynomially large dimensional Lie algebra $\fh$ associated to the problem.
Obviously, the complexity of this problem increases as $p$ does. Again, the
idea is to keep only the nonzero expectations of the transformed operator
$\hW'=U^\dagger  \hW^p  \cdots \hW^1 U = \hW'^p \cdots \hW'^1$,
where $\ket{\phi}=U \ket{\sf HW}$. For this purpose, all the raising operators
$\hE_{\alpha_j}$ appearing in $\hW'$ must be taken to the right side, by using
the commutation relations of the algebra, so that they destroy the state
$\ket{\sf HW}$. This problem is analogous to the one given by Wick's
theorem~\cite{Wic50} for fermionic operators. In Appendix~\ref{appaux2}, I
describe an efficient method for the classical computation of
Eq.~\ref{ho01}.



\section{Generalized Entanglement and Quantum Computation}
\label{sec4-4}
In Sec.~\ref{sec4-3} I presented a class of classical algorithms that allows 
one to efficiently obtain certain correlation functions of quantum systems in a
conventional computer. Here, I analyze some advantages of having a QC. For
example, assume then that one is interested in the evaluation of the
correlation function of Eq.~\ref{exp1}, when there is no polynomially large
dimensional Lie algebra $\fh$ associated with the problem. That is, either
$\ket{\phi}$ is not a GCS of $\fh$ or $\hW$ is not an element or group element
induced by $\fh$.  Then,  if a {\em bounded polynomial quantum algorithm}
(BPQA) like the ones introduced in
Sec.~\ref{sec2-2} exists in this case, Eq.~\ref{exp1} can be
efficiently computed using a QC. Usually, no known classical algorithm can be
used to compute such a correlation efficiently and this  represents, in some
cases, an exponential speed-up of the quantum simulation with respect to the
classical one.

In Sec.~\ref{sec2-5}, I discussed the complexity of simulating physical systems
using QCs. Although certain QSs have been shown to be efficient, others, like
the obtention of the ground state properties in the two-dimensional Hubbard
model, remained inefficient. The problem relied in the complexity of preparing
the desired initial state $\ket{\phi}$. When simulating quantum systems, this
state is usually not completely known (e.g., it is the ground state of some
non-solvable Hamiltonian) and an approximated state, having a non-zero overlap
with $\ket{\phi}$, needs to be prepared. The purpose of the following
section is then to show a wide class of quantum states that can be prepared
efficiently by quantum networks.  Again, the results obtained are independent
of the physical representation of the QC whenever an efficient
mapping between the operators used to describe the system to be simulated and
the operators describing the computational model, exists.

\subsection{Efficient Initial State Preparation}
\label{sec4-4-1}
I start by describing the simple case when the initial state to be prepared is given
by
\begin{equation}
\label{iniprep1}
\ket{\phi} = U \ket{\sf HW},
\end{equation}
where $\ket{\sf HW}$ is the highest weight of some Lie algebra $\fh$, and is
considered to be the boot-up state of the computer, which can be efficiently
initialized. If the operator $U$ of Eq.~\ref{iniprep1} is a group operation
induced by
$\fh$, that is, $U=e^{i\hA}$ and $\hA\in \fh=\{\hO_1, \cdots, \hO_M\}$, and 
$M \le \mbox{poly}(N)$, then
$\ket{\phi}$ can be prepared efficiently with arbitrary accuracy by using, for
example, a first order Trotter decomposition~\cite{Suz93}. 
In this case, the state $\ket{\phi}$ is
generalized unentangled and GCS of $\fh$. 
Then, $\hA = \sum_{j=1}^M
\zeta_j \hO_j$ and
\begin{eqnarray}
e^{i \hA} = \prod e^{i \hA \Delta t}, \\
e^{i \hA \Delta t} \approx \prod\limits_{j=1}^M e^{i \zeta_j \hO_j \Delta t}.
\end{eqnarray}
Assuming that every operation $e^{i \zeta_j \hO_j \Delta t}$ is either an elementary
gate or can be performed by applying a small number of them, the evolution
$e^{i \hA}$
can be then approximated by applying a polynomially large number of these gates. 
Particular examples were discussed in Sec.~\ref{sec2-3}.

Another interesting case is when the state $\ket{\phi}$ is not given in the
form of Eq.~\ref{iniprep1} but some information, like the expectation values of
a set of observables (i.e., its reduced state), $\langle \hO_j \rangle$, is
known or can be easily obtained. If $\fh =\{\hO_1, \cdots, \hO_M\}$ is a
semi-simple Lie algebra, with $M \le \mbox{poly}(N)$, and $\ket{\phi}$ is shown
to be a GCS of $\fh$ by calculating, for example, the $\fh$-purity
(Sec.~\ref{sec3-2-2}), then it can be prepared efficiently with a QC.
To show this, one needs to obtain first the operator
$\hA=\sum_{j=1}^M \zeta_j \hO_j$ such that $U= e^{i \hA}$
is the transformation of Eq.~\ref{iniprep1}.

For this purpose, I define the fictitious Hamiltonian
\begin{equation}
H_F= -\sum\limits_{j=1}^M \langle \hO_j \rangle \hO_j,
\end{equation}
whose expectation value over the state $\ket{\psi}$ is
\begin{equation}
\label{fictexpec}
\bra{\psi} H_F \ket{\psi}= -\sum\limits_{j=1}^M \langle \hO_j \rangle^2.
\end{equation}
If $\ket{\tilde{\psi}}= V \ket{\psi}$ is another GCS of $\fh$, with
$V$ a unitary group operation that transforms as
\begin{equation}
V^\dagger \hO_j V = \sum_{j'=1}^M \nu_{jj'} \hO_{j'},
\end{equation}
then
\begin{equation}
\bra{\tilde{\psi}} H_F \ket{\tilde{\psi}}= -\sum\limits_{j,j'=1}^M
\langle \hO_j \rangle \nu_{jj'}\langle \hO_{j'} \rangle.
\end{equation}
Since the $M \times M$ matrix defined by the real coefficients $\nu_{jj'}$ is
orthogonal, one obtains (Schur's inequality)
\begin{equation}
\bra{\tilde{\psi}} H_F \ket{\tilde{\psi}} \ge \bra{\psi} H_F \ket{\psi}.
\end{equation}
Therefore, $\ket{\psi}$ is the unique GCS and ground state
that minimizes such expectation value.

Assume now that $H_F$ can be efficiently diagonalized with a conventional
computer, obtaining
\begin{equation}
H_F = U H_0 U^\dagger,
\end{equation}
where $U$ is a unitary group operation and 
\begin{equation}
\label{unperturbed}
H_0 = \sum\limits_{k=1}^r \gamma_k \hh_k \ ,
\end{equation}
is an unperturbed Hamiltonian (i.e., $\hh_k$ are elements in the CSA of $\fh$).
Such a diagonalization algorithm is described in Sec.~\ref{sec5-2-1}. 
Defining $\ket{\bar{\psi}} = U^\dagger \ket{\psi}$, one obtains
\begin{equation}
\bra{\psi} H_F \ket{\psi} = \bra{\bar{\psi}} H_0 \ket{\bar{\psi}},
\end{equation}
which takes its minimum value, by definition, 
for a certain weight state $\ket{\phi}$. In particular, the coefficients $\gamma_k$ of
Eq.~\ref{unperturbed} can be chosen such that $\ket{\bar{\psi}}= \ket{\sf
HW}$ (see Chap.~\ref{chapter5}).
Therefore, $\ket{\psi} = U \ket{\sf HW}$, and $U$ is the desired group
operation. Again, $U$ can be approximated by applying a polynomially large
number of elementary gates so $\ket{\psi}$ can be prepared efficiently with a
QC.

\section{Summary}
In this chapter I have addressed several issues relating the concept of GE with
quantum and classical complexity. I have shown that the creation of GE relative
to every polynomially large dimensional
Lie algebra is a necessary requirement to gain
efficiency over the known classical algorithms. Otherwise, efficient classical
simulations exist, which realize the computation by working in low-dimensional
representations of the algebra. This result goes beyond the requirement
of creating entangled states in the usual sense. For example, a quantum
algorithm involving gates induced by the Lie algebra $\fh=\fso(2N)$,  which is
presented in detail in the next chapter, creates entangled states in the
standard sense (i.e., non-separable).
However, since the dimension of $\fh$ is polynomially large, such quantum
computation can be efficiently simulated with a CC.

The efficient preparation of a wide class of states by quantum networks has also
been studied. Again, any GCS or generalized unentangled state of a polynomially
large dimensional Lie algebra can be efficiently prepared on a QC.

\chapter{Generalized Entanglement and Many-Body Physics}
\label{chapter5}
\begin{quote}
{\em Information is inevitably physical.

Rolf Landauer}
\end{quote}

Every QC has associated certain physical representation (e.g., spin-1/2 system,
etc.). The fact that information is not just an abstract entity and is always
linked to a physical system, implies that it must be governed by the laws of
physics. In Quantum Information Theory, for example, the manipulation and
control of information is based on the foundations and laws of the quantum
mechanics. On the other hand, many concepts, such as QE, have been developed
only from an information-theory point of view. It is expected then that these
concepts are also useful for the analysis of physical phenomena. In this
chapter, I apply the notion of GE (Chap.~\ref{chapter3}) to the study of
different problems in many-body physics.

First, I will focus on the characterization of quantum phase
transitions (QPTs) in matter. QPTs are qualitative changes occurring in the
properties of the ground state of a many-body system due to
modifications either in the interactions among its constituents or in
their interactions with an external probe~\cite{Sac99}, while the system
remains at zero temperature.  Typically, such changes are induced as a
parameter $g$ in the system Hamiltonian $H(g)$ is varied across a
point at which the transition is made from one quantum phase to a
different one.  Often some correlation length diverges at this point,
in which case the latter is called a {\em quantum critical point}.
Because thermal fluctuations are inhibited, QPTs are purely driven by
quantum fluctuations.  Thus, these 
are purely quantum phenomena: A classical system in a pure state cannot
exhibit quantum correlations. Prominent examples of QPTs are the quantum
paramagnet to ferromagnet transition occurring in Ising spin systems
under an external transverse magnetic field~\cite{LSM61,Pfe70,BM71}, the
superconductor to insulator transition in high-$T_c$ superconducting
systems, and the superfluid to Mott insulator transition originally
predicted for liquid helium and recently observed in ultracold atomic
gases~\cite{GME02}.

Since entanglement is a property inherent to quantum states and
intimately related to quantum correlations (Chap.~\ref{chapter3}), 
one would expect that, in some
appropriately defined sense, the entanglement present in the ground state
undergoes a substantial change across a point where a QPT occurs. 
Thus, the concept of GE becomes specially well suited for
this study because it is directly applicable to any algebraic language
associated to the system under study.

Another important problem in quantum mechanics is to exactly diagonalize and
obtain the spectrum of a Hamiltonian of a many-body system. In this case, no
approximate methods (e.g., mean field theory) are needed and one has complete
knowledge of the physical properties of the system through algebraic methods.
By using information-theory concepts such as the ones described for the efficient
simulation of physical systems (Sec.~\ref{sec4-3}), I will show that whenever
there is a Lie algebraic structure behind a problem in quantum physics, and the
dimension of the associated Lie algebra is small enough, such problems can be
solved easily by using a CC. This
constitutes the final part of my thesis after which come the conclusions 
(Chap.~\ref{chapter6}).

\section{Entanglement and Quantum Phase Transitions}
\label{sec5-1}
In
this section, I characterize the QPTs present
in the Lipkin-Meshkov-Glick (LMG) model and in the anisotropic XY model in an external
magnetic field through the GE notion, relative to a particular subset
of observables which will be appropriately chosen in each case.
Interestingly, for both of these models the ground states can be computed
exactly by mapping the set of observable operators involved in the
system Hamiltonian to a new set of operators which satisfy the same
commutation relations; thus, preserving the underlying algebraic
structure. In the new operator language, the models are seen to
contain some symmetries that make them exactly solvable, allowing one
to obtain the ground state properties in a number of operations that
scales polynomially with the system size.
It is possible then to understand which quantum
correlations give rise to the QPTs in these cases.

Several issues should be considered when looking for an algebra $\fh$
of observables that may make the corresponding relative purity a good
indicator of a QPT. First, one observes that in each of
these cases a preferred Lie algebra exists, where the respective
ground state would have maximum $\fh$-purity independently of the
interaction strengths in the Hamiltonian. The purity relative to such
an algebra remains constant, therefore it does not identify the
QPT.  (In these cases, this algebra is in fact the Lie algebra generated
by the parametrized family of model Hamiltonians, as the parameters are
varied.)  Thus, one needs to extract a subalgebra relative to which the
ground state may be generalized entangled, depending on the parameters
in the Hamiltonian.  A second, closely related observation is that the
purity must contain information about quantum correlations which
undergo a qualitative change as the transition point is crossed: thus,
the corresponding degree of entanglement, as measured by the purity,
must depend on the interaction strengths governing the phase
transition.  Finally, whenever a degeneracy of the ground state exists
or emerges in the thermodynamic limit, a physical requirement is that 
the purity be the same for all ground states.

Although these restrictions together turn out to be sufficient for
choosing the relevant algebra of observables in the following two
models, they do not provide an unambiguous answer when solving a
non-integrable model whose exact ground state solution cannot be
computed efficiently.  Typically, in the latter cases the ground
states are GCSs of Lie algebras for each of which the dimension increases
exponentially with the system size.  Choosing the observable
subalgebra that contains the proper information on the QPTs 
(such as information on critical exponents) then becomes, 
in general, a difficult task.

A concept of {\em general mean-field Hamiltonians} (GMFH)
emerges from these considerations. Given a Hilbert space ${\cal H}$ of
dimension $p^N$ (with $p$ an integer $>1$), I define a GMFH
as the Hermitian operator
\begin{equation}
H_{\sf MF}=\sum_\alpha \kappa_j \hO_j \: , \hspace{5mm}
\kappa_j \in \mathbb{R} \:,
\end{equation}
which is an element of an irreducibly represented Lie algebra of 
Hermitian operators
$\fh=\{\hO_1,\cdots,\hO_M\}$, whose dimension scales at most polynomially in $N$
that is, $M\le\mbox{poly}(N)$. When the ground state of 
$H_{\sf MF}$ is non-degenerate, it turns out to be a GCS 
of $\fh$~\cite{BKO03}, while the
remaining eigenstates 
(some of which may also be GCSs) and energies can be efficiently
computed. The connection between Lie-algebraic mean-field Hamiltonians
and their efficient solvability deserves careful analysis in its
own right, and will be presented in the following section.

\subsection{Lipkin-Meshkov-Glick Model}
\label{sec5-1-1}
Originally introduced in the context of nuclear physics~\cite{LMG65},
the Lipkin-Meshkov-Glick (LMG) model is widely used as a testbed for
studying critical phenomena in (pseudo) spin systems. This model
was shown to be exactly-solvable in Ref.~\cite{OSD05}. In this
section, I investigate the critical properties of this model by
calculating the purity relative to a particular subset of observables,
which will be chosen by analyzing the {\em classical} behavior of the ground
state of the system. For this purpose, I first need to map the model
to a {\em single} spin, where it becomes solvable and where the
standard notion of entanglement is not immediately applicable.

The model is constructed by considering $N$ fermions distributed in two
$N$-fold  degenerate levels (termed upper and lower shells).  The
latter are separated by  an energy gap $\epsilon$, which will be set
here equal to 1.  The quantum number  $\sigma =\pm 1$ ($\uparrow$ or
$\downarrow$) labels the level while the quantum  number $k$ denotes
the particular degenerate state in the level (for both shells,  $k \in
\{k_1,\ldots,k_N\}$).  In addition, I consider a ``monopole-monopole''
interaction that scatters pairs of particles between the two levels
without  changing $k$. The model Hamiltonian may  be written as 
\begin{eqnarray}
\label{lmghamilt1}
H=H_0 + \hat{V} + \hat{W} =
\frac{1}{2} \sum\limits_{k,\sigma} \sigma c^\dagger_{k \sigma}c^{\;}_{k \sigma}
&+ & \frac{V}{2N} \sum\limits_{k,k',\sigma} c^\dagger_{k \sigma}
c^\dagger_{k' \sigma} c^{\;}_{k'\overline{\sigma} } 
c^{\;}_{k \overline{\sigma}}  \\
\nonumber
&+&\frac{W}{2N}
\sum\limits_{k,k',\sigma} c^\dagger_{k \sigma}
c^\dagger_{k' \overline{\sigma}} c^{\;}_{k' \sigma} 
c^{\;}_{k \overline{\sigma}} \;, 
\end{eqnarray}
where $\overline{\sigma} = -\sigma$, and the fermionic operators 
$c^\dagger_{k \sigma}$ ($c^{\;}_{k \sigma}$) create (annihilate) a
fermion in the  level identified by the quantum numbers $(k,\sigma)$
and satisfy the fermionic  commutation relations given in Sec.~\ref{sec2-3-1}.
Thus, the  interaction $\hat{V}$ scatters a
pair of particles belonging to one of the  levels, and the interaction
$\hat{W}$ scatters a pair of particles belonging to   different levels.
Note that the factor $1/N$ must be present in the interaction  terms
for stability reasons, as the energy per particle must be finite in 
the thermodynamic limit.

Upon introducing the pseudospin operators
\begin{eqnarray}
\label{pseudospin1}
J_+ &=& \sum\limits_k c^\dagger_{k \uparrow} c^{\;}_{k \downarrow}\;, \\
\label{pseudospin2}
J_- &=& \sum\limits_k c^\dagger_{k \downarrow} c^{\;}_{k \uparrow}\;, \\
\label{pseudospin3}
J_z &=& \frac{1}{2} \sum\limits_{k,\sigma} \sigma c^\dagger_{k \sigma} 
c^{\;}_{k \sigma}
= \frac{1}{2} \Big( n_\uparrow -n_\downarrow \Big) \;,
\end{eqnarray}
which satisfy the $\fsu(2)$ commutation relations of the angular momentum 
algebra, 
\begin{eqnarray}
\left[ J_z, J_{\pm} \right] &=& \pm J_{\pm}\;, \label{angmom0}\\
\left[ J_+ , J_- \right] &=& J_z \;,
\label{angmom}
\end{eqnarray}
the Hamiltonian of Eq.~\ref{lmghamilt1} may be rewritten as
\begin{equation}
\label{lmghamilt2}
H=J_z + \frac{V}{2N} (J_+^2 + J_-^2 ) + \frac{W}{2N} (J_+ J_- + J_- J_+)\;.
\end{equation}
As defined by Eq.~\ref{lmghamilt2}, $H$ is invariant under the
$\mathbb{Z}_2$ inversion symmetry operation $K$ that transforms
$(J_x,J_y,J_z) \mapsto (-J_x,-J_y,J_z)$, and it also commutes with the
(Casimir) total angular momentum operator ${\bf J}^2 = J_x^2
+J_y^2+J_z^2$. Therefore, the non-degenerate eigenstates of $H$ are
simultaneous eigenstates of both $K$ and ${\bf J}^2$, and they may be
obtained by diagonalizing matrices of dimension $2J+1$ (whereby the
solubility of the model).  Notice that, by definition of $J_z$ as in
Eq.~\ref{pseudospin3}, the maximum eigenvalue of $J_z$ and $J=|{\bf
J}|$ is $N/2$.  In particular, for a system with $N$ fermions as
assumed, both the ground state $\ket{g}$ and the first excited state
$\ket{e}$ belong to the largest possible angular momentum eigenvalue
$J=N/2$~\cite{LMG65} (so-called half-filling configurations); thus,
they can be computed by diagonalizing a matrix of dimension $N+1$.

\begin{figure}[htb]
\begin{center}
\includegraphics*[height=5cm]{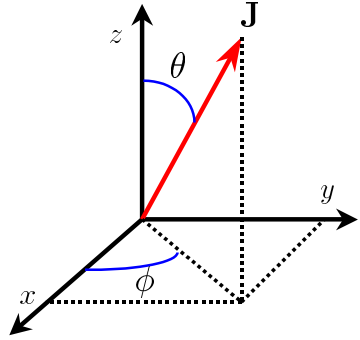}
\end{center}
\caption{Angular momentum coordinates in the three-dimensional space.}
\label{fig5-1}
\end{figure}

The Hamiltonian of Eq.~\ref{lmghamilt2} does not exhibit a QPT for finite
$N$. It is  important to remark that some critical properties of the
LMG model in the thermodynamic limit $N\rightarrow \infty$ can be
understood by using a semiclassical approach (note that 
the critical behavior is essentially mean-field): first, I
replace the angular momentum  operators in $H/N$ (with $H$ given in
Eq.~\ref{lmghamilt2}) by their classical  components (Fig.~\ref{fig5-1}); that is
\begin{eqnarray}
\label{angmomtrans}
{\bf J}=(J_x,J_y,J_z) &\rightarrow &\left(  J\sin \theta \cos \phi , J \sin
\theta \sin \phi, J \cos \theta \right) \;, \\
H/N &\rightarrow & h_c(j,\theta,\phi)\;, 
\end{eqnarray}
where $h_c$ is the resulting classical Hamiltonian and $j=J/N$,
$j=0,\ldots,1/2$.   In this way, one can show that in the thermodynamic
limit (Appendix~\ref{appE}) 
\begin{equation}
\label{lmgenergy}
\lim_{N \rightarrow \infty} \frac{\langle g | H | g  \rangle}{N}  =
\lim_{N \rightarrow \infty} \frac{E_g}{N} = \min_{j ,\theta,\phi}
h_c(j,\theta,\phi)\;,
\end{equation}
so the ground state energy per particle $E_g/N$ can be easily evaluated 
by minimizing 
\begin{equation}
\label{classichamilt}
h_c (j, \theta, \phi) = j \cos \theta  + \frac{V}{2}j^2 \sin^2 \theta  \cos(2 \phi)
+Wj^2 \sin^2  \theta \;.
\end{equation} 

\begin{figure}[hbt]
\begin{center}
\includegraphics*[width=9cm]{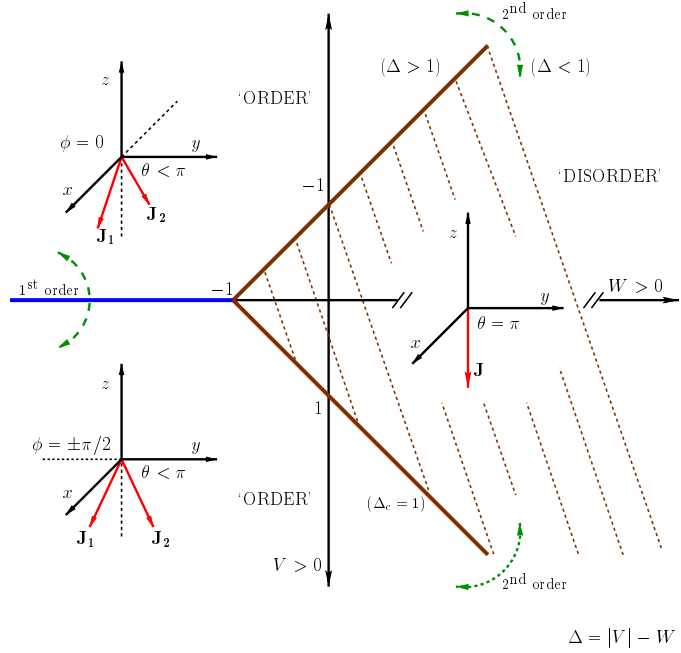}
\end{center}
\caption{Representation of the classical ground state of the LMG model.}
\label{fig5-2}
\end{figure}

As mentioned, the ground and first excited states have maximum angular
momentum  $j=1/2$. In Fig.~\ref{fig5-2} I show the orientation
of the angular  momentum in the ground states of the classical
Hamiltonian $h_c$,  represented by the vectors ${\bf J}$, ${\bf J_1}$,
and  ${\bf J_2}$, for different values of $V$ and $W$. When $\Delta
=|V|-W \leq 1$ we have $\theta = \pi$ and  the classical angular
momentum is oriented in the negative $z$-direction. However, when
$\Delta > 1$ we have $\cos \theta  = -\Delta^{-1}$  and the classical
ground state becomes two-fold degenerate (notice that $h_c$ is
invariant under the  transformation $\phi \mapsto - \phi$). In this
region and for $V<0$ the angular momentum is oriented in the XZ
plane  ($\phi=0$) while for $V>0$ it is oriented in the YZ plane
($\phi = \pm\pi/2$). The model has a gauge symmetry in the line $V=0$,
$W<-1$, where $\phi$ can take any possible value.

\subsubsection{First and Second Order QPTs, and Critical Behavior}
For the Hamiltonian of Eq.~\ref{lmghamilt1}, the
quantum  system undergoes a second order QPT at the critical boundary
$\Delta_c=|V_c|-W_c=1$,  where for $\Delta>\Delta_c$ the ground and
first excited states $\ket{g}$ and $\ket{e}$ become degenerate in the
thermodynamic limit, and the inversion symmetry $K$ breaks.   The order
parameter is given by the mean number of fermions in the upper shell 
$\langle n_\uparrow \rangle = 1/2+\langle J_z \rangle /N$, which  in
the thermodynamic limit converges to its classical value, 
\begin{equation}
\label{lmgop}
\lim_{N\rightarrow \infty} 
\langle n_\uparrow \rangle= \frac{1+\cos\theta}{2}\;.
\end{equation}
Obviously, for $\Delta \leq \Delta_c$ one has $\langle n_\uparrow
\rangle=0$, and  $\langle n_\uparrow \rangle > 0$, otherwise 
(Fig.~\ref{fig5-2}). The critical exponents of the order parameter
are easily computed by making a Taylor expansion near the  critical
points ($\Delta \rightarrow 1^+$). Defining the quantities $x=V_c-V$ 
and $y=W_c-W$, one obtains
\[
\lim_{\Delta \rightarrow 1^+} \langle n_\uparrow \rangle =
\left \{
\begin{array}{l}
(y^\alpha-x^\beta)/2 \hspace{3mm}\mbox{ for }V>0\\ 
(y^\alpha+x^\beta)/2 \hspace{3mm}\mbox{ for }V<0 \end{array}
\right. ,
\]
where the critical exponents are $\alpha=1$ and $\beta=1$.

\begin{figure}[hbt]
\begin{center}
\includegraphics[width=10cm]{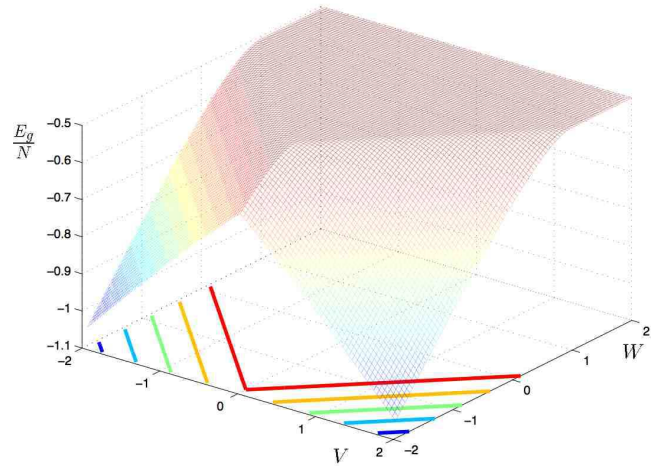}
\end{center}
\caption{Ground state energy per particle in the LMG model.}
\label{fig5-3}
\end{figure}

In Fig.~\ref{fig5-3} I show the exact ground state energy per
particle $E_g/N$ (with $E_g = \langle g | H | g\rangle$) as a function
of $V$ and $W$ in the  thermodynamic limit (Eqs.~\ref{lmgenergy}).
One can see that also in the broken symmetry region ($\Delta>1$)
the system undergoes a first order QPT at $V=0$; that is, the first
derivative of the ground state energy with respect to $V$ is not
continuous in this line.

\subsubsection{Purity as an Indicator of the QPTs in the LMG Model}
The standard notion of entanglement is not directly applicable to the
LMG model as described by Eq.~\ref{lmghamilt2}, as this is a single
spin system and no physically natural partition into subsystems is
possible.   Therefore, using the $\fh$-purity as a measure of
entanglement becomes an advantage from this point of view, since the
latter only depends on a particular subset of observables and no
partition of the system is necessary.  The first required step is the
identification of a relevant Lie algebra of observables relative to
which the purity has to be calculated.

Since both the ground and first excited states of the quantum LMG model
may be understood as states of a system carrying total angular
momentum $J=N/2$, a  first possible algebra to consider is the
$\fsu(N+1)$ algebra acting on the  relevant $(N+1)$-dimensional
eigenspace.  Relative to this algebra, $|g\rangle$ is generalized
unentangled for arbitrary values of $V$ and $W$, thus the corresponding purity
remains constant and does not signal the QPTs.  However, the 
family of Hamiltonians of Eq.~\ref{lmghamilt2} do not generate this 
Lie algebra, but rather an $\fsu(2)$ algebra.

Thus a natural choice, suggested by the commutation
relationships of Eqs.~\ref{angmom0} and \ref{angmom}, is to study
the purity relative to the spin-$N/2$  representation of the angular
momentum Lie algebra  $\fh=\fsu(2)= \{ J_x,J_y,J_z \}$:
\begin{equation}
P_{\fh} (\ket{\psi}) = \frac{4}{N^2}\Big[ \langle J_x \rangle ^2 
+\langle J_y \rangle ^2 + \langle J_z \rangle ^2 \Big] \;,
\end{equation}
where the normalization factor ${\sf K}=N^2/4$ is chosen to ensure
that the maximum of $P_{\fh}$ is equal to 1 (Eq.~\ref{purity}).  
With this normalization
factor, $P_{\fh}$ can be calculated exactly in the thermodynamic limit
by relying on the semi-classical approach described earlier
(Appendix~\ref{appE} and Eq.~\ref{angmomtrans}).  For $V=0$ and
arbitrary $W>0$, $|g\rangle= | J_z =-N/2\rangle$ which is a GCS of
$\fsu(2)$ and has $P_{\fh}=1$. For generic interaction values such
that $\Delta \leq 1$, the classical angular momentum depicted in Fig.
\ref{fig5-2} is oriented along the $z$ direction and is not
degenerate: Because $\langle J_x \rangle= \langle J_y \rangle= 0$,
only $\langle J_z \rangle$ contributes to $P_{\fh}$. By recalling that
$ \lim_{N \rightarrow \infty}\langle J_z/N \rangle= -1/2$, this gives
$P_{\fh}=1$, so that {\em as far as relative purity is concerned the
ground state behaves asymptotically like a coherent state in the
thermodynamic limit}.  Physically, this means that with respect to the
relevant fluctuations, GCSs of $\fsu(2)$ are a good approximation of
the quantum ground state for large particle numbers, as is well
established for this model~\cite{GF78}.  However, in the region
$\Delta > 1$ the ground state (both classical and quantum) is two-fold
degenerate in the $N \rightarrow \infty$ limit, and the value of
$P_{\fh}$ depends in general on the particular linear combination of
degenerate states.  This can be understood from
Fig.~\ref{fig5-2}, where different linear combinations of the
two degenerate vectors ${\bf J_1}$ and ${\bf J_2}$ imply different
values of $\langle J_x \rangle $ for $V<0$ and different values of
$\langle J_y \rangle $ for $V>0$, while $\langle J_z \rangle$ remains
constant.  With these features, the purity relative to the $\fsu(2)$
algebra will not be a good indicator of the QPT.

An alternative option is then to look at a subalgebra of $\fsu(2)$. In 
particular, if I only consider the purity relative to the single
observable  $\fh=\fso(2)=\{J_z\}$ (i.e., a particular CSA of
$\fsu(2)$), and retain the  same normalization as above, then 
\begin{equation}
\label{jzpuritylmg}
P_{\fh}(\ket{\psi}) =\frac{4}{N^2} \langle J_z \rangle ^2 \; .
\end{equation}
This new purity will be a good indicator of the QPT, since $P_{\fh}=1$
only for $\Delta \leq 1$ in the thermodynamic limit, and in addition
$P_{\fh}$ does not  depend on the particular linear combination of the
two-fold degenerate states in  the region $\Delta > 1$, where
$P_{\fh}<1$. Obviously, in this case $P_{\fh}$ is straightforwardly
related to the order parameter (Eq.~\ref{lmgop}); the critical
exponents of $P_{\fh}-1$ are indeed the same ($\alpha=1$ and
$\beta=1$).

In the region $\Delta <1 $ where $P_{\fh} =1$, the quantum
ground state of the LMG model (Eq.~\ref{lmghamilt2}) does not have a
well defined $z$-component of angular momentum except at $V= 0$ ($[ H,
J_z ] \neq 0$ if $V \neq 0$), thus in general it does not lie on a
coherent orbit of this algebra for finite $N$.  However, as discussed
above, it behaves asymptotically (in the infinite $N$ limit) as a GCS
(in the sense that $P_{\fh}\rightarrow 1$).  

\begin{figure}[hbt]
\begin{center}
\includegraphics*[width=9cm]{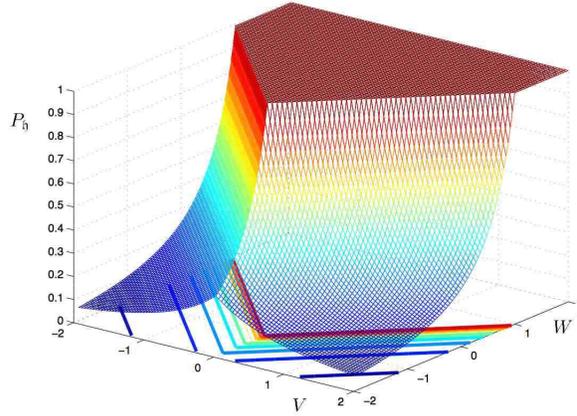}
\end{center}
\caption{Purity relative to the observable $J_z$ in the ground state 
of the LMG model.}
\label{fig5-4}
\end{figure}

In Fig.~\ref{fig5-4} I show the behavior of $P_{\fh}$ as a
function of  the parameters $V$ and $W$. The purity
relative to $J_z$ is then a good indicator not only of the second order QPT
but also of the first order QPT (the line $V=0$, $W<-1$).  

\subsection{Anisotropic XY Model in a Transverse Magnetic Field}
\label{sec5-1-2}
In this section, I exploit the purity relative to the $\fu(N)$
algebra (introduced in Sec.~\ref{sec3-3-3}) as a measure of GE which is
able to identify the paramagnetic-to-ferromagnetic QPT in the
anisotropic one-dimensional spin-1/2 XY model in a transverse magnetic
field and classify its universality properties.

\begin{figure}[hbt]
\begin{center}
\includegraphics*[width=9cm]{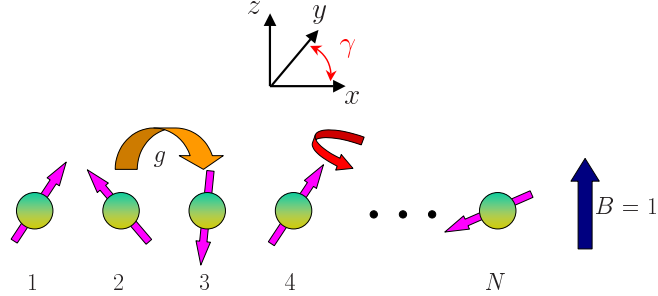}
\end{center}
\caption{Anisotropic one-dimensional XY model in an external transverse
magnetic field $B$.}
\label{fig5-5}
\end{figure}

The model Hamiltonian for a chain of $N$ sites is given by 
(Fig.~\ref{fig5-5})
\begin{equation} 
\label{Ham5-1} 
H =-g \sum\limits_{i=1}^N  \Big[(1+\gamma)  \sigma_x^i \sigma_x^{i+1}+ (1-\gamma) 
\sigma_y^i \sigma_y^{i+1}\Big] + \sum\limits_{i=1}^N \sigma_z^ i \;,
\end{equation}
where the operators  $\sigma_{\alpha}^i$ ($\alpha =x,y,z$)  are the
Pauli spin-1/2 operators on site $i$ (Sec.~\ref{sec2-1}),
$g$ is the parameter one may tune to drive the
QPT, and $0 < \gamma \leq 1 $ is the amount of anisotropy in the
XY plane. In particular, for $\gamma=1$, Eq.~\ref{Ham5-1} reduces to
the Ising  model in a transverse magnetic field~\cite{Pfe70}, while for $\gamma\rightarrow 0$
the model becomes isotropic. Periodic boundary conditions are
considered here, that is  $\sigma_{\alpha}^{i+N} =
\sigma_{\alpha}^{i}$, for all $i$ and $\alpha$.

When $g \gg 1 $ and $\gamma=1$ the model is Ising-like.
In this limit, the spin-spin
interactions are the dominant  contribution to the Hamiltonian
(Eq.~\ref{Ham5-1}), and the ground state becomes  degenerate in the
thermodynamic limit, exhibiting ferromagnetic long-range order 
correlations in the $x$ direction:  $M_x^2=\lim_{N\rightarrow \infty}
\langle \sigma_x^1 \sigma_x^{N/2} \rangle > 0$, where $M_x$ is the
magnetization in the $x$ direction. In the opposite limit  where $g
\rightarrow 0$, the external magnetic field becomes important, the
spins  tend to align in the $z$ direction, and the magnetization in the $x$
direction vanishes:  $M_x^2 =\lim_{N\rightarrow \infty} \langle
\sigma_x^1 \sigma_x^{N/2} \rangle = 0$. Thus, in the thermodynamic
limit the model is subject to a paramagnetic-to-ferromagnetic  second
order QPT at a critical point $g_c$ that will be determined later, 
with critical behavior belonging to the two-dimensional Ising
model universality class. 

This model can be exactly solved using the Jordan-Wigner
transformation~\cite{JW28},
which maps the Pauli (spin 1/2) algebra into the
canonical fermion  algebra through (Sec.~\ref{sec2-3-1})
\begin{equation}  
c^{\dagger}_j=  \prod\limits_{l=1}^{j-1} (-\sigma_z^l)
 \sigma_+^j \;. 
\end{equation} 

In order to find the exact ground state, I first need to write the
Hamiltonian given in Eq.~\ref{Ham5-1} in terms of these fermionic
operators, 
\begin{equation}
\label{Ham5-2}
H = -2g \sum\limits_{i=1}^{N-1} ( c^{\dagger}_i c^{\;}_{i+1}   + \gamma
c^{\dagger}_i c^{\dagger}_{i+1} + h.c.)+ 2g K (c^{\dagger}_N c^{\;}_1 + \gamma
c^{\dagger}_N c^{\dagger}_1 + h.c.) + 2 \hat{N}\;,
\end{equation}
where $K=\prod\limits_{j=1}^N (-\sigma_z^j)$ is an operator that
commutes with the Hamiltonian, and $\hat{N}=\sum\limits_{i=1}^N
c^\dagger_i c^{\;}_i$ is the total number operator (here, I choose
$N$ to be even). Then, the eigenvalue of $K$ is a good quantum number,
and noticing that $K= e^{i \pi \hat{N} }$, one obtains $K=+1 (-1)$
whenever the (non-degenerate) eigenstate of $H$ is a linear
combination of states with an even (odd) number of fermions.  In
particular, the numerical solution of this model in finite systems
(with $N$ even) indicates that the ground state has eigenvalue $K=+1$,
implying anti-periodic boundary conditions in Eq.~\ref{Ham5-2}.

The second step is to rewrite the Hamiltonian in terms of the
fermionic operators $\tilde{c}^{\dagger}_{k}$
($\tilde{c}^{\;}_{k}$), defined by the   Fourier transform of the
operators $c^{\dagger}_j$ ($c^{\;}_j$):
\begin{equation}
\tilde{c}^{\dagger}_{k}= \frac{1} {\sqrt{N}} \sum \limits_{j=1}^N e^{-ikj} 
c^{\dagger}_j \;,
\end{equation}
where the set $V$ of possible $k$ is determined by the anti-periodic
boundary conditions in the fermionic operators:  $V=V_+ + V_-=[ \pm
\frac{\pi}{N}  , \pm \frac {3 \pi}{N}, \cdots, \pm \frac  {(N-1)\pi}{N}
]$. Therefore,
\begin{equation}
\label{Ham5-3}
H +N = -2 \sum\limits_{k \epsilon V} (-1+2g \cos k) \tilde{c}^{\dagger}_k  
\tilde{c}^{\;}_k +  
i g \gamma \sin k (\tilde{c}^{\dagger}_{-k} \tilde{c}^{\dagger}_k 
+  \tilde{c}^{\;}_{-k} \tilde{c}^{\;}_k )\;.
\end{equation}

The third and final step is to diagonalize the Hamiltonian of Eq.~\ref{Ham5-3} 
using the Bogolubov canonical transformation~\cite{BR86}
\[ \left \{
\begin{array}{l}
\gamma_k = u_k \tilde{c}^{\;}_k - i v_k \tilde{c}^{\dagger}_{-k} \\
\gamma^{\dagger}_{-k} = u_k \tilde{c}^{\dagger}_{-k} - i v_k \tilde{c}^{\;}_k
\end{array}
\right. \;,
\]
where the real coefficients $u_k$ and $v_k$ satisfy the relations
\begin{equation}
u_k = u_{-k} \mbox{, } v_k = - v_{-k} \;, \mbox{  and  } \; u_k^2 + v_k^2 =1\;,
\end{equation}
with
\begin{equation}
\label{relation1}
u_k= \cos \Big(\frac{\phi_k}{2}\Big) \;, \mbox{    } 
v_k= \sin \Big(\frac{\phi_k}{2}\Big) \;,
\end{equation}
and $\phi_k$ given by
\begin{equation}
\label{relation2}
\tan (\phi_k) = \frac { 2 g \gamma \sin k} { -1 + 2g \cos k}\;.
\end{equation}

In this way, the quasiparticle creation and annihilation operators
$\gamma^\dagger_k$  and $\gamma^{\;}_k$, also satisfy  the canonical
fermionic  anti-commutation relations of Eq.~\ref{fermcom}, and the
Hamiltonian  may be finally rewritten as 
\begin{equation}
\label{diago}
H=  \sum\limits_{k \epsilon V} \xi_k (\gamma^{\dagger}_k \gamma^{\;}_k - 1/2)\;,
\end{equation}
where $\xi_k =2\sqrt {(-1+2g \cos k)^2 + 4 g^2 \gamma^2 \sin^2 k }$ is
the quasiparticle energy.  Since in general $\xi_k > 0$,  the ground
state is the quantum state with no quasiparticles (BCS state~\cite{Tak99}),
such that $\gamma_k \ket{\sf BCS} =0$.  
Thus, one finds
\begin{equation}
\label{Ground}
\ket{\sf BCS} = \prod \limits_{k \epsilon V_+} (u_k + i v_k \tilde{c}^{\dagger}_k
\tilde{c}^{\dagger}_{-k} ) \ket{\sf vac}\;,
\end{equation}
where $\ket{\sf vac}$ is the state with no fermions ($\tilde{c}_k
\ket{\sf vac}=0$).  

Excited states with an even number of fermions ($K=+1$) can be
obtained applying pairs of quasiparticle creation operators
$\gamma^\dagger_k$ to the $\ket{{\sf BCS}}$ state. However, one should
be more rigorous when obtaining excited states with an odd number of
particles, since $K=-1$ implies periodic boundary conditions in
Eq.~\ref{Ham5-2}, and the new set of possible $k$'s (wave vectors) is
$\overline{V}=[ -\pi, \cdots, -\frac{2 \pi}{N}, 0, \frac{2 \pi}{N},
\cdots, \frac{2(N-1) \pi}{N} ]$ (different from $V$).

\subsubsection{QPT and Critical Point}
\begin{figure}[hbt]
\begin{center}
\includegraphics*[width=9cm]{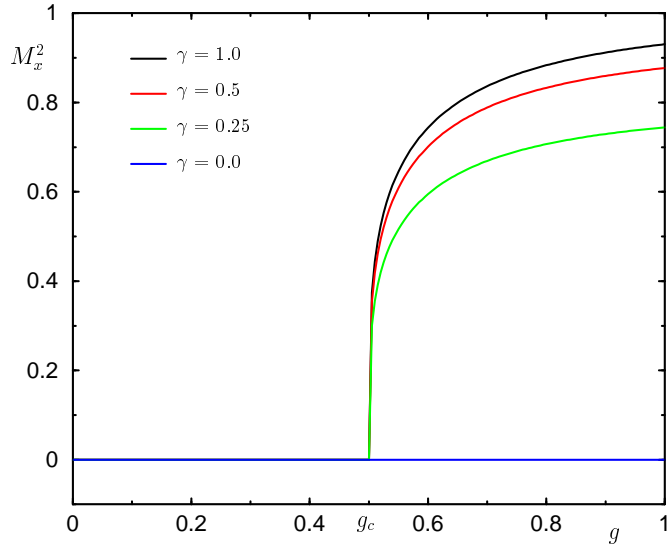}
\end{center}
\caption{Order parameter $M_x^2$ in the thermodynamic limit as a function of 
$g$ for different anisotropies $\gamma$.  The critical point is at $g_c=1/2$. }
\label{fig5-6}
\end{figure}

In Fig.~\ref{fig5-6} I show the order parameter $M_x^2=
\lim_{N \rightarrow \infty} \langle \sigma_x^1 \sigma_x^{N/2} \rangle$
as a function of $g$ in the thermodynamic limit and for different
anisotropies $\gamma$~\cite{BM71}. Then, $M_x^2=0$ for $g 
\leq g_c$ and $M_x^2 \neq 0$ for $g>g_c$, so the critical point is
located at $g_c=1/2$, regardless of the value of $\gamma$.  The value
of $g_c$ can also be obtained by setting $\xi_k=0$ in Eq.~\ref{diago}, 
where the gap vanishes.   

The Hamiltonian of Eq.~\ref{Ham5-1} is invariant under the
transformation that maps $(\sigma_x^i;\sigma_y^j;\sigma_z^k) \mapsto (
-\sigma_x^i;-\sigma_y^j; \sigma_z^k)$ ($\mathbb{Z}_2$ symmetry), implying
that $\langle \sigma_x^i \rangle = 0$ for all $g$. However, since in
the thermodynamic limit the ground state becomes two-fold degenerate,
for $g>g_c$ , it is possible to build up a ground state where the discrete
$\mathbb{Z}_2$ symmetry is broken, i.e. $\langle \sigma_x^i \rangle \neq
0$.  This statement can be easily understood if we consider the case
of $\gamma=1$, where for $0\leq g < g_c$ the ground state has no
magnetization in the $x$ direction: For $g=0$, the spins align with
the magnetic field, while an infinitesimal spin interaction disorders
the system and $M_x=0$. On the other hand, for $g \rightarrow \infty$
the states $\ket{g_1}= \frac{1}{\sqrt{2}}[\ket{ \rightarrow , \cdots ,
\rightarrow} + \ket{ \leftarrow, \cdots, \leftarrow}]$ and $\ket{g_2}=
\frac{1}{\sqrt{2}}[\ket{ \rightarrow, \cdots , \rightarrow} - \ket{
\leftarrow, \cdots, \leftarrow}],$ with $\ket{\rightarrow}=
\frac{1}{\sqrt{2}} [\ket{\uparrow} +\ket{\downarrow}]$ and
$\ket{\leftarrow}= \frac{1}{\sqrt{2}} [\ket{\uparrow}
-\ket{\downarrow}]$, become degenerate in the thermodynamic limit, and
a ground state with $\langle \sigma_x^i \rangle \neq 0$ can be
constructed from a linear combination.

Remarkably, this paramagnetic-to-ferromagnetic QPT does not exist in the
isotropic limit ($\gamma=0$). In this case, the Hamiltonian of Eq.~\ref{Ham5-1}
has a continuous $\fu(1)$ symmetry; that is, it is invariant under any
$\hat{z}$ rotation of the form $\exp [i \theta \sum_j \sigma_z^j]$. Since the
model is one-dimensional, this symmetry cannot be spontaneously broken,
regardless of the magnitude of the coupling constants. Nevertheless, a simple
calculation of the ground state energy indicates a divergence in its second
derivative at the critical point $g_c=1/2$, thus, a second order non-broken
symmetry QPT. For $g < g_c$ all the spins (in the ground state) are aligned
with the external magnetic field, with total magnetization in the $\hat{z}$
direction $M_z=\sum_j \langle \sigma_z^j \rangle = -N$, and the quantum phase
is gapped. For $g \ge g_c$, the total magnetization in the $\hat{z}$ direction
is $M_z \ge -N$, the gap  vanishes, and the quantum phase becomes critical
(i.e., the spin-spin correlation functions decay with a power law), with an
emergent  $\fu(1)$ gauge symmetry~\cite{BO04}. Then, in terms of fermionic
operators (Eq.~\ref{Ham5-2}), an insulator-metal (or superfluid)  like second
order QPT exists at $g_c$ for the isotropic case, with no symmetry order
parameter. It is a Lifshitz transition.
\subsubsection{$\fu(N)$-Purity in the BCS State and Critical Behavior}
The $\ket{\sf BCS}$ state of Eq.~\ref{Ground} is a GCS of the
algebra of observables $\fh=\fso (2N)$. This is spanned by an orthonormal 
Hermitian basis which is constructed by adjoining to the basis of 
$\fu(N)$, given in Eq.~\ref{uNbasis}, the following set $\fr$ of 
number-non-conserving fermionic operators:
\begin{equation}
\fr=\left\{
\begin{array}{ll}
(c^{\dagger}_j c^\dagger_{j'} + c^{\;}_{j'} c^{\;}_j) & \mbox{  with } 1\leq j<j' \leq N \cr
i(c^{\dagger}_j c^\dagger_{j'} - c^{\;}_{j'} c^{\;}_j) & \mbox{  with } 1\leq j<j' \leq N \cr
\end{array}
\right. \:, \hspace{5mm}  \fso(2N) = \fu(N) \oplus \fr \:.
\label{so2Nbasis}
\end{equation}
Then, the $\ket{\sf BCS}$ state is generalized unentangled with
respect  to the $\fso (2N)$ algebra and its purity $P_{\fh}$
contains no information about the phase transition:
$P_{\fh}=1\mbox{ } \forall g,\gamma$. Therefore, in order to
characterize the QPT one needs to look at the possible subalgebras of
$\fso(2N)$.  A natural choice is to restrict to operators which
preserve the total fermion number, that is, to consider the $\fu(N)$
algebra  defined in Sec.~\ref{sec3-3-3}, relative to which
the $\ket{\sf BCS}$  state may become generalized entangled. (Note that
as mentioned in Sec.~\ref{sec3-3-3}, the  $\fu(N)$ algebra can
also be written in terms of the fermionic operators 
$\tilde{c}^\dagger_{k}$ and $\tilde{c}_{k}$, with $k$ belonging to the
set $V$.)

In the $\ket{\sf BCS}$ state, $\langle \tilde{c}^\dagger_{k}
\tilde{c}^{\;}_{k'}  \rangle \neq 0$ only if $k=k'$, thus using Eq.
\ref{unpurity} the purity  relative to $\fh=\fu(N)$ is:
\begin{equation}
\label{unpurity2}
P_{\fh} (\ket{\sf BCS}) = 
\frac{4}{N} \sum\limits_{k \epsilon V} \langle \tilde{c}^{\dagger}_k \tilde{c}^{\;}_k -1/2 \rangle^2
= \frac{4}{N} \sum\limits_{k \epsilon V} (v_k^2 -1/2)^2 \;,
\end{equation}
where the coefficients $v_k$ can be obtained from Eqs.
\ref{relation1} and \ref{relation2}.  In particular, for $g=0$ the
spins are aligned with the magnetic field and the fully polarized
$\ket{\sf BCS}_{g=0}= \ket{\downarrow,  \downarrow, \ldots ,\downarrow}$
state is generalized unentangled in this limit (a GCS of $\fu(N)$ with
$P_{\fh}=1$). In the thermodynamic limit, the purity relative to the
$\fu(N)$ algebra can be obtained by integrating Eq.~\ref{unpurity2}:
\begin{equation}
P_{\fh}(\ket{\sf BCS})=\frac{2}{\pi}\int\limits_0^{2 \pi} (v_k^2 -1/2)^2 dk\:,
\end{equation}
leading to the following result:
\begin{equation}
\label{purity5}
P_{\fh} (\ket{\sf BCS}) = \left\{ \begin{array}{cl} \frac {1}{1- \gamma ^2} 
\Big[ 1 - \frac{\gamma^2}{\sqrt{1 - 4g^2 (1-\gamma^2)}} \Big] 
& \mbox{ if } g \leq 1/2 \\  
\frac{1}{1+ \gamma} & \mbox{ if } g>1/2 \end{array} \right.\; .
\end{equation}
Although this function is continuous, its derivative is not and has a
drastic change at $g=1/2$, where the QPT occurs. Moreover, $P_{\fh}$ is
minimum for $g>1/2$ implying maximum entanglement at the transition
point and in the ordered  (ferromagnetic) phase.  Remarkably, for
$g>1/2$ and $N\rightarrow \infty$, where  the ground state of  the
anisotropic XY model in a transverse magnetic field is  two-fold
degenerate, $P_{\fh}$ remains invariant for arbitrary linear 
combinations of the two degenerate states.

\begin{figure}[hbt]
\begin{center}
\includegraphics*[width=11cm]{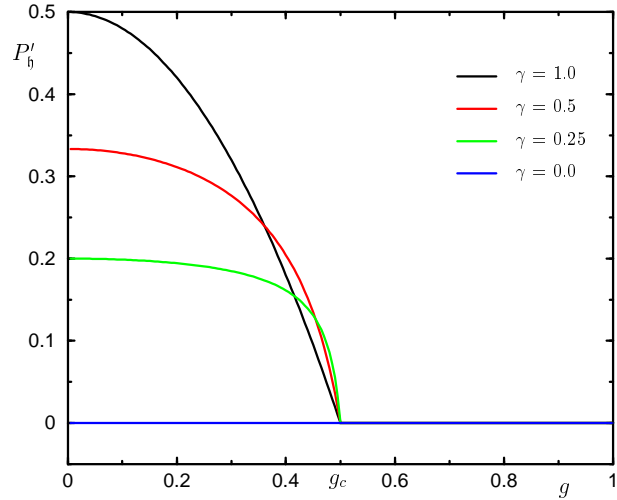}
\end{center}
\caption{Shifted purity $P'_{\fu(N)}$ of the $\ket{\sf BCS}$ as a
function of $g$  for different anisotropies $\gamma$, Eq.
\ref{disorder0}. $P'_{\fu(N)}$ behaves like a disorder  parameter for
this model, sharply identifying the QPT at $g_c=1/2$. }
\label{fig5-7}
\end{figure}

As defined, for large $g$ the purity $P_{\fh}$ approaches a constant
value  which depends on $\gamma$.  It is convenient to remove such
dependence in the  ordered phase by introducing a new quantity
$P'_{\fh}= P_{\fh} - \frac{1}{1+\gamma}$ (shifted purity).  Thus,
\begin{equation}
P'_{\fh}(\ket{\sf BCS}) = \left\{ \begin{array}{cl} \frac{\gamma}{1- \gamma^2} 
\Big[ 1 - \frac{\gamma} {\sqrt{1 - 4g^2 (1- \gamma^2)}} \Big]& 
\mbox{ if } g \leq 1/2 \\  0 & \mbox{ if } g>1/2  \end{array} \right. \;.
\label{disorder0}
\end{equation}
The new function $P'_{\fh}$ behaves like a {\it disorder parameter} for
the system,  being zero in the ferromagnetic (ordered) phase and
different from zero in the  paramagnetic (disordered) one.  The behavior
of $P'_{\fh}$ as a function of $g$ in  the thermodynamic limit is
depicted in Fig.~\ref{fig5-7} for different values of  $\gamma$.  In
the special case of the Ising model in a transverse magnetic field 
($\gamma=1$), one has the simple behavior $P'_{\fh}= 1/2 -2g^2$  for $g
\leq 1/2$  and $P'_{\fh}=0$ if $g>1/2$.

\begin{figure}[hbt]
\begin{center}
\includegraphics[width=9cm]{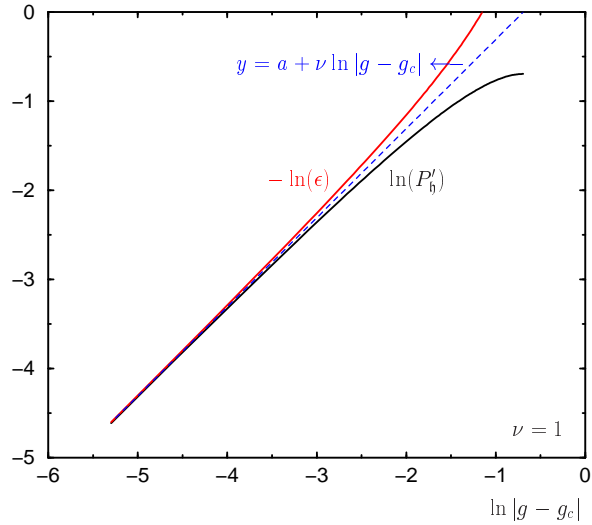}
\end{center}
\caption{Scaling properties of the disorder parameter for anisotropy
$\gamma=1$. The exponent $\nu=1$ belongs to the Ising universality class.}
\label{fig5-8}
\end{figure}

The critical behavior of the system is characterized by a power-law
divergence  of the {\it correlation length} $\epsilon$, which is
defined such that for $g<1/2$, 
\begin{equation}
\lim_{|i-j| \rightarrow \infty} |
\langle \sigma_x^i \sigma_x^j \rangle | \sim \exp
(-\frac{|i-j|}{\epsilon}). 
\end{equation}
Thus, $\epsilon \rightarrow \infty$ signals
the emergence of long-range correlations in the ordered region $g>1/2$.
Near the critical point ($g \rightarrow 1/2^-$), the correlation length
behaves  as $\epsilon \sim (g_c - g)^{- \nu}$, where $\nu$ is a
critical exponent and the value $\nu =1$ corresponds to the Ising
universality class.  Let the  parameter $\lambda_2 = e^{-1/\epsilon}$. 
The fact that the purity contains  information about the critical
properties of the model follows from the  possibility of expressing
$P'_{\fh}$ for $g <1/2$ as a function of the  correlation length, 
\begin{equation}
\label{disorder}
P'_{\fh} (\ket{\sf BCS}) = 
\frac{\gamma} {1- \gamma^2} \left [ 1 + \frac {\gamma} {2g \lambda_2 (1 -
\gamma) -1 } \right ] \ ,
\end{equation}
where a known relation between $g$, $\gamma$, and $\lambda_2$ has been 
exploited~\cite{BM71}.  Performing a Taylor expansion of Eq.
\ref{disorder}  in the region $g \rightarrow 1/2^-$, one obtains
(Fig.~\ref{fig5-8})
\begin{equation}
P'_{\fh} \sim 2  (g_c - g )^\nu /\gamma \ \  \nu=1,\gamma >0. 
\end{equation}
Thus, the name disorder  parameter for $P'_{\fh}$
is consistent.

Some physical insight into the meaning of the ground-state purity may be
gained by  noting that Eq.~\ref{unpurity2} can be written in terms
of the fluctuations  of the total fermion operator $\hat{N}$ 
\begin{equation}
\label{fluct1}
P_{\fh}(\ket{\sf BCS}) =1 - \frac{2}{N} \Big(\langle \hat{N}^2 \rangle - 
\langle \hat{N}\rangle ^2 \Big)\;,
\end{equation}
where the $\ket{\sf BCS}$-property  $\langle \tilde{c}^\dagger_k
\tilde{c}_{k'}\rangle= \delta_{k,k'} v_k^2$ has been used.   In
general, the purity relative to a given algebra can be written in terms
of fluctuations of observables~\cite{BKO03}. 

\begin{figure}[hbt]
\begin{center}
\includegraphics*[width=9cm]{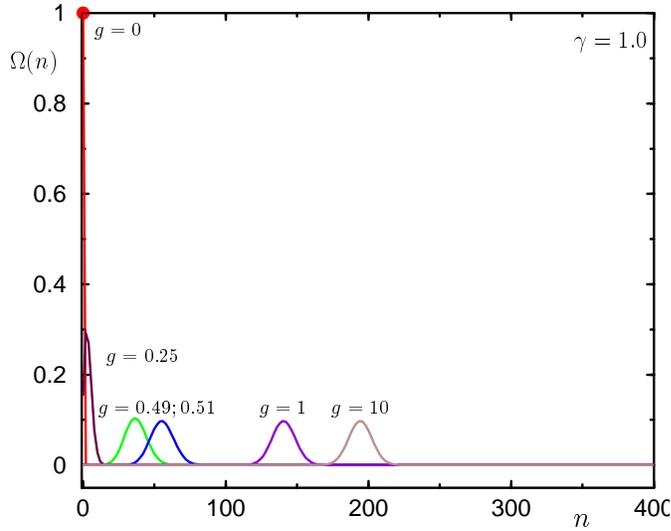}
\label{distribution}
\end{center}
\caption{Distribution of the fermion number in the $\ket{\sf BCS}$ state
for a chain of $N=400$ sites and anisotropy $\gamma=1$.}
\label{fig5-9}
\end{figure}

Since fluctuations of observables are at the root of QPTs, it is not
surprising  that this quantity succeeds at identifying the critical
point. Interestingly,  by recalling that $P_{\fso(2N)}(\ket{\sf
BCS})=1$, the $\fu(N)$-purity can also be formally expressed as
\begin{equation}
\label{fluct2}
P_{\fu(N)}(\ket{\sf BCS}) =1 - \sum_{A_\alpha \in \fr} 
\langle A_\alpha \rangle ^2 \:,
\end{equation}
where the sum only extends to the non-number-conserving $\fso(2N)$-generators
belonging to the set $\fr$ specified in Eq.~\ref{so2Nbasis}. 
Thus, the purity is entirely contributed by expectations of operators
connecting different $\fu(N)$-irreps, the net effect of correlating
representations with a different particle number resulting in the
fluctuation of a {\it single} operator, given by $\hat{N}=\sum_k
\tilde{c}_k^\dagger \tilde{c}_k$.  In Fig.~\ref{fig5-9}, I  show
the probability $\Omega(n)$ of having $n$ fermions in a chain of
$N=400$ sites  for $\gamma=1$. For $g>1/2$ the
fluctuations remain almost constant,  and so does the purity.

Again, the isotropic case ($\gamma=0$) is particular in the sense that
$P_{\fh}=1$ (or $P'_{\fh}=0$, see Fig.~\ref{fig5-7}), without
identifying the corresponding metal-insulator QPT. The reason is that
in this limit, the Hamiltonian of Eq.~\ref{Ham5-2} contains only
fermionic operators that preserve the number of particles (i.e., $H
\in \fu(N)$), and the ground state of the system is always a GCS of
the $\fu(N)$ algebra. Therefore, in order to obtain information about
this QPT, one should look at algebras other than $\fu(N)$, relative 
to which the ground state is generalized entangled. A more detailed analysis can
be found in Ref.~\cite{SOB04}.

\section{General Mean-Field Hamiltonians}
\label{sec5-2}
A Lie-algebraic analysis of many-body problems leads to a powerful tool for
finding the spectrum and eigenstates of the Hamiltonian that describes the
interactions in the system:
\begin{equation}
\label{gmfh1}
H = \sum_{j=1}^M \kappa_j \hO_j, \ \kappa_j \in \mathbb{R},
\end{equation}
where the Hermitian operators  $\hO_j$ are linearly independent
Schmidt-orthogonal elements of a Lie algebra  $\fh= \{\hO_1, \cdots \hO_M\}$
(Sec.~\ref{sec3-2-2}). Each operator $\hO_j$ corresponds to a $d\times d$
matrix acting on the Hilbert space $\cH$ (of dimension $d$) associated with the
physical system. In general, $d$ is infinite or scales exponentially with
a certain variable $N$, like the volume of the system, etc.  For example,  for a
$N$-spin-1/2 system one obtains $d=2^N$. However,  if the system is bosonic or
is composed of harmonic oscillators, then $d \rightarrow \infty$. Throughout
this section, I will assume that $N  \le \mbox{poly} [\log (d)]$.

Without loss of generality, one can consider $\fh$ to be a semi-simple Lie
algebra acting irreducibly on $\cH$: The Casimir elements in $H$ behave as
constants (symmetries) and do not change the physics.   If the dimension of
$\fh$ satisfies $M \le \mbox{poly}(N)$, then $H$ is defined to be a {\em
general mean field Hamiltonian} and is denoted as $H_{MF}$.
Since $\fh$ admits a
CW decomposition (see Sec.~\ref{sec3-2-2}), one obtains
\begin{equation}
\label{Ham5-5}
H_{MF}=\sum\limits_{k=1}^r \gamma_k \hh_k + \sum\limits_{j=1}^{l} \iota_j
\hE_{\alpha_j} + \iota^*_j \hE_{-\alpha_j},
\end{equation}
where $\gamma_k \in \mathbb{R}$ and $\iota_j \in \mathbb{C}$ are known (or can
easily be computed) coefficients, $\hh_k \in \fh_D$ (i.e., the CSA of $\fh$),
$\hE_{\pm \alpha_j}$ are ladder operators, and $(r,l) \le \mbox{poly}(N)$.

The eigenstates $\ket{\phi^p}$  of $\fh_D$ (i.e., the weight states of the
standard representation given by $\cH$) and the corresponding eigenvalues
$e_k^p$, satisfying
\begin{equation}
\hh_k \ket{\phi^p} = e_k^p \ket{\phi^p},
\end{equation}
are assumed to be known with high accuracy. In particular, every weight state can
always be obtained by the (efficient) successive application of lowering
operators to a highest weight state $\ket{{\sf HW}} \in \cH$:
\begin{equation}
\label{gmfhws}
\ket{\phi^p} = N_p \prod\limits_{j=1}^{l} \hE_{-\alpha_j}^{n_j} \ket{\sf HW},
\end{equation}
where $N_p \in \mathbb{R}$ is a normalization factor ($n_j \in \mathbb{Z}$). If
$(e_1, \cdots, e_r)$ is the highest
weight vector (associated with $\ket{\sf HW}$), then
\begin{equation}
\label{hweigen}
\hh_k \ket{\phi^p} = \left[e_k - \sum\limits_{j=1}^l n_j \alpha_j^k \right]
\ket{\phi^p}.
\end{equation}
Therefore, all the information about the weight states of $\fh$ can be obtained
from $\ket{\sf HW}$, its weights, and the root vectors $\alpha_j$.
This applies to every irreducible representation of $\fh$.

The diagonal form $H_D \in \fh_D$ of Eq.~\ref{Ham5-5} is the Hermitian operator
\begin{equation}
\label{Ham5-6}
H_D = \sum\limits_{k=1}^r \varepsilon_k \hh_k = U H U^\dagger,
\end{equation}
where $U=e^{iF}$ is a unitary operator ($F = F^\dagger$). In principle, $U$ and
$H_D$ can be obtained by a simple matrix diagonalization algorithm working in
some low-dimensional faithful representation of $\fh$. However, this is not
sufficient to show that the eigenvectors and eigenvalues obtained correspond
to a group
operation and an element in $\fh_D$, respectively. Nevertheless, an important
result in Lie theory states that an operator such as 
$U$ can always be chosen to be a
group operator induced by $\fh$, that is
\begin{equation}
\label{group1}
F=  \sum\limits_{j=1}^M \zeta_j \hO_j \  , \in \fh.
\end{equation}

Because $H_D \in \fh_D$, its eigenstates are the weight states $\ket{\phi^p}$
of Eq.~\ref{gmfhws}.
Then,
\begin{equation}
\ket{\tilde{\phi}^p} = U^\dagger \ket{\phi^p} = N_p U^\dagger \prod_{j=1}^{l}
\hE^{n_j}_{-\alpha_j} \ket{\sf HW}
\end{equation}
are the eigenstates of $H$. In particular, $F$ can be chosen such that
$\ket{{\sf HW}}$ is the ground state of Eq.~\ref{Ham5-6}\footnote{Equivalently,
one can choose $F$ such that the lowest weight state $\ket{\sf LW}$ is a ground
state of $H_D$.} and
\begin{equation}
\ket{G}=U^\dagger \ket{\sf HW}
\end{equation}
is the ground state of Eq.~\ref{Ham5-5}.
Therefore, $\ket{G}$ is
generalized unentangled and GCS of $\fh$.
Moreover, considering that $U$ is a similarity transformation, the
spectrum of Eq.~\ref{Ham5-5} is then given by
\begin{equation}
H \ket{\tilde{\phi}^p} = \left[\sum\limits_{k=1}^r \varepsilon_k e_k^p
\right]\ket{\tilde{\phi}^p},
\end{equation}
where the weights $e_k^p$ can be obtained from the highest weight vector as in
Eq.~\ref{hweigen}.

In brief, Eq.~\ref{Ham5-5} is diagonalized if the coefficients $\varepsilon_k$
of Eq.~\ref{Ham5-6} and $\zeta_j$ of Eq.~\ref{group1} are obtained. This
can be done in any
faithful representation of $\fh$.  To show this, assume that $\bar{O}_j$ is any
faithful matrix representation associated to the operator $\hO_j$. From the
definition of the exponential mapping, one obtains
\begin{equation}
\label{expserie}
e^{iF} H e^{-iF} = H + i \left[F,H \right] -\frac{1}{2} \left[F, \left[F,H
\right] \right]+ \cdots = \sum\limits_{k=1}^r \varepsilon_k \hh_k,
\end{equation}
and, equivalently,
\begin{equation}
e^{i\bar{F}} \bar{H} e^{-i\bar{F}}= \bar{H} + i \left[\bar{F}, \bar{H} \right] 
-\frac{1}{2} \left[\bar{F}, \left[\bar{F},\bar{H}
\right] \right]+ \cdots = \sum\limits_{k=1}^r \varepsilon_k \bar{h}_k,
\end{equation}
which is the matrix representation of Eq.~\ref{expserie}.  Since $M \le
\mbox{poly}(N)$, then $H_{MF}$ can be exactly diagonalized\footnote{By exact
diagonalization, I mean efficient diagonalization.} using the  classical
algorithm given in the following section.

\subsection{Diagonalization Procedure}
\label{sec5-2-1}
Any element $H$ of a real semi-simple Lie algebra $\fh$ can be diagonalized
through a similarity group transformation induced by $\fh$. Based on
Ref.~\cite{Wil93},  in this section I introduce a classical  algorithm for the
diagonalization when $\fh$ is also a compact Lie algebra. This algorithm is a
generalization of the well known Jacobi's algorithm~\cite{PTV92} used for
general matrix diagonalization.

Assume then that $\Phi$ is a faithful matrix representation of $\fh =\{\hO_1, \cdots,
\hO_M \}$, satisfying
\begin{equation}
\label{inner2}
{\sf Tr} \left[ \bar{O}_j \bar{O}_{j'} \right] = \delta_{jj'},
\end{equation}
with $\bar{O}_j = \Phi(O_j)$. Equation~\ref{inner2} is always satisfied
in the case of compact Lie algebras if working with the adjoint representation.
In general, an orthogonal basis is given
by the Hermitian matrices $\bar{h}_k$, $[\bar{E}_{\alpha_j} +
\bar{E}_{-\alpha_j}]$, and $i[\bar{E}_{\alpha_j} -\bar{E}_{-\alpha_j}]$ (up to
normalization factors), which are representations of the operators in the CSA
of $\fh$ and in the set of ladder operators, respectively. Therefore, the trace
inner product of Eq.~\ref{inner2} allows one to project any operator in $\fh$
onto the sets $\fh_D$, $\fh_+$, and $\fh_-$. That is, the coefficients
$\gamma_k$ and $\iota_j$ of Eq.~\ref{Ham5-5} can be obtained by projecting the
matrix $\bar{H}_{MF}$, associated to $H_{MF}$, onto the representations of the
corresponding operators. 

To find the group operation $U$ of Eq.~\ref{Ham5-6}, one starts by
searching classically the index $t$ such that $|\iota_t| \ge |\iota_{j}| \
\forall j \in[1 \cdots l]$.
Second, one needs to perform a particular group (unitary)
operation $U_t$ such that
the transformed operator $H^1 = U_t H^0 U_t^\dagger$, with $H^0 \equiv H_{MF}$,
is
\begin{equation}
\label{Ham4}
H^1 = \sum\limits_{k=1}^r \gamma^1_k \hh_k + \sum\limits_{j=1}^l
\iota^1_{j} \hE_{\alpha_{j}} + (\iota^1_{j})^* \hE_{-\alpha_{j}}\ 
; \ \iota^1_t=0.
\end{equation}
The operator $U_t$ can be obtained by noticing the existence in $\fh$
of the $\fsu^t(2)$ Lie algebra
\begin{equation}
\fsu^t(2)
\left\{
\matrix{
S_+^t &=& \left[\sum_k (\alpha^k_t)^2\right]^{-1} \hE_{\alpha_t}, \cr
S_-^t& = &\left[\sum_k (\alpha^k_t)^2\right]^{-1} \hE_{-\alpha_t}, \cr
S_z^t &=& \left[\sum_k (\alpha^k_t)^2\right]^{-2} \left[\sum_k \alpha^k_t
\hh_k\right],
}
\right.
\end{equation}
where $\alpha_t^k$ are the components of the associated root vector. The
operators $S_x^t$ and  $S_y^t$ are obtained from the relations $S_\pm^t
=\frac{1}{2}[S_x^t \pm i S_y^t]$. Since  $\hh_k$, $\hE_{\alpha_j}$, and
$\hE_{-\alpha_j}$ satisfy the commutation relations defined in
Sec.~\ref{sec3-2-2}, the operators  $S_x^t$, $S_y^t$, and $S_z^t$ satisfy the
$\fsu(2)$ commutation relations of Eqs.~\ref{comm1}  (i.e., spin operators).
Equation~\ref{Ham5-5} can then be written as
\begin{equation}
H= H_t + H_t^\perp,
\end{equation}
where $H_t=\nu^t_x S^t_x + \nu^t_y S^t_y + \nu^t_z S^t_z\in \fsu^t(2) , \
 \nu^t_\alpha \in \mathbb{R}$, and $H_t^\perp$ is an element of the orthogonal
complement $\fsu^t(2)^\perp$ as defined by  Eq.~\ref{inner2}.
 
Therefore, $U_t$ is defined as the operator in the group $SU^t(2) \equiv e^{i
\fsu^t(2)}$ that diagonalizes $H_t \in \fsu^t(2)$. That is,
\begin{equation}
\label{su2op}
U_t = \exp \left[i(\mu^t_x S^t_x + \mu^t_y S^t_y+ \mu^t_z S^t_z) \right],
\end{equation}
where $\mu^t_\alpha \in \mathbb{R}$, and
\begin{equation}
U_t H_t U_t^\dagger \propto S_z^t \in \fh_D.
\end{equation}
This is a diagonalization problem in $\fsu(2)$ and the defining
coefficients $\mu^t_\alpha$ of Eq.~\ref{su2op} can be obtained through the
diagonalization of a $2 \times 2 $ matrix\footnote{In fact, different sets of
coefficients
$\mu_\alpha^t$ can be used to diagonalize this problem allowing one to choose
the sign of the proportionality coefficient}.

Remarkably, $U_t$ leaves invariant the decomposition
\begin{equation}
\label{algdec}
\fh= \fsu^t(2) \oplus \fh_D^{t,\perp} \oplus \fh_\pm^{t,\perp},
\end{equation}
where $\fh_D^{t,\perp}$ and $\fh_\pm^{t,\perp}$
are the sets of operators in $\fh_D$ and $(\fh_+ \oplus
\fh_-)$, respectively,
which are orthogonal to $\fsu^t(2)$. To show this, I first notice that
\begin{equation}
\label{invprop1}
\left[ \fsu^t(2) , \fh_D^{t,\perp} \right] =0,
\end{equation}
or otherwise a group operation in $SU^t(2) \equiv e^{i \fsu^t(2)}$ could transform
operators in $\fh_D^{t,\perp}$ into operators in $\fsu^t(2)$. This is not
allowed due to the orthogonalization property of Eq.~\ref{inner2}. 
For the same reason,
\begin{equation}
\label{invprop2}
\left[ \fsu^t(2) , \fh_\pm^{t,\perp} \right] \subset \fh_\pm^{t,\perp}.
\end{equation}
Equations~\ref{invprop1} and \ref{invprop2} are sufficient to guarantee the
invariance of the decomposition in Eq.~\ref{algdec} under the action of $U_t$.
Then, such $U_t$ anihilates the $t$-component of $H_{MF}$ and transforms it as in
Eq.~\ref{Ham4}.

The same $\fsu(2)$-diagonalization procedure is applied to $H^1$ and so on, to
obtain $H^2, \cdots, H^p$. In each step, a certain component $t$ is eliminated as
described above. Therefore, $H^p$ gets closer to a diagonal form and
\begin{equation}
\lim_{p\rightarrow \infty} H^p \subset \fh_D.
\end{equation}
To show this, I first write $H^p$ as
\begin{equation}
\label{pham}
H^p = \sum\limits_{k=1}^r \gamma^p_k \hh_k + \sum\limits_{j=1}^l
\iota^p_j \hE_{\alpha_j} + (\iota^p_j )^* \hE_{-\alpha_j} ,
\end{equation}
where $H^0 \equiv H_{MF}$ and $\gamma^0_k = \gamma_k$, $\iota^0_j = \iota_j$.
The square distance of $H^p$ to $\fh_D$ is defined as
\begin{equation}
\label{sqdistance}
d_C(H^p) = \sum\limits_{j=1}^l |\iota^p_j|^2,
\end{equation}
which is equivalent to count for the off-diagonal elements as in the 
Jacobi's diagonalization algorithm for symmetric matrices~\cite{PTV92}. 
In particular,
$d_C(H^p)=0$ when $H^p \in \fh_D$. 
If $t$ denotes the index of the biggest
$|\iota^{p-1}_j|$ in the $p-1$ iteration, then
\begin{equation}
\sum\limits_{j=1}^l |\iota^{p-1}_j|^2 \le l |\iota^{p-1}_t|^2.
\end{equation}
Moreover, since
$U_t$ leaves $\fh_\pm^{t,\perp}$ invariant (Eq.~\ref{algdec}), then
\begin{equation}
d_C(H^p) = \sum\limits_{j=1 (\ne t)}^l |\iota^p_j|^2=
\sum\limits_{j=1 (\ne t)}^l |\iota^{p-1}_j|^2.
\end{equation}
Therefore,
\begin{equation}
d_c(H^p)= \sum\limits_{j=1 ( \ne t)}^l
|\iota^{p-1}_j|^2 = \sum_{j=1}^l |\iota^{p-1}_j|^2 -
|\iota^{p-1}_t|^2 \le \left( \frac{l-1}{l} \right )
\sum\limits_{j=1}^l
|\iota^{p-1}_j|^2;
\end{equation}
that is,
\begin{equation}
\label{scaling0}
d_c(H^p) \le \left( \frac{l-1}{l} \right ) d_C(H^{p-1}).
\end{equation}
Since $(\frac{l-1}{l}) <1$, then
\begin{equation}
\lim_{p \rightarrow \infty} d_C(H^p) = 0.
\end{equation}

The group operation $U$ of Eq.~\ref{Ham5-6} is then
\begin{equation}
U = \lim_{P \rightarrow \infty} \prod\limits_{p=1}^P U_{t(p)},
\end{equation}
where $t(p)$ is the index for the biggest $|\iota^p_j|$ at step $p$.
Nevertheless, Eq.~\ref{scaling0} assures a rapid convergence and $U$ can be
well approximated by a finite product of operators (i.e., $P < \infty$). 
If the desired accuracy is denoted by $\epsilon$, 
the number $P$ of required iterations
is defined by
\begin{equation}
\label{acc1}
d_C(H^P) \le \epsilon,
\end{equation}
which is satisfied if (see Eq~\ref{scaling0})
\begin{equation}
\label{scaling1}
\left[\frac{l}{l-1} \right]^P d_C(H_{MF}) \le \epsilon.
\end{equation}
Naturally, the larger the dimension $M$ of $\fh$ is (i.e., the larger $l$ is), 
the larger the number $P$
of iterations required to obtain such an accuracy. 

To obtain $P$ as a function of $M$, I start by
rewriting Eq.~\ref{scaling1} as ($M=2l+r$)
\begin{equation}
\label{scalin2}
\log \epsilon \ge P \log (1-1/l) + \log (d_C(H_{MF})) \sim -P /l+ \log
(d_C(H_{MF})).
\end{equation}
Since $H_{MF} \in \fh$, the following approximation can be performed: 
\begin{equation}
d_C(H_{MF}) \sim M
d^0_C \ \ \ (M \gg1),
\end{equation}
where $d^0_C \in \mathbb{R}$ referres to some characteristic reference
distance. Therefore, Eq.~\ref{scaling1} reads
\begin{equation}
\label{scaling3}
P \ge l \left[ \log M - \log \frac{\epsilon}{d^0_C} \right];
\end{equation}
that is, the desired accuracy $\epsilon$ in the diagonalization
is guaranteed if $P$ satisfies 
Eq. \ref{scaling3}. In particular, if $M \gg1$, the integer $P$ is bounded above
by a polynomial of second order in $M$ (i.e., $M > \log M$ for $M \gg1$).

Equation~\ref{scaling3} is necessary but not sufficient  to assure the
efficiency of this diagonalization method. It remains to be shown then that
$H^P$ can be obtained efficiently. For this purpose, I consider a simple
classical algorithm based mainly on standard matrix multiplication. The idea is to
work in a certain $q \times q$ dimensional and faithful representation of $\fh$, such
as the adjoint representation (Appendix~\ref{appC}). 
In brief, the algorithm is based on two
main steps: The search for the biggest $|\iota^p_t|$ and the diagonalization in
$\fsu^t(2)$. These steps are repeated $P$ times to diagonalize $H$.
 If $\bar{H}^p$ (where $p \in [1 \cdots P]$) is the matrix
representation of $H^p$, the biggest $|\iota^p_t|$ is found by projecting
the corresponding matrix onto the matrices $\bar{E}_{\alpha_j}$ and 
$\bar{E}_{-\alpha_j}$, with $j \in [1\cdots l]$ (i.e., the representations
of the ladder operators). This can be done with a conventional computer
in ${\cO}(q^2 l)$ computational operations (i.e., addition and
multiplication of complex numbers). Once 
$\iota^p_t$ has been found, the corresponding $\fsu^t(2)$-rotation has to be
performed. This operation is also represented by a $q \times q$ matrix and the
representation $\bar{H}^{p+1}$ of $H^{p+1}$ can be obtained with ${\cO
}(q^2)$ computational operations.

In brief, the coefficients defining $H^P$ in Eq.~\ref{pham} can be obtained with
computational accuracy in ${\cO}(Pl q^2)$ computational operations. From
Eq.~\ref{scaling3} and if $M \le \mbox{poly}(N)$  and $q$  satisfies $q \le M$,
then $H^P$ can be efficiently obtained\footnote{This evaluation is done at the
computational accuracy given by the CC. Such an accuracy decreases polynomially
in the
number of computational operations, so the method is efficient.} in, at most, 
$\mbox{poly}(N)$
operations.

\subsection{Example: Fermionic Systems}
\label{sec5-2-2}
Consider a fermionic lattice system with Hamiltonian given by
\begin{equation}
H = \sum\limits_{j,j'=1}^N \lambda_{jj'} c^\dagger_j c_{j'},
\end{equation}
where $N$ is the size of the lattice, $\lambda_{jj'}=(\lambda_{j'j})^*$, and
$c^\dagger_j$ ($c_j$) the fermionic creation (annihilation) operators for site
$j$. Such a Hamiltonian is known to be exactly solvable and can be diagonalized
through a Bogolubov transformation~\cite{BR86}. Nevertheless, it constitutes a
simple example to apply the methods described above. 

For this purpose, I start by noticing that $H \in \fu(N)$, where
$\fu(N)=\{c^\dagger_j c_{j'} ; \ j,j' \subset[1 \cdots N]\}$ is the Lie algebra
of dimension $M=N^2$ introduced in Sec.~\ref{sec3-3-3}. Therefore, it can be
efficiently diagonalized by working in a low-dimensional, faithful,
representation of $\fu(N)$ such as the one given by
\begin{equation}
c^\dagger_j c_{j'} \leftrightarrow T_{jj'},
\end{equation}
where $T_{jj'}$ are the $N\times N$ matrices with +1 in the $j$th row and
$j'$th column, and zeros otherwise. In this representation, $H$ has associated
the following matrix:
\begin{equation}
\bar{H}= \pmatrix{ \lambda_{11} & \lambda_{12} &\cdots & \lambda_{1N} \cr
\vdots &\vdots & & \vdots \cr \lambda_{N1} &\lambda_{N2}
&\cdots & \lambda_{NN} } ,
\end{equation}
which can be easily diagonalized in a conventional computer, obtaining
\begin{equation}
\label{fermdiag}
\bar{H}_D =\bar{U} \bar{H} \bar{U}^\dagger = \pmatrix { \varepsilon_1 & 0 & \cdots &0 \cr
0 & \varepsilon_2 & \cdots & 0 \cr \vdots & \vdots & & \vdots \cr 0 & 0 & \cdots
& \varepsilon_N},
\end{equation}
where $\bar{U}$ is a $N \times N$ unitary matrix. In this representation,
$T_{jj'}$ span the whole set of $N \times N$ matrices, and any possible
$\bar{U}$ is
directly associated with a group element induced by $\fu(N)$. For this reason,
the diagonalization
procedure described in the previous section to assure a group transformation 
needs not be performed. Therefore, 
Eq.~\ref{fermdiag} defines the diagonal form of $H$ through
\begin{equation}
\bar{H}_D = \sum\limits_{k=1}^N \varepsilon_k T_{kk} \leftrightarrow
\sum\limits_{k=1}^N \varepsilon_k c^\dagger_k c_k \equiv H_D,
\end{equation}
where $\varepsilon_k$, the eigenvalues of $\bar{H}$, are now the single-fermion
energies.

In this case, the diagonalization method used turns out to be exactly
equivalent to the one based on the Bogolubov transformation for fermionic
quadratic Hamiltonians. It is also directly applicable to diagonalize a bigger
family defined by the fermionic operators of the algebra $\fso(2N)$
(Sec.~\ref{sec5-1-2}). Additionally, this method can also be used to diagonalize
more complex problems in which a Bogolubov transformation is not
straightforward to implement.

\section{Summary}
In this chapter, I have addressed two important problems in condensed matter
theory through a quantum information theory point of view:  The
characterization of quantum phase transitions and the efficient diagonalization
of many-body Hamiltonians. It has been shown that the concept of GE, which
becomes very useful when there is a Lie algebraic structure associated with the
problem, plays a
significant role in these cases. 

First, since QE is related to the existence of quantum correlations and these
dominate the different phases, I have shown that by choosing a relevant set of
observables, the relative purity contains information about the critical
exponents of the phase transitions in two models of interest. Second,
as motivated by the results of Chap.~\ref{chapter4}, I have shown that
whenever an interaction Hamiltonian is an element of a low-dimensional
semi-simple Lie algebra, it can be diagonalized efficiently through algebraic
methods. This constitutes a major result in condensed matter theory and, in
principle, it could be extended for the case of infinite dimensional Lie
algebras which are generated by a finite set of functions~\cite{OSD05}.

\chapter{Conclusions}
\label{chapter6}
When is a quantum computer useful?; Which problems can be solved more
efficiently with a quantum computer than with a conventional one?  Although not
known yet,  finding the answer to these questions constitutes the main reason
that makes the science of quantum information a prospering and rapid-growing
field. In this thesis, I have addressed various subjects in quantum information
theory,
including the quantum simulations of physical systems,
quantum entanglement, and quantum complexity, to prove some of the capabilities
of quantum computation. The main idea can be briefly stated as follows: To
efficiently simulate a physical system with a quantum computer, the laws of
quantum mechanics need to be exploited to our advantage. Quantum entanglement,
a non-classical property, is at the core of many tasks in quantum information
and, if entangled states are created in a quantum simulation, such phenomena
cannot be easily reproduced with a conventional computer. Therefore, it is 
expected that a  computer which imitates the system to be simulated, i.e., a
quantum computer, will be the most efficient device for this purpose. 
In the following, I present the principal results and conclusions from each
chapter of this thesis.
\\

In Chap.~\ref{chapter2}, I addressed several broad issues associated with
the simulation of physical phenomena by quantum networks. I first introduced
the conventional model of quantum computation (Sec.~\ref{sec2-1-1}) as the main model used to
perform these simulations. I studied the implementation of deterministic
quantum algorithms (Sec.~\ref{sec2-2}) which allow one to obtain relevant physical properties,
usually related with the evaluation of some correlation function of the system
under study. In general, the physical system one is interested in simulating is
expressed by some operator algebra that may differ from the operator
algebra associated with the quantum computer (i.e., Pauli operators). I pointed
out that efficient mappings between these two sets of  operators exist in many
cases (Sec.~\ref{sec2-3}) and are sufficient to establish the equivalence of the different physical
models to a universal model of quantum computation such as the conventional
model. In Sec.~\ref{sec2-3}, I explained how these mappings can be used to perform
quantum simulations of fermionic, anyonic, and bosonic systems, respectively,
in a quantum computer made of qubits.

I also explored various issues associated with efficient quantum simulations. A
simulation is said to be efficient when the amount of resources required is
bounded above by a polynomial in some variable $N$, such as the size of the
physical system to be simulated. The main topics I addressed were how to reduce
the number of qubits and  elementary quantum gates needed for the simulation
and how to increase the amount of physical information measurable. I showed
that the evaluation of some correlation functions in efficiently prepared
initial quantum states, like fermionic product states (Sec.~\ref{sec2-3-1}),
can be efficiently done on a quantum computer.  In some cases, this presents an exponential
speed-up with respect to the corresponding classical simulation
(Sec.~\ref{sec2-5}), where no known efficient algorithms exist (i.e., it would
require an exponentially large amount of resources). However, it remains to be
shown wether other tasks related with physical simulations on quantum computers can be performed
efficiently or not. For example, there is no known efficient quantum algorithm
to obtain the ground state energy of  the two-dimensional Hubbard Hamiltonian.
This is due to an exponential decay of the overlap between the efficiently
prepared initial state and the actual ground state of the model
(Sec.~\ref{sec2-4}). 

As a proof of principles, in Sec.~\ref{sec2-6} I presented the simulation of a
quantum many-fermion system, the Fano-Anderson model, using a liquid-state NMR
based quantum information processor. Relevant correlation functions were obtained by executing the
quantum algorithms described in Sec.~\ref{sec2-3}. For this purpose, a pulse
sequence consisting of rf pulses acting on an ensemble of trans-crotonic acid
molecules was performed. Moreover, different approximation and refocusing
schemes were used to optimize such pulse sequence and minimize the errors of
the simulation. The results obtained were very accurate because the overall
duration of the simulation was much smaller than the decoherence time of the
system. This quantum simulation was performed efficiently, i.e.,
with polynomial complexity in the system size.

Although the studies on efficiency were done by considering the conventional
model of quantum computation,  the results obtained are independent of the
actual physical representation of the quantum computer. A generalization of
these results can be obtained by means of the notion of generalized
entanglement, which was presented in Chap.~\ref{chapter3}. This generalization
of entanglement goes beyond the standard subsystem-based approach, and is a
feature of quantum states relative to a preferred set of observables of the
system under study: it is an observer-dependent concept. To understand the
main differences with the usual notion, I described the main properties of
quantum entanglement in Sec.~\ref{sec3-1}. In Sec.~\ref{sec3-2-2} I tied together the
theory of entanglement and the theory of coherent states whenever the preferred
set of observables is a semi-simple Lie algebra. Important results were obtained
in this case, where generalized unentangled states were defined as the extremal
states of the algebra or generalized coherent states. Furthermore, these states
present least uncertainty and can be considered as the most classical states in
some sense. Some examples were presented in Sec.~\ref{sec3-3}.

The main conclusion of Chap.~\ref{chapter3} is then that conventional
entanglement is a special case of a much more general theory, and this should
be deeply analyzed to take advantage of the quantum world in different quantum
information protocols like quantum teleportation, quantum computation, etc. 
For this purpose, in Chap.~\ref{chapter4}, I studied some of the capabilities of
generalized entanglement. Since traditional entanglement is known to be a
resource for several tasks in quantum information (Sec.~\ref{sec4-1}),
including quantum computation (Sec.~\ref{sec4-2}), one would expect that a more
general theory of
entanglement  would allow one to better understand the reasons lying behind
the power of quantum computers. Therefore, in Sec.~\ref{sec4-3} I presented a wide class of
problems that can be solved efficiently on
both, a quantum computer and a conventional one.  These problems are related with the
evaluation of a particular type of correlation function. In these special cases, the corresponding
quantum simulation involves only generalized unentangled states relative to a
certain polynomially large (or polynomially bounded)
set of observables (Sec.~\ref{sec4-4}). I showed that if no generalized
entangled states, with respect to these sets, is created at some step of the
quantum simulation, this task can be efficiently reproduced with a conventional
computer.

This important result indicates that although entangled states (in the usual
sense) could be involved in the quantum simulation, this is not a sufficient
condition to state that a quantum computer is more powerful than a classical
one for these simulations. Nevertheless, if generalized entangled states
relative to all polynomially large sets of observables are created in the
quantum simulation, such phenomena cannot be easily reproduced with a
conventional computer, and no known efficient classical algorithms exist in
this case. This represents, again, an exponential speed-up with respect to the
classical simulation.

In Chap.~\ref{chapter2}, I discussed some issues related with efficient initial
state preparation when simulating a physical system with a quantum computer made of qubits. 
These results have been generalized in Sec.~\ref{sec4-4-1} from the point of
view of Lie algebras. Again, if the initial state to be prepared is the
generalized coherent state of a Lie algebra with polynomially large (or
polynomially bounded) dimension, such a state can be prepared efficiently; that
is, it can be prepared by applying a polynomially large number of elementary
gates. The type of gates depend on the physical representation of the quantum
computer but
the existence of one-to-one mappings makes this result independent of
such representation.

In Chap.~\ref{chapter5}, I addressed two important topics in condensed matter
theory: The characterization of broken-symmetry quantum phase transitions and
the exact diagonalization of Hamiltonians, from the point of view of
generalized entanglement.
In Sec.~\ref{sec5-1}, I showed that the relative purity, which
constitutes a measure of generalized entanglement in the Lie algebraic
case, successfully distinguishes between the different phases present in
the LMG model and in the anisotropic XY model in a transverse magnetic field.
In these cases, the corresponding ground states can be exactly obtained, so
choosing the preferred set of observables that contains the relevant
correlations in the different phases becomes relatively easy. Nevertheless,
applying these concepts to a more general case can be done, in principle, by
following the same strategy. However, determining in a systematic way the
minimal set of observables whose relative purity is able to signal and
characterize the quantum phase transition requires, in general, an elaborate
analysis.

Finally, in Sec.~\ref{sec5-2} I introduced the general mean-field Hamiltonians
as those operators that are elements of polynomially large (or polynomially
bounded) dimensional Lie algebras. This is a generalization of the known
mean-field Hamiltonians such as the one composed by quadratic fermionic or
bosonic operators. I pointed out that the existence of low-dimensional faithful
representations guarantees the existence of efficient classical algorithms for
their diagonalization (Sec.~\ref{sec5-2-1}). In particular, the Bogolubov
transformation is a special case of this type (Sec.~\ref{sec5-2-2}). 

Much remains to be done to really understand the power of quantum computers. As is the case
for other
investigators in the field, I believe that a complete understanding of quantum
entanglement for pure and mixed states is the key that will unlock such power.
I hope that this thesis has been an interesting approach to the subject.
In the following section, I list a set of problems that may deserve further
investigation but are out of the scope of the present work.

\section{Future Directions}
Most of the results about efficiency in quantum simulations of  physical
systems (Chap.~\ref{chapter2}) were based on the implementation of a particular
type of algorithms. Nowadays, adiabatic quantum computation~\cite{FGG00} has
emerged as an important topic in which it is expected that certain quantum states,
such as the ground state of the two-dimensional Hubbard model, could be
efficiently prepared by slowly changing some Hamiltonian interactions. It is
important then to investigate wether there is a connection between the quantum
complexity associated to both types of algorithms.

Regarding the theory of generalized entanglement, several issues related to
mixed-state entanglement also need further investigation. First, the natural
extension makes any measure, such as the relative purity,  very hard to compute
and more efficient ways for its evaluation need to be studied. Second, it is
important to extend the results about efficiency to the mixed-state case. For
example, the simplest case would be to consider mixed states which are a finite
convex combination of generalized unentangled states relative to a polynomially
large dimensional Lie algebra.  Here, the computation of some correlation
functions would still be tractable (efficient) on a conventional computer. 

The classical algorithms described in Chap.~\ref{chapter4} were used to show
that certain tasks, like the evaluation of some correlation functions, can be
done with polynomial complexity on a conventional computer. However, it may be
possible to find even more efficient classical algorithms for this purpose, and
it would be worthwhile to compare their complexity with that of the 
corresponding quantum
one. In fact, many quantum algorithms found in the literature, like Grover's
algorithm, do not have an exponential speed-up with respect to their classical
simulation.

In Chap.~\ref{chapter5}, I analyzed some quantum phase transitions from the
point of view of generalized entanglement. It yet remains to be understood how
to choose the set of observables that captures the most relevant correlations,
which distinguish different quantum phases, in a more general case. That is, by
just performing a Lie algebraic analysis of the interaction in the Hamiltonian,
can we always distinguish the preferred set of observables that characterizes
the phase transition?

Finding the solution to these and other problems, like characterizing a bigger
set of exactly-solvable systems in a Lie algebraic framework, constitute,
together with the results obtained throughout this thesis, an advance towards the
unification of Physics and Information Processing Theory. At the end,
information is physical.

\appendix
\setcounter{chapter}{0}
\roman{chapter}

\chapter{Discrete Fourier Transforms}
\label{appA}
In practice, to evaluate the discrete Fast Fourier Transform (DFFT) 
one uses discrete samples, therefore Eq.~\ref{FFT} must be modified 
accordingly. From Fig.~\ref{fig2-12} it is observed that instead of having 
$\delta$-functions (Dirac's functions), there are finite peaks in some 
range of energies, close to the eigenvalues of the Hamiltonian.
Accordingly, one cannot determine the eigenvalues with the same 
accuracy as other numerical calculations. However, there are some 
methods that give the results more accurately than the DFFT.

As a function of the frequency $\eta_l$, the DFFT
($\tilde{S}(\eta_l)$) is given by:
\begin{equation}
\label{DFTS}
\tilde{S}(\eta_l)=\Delta t \sum\limits_{j=1}^{M} S(t_j) 
e^{i \eta_l t_j} \ ,
\end{equation}
where $t_j=j \Delta t$ are the different times at which the  function $S$ is
sampled (Eq.~\ref{spectra2}),  $\eta_l=\frac{2 \pi l}{M \Delta t}$ are the
possible frequencies to evaluate the FFT of $S(t)$\footnote{Only a discrete set
of frequencies can be obtained from the evaluation of the DFT over a discrete
sample. In this case, the Nyquist critical frequency is given by $\nu_c=
\frac{2 \pi}{\Delta t}$.}, and $M$ is the number of samples. 

Since one is interested in $S(t)=\sum_n|\gamma_n|^2 \
e^{-i\lambda_nt}$ (Eq.~\ref{spectra2}), then
\begin{equation}
\tilde{S}(\eta_l)=\Delta t\sum_n 
|\gamma_n|^2 \sum\limits_{j=0}^{N-1}
e^{i[\eta_l-\lambda_n]t_j} \ ,
\end{equation}
and
\begin{equation}
\label{serie}
\tilde{S}(\eta_l)=\Delta t\sum_n |\gamma_n|^2 
\ \frac{e^{i (\eta_l -
\lambda_n)\Delta t N}-1}{e^{i (\eta_l -\lambda_n)\Delta t}-1} \ .
\end{equation}
If $\eta_l$ is close to one of the eigenvalues $\lambda_n$, and these are
sufficiently far apart to be well resolved, all terms in the sum of
Eq.~\ref{serie}, other than $n$, can be neglected. Taking $\eta_l$ and
$\eta_{l+1}=\eta_l+\frac{2\pi}{M \Delta t}$, 
both close to $\lambda_n$ in such a way that
$|\tilde{S}(\eta_l)|, |\tilde{S}(\eta_{l+1})|\gg 0$, then 
\begin{equation}
\label{coef}
\frac{\tilde{S}(\eta_{l+1})}{\tilde{S}(\eta_l)} \approx
\frac{e^{i (\eta_l-\lambda_n)
\Delta t}-1}
{e^{i (\eta_{l+1}-\lambda_n)\Delta t}-1} \ .
\end{equation}
After simple algebraic manipulations (and approximating  $e^{i ( \eta_l -
\lambda_n ) \Delta t} \approx 1+i (\eta_l - \lambda_n ) \Delta t$ and the
same for the denominator in Eq.~\ref{coef}) the correction to
the energy is
\begin{equation}
\lambda_n=\eta_l+\Delta\lambda_n
\end{equation}
with
\begin{equation}
\Delta\lambda_n \approx -\frac {2\pi}{M \Delta t}
{\rm Re}
\biggl[\frac{\tilde{S}
(\eta_{l+1})}{\tilde{S}(\eta_l)-\tilde{S}
(\eta_{l+1})} \biggr].
\end{equation}

\chapter{Discrete Fourier Transform and Propagation of Errors}
\label{appB}
Theoretically, the function $S(t)$ of Eq. \ref{fanospectrum} is a
linear combination of two complex functions having different
frequencies: $S(t) = |\gamma_1|^2 e^{-i \lambda_1 t} + |\gamma_2|^2
e^{-i \lambda_2 t}$, where $\lambda_i$ are the eigenvalues of the
one-particle eigenstates, defined as $\ket{1{\sf P}_i}$,  in the
Fano-Anderson model with $n=1$ site and the impurity (Sec.~\ref{sec2-7}),
and $\lambda_i=|\langle \phi \ket{1{\sf P}_i}|^2$,
with $\ket{\phi}=\ket{\downarrow_1
\uparrow_2}$. However, the liquid-state NMR setting used to experimentally
measure $S(t)$ adds a set of errors that cannot be
controlled, and the function $S(t)$ shown in Fig.~\ref{fig2-21} is no
longer a contribution of only two different frequencies.

As mentioned in Sec.~\ref{sec2-7}, $S(t)$ was obtained experimentally
for a discrete set of values $t_j = j \Delta t$, with
$j=[1,\cdots,M=128]$ and $\Delta t= 0.1$ s. Its DFT is given by Eq.~\ref{DFTS}.
Since one is evaluating the spectrum of a physical (Hermitian) Hamiltonian, the
imaginary part of $\tilde{S}(\eta_l)$ is zero\footnote{Due to experimental
errors, the imaginary part of $\tilde{S}(\eta_l)$ could be different from zero.
However, I only consider its real part because it contains all the desired
information.}.  In Fig.~\ref{fig2-22}, I show $\tilde{S}(\eta_l)$ obtained from
the experimental points $S(t_j)$ of Fig.~\ref{fig2-21}. Its error bars (i.e.,
the size of the line in the figure) were calculated by considering the
experimental error bars of $S(t_j)$ in the following way: First, I rewrite Eq.
\ref{DFTS} as
\begin{equation}
\tilde{S}(\eta_l) = \sum\limits_{j=1}^M Q_{lj} ,
\end{equation}
with $Q_{lj} = M^{-1} [{\sf Re}(S(t_j)) \cos( \eta_l t_j) - {\sf
Im}(S(t_j)) \sin( \eta_l t_j)]$ (real).  Then, the approximate standard
deviation ${\sf E}\tilde{S}_l$ of $\tilde{S}(\eta_l)$ depends on
the errors ${\sf E}Q_{lj}$ of $Q_{lj}$ as (considering a normal
distribution \cite{Tay97})
\begin{equation}
\label{totalerror1}
[{\sf E}\tilde{S}_l ]^2 \approx  \sum\limits_{j=1}^M [{\sf E}Q_{lj}]^2 .
\end{equation}
On the other hand, ${\sf E}Q_{lj}$ is calculated as~\cite{Tay97}
\begin{equation}
\label{Qerror}
[{\sf E}Q_{lj}]^2 = \left| \frac{\partial Q_{lj}}{\partial{\sf
Re}(S(t_j))} \right|^2 {\sf E_R}^2 + \left|\frac{\partial
Q_{lj}}{\partial{\sf Im}(S(t_j))} \right|^2 {\sf E_I}^2 ,
\end{equation}
where ${\sf E_R}$ and ${\sf E_I}$ are the standard deviations of the real and
imaginary parts of $S(t_j)$ (see Fig.~\ref{fig2-21}),
respectively. For experimental reasons (Sec.~\ref{sec2-7-1})
these errors are almost constant, having ${\sf E_R}\sim{\sf
E_I}\sim{\sf E_S}$ independently of $t_j$ (see Fig.~\ref{fig2-21}),
where ${\sf E_S}$ is taken as the largest standard deviation.  Combining
Eqs. \ref{totalerror1} and \ref{Qerror}, we obtain
\begin{equation}
\label{totalerror}
{\sf E}\tilde{S}_l = \left [ M^{-2} {\sf E_S}^2 \sum\limits_{j=1}^M
[|\cos (\eta_l t_j)|^2 + |\sin (\eta_l t_j)|^2 ] \right] ^{1/2} =
\frac{{\sf E_S}}{\sqrt{M}}.
\end{equation}
In the experiment, $M=128$ and ${\sf E}_S \approx 0.04$, obtaining
${\sf E}\tilde{S}_l \approx 0.0035$, which
determines the (constant) error bars (i.e., the size of the
dots representing data points) shown in Fig.~\ref{fig2-22}. 

The standard deviation ${\sf E}\eta_l$
in the frequency domain is due to the resolution of the sampling
time $\Delta t$. This resolution is related to the error coming from the
implementation of the $z$ rotations in the refocusing procedure.
A bound for this error is given by the resolution of the spectrum;
that is,
\begin{equation}
{\sf E}\eta_l \le \frac{2 \pi}{M \Delta t}\approx 0.5 \ .
\end{equation}

\chapter{The Adjoint Representation}
\label{appC}
The adjoint representation of an $M$-dimensional abstract
Lie algebra $\fh=\{ \hO_1,
\cdots \hO_M \}$ is the transformation that maps every operator in $\fh$ into a
$M \times M$ dimensional complex matrix given by the structure factors of
$\fh$~\cite{Geo99}.
If
\begin{equation}
\left[ \hO_j, \hO_{j'} \right] = i \sum\limits_{k=1}^M f_{jj'}^k \hO_k,
\end{equation}
where $[,]$ is the antisymetric form (e.g., the commutator) and  $f_{jj'}^k$ are
the structure factors, the matrices $\bar{O}_j$ given by the elements
\begin{equation}
\label{adjoint}
|| \bar{O}_j||_{j'k} = -i f_{jj'}^k \; \ \ j,j',k=[1 \cdots M]
\end{equation}
define the adjoint representation of $\fh$. In other words
\begin{equation}
\left[ \bar{O}_j , \bar{O}_{j'} \right] = i \sum\limits_{k=1}^M f_{jj'}^k
\bar{O}_k
\end{equation}
where $[\bar{A}, \bar{B}] = \bar{A}\bar{B}-\bar{B}\bar{A}$ is the usual
commutator between matrices.
Since the operators $\hO_j$ are Hermitian, the factors $f_{jj'}^k$ are real and
the adjoint representation is pure imaginary.

The Killing form is the bilinear form $K:\fh \times \fh \rightarrow \mathbb{R}$
given by the following mapping:
\begin{equation}
K(\hO_j,\hO_{j'})={\sf Tr} [ \bar{O}_j \bar{O}_{j'}].
\end{equation}
This mapping defines a convenient inner product in $\fh$. In particular, if
$\fh$ is compact one can always choose a linear transformation of the operators
in $\fh$ such that
\begin{equation}
\label{ortadj}
K(\hO_j,\hO_{j'})= \delta_{jj'}.
\end{equation}
In this case, the observables $\hO_j$ are said to be Schmidt-orthogonal.
Moreover, if the algebra $\fh$ is semi-simple and Eq.~\ref{ortadj} is satisfied, the
adjoint representation is a faithful representation of $\fh$; that is, every
operator $\hO_j$ has associated a unique, linearly independent, $M \times M$
matrix.

A group operation induced by $\fh$ is a linear unitary transformation of the form
\begin{equation}
U=e^{i\hat{H}},
\end{equation}
where $\hat{H}=\hat{H}^\dagger=\sum\limits_{j=1}^M \zeta_j \hO_j \in \fh$, and
$\zeta_j \in \mathbb{R}$. The exponential mapping $e^{iH}$ is
defined by
\begin{equation}
e^{i\hat{H}} = \one + i\hH + \frac{(i\hH)^2}{2!} +\frac{(i\hH)^3}{3!}+ \cdots
\end{equation}
Then, its action over an element of the
algebra is given by
\begin{equation}
\label{groupaction}
\hO'_j=e^{-i\hH} \hO_j e^{i\hH} = \hO_j -i \left[\hH , \hO_j \right] -\frac{1}{2}
\left[ \hH, \left[\hH , \hO_j \right] \right] + \cdots,
\end{equation}
which also belongs to the algebra, that is, $\hO'_j = \sum\limits_{j'=1}^M
\nu_{jj'} \hO_{j'} \ \in \fh$. The real coefficients $\nu_{jj'}$ define
an $M \times M$ dimensional
matrix $\nu$, whose properties are given by the nature of the adjoint
representation of $\fh$. To see this, consider the following decomposition:
\begin{equation}
U=\prod U_\Delta = \prod e^{i \Delta \hH}
\end{equation}
where the infinitesimal group operation $U_\Delta$ can be approximated by
\begin{equation}
U_{\Delta} \approx \one + i \Delta \hH \ ;\ \Delta \rightarrow 0.
\end{equation}
Naturally, this infinitesimal transformation induces another infinitesimal
transformation over $\fh$ given by the matrix
\begin{equation}
\nu \approx \one + \Delta . \mu,
\end{equation}
where $\mu$ is an $M \times M$ dimensional matrix with real coefficients
$\mu_{jj'}$ .

Keeping the first-order terms in Eq.~\ref{groupaction}, one obtains
\begin{equation}
\hO_j -i \Delta .\left[ \hH , \hO_j \right] = \hO_j + \Delta .\sum_{j'=1}^M
\mu_{jj'} \hO_{j'}.
\end{equation}
The matrix $\mu$ is then given by the adjoint representation $\bar{H}$ of the
element $\hH$ as defined by Eq.~\ref{adjoint}:
\begin{equation}
\bar{H}= i \mu,
\end{equation}
and the infinitesimal transformation is given by
\begin{equation}
\one + \Delta . \mu \approx e^{i \Delta .\mu} \ .
\end{equation}
Since $U$ is obtained after the successive application of $U_\Delta$, one obtains
\begin{equation}
\label{adjoint2}
\pmatrix{\hO'_1 \cr \vdots \cr \hO'_M}=
e^{-i\hH} \pmatrix{\hO_1 \cr \vdots \cr \hO_M} e^{i \hH} = e^{i \bar{H}} \pmatrix{\hO_1 \cr \vdots \cr \hO_M},
\end{equation}
where the $M \times M$ dimensional matrix $e^{i \bar{H}}$ is obtained by
exponentiating the adjoint representation of $\hH$. Equation~\ref{adjoint2} is
usually referred to the adjoint action of the group. In many cases\footnote{That
is, only valid for simply connected Lie algebras.}, the matrix
$e^{i \bar{H}}$ defines the adjoint representation of the group induced by
$\fh$.

\chapter{Separability, Generalized Unentanglement, and Local Purities}
\label{appD}
For a quantum system ${\cal S}$ whose pure states $\ket{\psi}$ belong to 
a Hilbert space ${\cal H}$ of dimension dim$({\cal H})=d$, the purity 
relative to the (real) Lie algebra of all traceless observables $\fh=\fsu(d)$
spanned by an orthogonal, commonly normalized Hermitian basis 
$\{ \hA_1, \cdots ,\hA_{M} \}$, $M=d^2-1$, is, according to Eq.~\ref{purity}, 
given by:
\begin{equation}
\label{puritysubsystem}
P_{\fh} (|\psi\rangle ) = {\sf K} \sum\limits_{i=1}^{M} \langle \hA_i
\rangle^2 .
\end{equation}
The normalization factor ${\sf K}$ depends on $d$ and is determined so that 
the maximum purity value is 1. If {\sf Tr}$(\hA_i \hA_{i'})=\delta_{ii'}$
(as for the standard spin-$1$ Gell-Mann matrices), then
${\sf K}=d/(d-1)$, whereas in the case {\sf Tr}$(\hA_i \hA_{i'})= d
\delta_{ii'}$ (as for ordinary spin-$1/2$ Pauli matrices), 
${\sf K}=1/(d-1)$.  Recall that any quantum state $\ket{\psi} \in {\cal H}$ 
can be obtained by applying a group operator $U$ to a reference state 
$\ket{{\sf ref}}$ [a highest or lowest weight state of $\fsu (d)$]; that is
\begin{equation}
\ket{\psi} =U \ket{{\sf ref}} \:,
\end{equation}
with $U=e^{i \sum_i\zeta_i \hA_i}$,  
and $\zeta_i \in \mathbb{R}$.  Therefore, any quantum state $\ket{\psi}$ is 
a GCS of $\fsu (d)$, thus generalized unentangled relative to the 
algebra of all observables: $P_{\fh} (\ket{\psi}) =1$ for all $\ket{\psi}$.

Let's now assume that ${\cal S}$ is composed of $N$ distinguishable subsytems, 
corresponding to a factorization ${\cal H} = \bigotimes_{j=1}^N {\cal H}_j$, 
with dim$({\cal H}_j)=d_j$, $d=\prod_j d_j$. Then the set of all {\it local} 
observables on ${\cal S}$ becomes 
$\fh=\fh_{loc} = \bigoplus_j \fsu^j(d_j)$.  An orthonormal basis which is suitable
for calculating the local purity $P_{\fh}$ may be obtained by considering a 
collection of orthonormal bases 
$ \fsu^j(d_j)=\{  \hA^j_1, \cdots , \hA^j_{L_j} \}$, 
$L_j=d_j^2-1$, each acting on the $j$th subsystem; that is, 
\begin{equation}
\label{Basis}
\hA^j_i = \overbrace{ \one^1 \otimes \one^2 \otimes \cdots \otimes
\underbrace{\hA_i}_{j^{th}\ \mbox{factor}} \otimes \cdots
\otimes \one^N}^{N\ \mbox{factors}} \: ,
\end{equation}
where $\one^j =\one/\sqrt{d_j}$. Then for any pure state $|\psi\rangle 
\in {\cal H}$ one may write
\begin{equation}
\label{puritysub}
P_{\fh} (|\psi\rangle )= {\sf K'} \sum\limits_{j=1}^N \Big[
\sum\limits_{i=1}^{L_j} 
\langle \hA^j_{i} 
\rangle^2 \Big] \:.
\end{equation}
By letting $\fh_j=\mbox{span}\{ {\hA}^j_{i} \}$ be the Lie algebra of traceless
Hermitian operators acting on ${\cal H}_j$ alone, the above equation is also 
naturally rewritten as
\begin{equation}
\label{puritysub1}
P_{\fh} (|\psi\rangle )= {\sf K'} \sum\limits_{j=1}^N 
{1 \over {\sf K}_j} P_{\fh_j} (|\psi\rangle )\:, 
\hspace{5mm} {\sf K}_j ={d_j \over {d_j-1}}\:.
\end{equation}
The $\fh_j$-purity $P_{\fh_j}$ may be simply related to the conventional 
subsystem purity.  Let $\rho_j ={\sf Tr}_{j' \not = j} 
(\{ |\psi\rangle\langle \psi|\})$ be the reduced density operator 
describing the state of the $j$th subsystem. Because the 
latter can be represented as 
\begin{equation}
\rho_j = \frac{\one}{d_j} + \sum\limits_{i=1}^{L_j} 
\langle \hA^j_{i} \rangle \hA^j_{i} \:,
\end{equation}
Eq.~\ref{puritysub} can be equivalently expressed as
\begin{equation}
\label{puritysub2}
P_{\fh} (|\psi\rangle )= {\sf K'} \sum\limits_{j=1}^N \Big[
{\sf Tr} \rho_j^2 - {1 \over d_j}  \Big] \:,
\end{equation}
that is, $P_{\fh_j} (|\psi\rangle ) = (d_j {\sf Tr} \rho_j^2 - 1)/(d_j-1) $. 
Clearly, the maximum value of either Eqs. ~\ref{puritysub1} or \ref{puritysub2} 
will be attained if, and only if, each of the conventional purities 
${\sf Tr}\rho_j^2=1$ $\leftrightarrow$ $P_{\fh_j} =1$ for all $j$, 
which allows one to determine the ${\sf K}'$-normalization factor as
\begin{equation}
{\sf K}'= \frac{1}{\sum_j {1 \over {\sf K}_j} } =
\frac{1}{ N - \sum_j {1 \over d_j}  } = 
\frac{1}{ N \Big( 1 - {1\over N} \sum_j {1 \over d_j} \Big) }  \:.
\end{equation}
Accordingly, 
\begin{equation}
P_{\fh_{loc}} (\ket{\psi}) = \mbox{max} =1 \leftrightarrow \ket{\psi} = 
\ket{\phi_1} \otimes \cdots \otimes \ket{\phi_N} \:,
\end{equation}
and the equivalence with the standard notions of separability and 
entanglement are recovered. 
Note that for the case of $N$ qubits considered in Sec.~\ref{sec3-3-2}, the
above  value simplifies to ${\sf K}'=2/N$ which in turn gives the purity
expression  of Eq.~\ref{purity3} once the standard unnormalized Pauli matrices are  used
($\hA^j_{i} = \sigma^j_{i} /\sqrt{2}$, thus removing the  overall
factor 2).

\chapter{Approximations of the exponential matrix}
\label{appaux1}

To classically compute the correlation function
\begin{equation}
\label{correlation1}
\bra{\sf HW} e^{-i\hH} \hW e^{i\hH} \ket{\sf HW},
\end{equation}
one  works in a low dimensional representation  of the algebra $\fh$, such as
the  adjoint representation. Here, $\ket{\sf HW}$ is a highest weight state of
$\fh$ in some representation associated with the Hilbert space $\cH$, and $\hH$
and $\hW 
\in \fh$ are operators that map states in $\cH$ into states in the same space.
As before, I consider such an algebra to be compact semi-simple. Then, if $\fh =
\{ \hO_1, \cdots , \hO_M \}$, there is a Killing form (for the adjoint
representation) such that
\begin{equation}
K(\bar{O}_i,\bar{O}_j)= \delta_{ij},
\end{equation}
where $\bar{O}_k$ is the matrix representation of the operator $O_k$.

In order to approximate a matrix, it is necessary to define a suitable
norm. Here, I use the second norm defined by 
\begin{equation}
|A| = \max |Ax|
\end{equation}
where $A$ is a $n \times n$ matrix and $x \in \mathbb{C}^n$ is a unit vector.  If $A$ can be
diagonalized, this norm is equivalent to the largest eigenvalue. 
Nevertheless, the results and proofs of this appendix apply to any definition of
matrix norm.

The first step is to obtain, with the best possible method, a good approximation to
the exponential matrix $e^{i\bar{H}}$, where $\bar{H}$ is the representation of
the operator $\hH$. In Ref.~\cite{ML03}, the
authors state that the best method to evaluate such an exponential is the so
called {\em scaling and squaring method}, which uses Pad\'{e} approximants.
This method is the one used in software like MatLab, etc. (However, if one is
interested in evaluating $e^{it\bar{H}}$ for different values of $t$, the
method of obtaining the eigenvalues and eigenvectors of the Hermitian matrix
$\bar{H}$ might be more efficient.)

A Pad\'{e} approximation to $e^{A}$ ($A \in \mathbb{C}^{2n}$) is defined by
\begin{equation}
e^A \sim R_{pq}(A) = [D_{pq}(A)]^{-1} N_{pq}(A)
\end{equation}
with
\begin{equation}
N_{pq}(A) = \sum\limits_{j=0}^p \frac{(p+q-j)! p!}{(p+q)! j! (p-j)!} A^j
\end{equation}
and
\begin{equation}
D_{pq}(A) = \sum\limits_{j=0}^q \frac{(p+q-j)! q!}{(p+q)! j! (q-j)!}(- A)^j.
\end{equation}

Interestingly, Pad\'{e} approximants can be used if $|A|$ is not too large. In
this case, choosing $p=q$ gives the best approximation. To calculate the
matrices $N_{pq}$ or $D_{pq}$ takes the order of $qn^3$ flop operations, defined
as the computational operation $a \rightarrow
ax + y$. 
The idea is then to use the property
\begin{equation}
e^A = (e^{A/m})^m.
\end{equation}
Therefore, $m$ must be chosen such that it is a power of two and for which $e^{A/m}$ can
be efficiently computed. Then, the final matrix is obtained by $m$ matrix
multiplications.

A common criteria for choosing $m$ is given by
\begin{equation}
\label{bound1}
|A|/m \le 1.
\end{equation}
This is a criteria that might be too restrictive, but I will use it here. Then,
$m=2^s$, where $s$ will be given by Eq.~\ref{bound1}. In this way, the matrix
$e^{A/m}$ can be efficiently computed by using a Taylor expansion or a Pad\'{e}
approximant.

In particular, if $e^{A/2^s}$ is approximated by $R_{qq} (A/2^s)$, then $q$
must be chosen such that the approximation has a small error. In the following,
I present some proofs obtained in the Appendix of the mentioned paper and
which I do not describe in detail here.

First, if $A$ is a matrix with $|A|<1$, then 
\begin{equation}
|\log(\one +A)| \le \frac{|A|}{1-|A|}. 
\end{equation}
Second, if $|A| \le 1/2$ and $p>0$, then 
\begin{equation}
|D_{pq}(A)^{-1}| \le (q+p)/p. 
\end{equation}
Third, if $|A| \le 1/2$, then 
\begin{equation}
R_{pq}(A)= e^{A+F}, 
\end{equation}
where 
\begin{equation}
|F| \le 8 |A|^{p+q+1} \frac{p! q!}{(p+q)! (p+q+1)!}.
\end{equation}
The $n \times n$ matrix $F$ can be shown to commute with the matrix $A$.  This
is because $F$ must be a function of $A$, since the Pad\'{e} approximants are
functions of $A$, too. 

Then, if $p=q$, $|A|/2^s \le 1/2$, I obtain 
\begin{equation}
[R_{qq}(A/2^s)]^{2^s} = e^{A +E}, 
\end{equation}
with
\begin{equation}
\label{approx1}
\frac{|E|}{|A|} \le 8 \left( \frac{|A|}{2^s} \right)^{2q} \frac{(q!)^2} {(2q)!
(2q+1)!} \le \left( \frac{1}{2} \right)^{2q-3} \frac{(q!)^2} {(2q)!(2q+1)!}.
\end{equation}

Naturally, a low ratio between norms would give a good approximation. In the
following, I relate these results with the specific problem of evaluating the
correlation functions of Eq.~\ref{correlation1}.

As mentioned before, I am interested in the approximation of the matrix $e^{i
\bar{H}}$ (i.e., $A = i \bar{H}$), where $\bar{H}$ is some low-dimensional
matrix representation of the operator
\begin{equation}
\hH = \sum_{j=1}^M \zeta_j \hO_j \ ; \
\zeta_j \in \mathbb{R}, \ \hH \in \fh.
\end{equation}
Here, $M$ denotes the dimension of the compact semi-simple Lie algebra $\fh$.
If a Killing form exists for the representation, I obtain
\begin{equation}
|A|=|\bar{H}| \le \sqrt{{\sf Tr} (\bar{H}^2)} = \sqrt{\sum_{j=1}^M \zeta_j^2}.
\end{equation}
Then, if each $|\zeta_j| \le d_0$, where $d_0$ is some bound for the
coefficients that build the operator $\hH$, I obtain
\begin{equation}
\label{norm2}
|A|=|\bar{H}| \le \sqrt{\zeta_j^2} \le \sqrt{M} d_0.
\end{equation}

Assuming that $|A|/2^s \le 1/2$, then
\begin{equation}
{\sqrt{M} d_0}{2^s} \le 1/2,
\end{equation}
yielding to
\begin{equation}
s \ge \log_2 (\sqrt{M} d_0) +1.
\end{equation}
Equivalently, 
\begin{equation}
\label{boundm}
m=2^s \ge 2 \sqrt{M} d_0.
\end{equation}
Equation~\ref{boundm} tells one that the exponential can be approximated  by an
amount of products of matrices $R_{qq} (i\bar{H}/m) \sim e^{i\bar{H}/m}$
efficiently in $M$.

The error of the approximation is given by
\begin{equation}
\sigma = | [R_{qq}(i \bar{H}/2^s)]^{2^s} \bar{W} 
[R_{qq}(-i \bar{H}/2^s)]^{2^s} - \bar{W} | =
| e^{i \bar{H}} ( e^E  \bar{W} e^{E^\dagger} - \bar{W} ) e^{-i \bar{H}}|.
\end{equation}
Because of the properties of the norm (i.e., $|AB|\le |A||B|$ and
$|A+B|\le |A|+|B|$),
one obtains
\begin{equation}
\sigma \le |E| e^{|E|} |\bar{W}| e^{|E^\dagger|} +
e^{|E|} |\bar{W}| |E^\dagger| e^{|E^\dagger|},
\end{equation}
and considering that
\begin{equation}
\hW = \sum\limits_{j=1}^M \varsigma_j \hO_j
\end{equation}
with $\varsigma_j \le d_0$, then
\begin{equation}
|\bar{W}| \le \sqrt{ \sum_{j=1}^M \varsigma_j^2} \le \sqrt{M} d_0.
\end{equation}
Evenmore, because $E$ can be diagonalized (it commutes with $H$) we obtain $|E|
\equiv |E^\dagger|$. (This property might be satisfied for every matrix).
Then,
\begin{equation}
\label{approx4}
\sigma \le |E| e^{2|E|} \sqrt{M} d_0.
\end{equation}

Equation~\ref{approx4} tells one that for small values of $|E|$, the approximation can
be performed with high accuracy. Then, if
\begin{equation}
\frac{|E|}{|A|} = \frac{|E|}{|\bar{H}|} \le \epsilon_1,
\end{equation}
I obtain
\begin{equation}
\label{approx5}
\sigma \le \epsilon_1 |\bar{H}| e^{2\epsilon_1 |\bar{H}|} \sqrt{M} d_0
\le \epsilon_1 M d_0^2 e^{2\epsilon_1 \sqrt{M} d_0} 
\le \epsilon_1 M d_0^2 e^{2\epsilon_1 M d_0} 
\le \epsilon,
\end{equation}
where $\epsilon$ is the maximum tolerable error in the approximation. This
error determines $\epsilon_1$, which determines the integer $q$ of
Eq.~\ref{approx1}. In fact, for a constant error $\epsilon$, the higher $M$ is,
the smaller $\epsilon_1$. Then $q$ increases and the approximation needs to be
done by higher order Pad\'{e} approximants.

Since I am interested in the case when $M \le \mbox{poly}(N)$, where $N \le
\log(d)$ is an integer that depends on the dimension $d$ of the Hilbert space
$\cH$, it remains to be shown that for fixed $\epsilon$, the integer $q$ scales
at most polynomially with $M$ or $N$. First, Eq.~\ref{approx5} tells one that for
fixed $\epsilon$, I can consider
\begin{equation}
\epsilon_1 M = \tau_{\epsilon,d_0}
\end{equation}
where $\tau_{\epsilon,d_0} >0$ is the coefficient of proportionality.
Then (see Eq.~\ref{approx1})
\begin{equation}
\label{approx6}
\left( \frac{1}{2} \right)^{2q-3} \frac{(q!)^2} {(2q)!(2q+1)!}
\le \epsilon_1 = \frac{\tau_{\epsilon,d_0}}{M}.
\end{equation}
[The previous analysis did not consider roundoff errors and these might be
taken into account if needed. Nevertheless, one can consider that every step was done
at the accuracy given by the number of bits of our computer (Turing machine). 
Also, some bounds to the errors can be improved.]

\section{Scaling of the Method}
In this section, I am interested in obtaining the number of operations required
to obtain $\sigma \le \epsilon$ as a function of norms of operators, etc.
First, Eq.~\ref{approx1} can be bounded as follows:
\begin{equation}
\label{approx7}
\frac{|E|}{|\bar{H}|} \le \frac{8}{(2q)^{2q}}.
\end{equation}
Since
\begin{equation}
\label{approx8}
\sigma \le 2|E|e^{2|E|}  |\bar{W}| ,
\end{equation}
I obtain
\begin{equation}
\label{approx9}
\sigma \le \frac{16 |\bar{H}|}{(2q)^{2q}} e^{\frac{16 |\bar{H}|}{(2q)^{2q}}}|\bar{W}|
\le \frac{16 |\bar{H}|}{(2q)^{2q}} e |\bar{W}| \le \epsilon,
\end{equation}
where I have assumed that $0<\frac{16 |\bar{H}|}{(2q)^{2q}} \le 1$.
Since
\begin{equation}
\label{approx10}
2q \log_2(2q) \ge 2q ,
\end{equation}
I obtain
\begin{equation}
2q \ge 4+ \log_2 e + \log_2 |\bar{H}| + \log_2 |\bar{W}| + \log_2 (1/\epsilon)
\end{equation}
for the desired accuracy. 

Because calculating $N_{pq}$ or $D_{pq}$ takes $qM^3$ flop operations in the
adjoint representation, calculating the approximated exponential matrix takes
$(2q+m)M^3$ flops. Then, to obtain the approximated 
matrix $e^{-i\bar{H}} \bar{W}
e^{i\bar{H}}$ one needs $n\sim{\cal O}[(2q+m+2)M^3]$ operations. That is,
\begin{equation}
\label{flops}
n \sim [(\log_2 1/\epsilon +\log_2 |\bar{H}| +\log_2 |\bar{W}|)+|\bar{H}|/2]M^3
\end{equation}
flops, where I have considered $\epsilon \ll1$.

\chapter{Efficient Classical Evaluation of High-Order Correlation Functions}
\label{appaux2}
In this case, I am interested in the evaluation of correlation functions of
the form
\begin{equation}
\label{hocorrel}
\bra{\sf HW} \hW'^p \cdots \hW'^1 \ket{\sf HW}
\end{equation}
where $\ket{HW}$ is the highest weight state of a compact semi-simple Lie
algebra $\fh = \{ \hO_1, \cdots, \hO_M \}$, where each Hermitian operator
$\hO_j$ maps states of the Hilbert space $\cH$ into states in the same space.

In a Cartan-Weyl decomposition, each operator $\hW'^i$ can be decomposed as
\begin{equation}
\label{operdef}
\hW'^i = \sum_{k=1}^r \gamma_k^i \hh_k + \sum_{j=1}^l \lambda_j^i \hE_{\alpha_j} +
(\lambda_j^i)^* \hE_{-\alpha_j}.
\end{equation}
Again, the
roots $\alpha_j$ are considered to be positive so that the state $\ket{\sf HW}$
satisfies
\begin{eqnarray}
\label{raishw}
\hE_{\alpha_j} \ket{\sf HW} &=&0 \forall j, \\
\label{carhw}
\hh_k \ket{\sf HW} &=& e_k \ket{\sf HW}.
\end{eqnarray}

Assume that $\oS=(s_1, \cdots, s_q)$ is a vector whose $q$ components can take
an integer value in the set $[1, \cdots, l]$ (where $l$ is the number of positive
roots). Then, I want to show that
\begin{eqnarray}
\nonumber
\hE_{\beta} \hE_{-\alpha_{s_1}} \cdots \hE_{-\alpha_{s_q}} \ket{\sf HW} \equiv
\left[ \sum\limits_{j_1,\cdots,j_q=1}^l x_{j_1,\cdots,j_q}^{\beta,\oS} 
\hE_{-\alpha_{j_1}} \cdots \hE_{-\alpha_{j_q}} + \right. \\
\label{action1}
\left.
\sum\limits_{j_1,\cdots,j_{q-1}=1}^l x_{j_1,\cdots,j_{q-1}}^{\beta,\oS}
\hE_{-\alpha_{j_1}} \cdots \hE_{-\alpha_{j_{q-1}}} + \cdots +
x^{\beta,\oS} \right] \ket{\sf HW},
\end{eqnarray}
where $\beta$ is also a positive root.
To show this I use an inductive method. For this purpose, some commutation
relations are needed. These are
\begin{eqnarray}
\label{carcom}
\left[ \hE_{\beta}, \hE_{-\alpha_{s_1}} \right] &=& \sum_{k=1}^r \alpha_{s_1}^k
\hh_k \mbox{ if }
\beta=\alpha_{s_1} , \\
\label{raiscomx}
\left[ \hE_{\beta}, \hE_{-\alpha_{s_1}} \right] &= &N_{\beta,-\alpha_{s_1}}
\hE_{\beta-\alpha_{s_1}} \mbox{ if } \beta \ne \alpha_{s_1}.
\end{eqnarray}
Then, for $q=1$, the desired result is satisfied:
\begin{eqnarray}
\hE_{\beta} \hE_{-\alpha_{s_1}}\ket{\sf HW} &=& \sum_{k=1}^r \alpha_{s_1}^k e_k 
\ket{\sf HW} \mbox{ if } \beta=\alpha_{s_1} , \\
\hE_{\beta} \hE_{-\alpha_{s_1}} \ket{\sf HW} &=& 0 \mbox{ if } \beta>\alpha_{s_1},
\\
\hE_{\beta} \hE_{-\alpha_{s_1}} \ket{\sf HW} &=& N_{\beta,-\alpha_{s_1}}
\hE_{-(\alpha_{s_1}-\beta)}\ket{\sf HW} \mbox{ if } \beta<\alpha_{s_1},
\end{eqnarray}
where Eqs.~\ref{raishw} and \ref{carhw} have been used here.
($\alpha_i > \alpha_j$ if they differ in a positive root.)

I now assume that Eq.~\ref{action1} is valid for some value of $q$. Then, I
need to show that it remains valid for $q+1$. In the latter case, I need to
obtain
\begin{eqnarray}
\nonumber
\hE_{\beta} \hE_{-\alpha_i} \hE_{-\alpha_{s_1}} \cdots 
\hE_{-\alpha_{s_q}} \ket{\sf HW}
\equiv
\left[ \sum\limits_{j_1,\cdots,j_{q+1}=1}^l x_{j_1,\cdots,j_{q+1}}^{\beta,\oT} 
\hE_{-\alpha_{j_1}} \cdots \hE_{-\alpha_{j_{q+1}}} + \right. \\
\label{action2}
\left.
\sum\limits_{j_1,\cdots,j_q=1}^l x_{j_1,\cdots,j_q}^{\beta,\oT}
\hE_{-\alpha_{j_1}} \cdots \hE_{-\alpha_{j_q}} + \cdots +
x^{\beta,\oT} \right] \ket{\sf HW},
\end{eqnarray}
where $\alpha_i$ is also a positive root and $\oT=(i,s_1,\cdots,s_q)$. Again,
one can show the validity of Eq.~\ref{action2} using the commutation relations
of Eqs.~\ref{raiscomx} and \ref{carcom}. 

Consider then that $\beta = \alpha_i$. Therefore,
\begin{equation}
\hE_{\alpha_i} \hE_{-\alpha_i} \hE_{-\alpha_{s_1}} \cdots 
\hE_{-\alpha_{s_q}} \ket{\sf
HW}= (\hE_{-\alpha_i} \hE_{\alpha_i} + \sum_{k=1}^r \alpha_i^k \hh_k)
\hE_{-\alpha_{s_1}} \cdots \hE_{-\alpha_{s_q}} \ket{\sf HW},
\end{equation}
where each operator $\hh_k$ behaves as the constant $(e_k - \sum_{j=1}^q
\alpha_{s_j}^k)=\tilde{e}_k$ when acting over the corresponding weight states.
The validity of Eq.~\ref{action2} is obtained by noticing that
the operator $\hE_{-\alpha_i}$ acting on the state of
Eq.~\ref{action1} increases the order $q$ in 1. For this case, the 
coefficients 
$x_{j_1,\cdots}^{\alpha_i,\oT}$ can be obtained from the coefficients 
$x_{j_1,\cdots}^{\alpha_i,\oS}$ in the following way:
\begin{equation}
\label{update1}
\left\{
\matrix{ 
x^{\beta,\oT}&= &\sum_{k=1}^r \tilde{e}_k ,\cr
x_i^{\beta,\oT}&= & x^{\beta,\oS}, \cr
x_{i,j_1,\cdots,j_n}^{\beta,\oT}&= & x_{j_1,\cdots,j_n}^{\beta,\oS} \ \forall
n\in[1 \cdots q],}
\right.
\end{equation}
while the other coefficients remain zero.

Consider now that $\beta >
\alpha_i$. Then,
\begin{equation}
\hE_{\beta} \hE_{-\alpha_i} \hE_{-\alpha_{s_1}} \cdots 
\hE_{-\alpha_{s_q}} \ket{\sf
HW}= (\hE_{-\alpha_i} \hE_{\beta} + N_{\beta,-\alpha_i} \hE_{\alpha_k} )
\hE_{-\alpha_{s_1}} \cdots \hE_{-\alpha_{s_q}} \ket{\sf HW},
\end{equation}
where $\alpha_k = \beta- \alpha_i$ is a positive root.
Because of Eq.~\ref{action1}, the first term on the rhs is
a linear combination of states with order $q+1$ in $\hE_{-\alpha}$, 
while the second term is of
order $q$. Then, Eq.~\ref{action2} is satisfied for this case, too. The
coefficients can now be obtained by the following recursion:
\begin{equation}
\label{update2}
\left\{
\matrix{
x^{\beta,\oT} &=&N_{\beta,-\alpha_i} x^{\alpha_k,\oS},\cr
x_i^{\beta,\oT} &=& x^{\beta,\oS}, \cr
x_{j_1, \cdots, j_n}^{\beta,\oT}&=& N_{\beta,-\alpha_i} 
x_{j_1, \cdots, j_n}^{\alpha_k,\oS} \mbox{ for } n\in[1\cdots q], j_1 \ne i, \cr
x_{i,j_1, \cdots, j_n}^{\beta,\oT}  &=&  x_{j_1, \cdots, j_n}^{\beta,\oS}
+ N_{\beta,-\alpha_i} x_{i,j_1,\cdots,j_n}^{\alpha_k,\oS} \mbox{ for }
n\in[1\cdots q], }
\right.
\end{equation}
while the other coefficients remain zero.

Similar results are obtained for the case $\beta <
\alpha_i$. Then,
\begin{equation}
\hE_{\beta} \hE_{-\alpha_i} \hE_{-\alpha_{s_1}} \cdots 
\hE_{-\alpha_{s_q}} \ket{\sf
HW}= (\hE_{-\alpha_i} \hE_{\beta} + N_{\beta,-\alpha_i} \hE_{-\alpha_k} )
\hE_{-\alpha_{s_1}} \cdots \hE_{-\alpha_{s_q}} \ket{\sf HW},
\end{equation}
where $\alpha_k = \beta- \alpha_i$ is a positive root. Again, Eq.~\ref{action1}
tells one that the first term on the rhs is
a linear combination of states of order $q+1$ in $\hE_{-\alpha}$ while the
second term is just a weight state of order $q+1$ in $\hE_{-\alpha}$.
Therefore, Eq.~\ref{action1} is satisfied and the coefficients can be obtained
as
\begin{equation}
\label{update3}
\left\{
\matrix{
x_i^{\beta,\oT} &=& x^{\beta,\oS}, \cr
x_{i,j_1,\cdots,j_n}^{\beta,\oT} &=& x_{j_1, \cdots, j_n}^{\beta,\oS} \mbox{
for } n\in[1,\cdots,q] , \cr
x_{k,s_1,\cdots,s_q}^{\beta,\oT}&=& N_{\beta,-\alpha_i},}
\right.
\end{equation}
while the others remain zero.

For a fixed value of $q$, the number $f$ of coefficients $x^{\beta,\oS}_{j_1,
\cdots, j_n} \ \forall \beta, \oS$ is given by
\begin{equation}
f= \frac{l^{q+1} -1}{l-1} l^{q+1},
\end{equation}
and considering that $q \in [1\cdots p-1]$, where $p$ is the order of the
correlation in Eq.~\ref{hocorrel}, the total number $F$ of coefficients
$x^{\beta,\oS}_{j_1,
\cdots, j_n} \ \forall \beta, \oS$
satisfies
\begin{equation}
F \le p \frac{l^{p+1} -1}{l-1} l^{p+1} \le \mbox{poly}(l).
\end{equation}
These coefficients can be computed easily, in polynomial time with respect to
$l$, by using Eqs.~\ref{update1},
\ref{update2}, and \ref{update3}. The calculation of every coefficient is
needed for the following results.

Remarkably, Eq.~\ref{action1} yields to the same results for the action of the
operators $\hW'^i$ as defined by Eq.~\ref{operdef}:
\begin{eqnarray}
\nonumber
\hW'^q \cdots \hW'^1 \ket{\sf HW} \equiv \left[ \sum\limits_{s_1, \cdots s_q=1}^l
z_{s_1,\cdots ,s_q}^q \hE_{-\alpha_{s_1}} \cdots \hE_{-\alpha_{s_q}} + \right.
\\
\label{action3}
\left.
\sum\limits_{s_1, \cdots s_{q-1}=1}^l z_{s_1,\cdots ,s_{q-1}}^q
\hE_{-\alpha_{s_1}} \cdots \hE_{-\alpha_{s_{q-1}}} + \cdots z^q \right] \ket{\sf HW}.
\end{eqnarray}
The idea is then to update the coefficients $z_{s_1, \cdots}^q$ whenever
one of the operators $\hW'^i$ is multiplied by the left of Eq.~\ref{action3} until
$q=p$ (see Eq.~\ref{hocorrel}). The result is then
\begin{equation}
\bra{\sf HW} \hW'^p \cdots \hW'^1 \ket{\sf HW} = z^p.
\end{equation}
For this purpose, one needs to present a recursive method to update these coefficients.
First, if the operator $\gamma_k^{q+1}\hh_k$ acts on the state
$\hW'^q \cdots \hW'^1 \ket{\sf HW}$, the coefficients can be easily updated
as
\begin{equation}
\left\{
\matrix{
z^{q+1} &\rightarrow &\gamma_k^{q+1}e_k z^q ,\cr
z^{q+1}_{s_1, \cdots , s_n} &\rightarrow & \gamma_k^{q+1}(e_k - \sum_{i=1}^n
\alpha_{s_i}^k) z^q_{s_1, \cdots , s_n} ,}
\right.
\end{equation}
where $n \in [ 1 \cdots q]$. For a fixed value of $q$, the number of
computational operations to update all the coefficients $z^q$ due to the action of
a single operator $\gamma_k^{q+1}\hh_k$ is given by
\begin{equation}
F^q_h=2 + \sum_{n=1}^q l^q (q+2) = 2 + \frac{l^{q+1}(q-1)}{l-1} +
\frac{l^{q+2}-l}{(l-1)^2}.
\end{equation}

However, if the operator $(\lambda_j^{q+1})^* \hE_{-\alpha_j}$ acts on the state
$\hW'^q \cdots \hW'^1 \ket{\sf HW}$, the coefficients need to be updated as
\begin{equation}
\left\{
\matrix{
z_j^{q+1} & \rightarrow & (\lambda_j^{q+1})^* z^q, \cr
z_{j,s_1,\cdots,s_n}^{q+1} & \rightarrow &(\lambda_j^{q+1})^*
z_{s_1,\cdots,s_n}^q ,}
\right.
\end{equation}
while the other coefficients need not be updated or remain zero. The number of
operations to update all the coefficients in this case is  
\begin{equation}
F^q_-= 1+ \frac{l^{q+1}-1}{l-1}.
\end{equation}

Finally, when the operator
$\lambda_j^{q+1} \hE_{\alpha_j}$ acts on the state 
$\hW'^q \cdots \hW'^1 \ket{\sf HW}$, 
the coefficients need to be updated, too. For this purpose, I first rewrite
Eq.~\ref{action3} as
\begin{equation}
\label{action4}
\hW'^q \cdots \hW'^1 \ket{\sf HW} \equiv \left[ (\sum_{m=1}^q 
\sum_{\oS_m} z^q_{S_m^1,
\cdots, S_m^m} 
\hE_{-\alpha_{S_m^1}} \cdots \hE_{-\alpha_{S_m^m}})+ z^q \right] \ket{\sf HW},
\end{equation}
where the vector $\oS_m =(S_m^1, \cdots S_m^m)$ has $m$ components which can
take values in the set $[1 \cdots l]$. Therefore,
\begin{equation}
\label{action5}
\matrix{
\hE_{\alpha_j}\hW'^q \cdots \hW'^1 \ket{\sf HW} \equiv
 \sum_{m=1}^q \sum_{\oS_n} z^q_{S_m^1,\cdots, S_m^m}
 [ \sum_{j_1,\cdots,j_m=1}^l x_{j_1,\cdots,j_m}^{\alpha_j,\oS_m}
\hE_{-\alpha_{j_1}} \cdots \hE_{-\alpha_{j_m}} + \cr
\sum_{j_1,\cdots,j_{m-1}=1}^l x_{j_1,\cdots,j_{m-1}}^{\alpha_j,\oS_m}
\hE_{-\alpha_{j_1}} \cdots \hE_{-\alpha_{j_{m-1}}} +\cdots + x^{\alpha_j,\oS_m}]
\ket{\sf HW}.
}
\end{equation}
Then, the coefficients are updated as
\begin{equation}
\label{update7}
\left\{
\matrix{
z^{q+1} & \rightarrow & \lambda_j^{q+1} \left[\sum_{m=1}^q \sum_{\oS_m} z_{S_m^1,
\cdots , S_m^m}^q x^{\alpha_j, \oS_n} \right], \cr
z^{q+1}_{s_1,\cdots,s_n}  & \rightarrow & \lambda_j^{q+1} \sum_{m \ge n}^q
\sum_{\oS_m} z^q_{s_m^1, \cdots, s_m^m} x_{s_1, \cdots, s_n}^{\alpha_j,\oS_m}.}
\right.
\end{equation}
Given that the coefficients $x_{j_1...}^{\alpha_j,\oS_m}$ are known, 
the calculation of the coefficients in Eq.~\ref{update7} take
$\frac{l^{q+1}-1}{l-1}+\frac{l^{q+1}-l^n}{l-1}$
computational operations. Therefore, the number of operations to
update all the coefficients in this case is given by
\begin{equation}
F^q_+ = \frac{(q+1)l^{q+1}-1}{l-1} - \frac{l^{q+1}-1}{(l-1)^2}.
\end{equation}

In brief, to update the coefficients due to the action of the operator
$\hW'^{q+1}$ over the state $\hW'^q \cdots \hW'^1 \ket{\sf HW}$ it takes
\begin{equation}
F^q = rF^q_k + l (F^q_- + F^q_+)
\end{equation}
computational operations (assuming that
the coefficients $x_{j_1...}^{\alpha_j,\oS_m}$ are known). 
Therefore, the total number of
computational operations $\cal{F}$ to obtain the coefficients
$z^p_{S^1_m...}$ 
for the state
$\hW'^p \cdots \hW'^1 \ket{\sf HW}$ satisfies
\begin{equation}
{\cal F} \le p F^p,
\end{equation}
that is, it takes at most a polynomially large amount of operations, with
respect to $l$ and $r$, to classically compute the correlation of
Eq.~\ref{hocorrel}. (Though, the same is not true with respect to $p$. In fact,
the complexity of the method described here increases exponentially with $p$,
the order of the correlation function.)

If $M=r=2l$ (i.e., the dimension of $\fh$) satisfies
\begin{equation}
M \le \mbox{poly}(N),
\end{equation}
and $N \sim \log(d)$, where $d$ is the dimension of the associated Hilbert
space, the method presented here allows one to compute Eq.~\ref{hocorrel}
efficiently on a conventional computer.

\chapter{Classical Limit in the LMG Model}
\label{appE}
As I mentioned in Sec.~\ref{sec5-1-1}, some critical properties of the
LMG, such as the order parameter or the ground state energy per particle in the
thermodynamic limit, may be obtained using a semi-classical approach. Here,
I sketch a rough analysis of why such approximation is valid (for a 
more extensive analysis, see Ref.~\cite{SOB04}).

Defining the collective operators 
\begin{equation}
E_{(\sigma,\sigma')} = \sum\limits_{k=1}^N c^\dagger_{k \sigma}c^{\;}_{k \sigma'},
\end{equation}
where $\sigma,\sigma' = \uparrow\mbox{ or }\downarrow$ and the fermionic
operators $c^\dagger_{k \sigma}$ ($c^{\;}_{k \sigma}$) have been defined in
Sec.~\ref{sec2-3-1}.  The collective operators satisfy the $\fu(2)$ commutation 
relations;  that is
\begin{equation}
\left[ E_{(\sigma,\sigma')}, E_{(\sigma'',\sigma''')} \right ] =
\delta_{\sigma' \sigma''} E_{(\sigma,\sigma''')} -
\delta_{\sigma \sigma'''} E_{(\sigma'',\sigma')} .
\end{equation}

If the number of degenerate levels $N$ is very large, it is useful to  define
the intensive collective operators $\hat{E}_{(\sigma,\sigma')} =
E_{(\sigma,\sigma')}/N$, with commutation relations
\begin{equation}
\left[ \hat{E}_{(\sigma,\sigma')}, \hat{E}_{(\sigma'',\sigma''')} \right ] =
\frac{1}{N} \left ( \delta_{\sigma' \sigma''} \hat{E}_{(\sigma,\sigma''')} -
\delta_{\sigma \sigma'''} \hat{E}_{(\sigma'',\sigma')} \right).
\end{equation}
Therefore, the intensive collective  operators commute in the limit $N
\rightarrow \infty$, they are effectively classical and can be simultaneously 
diagonalized. Similarly, the intensive angular momentum operators 
$J_x/N=(\hat{E}_{(\uparrow,\downarrow)} + \hat{E}_{(\downarrow,\uparrow)})/2$, 
$J_y/N=(\hat{E}_{(\uparrow,\downarrow)} -
\hat{E}_{(\downarrow,\uparrow)})/2i$,  and
$J_z/N=(\hat{E}_{(\uparrow,\uparrow)} - \hat{E}_{(\downarrow,\downarrow)})/2$ 
(with $J_\alpha$ defined in Eqs.~\ref{pseudospin1}, \ref{pseudospin2}, and
\ref{pseudospin3})  commute with each other in the thermodynamic limit, so they
can be thought of as the angular momentum operators of a classical system.

Since the intensive LMG Hamiltonian $H/N$, with $H$ given in Eq.
(\ref{lmghamilt2}), can be written in terms of the intensive angular momentum
operators, it can be regarded as the Hamiltonian describing a classical 
system. The ground state of the LMG model $\ket{g}$ is then an eigenstate of
such  intensive operators when $N \rightarrow \infty$:  $(J_\alpha /N) \ket{g}
= j_\alpha \ket{g}$, $j_\alpha$ being the corresponding eigenvalue. In other
words, when obtaining some expectation values of intensive operators such as
$J_\alpha /N$ or $H/N$ the ground state $\ket{g}$ can be pictured as a
classical angular momentum with fixed coordinates in the three-dimensional space 
(see Fig.~\ref{fig5-1}).
 
This point of view makes it clear why such operators ought to be intensive. 
Otherwise, such a classical limit is not valid and terms of ${\cO}(1)$ (order 1) 
would 
be important for the calculations of the properties of the LMG model.
Obviously, all these concepts can be extended to more complicated Hamiltonians
such as the extended LMG model, or even Hamiltonians including interactions of
higher orders as in~\cite{Gil81}.

\bibliography{thesis}

\end{document}